\documentclass[11pt,a4paper]{article}
\pdfoutput=1

\usepackage[utf8]{inputenc}
\usepackage{a4wide}

\usepackage{amsmath}
\usepackage{braket}
\usepackage{cite}
\usepackage[table]{xcolor}
\usepackage{amssymb}
\usepackage{amsfonts}
\usepackage[normalem]{ulem}
\usepackage{booktabs}
\usepackage{makecell}
\usepackage{multirow}
\usepackage[printonlyused,nohyperlinks,nolist]{acronym}
\usepackage{pdflscape}
\usepackage{textcomp}
\usepackage{gensymb}
\usepackage{bigstrut}
\usepackage{array}
\usepackage{enumerate}
\usepackage[small,bf]{caption}
\usepackage{subcaption}
\usepackage{diagbox}
\usepackage{graphicx}
\usepackage{hyperref}
\usepackage{mathbbol}
\usepackage{appendix}
\usepackage{verbatim}
\usepackage{geometry}
\usepackage{makecell}
\usepackage{listings}
\usepackage{mathtools}
\usepackage{dsfont}
\usepackage{longtable}
\usepackage{threeparttable}
\usepackage{floatpag}
\usepackage{arydshln}
\usepackage{cleveref} 
\usepackage{tikz}
\usetikzlibrary{fit,matrix}
\usepackage{colortbl}

\DeclareMathAlphabet{\mathdutchcal}{U}{dutchcal}{m}{n}

\newcommand{\id}{\mathds{1}}
\newcommand{\n}{N}
\newcommand{\M}{\mathsf{M}}
\newcommand{\vw}{v}
\newcommand{\LY}{\mathcal{L}}
\newcommand{\Lm}{\mathcal{L}_\text{m} }
\newcommand{\Mu}{\mathcal{M}_u}
\newcommand{\Md}{\mathcal{M}_d}
\newcommand{\Mq}{\mathcal{M}_q}
\newcommand{\T}{U}
\newcommand{\B}{D}
\newcommand{\yu}{Y_u}
\newcommand{\yd}{Y_d}
\newcommand{\yq}{Y_q}
\newcommand{\yU}{Z_u}
\newcommand{\yD}{Z_d}
\newcommand{\yQ}{Z_q}
\newcommand{\mb}{\overline{M}}
\newcommand{\mB}{M_Q}
\newcommand{\mBi}[1]{M_{Q#1}}
\newcommand{\mBd}{D_Q}
\newcommand{\Q}{\mathcal{Q}}
\newcommand{\U}{\mathdutchcal{U}}
\newcommand{\D}{\mathdutchcal{D}}
\newcommand{\Yu}{\mathcal{Y}_u}
\newcommand{\Yd}{\mathcal{Y}_d}
\newcommand{\Yq}{\mathcal{Y}_q}
\newcommand{\mq}{m_q}
\newcommand{\Mb}{M}
\newcommand{\V}{\mathcal{V}}
\newcommand{\Vu}{\V^u}
\newcommand{\Vd}{\V^d}
\newcommand{\Vq}{\V^q}
\newcommand{\VuL}{\Vu_L}
\newcommand{\VdL}{\Vd_L}
\newcommand{\VqL}{\Vq_L}
\newcommand{\VuR}{\Vu_R}
\newcommand{\VdR}{\Vd_R}
\newcommand{\VqR}{\Vq_R}
\newcommand{\Dq}{\mathcal{D}_q}
\newcommand{\Du}{\mathcal{D}_u}
\newcommand{\Dd}{\mathcal{D}_d}
\newcommand{\msq}{m^q}
\newcommand{\msu}{m^u}
\newcommand{\msd}{m^d}
\newcommand{\Msq}{M^q}
\newcommand{\MT}{M_{T'}}
\newcommand{\MB}{M_{B'}}
\newcommand{\Aq}{A^q}
\newcommand{\AqR}{A_R^q}

\newcommand{\Bq}{B^q}
\newcommand{\BqR}{B_R^q}
\newcommand{\BqL}{B_L^q}
\newcommand{\BuR}{B_R^u}

\newcommand{\BdR}{B_R^d}

\newcommand{\CKM}{V_\text{CKM}}
\newcommand{\Vr}{V} 
\newcommand{\VL}{\Vr_L}
\newcommand{\VLa}{\Vr^L} 
\newcommand{\VLac}{\Vr^{L*}} 
\newcommand{\Vch}{\hat{\Vr}_\chi}
\newcommand{\VLh}{\hat{\Vr}_L}
\newcommand{\Vha}{\hat{\Vr}^L} 
\newcommand{\Vhac}{\hat{\Vr}^{L*}} 
\newcommand{\VR}{\Vr_R}
\newcommand{\VRa}{\Vr^R} 
\newcommand{\VRac}{\Vr^{R*}} 
\newcommand{\VRh}{\hat{\Vr}_R}
\newcommand{\VRha}{\hat{\Vr}^R} 
\newcommand{\VRhac}{\hat{\Vr}^{R*}} 
\newcommand{\Vhda}{\hat{\Vr}^{L(d)}}
\newcommand{\Vhua}{\hat{\Vr}^{L(u)}}
\newcommand{\dVhda}{\delta\Vhda}
\newcommand{\dVhua}{\delta\Vhua}
\newcommand{\Fr}{F} 
\newcommand{\Fu}{\Fr_u}
\newcommand{\Fua}{\Fr^u} 
\newcommand{\Fuac}{\Fr^{u*}} 
\newcommand{\Fd}{\Fr_d}
\newcommand{\Fda}{\Fr^d} 
\newcommand{\Fq}{\Fr_q}
\newcommand{\Fud}{\Fr_{u,d}}
\newcommand{\Fuh}{\hat{\Fr}_u}
\newcommand{\Fuha}{\hat{\Fr}^u} 
\newcommand{\Fuhac}{\hat{\Fr}^{u*}} 
\newcommand{\Fdh}{\hat{\Fr}_d}
\newcommand{\Fdha}{\hat{\Fr}^d} 
\newcommand{\Fqh}{\hat{\Fr}_q}
\newcommand{\Fqha}{\hat{\Fr}^q} 
\newcommand{\h}{h}
\newcommand{\WL}{\mathcal{W}_L}
\newcommand{\WR}{W_R}
\newcommand{\WRq}{W_R^q}
\newcommand{\WRu}{W_R^u}
\newcommand{\WRd}{W_R^d}
\newcommand{\WRp}{\varphi_R}
\newcommand{\Ydu}{\hat{Y}_u}
\newcommand{\Ydd}{\hat{Y}_d}
\newcommand{\Ydq}{\hat{Y}_q}
\newcommand{\VWB}{\hat{V}}
\newcommand{\UW}{\hat{U}}
\newcommand{\VUW}{\UW_{uL}}
\newcommand{\WUW}{\UW_{uR}}
\newcommand{\VDW}{\UW_{dL}}
\newcommand{\WDW}{\UW_{dR}}
\newcommand{\VQW}{\UW_{qL}}
\newcommand{\WQW}{\UW_{qR}}
\newcommand{\dU}{\delta U}
\newcommand{\ydp}{Y_d'}
\newcommand{\yup}{\boldsymbol{z}_u}
\newcommand{\yub}{\hat{\boldsymbol{z}}_u}
\newcommand{\ydb}{\boldsymbol{z}_d}
\newcommand{\yqp}{\boldsymbol{z}_q}
\newcommand{\ydbh}{\hat{\boldsymbol{z}}_d}
\newcommand{\yqh}{\hat{\boldsymbol{z}}_q}
\newcommand{\PWB}{P}
\newcommand{\VWT}{\tilde{V}}
\newcommand{\0}{\mathcal{O}}
\newcommand{\K}{\mathcal{K}}
\newcommand{\hi}{z}
\newcommand{\hiu}{\hi_u}
\newcommand{\hic}{\hi_c}
\newcommand{\hit}{\hi_t}
\newcommand{\hid}{\hi_d}
\newcommand{\his}{\hi_s}
\newcommand{\hib}{\hi_b}
\newcommand{\Rf}{R_4}
\newcommand{\uu}{\mathcal{U}}
\newcommand{\hh}{\mathdutchcal{H}}
\newcommand{\bi}{\mathcal{B}}
\newcommand{\tri}{\mathcal{T}}
\newcommand{\qua}{\mathcal{Q}}
\newcommand{\hbi}{\hat{\bi}}
\newcommand{\htr}{\hat{\tri}}
\newcommand{\hqu}{\hat{\qua}}
\newcommand{\Qh}{\hat{Q}}
\newcommand{\HqL}{\mathcal{H}_q}
\newcommand{\HuL}{\mathcal{H}_u}
\newcommand{\HdL}{\mathcal{H}_d}
\newcommand{\hq}{h_q}
\newcommand{\hu}{h_u}
\newcommand{\hd}{h_d}
\newcommand{\C}{H}
\newcommand{\HR}{\mathcal{H}_{R}}
\newcommand{\HqR}{\mathcal{H}_{qR}}
\newcommand{\hqR}{h_{qR}}
\newcommand{\CR}{\C_R}

\newcommand{\Iqp}[1]{I_{#1}^{q\prime}}
\newcommand{\Ip}[1]{I_{#1}'}
\newcommand{\Iqs}[1]{I_{#1}^{q}}
\newcommand{\Is}[1]{I_{#1}}
\newcommand{\Js}[1]{J_{#1}}
\newcommand{\MBS}{M_{\text{BSM}}}

\renewcommand{\approx}{\simeq}

\definecolor{darkgreen}{rgb}{0,0.35,0}

\definecolor{light-gray}{gray}{0.9}
\definecolor{lighter-gray}{gray}{0.95}

\allowdisplaybreaks
\numberwithin{equation}{section}

\DeclareMathOperator{\br}{Br}
\DeclareMathOperator{\diag}{diag}
\DeclareMathOperator{\tr}{Tr}
\DeclareMathOperator{\im}{Im}
\DeclareMathOperator{\re}{Re}

\makeatletter
\newcommand*{\org@overidelabel}{}
\let\org@overridelabel\@verridelabel
\@ifpackagelater{acronym}{2015/03/21}{%
  \renewcommand*{\@verridelabel}[1]{%
    \@bsphack
    \protected@write\@auxout{}{\string\AC@undonewlabel{#1@cref}}%
    \org@overridelabel{#1}%
    \@esphack
  }%
}{%
  \renewcommand*{\@verridelabel}[1]{%
    \@bsphack
    \protected@write\@auxout{}{\string\undonewlabel{#1@cref}}%
    \org@overridelabel{#1}%
    \@esphack
  }%
}
\makeatother

\newcommand*{\belowrulesepcolor}[1]{%
  \noalign{%
    \kern-\belowrulesep
    \begingroup
      \color{#1}%
      \hrule height\belowrulesep
    \endgroup
  }%
}
\newcommand*{\aboverulesepcolor}[1]{%
  \noalign{%
    \begingroup
      \color{#1}%
      \hrule height\aboverulesep
    \endgroup
    \kern-\aboverulesep
  }%
}

\usepackage{pifont}

\begin{document}

\begin{titlepage}

\vspace*{-15mm}
\begin{flushright}
CFTP/24-003 \\
TTP24-047 \\
P3H-24-103
\end{flushright}
\vspace*{1cm}

\begin{center}
{\bf\LARGE 
Vector-like quark doublets,\\[2mm]
weak-basis invariants and CP violation
}\\[1.2cm]

\renewcommand*{\thefootnote}{\fnsymbol{footnote}}\large
F.~Albergaria\(^{\,a,}\)\footnote{\texttt{francisco.albergaria@tecnico.ulisboa.pt}},
J.~F.~Bastos\(^{\,a,}\)\footnote{\texttt{jose.bastos@tecnico.ulisboa.pt}},
B.~Belfatto\(^{\,b,}\)\footnote{\texttt{benedetta.belfatto@kit.edu}},
G.~C.~Branco\(^{\,a,}\)\footnote{\texttt{gbranco@tecnico.ulisboa.pt}},\\[2mm]
J.~T.~Penedo\(^{\,c,}\)\footnote{\texttt{jpenedo@roma3.infn.it}},
A.~Rodríguez-Sánchez\(^{\,d,}\)\footnote{\texttt{anrosanz@ific.uv.es}},
J.~I.~Silva-Marcos\(^{\,a,}\)\footnote{\texttt{juca@cftp.tecnico.ulisboa.pt}}\\
 \vspace{7mm}
\renewcommand*{\thefootnote}{\arabic{footnote}}
\normalsize
\(^{a}\)\,{\it CFTP, Departamento de Física, Instituto Superior Técnico, Universidade de Lisboa,\\
Avenida Rovisco Pais 1, 1049-001 Lisboa, Portugal} \\
\vspace{2mm}
\(^{b}\)\,\it Institut f\"ur Theoretische Teilchenphysik, Karlsruhe Institute of Technology, Engesserstra{\ss}e 7, D-76128 Karlsruhe, Germany \\
\vspace{2mm}
\(^{c}\)\,{\it INFN, Sezione di Roma Tre, Via della Vasca Navale 84, 00146, Rome, Italy} \\
\vspace{2mm}
\(^{d}\)\,{\it Instituto de Física Corpuscular, Universitat de València - CSIC,\\
Parque Científico, Catedrático José Beltrán 2, E-46980 Paterna, Spain} \\

\end{center}
\vspace{6mm}

\begin{abstract}
We study Standard Model extensions with isodoublet vector-like quarks with standard charges. Their presence induces right-handed charged and neutral currents. We identify minimal sets of independent parameters characterizing these extensions, describe useful weak bases, and provide parameterizations for all quark mixing. 
We analyze the intricacies of CP violation in such scenarios, finding a complete set of CP-odd invariants for the single doublet case.
Crucially, we uncover a connection between weak-basis invariants and effective rephasing invariants involving only standard quarks.
These results allow us to explore the phenomenology of doublet vector-like quarks through a rephasing-invariant analysis, with an emphasis on CP violation, including the potential role of these fields in explaining the Cabibbo angle anomalies.

\end{abstract}

\end{titlepage}
\setcounter{footnote}{0}
\setcounter{page}{2}
%
\pagestyle{empty}
\tableofcontents
\vskip 1cm
\hrule
\vskip 1cm

\vfill
\clearpage

\pagestyle{plain}
\setcounter{page}{1}

\section{Introduction}
\label{sec:intro}

\Acp{VLQ} are one of the simplest and best-motivated extensions of the \ac{SM}, which may play an important role in solving some of the open questions in particle physics. They appear in models of grand unification~\cite{Gursey:1975ki,Achiman:1978vg, Berezhiani:1989bd,Barbieri:1994kw,Berezhiani:1995dt}, extra dimensions~\cite{Randall:1999ee}, 
and emerge in models addressing the electroweak hierarchy problem 
in which the light Higgs arises as a pseudo-Goldstone boson of a global symmetry~\cite{Arkani-Hamed:2001nha,Arkani-Hamed:2002sdy,Perelstein:2003wd,Han:2003wu,Fajfer:2013wca}.
They play a fundamental part in models with inter-family symmetries which explain the origin of fermion mass hierarchies and mixings~\cite{Berezhiani:1983hm,Dimopoulos:1983rz,Berezhiani:1985in,Berezhiani:1991ds,Berezhiani:1992pj,Berezhiani:2000cg,Berezhiani:1990be}, and provide a natural realization of the minimal flavour violation scenario \cite{Berezhiani:1996ii,Anselm:1996jm,Berezhiani:2001mh}. They can also be relevant to axion models~\cite{Kim:1979if,Berezhiani:1989fp} as well as axion-less models of the Nelson-Barr type solving the strong CP problem~\cite{Nelson:1983zb,Barr:1984qx,Babu:1989rb,Berezhiani:1990vp,Berezhiani:1992pq,Vecchi:2014hpa,Kuchimanchi:2023imj}. 

These extensions also present a very rich phenomenology that can affect many observables, 
including flavour-changing neutral current phenomena
(neutral meson mixing parameters, meson decay amplitudes) as well as flavour-conserving processes as electroweak precision measurements, and weak charged currents, by inducing modifications of the \ac{CKM} matrix or generating right-handed mixing
(see e.g.~\cite{Branco:1986my,delAguila:1989rq,Nir:1990yq,delAguila:2000rc,Barenboim:2001fd,Cacciapaglia:2010vn,Botella:2012ju,Ishiwata:2015cga,Wang:2016mjr,Biekotter:2016kgi,Bobeth:2016llm,Botella:2016ibj,Cacciapaglia:2018lld,Belfatto:2019swo,Belfatto:2021jhf,Branco:2021vhs,Balaji:2021lpr,Botella:2021uxz,Belfatto:2023tbv,Cepedello:2024qmq}).
Remarkably, vector-like quarks have been proposed as viable explanations for Cabibbo angle anomalies,
\cite{Belfatto:2019swo,Cheung:2020vqm,Endo:2020tkb,Belfatto:2021jhf,Branco:2021vhs,Crivellin:2020ebi,Botella:2021uxz,Crivellin:2022rhw,Dcruz:2022rjg},  that is, 
the tensions emerging after three different precise determinations of the Cabibbo angle.
Moreover, since they introduce extra --- and in general complex --- couplings in the Yukawa Lagrangian, new sources of \ac{CPV} arise which may be crucial in explaining the \ac{BAU}, given that \ac{CPV} within the \ac{SM} is too suppressed to generate a large enough \ac{BAU}, consistent with observations. 

If the scalar sector of a model only includes $SU(2)_L$ doublets, \acp{VLQ} mixing with standard model quarks can only appear in seven types of $SU(2)_L$ multiplets: singlets with $Y = - 1/3$ or $Y = 2/3$ (see~\cite{Alves:2023ufm} for a recent review); doublets with $Y = - 5/6$, $Y = 1/6$ or $Y = 7/6$; or triplets with $Y = - 1/3$ or $Y = 2/3$ (see e.g.~\cite{delAguila:2000rc,delAguila:2000aa,Aguilar-Saavedra:2013qpa,Ishiwata:2015cga}).

In this paper, we are specifically interested in studying models
of \ac{VLQ} isodoublets
with non-exotic charges, i.e.~\acp{VLQ} with $Y=1/6$, similarly to the quark doublets present in the \ac{SM}.
These appear to be capable of addressing several current anomalies.
Most notably, it was shown that \ac{VLQ} doublets
mixing with the light generations of \ac{SM} quarks 
are favoured candidates in explaining the \acp{CAA}~\cite{Belfatto:2021jhf,Crivellin:2022rhw,Belfatto:2023tbv,Cirigliano:2023nol}.
Indeed, models of \ac{VLQ} doublets
i) induce right-handed currents, which can potentially accommodate all \ac{CAA} tensions, and ii) are in agreement with electroweak precision observables, e.g.~the $W$-boson mass. Note that, unlike in the case of singlet extensions, there are no significant deviations from \ac{CKM} unitarity at leading order in these models.

We study these new-physics effects and sources of CP violation using a \ac{WBI} description.
In contrast to extensive surveys of vector-like quark singlet models (parameterizations, construction of weak-basis invariants, CP-odd invariants), to our knowledge,
the literature lacks studies of vector-like quark doublets based on \acp{WBI}.
The usefulness of \ac{WBI} quantities lies in the fact that they remain unmodified under weak-basis changes.
Thus, \ac{WBI} quantities can be used to identify the physical information contained in any arbitrary parameterization of the Yukawa matrices.
This allows for an unambiguous
connection to physical observables such as quark masses and mixings~\cite{Bernabeu:1986fc,Gronau:1986xb,Olechowski:1989ny,delAguila:1996pa,delAguila:1997vn,Branco:2011aa,Albergaria:2022zaq,Bento:2023owf,deLima:2024vrn}.
The existence of complex \acp{WBI} signals the presence of physical phases and CP violation.
The imaginary parts of such invariants --- the so-called CP-odd \acp{WBI} --- are one of the focuses of this work.

In the \ac{SM}, the only independent CP-odd \ac{WBI} is of mass dimension $\M=12$ \cite{Branco:1999fs} and is proportional to \ac{CKM} quartets $V_{\alpha i}V_{\beta j}V^*_{\alpha j}V^*_{\beta i}$, resulting in a large suppression of CP violation effects.
In the extensively-studied isosinglet \ac{VLQ} case, the potential for enhancement is well-known, given the presence of a larger number of independent CP-odd \acp{WBI}, including some of dimension $\M=8$~\cite{delAguila:1997vn,Albergaria:2022zaq,deLima:2024vrn}. However, as we emphasize in this work, the doublet representation can allow for lower-dimension \acp{WBI}, starting from $\M=6$ invariants, which suggests an even greater enhancement.%
\footnote{It has been pointed out that models with isodoublet \acp{VLQ}
are uniquely capable of generating significant CP violation in the coupling of third-generation quarks to the Higgs boson, whereas for all other \ac{VLQ} realizations CP violation is suppressed in these couplings~\cite{Chen:2015uza}.}
This is directly related to the fact that in the doublet scenario new kinds of rephasing invariants arise from the interplay between the two types of charged currents. In particular, bilinears of the form $\VRa_{\alpha i}\VLac_{\alpha i}$ are now allowed, where $\Vr_{L(R)}$ represents the mixing matrix for the left(right)-handed charged currents. In this way, the number of insertions of the mixing elements needed to construct CP-odd \acp{WBI} can be brought to a minimum, potentially allowing for strong \ac{CPV} effects.

\vskip 2mm
This paper is divided into the following sections, each of which is as self-contained as possible.
In~\cref{sec:setup}, we
set up the framework and notation (see also~\cref{app:notation}) and
display the gauge and Higgs interactions of the model.
In~\cref{sec:WB}, we discuss
various convenient weak bases, count the physical parameters and present the most relevant rephasing invariants.
In~\cref{sec:rotating}, we present
approximate expressions for the quark masses and the mixings as a function of the Lagrangian parameters, in the realistic limit in which the \acp{VLQ} mass scale is at least a few times larger than the electroweak scale. 
We also derive an exact parameterization for the quark mixing matrices for any number $\n$ of \ac{VLQ} doublets (see~\cref{sec:param}).
In~\cref{sec:CPeven}, we explain how to construct \acfp{WBI} and how they are related to physical parameters. There, we also identify a set of \acp{WBI} characterizing the single \ac{VLQ} doublet scenario 
(see~\cref{sec:mapping} for the case $\n > 1$).
In~\cref{sec:CPodd}, we delve into the topic of CP violation, presenting CP-odd \acp{WBI}, their connection to effective rephasing invariants (see also~\cref{sec:EFT} for the \ac{EFT} description of these models), and conditions for CP conservation with one \ac{VLQ} doublet.
In~\cref{sec:pheno}, we analyze some of the phenomenology of these models using a \ac{WBI} description, particularly in relation to the \acp{CAA} and \ac{CPV} effects. 
Finally, we summarize our results in~\cref{sec:summary}.

\vfill
\clearpage

\section{Setup}
\label{sec:setup}

We extend the \ac{SM} by adding $\n$ \acf{VLQ} isodoublets $Q_{L,R} =(\T,\B)_{L,R}$ in the same representation of $SU(3)_c\times SU(2)_L \times U(1)_Y$ as standard \ac{LH} quarks, i.e.~with the \ac{SM} quantum numbers $(\mathbf{3},\mathbf{2})_{1/6}$. Thus, $\T$ and $\B$ have electric charges $+2/3$ and $-1/3$, respectively. We define
\begin{equation} \label{eq:doublets}
Q^0_{L \alpha} = \begin{pmatrix}
    \T^0_{L \alpha} \\[2mm] \B^0_{L \alpha}
\end{pmatrix}\,,\quad
Q^0_{R \alpha} = \begin{pmatrix}
    \T^0_{R \alpha} \\[2mm] \B^0_{R \alpha}
\end{pmatrix}
\qquad (\alpha = 1,\ldots,\n)\,.    
\end{equation}
The `0' superscript indicates that fields are in a flavour basis, as opposed to the mass basis. \ac{SM} quark doublets and singlets are denoted by $q^0_{Li} = (u^0_{Li},\,\,d^0_{Li})^T$, and $u^0_{Ri}$ and $d^0_{Ri}$, respectively, using Latin family indices ($i=1,2,3$).

Along with the standard Yukawa terms, the relevant Lagrangian includes new couplings and mass terms. Prior to \ac{EWSB}, it reads
\begin{equation} \label{eq:LY1}
\begin{aligned}
-\LY \,=\,\,
 & \left(\yu\right)_{ij} \, \overline{q^0_{L}}_i \, \tilde{\Phi} \, u^0_{Rj} 
 \,+\, \left(\yd\right)_{ij} \,  \overline{q^0_{L}}_i \, \Phi \, d^0_{Rj} 
 \,+\, \text{h.c.}
\\[2mm]
\,+\, &\left(\yU\right)_{\alpha j} \, \overline{Q^0_{L}}_\alpha \, \tilde{\Phi} \, u^0_{Rj}
\,+\, \left(\yD\right)_{\alpha j} \, \overline{Q^0_{L}}_\alpha \,\Phi\,  d^0_{Rj} \,+\, \text{h.c.}
\\[2mm]
\,+\, &\,\left(\mb\right)_{i\beta} \,\,\overline{q^0_{L}}_i \,  Q^0_{R\beta} 
\,+\, \,\left(\mB\right)_{\alpha\beta} \,\, \overline{Q^0_{L}}_\alpha \,Q^0_{R\beta} \,+\, \text{h.c.}
\,,\,
\end{aligned}
\end{equation}
where the first line contains the standard terms, and $i,j = 1,2,3$ and $\alpha,\beta = 1,\ldots,\n$. Here, $\Phi$ denotes the Higgs doublet, while $\tilde\Phi \equiv \epsilon\Phi^*$. The last line in $\LY$ contains bare mass terms, controlled by $\mb$ and $\mB$. 
It may also prove useful to collect all \ac{LH} $SU(2)_L$ doublet quark fields in a single $(3+\n)$-dimensional flavour vector
\begin{equation}
\Q_L^0 = \begin{pmatrix}
    q^0_{L} \\[2mm] Q^0_{L}
\end{pmatrix}\,,
\end{equation}
which allows us to rewrite the Lagrangian of~\cref{eq:LY1} in a compact form:
\begin{equation} \label{eq:LY2}
-\LY \,=\,\,
      \overline{\Q^0_{L}} \, \tilde{\Phi} \, \Yu \, u^0_{R} 
\,+\, \overline{\Q^0_{L}} \, \Phi         \, \Yd \, d^0_{R} 
\,+\, \overline{\Q^0_{L}} \, \Mb \,  Q^0_{R} 
\,+\, \text{h.c.}
\,,
\end{equation}
where the $(3+\n)\times 3$ matrices $\Yu$ and $\Yd$ contain all Yukawa couplings to the Higgs doublet and the $(3+\n)\times \n$ matrix $\Mb$ contains all bare mass terms.
Given that the $3+\n$ species of \ac{LH} doublets have identical quantum numbers, one can consider a basis in field space where $\mb$ vanishes and $\mB$ is real and diagonal ($\mB\to \mBd$; see also~\cref{sec:minimalweak}).
From~\cref{sec:WB} onwards, we will take a Lagrangian with $\mb=0$ and $\mB = \mBd$ as our starting point, without loss of generality, cf.~\cref{eq:L1,eq:LM}.

Following \ac{EWSB}, one has $\Phi \to (0,\,\, \vw + h/\sqrt{2})^T$, with the \ac{VEV} $\vw \approx 174$ GeV, and obtains the tree-level mass terms
\begin{equation} \label{eq:LM1}
\begin{aligned}
-\Lm
\,&=\,
\begin{pmatrix}
    \overline{u^0_L}  &     \overline{\T^0_L} 
\end{pmatrix}
\begin{pmatrix}
    \vw\, \yu  &    \mb \\[2mm]     
    \vw\, \yU  &    \mB
\end{pmatrix}
\begin{pmatrix}
    u^0_R  \\[2mm]     \T^0_R 
\end{pmatrix}
+
\begin{pmatrix}
    \overline{d^0_L}  &     \overline{\B^0_L} 
\end{pmatrix}
\begin{pmatrix}
    \vw\, \yd  &    \mb \\[2mm]     
    \vw\, \yD  &    \mB
\end{pmatrix}
\begin{pmatrix}
    d^0_R  \\[2mm]     \B^0_R
\end{pmatrix}
+ \text{h.c.}
\\[2mm]
\,&\equiv\, \overline{\U^0_L} \Mu \U^0_R \,+\, \overline{\D^0_L} \Md \D^0_R 
+ \text{h.c.}
\,,
\end{aligned}
\end{equation}
where the $\Mq$ are $(3+\n)$-dimensional square matrices, with
\begin{equation} \label{eq:defsM}
\begin{aligned}
\Mq =  \left(\begin{array}{c:c} 
{ } & { } \\
\,\mq\,{ } & \,\,\Mb\,\,{ }\\   
{ } & { } \\
\end{array}\right)\,,
\qquad
\mq \equiv \vw\,\Yq = \vw
\begin{pmatrix}
    \yq  \\[2mm]     
    \yQ 
\end{pmatrix}
\,,
\qquad
\Mb \equiv \begin{pmatrix}
       \mb \\[2mm]     
       \mB
\end{pmatrix}\,,
\end{aligned}
\end{equation}
and $q=u,d$.
We have also defined the vectors
\begin{equation}
\U^0_{L,R} = \begin{pmatrix}
    u^0_{L,R}  \\[2mm]     \T^0_{L,R}
\end{pmatrix}\,,\qquad
\D^0_{L,R} = \begin{pmatrix}
    d^0_{L,R}  \\[2mm]     \B^0_{L,R}
\end{pmatrix}\,,
\end{equation}
in flavour space.
The mass matrices $\Mu$ and $\Md$ can be diagonalized via bi-unitary transformations (a singular value decomposition). To this end, one defines the mass basis fields via
\begin{equation} \label{eq:rot}
    \U^0_{L,R}
    \,=\, \Vu_{L,R}\, \U_{L,R}
    \,=\, \Vu_{L,R}
    \begin{pmatrix} u_{L,R} \\[2mm] \T_{L,R} \end{pmatrix}\,,
    \quad\,\,
    \D^0_{L,R}
    \,=\, \Vd_{L,R}\, \D_{L,R}
    \,=\, \Vd_{L,R}
    \begin{pmatrix} d_{L,R} \\[2mm] \B_{L,R} \end{pmatrix}\,,
\end{equation}
where the four $\V$ matrices are $(3+\n)$-dimensional unitary matrices. One then has
\begin{equation} \label{eq:LM2}
\begin{aligned}
-\Lm \,=\,
\overline{\U}_L \,\Du\, \U_R \,+\, \overline{\D}_L \,\Dd\, \D_R 
+ \text{h.c.}
\end{aligned}
\end{equation}
and
\begin{equation} \label{eq:diag}
{\VqL}^\dagger \, \Mq \, \VqR 
\, = \, 
\Dq
\, \equiv \, 
\diag(\msq_1,\,\msq_2,\,\msq_3,\,\Msq_1,\,\ldots,\,\Msq_\n) \qquad (q=u,d)
\,,
\end{equation}
where the $\msq_i$ and $\Msq_\alpha$ are the (non-negative) physical quark masses in each sector, henceforth assumed positive for simplicity.
In the case of a single \ac{VLQ} doublet ($\n=1$), the $\Msq_1$ are simply denoted $\MT$ and $\MB$, for $q=u,d$ respectively.
As an alternative, one can denote light (standard) quark masses by $m_\alpha$ ($\alpha = u,c,t$) and $m_i$ ($i=d,s,b$), where up- and down-type quarks are labelled by Greek and Latin indices respectively. 

Note that~\cref{eq:diag} describes the singular value decompositions of the $\Mq$. As with any such decomposition, one can find the unitary matrices $\VqL$ and $\VqR$
via the usual (diagonalization) relations 
\begin{equation} \label{eq:svd2}
{\VqL}^\dagger\, \Mq \Mq^\dagger\, \VqL
\,=\,
{\VqR}^\dagger\, \Mq^\dagger \Mq\, \VqR
\,=\, \Dq^2 \,.    
\end{equation}
Before proceeding, it also proves convenient to decompose each $\V$ matrix into its upper and lower blocks, according to
\begin{equation} \label{eq:AB}
\Vq_{L,R} \,=\,
\setlength{\extrarowheight}{1.2pt}
    \left(\begin{array}{c}
     { }\\[-4mm]
      \qquad \Aq_{L,R} \qquad 
      \\[2mm] \hdashline[2pt/2pt]
       { }\\[-4mm]
      \qquad \Bq_{L,R} \qquad 
      \\[2mm]
    \end{array}\right) 
    \setlength{\extrarowheight}{6pt}
\,,
\end{equation}
where, independently of $q$, the $\Aq$ are $3 \times (3+\n)$ matrices while the
$\Bq$ are $\n \times (3+\n)$ matrices.
\Cref{eq:diag} directly implies the relations
\begin{equation} \label{eq:mqM}
\mq \,=\, \VqL \,\Dq \, {\AqR}^\dagger
\,,\qquad
\Mb \,=\, \VqL \,\Dq \, {\BqR}^\dagger
\,.
\end{equation}
Finally, for each chirality, the matrices $\Aq$ and $\Bq$ obey the relations
\begin{equation} \label{eq:ABunit}
\Aq {\Aq}^\dagger = \id_3 \,, \quad
\Bq {\Bq}^\dagger = \id_\n \,, \quad
\Aq {\Bq}^\dagger = 0 \,, \quad
{\Aq}^\dagger \Aq + {\Bq}^\dagger \Bq = \id_{3+\n} \,, 
\end{equation}
which follow from the unitarity of $\Vq$.

\subsection{Gauge and Higgs interactions}
\label{sec:gauge}
The introduction of doublet \acp{VLQ} will also modify the form of the \ac{EW} gauge interactions. The interactions of quarks with the $W$ boson are encoded in the charged-current Lagrangian, which reads
\begin{equation} \label{eq:LW}
\begin{aligned}
    \mathcal{L}_W \,=\,
   &- \frac{g}{\sqrt{2}} \,W_\mu^+ \left[
   \overline{\U^0_L}  \,\gamma^\mu\,  \D^0_L
   \,+\, \overline{\T^0_R}  \,\gamma^\mu\,  \B^0_R
   \right]
+ \text{h.c.}
   \\[2mm]
   \,=\,
   &- \frac{g}{\sqrt{2}} \,W_\mu^+ \left[
   \overline{\U_L}  \,\gamma^\mu\, \VL \, \D_L
   \,+\, \overline{\U_R}  \,\gamma^\mu\, \VR \,  \D_R
   \right]
+ \text{h.c.} \,,
\end{aligned}
\end{equation}
where we have defined the (3+$\n$)-dimensional unitary matrix $\VL$, and the (3+$\n$)-dimensional \ac{RH} current matrix $\VR$, with
\begin{equation} \label{eq:VLVR}
	\VL \,=\,
    {\VuL}^\dagger \VdL\,,
\qquad
    \VR \,=\,
    {\BuR}^\dagger \BdR
    \,=\,
     {\VuR}^\dagger\, \diag(0,0,0,1,\ldots,1)\, \VdR\,. 
\end{equation}
Important differences arise with respect to an extension of the \ac{SM} with \ac{VLQ} isosinglets (see e.g.~\cite{Alves:2023ufm}). Namely, there are two physical mixing matrices, one for each chirality, with $\VR$
originating from the $W_\mu^+ \overline{\T_R^0} \gamma^{\mu} \B_R^0$ term after performing the rotation to the mass eigenbasis, see~\cref{eq:LW}.
The charged-current couplings are determined by
the elements of these matrices. In the simplest case of $\n=1$, these are $4\times 4$ square matrices, with 
\begin{equation} \label{eq:VLVR1}
    \VR = {\VuR}^\dagger \diag(0,0,0,1)\, \VdR \,, \qquad (\text{for }\n=1)\,.
\end{equation}
While the enlarged \ac{LH} quark mixing matrix $\VL$ is necessarily unitary,
taking the form
\begin{equation} \label{eq:CKM}
\VL \,=\, 
  \tikz[baseline=(M.west)]{%
    \node[matrix of math nodes,matrix anchor=west,left delimiter=(,right delimiter=),ampersand replacement=\&] (M) {%
\quad V_{ud}\quad  \& \quad V_{us}\quad  \& \quad V_{ub}\quad  \& \quad \cdots \,\,  \\[1mm]
\quad V_{cd}\quad  \& \quad V_{cs}\quad  \& \quad V_{cb}\quad  \& \quad \cdots \,\,  \\[1mm]
\quad V_{td}\quad  \& \quad V_{ts}\quad  \& \quad V_{tb}\quad  \& \quad \cdots \,\,  \\[1mm]
\vdots \& \vdots \& \vdots \& \quad \ddots \,\,  \\\\
    };
    \node[draw,fit=(M-3-1)(M-1-3),inner sep = -1pt,label={ $\CKM$}] {};
  }
\,,
\end{equation}
the new \ac{RH} mixing matrix $\VR$ cannot be unitary, since $\det \VR = 0$. Here, the $3\times 3$ upper-left block of $\VL$ can be identified as the \ac{CKM} quark mixing matrix $\CKM$, which is no longer unitary in general.
Nevertheless, at leading order in \ac{VLQ} corrections and unlike in \ac{SM} extensions with isosinglet \acp{VLQ}, the matrix $\CKM$ is still unitary to a good approximation (see also~\cref{sec:LH,sec:ESWB}).
Finally, note that factorizable phases in $\VL$ are not necessarily unphysical as in the \ac{SM}, where one can exploit the freedom of identically rephasing \ac{LH} and \ac{RH} quark fields while keeping the (diagonalized) mass terms invariant. Indeed, the mixing of \ac{RH} quarks here means that any such rephasing will modify $\VR$. Hence, in any physical basis, at least one of $\VL$ or $\VR$ will contain factorizable phases that cannot be eliminated.

\vskip 2mm
 
Additionally, the mixing of \ac{SM} quarks with the \acp{VLQ} induces couplings to the Higgs and $Z$ bosons which are flavour-non-diagonal in the mass basis and originate flavour-changing phenomena.
The weak neutral current Lagrangian describing  the interactions of quarks with the $Z$ boson is
\begin{equation} \label{eq:LZ}
\begin{aligned}
    \mathcal{L}_Z \,=\,
   &- \frac{g}{2 c_W} \,Z_\mu  \left[
    \overline{\U^0_L} \,\gamma^\mu\, \U^0_L
-   \overline{\D^0_L} \, \gamma^\mu\, \D^0_L
+ \overline{\T^0_R} \, \gamma^\mu  \, \T^0_R 
- \overline{\B^0_R} \, \gamma^\mu \, \B^0_R 
-2 s_W^2 J^\mu_\text{em}
\right]
   \\[2mm]
    \,=\, 
    &- \frac{g}{2 c_W} \,Z_\mu \left[
  \overline{\U_L} \, \gamma^\mu \, \U_L
- \overline{\D_L} \, \gamma^\mu \, \D_L
+\overline{\U_R} \, \gamma^\mu \, \Fu \, \U_R
-\overline{\D_R} \, \gamma^\mu \, \Fd \, \D_R
-2 s_W^2 J^\mu_\text{em}
\right]\,,
\end{aligned}
\end{equation}
with $c_W= \cos \theta_W$ and $s_W= \sin \theta_W$, while the electromagnetic current reads
\begin{equation}
J^\mu_\text{em} \,=\, + \frac{2}{3}  \, \overline{\U} \, \gamma^\mu \, \U - \frac{1}{3}\, \overline{\D} \, \gamma^\mu \, \D\,,
\end{equation}
with $\U = \U_L + \U_R$ and $\D = \D_L + \D_R$,
and preserves its diagonal form in moving from the flavour to the mass basis.
In~\cref{eq:LZ} we have introduced the $(3+\n)$-dimensional Hermitian matrices
\begin{equation} \label{eq:vnc}
\begin{aligned}
	\Fu \,&=\,
    {\BuR}^\dagger \BuR
    \,=\,
	\VR^{}\, \VR^\dagger
    \,=\,
     {\VuR}^\dagger\, \diag(0,0,0,1,\ldots,1)\, \VuR\,,
\\
    \Fd \,&=\,
    {\BdR}^\dagger \BdR
    \,=\,
	\VR^\dagger\, \VR^{} 
    \,=\,
     {\VdR}^\dagger\, \diag(0,0,0,1,\ldots,1)\, \VdR\,. 
\end{aligned}
\end{equation}
Note that \ac{LH} couplings remain diagonal at tree level, as in the \ac{SM}, while 
\acp{FCNC} appear only in the \ac{RH} sector, controlled by these matrices $\Fq$.
This is in contrast with the case of \ac{VLQ} singlets, for which only \ac{LH} $Z$-mediated \acp{FCNC} are present.
In the simplest case of $\n=1$, the matrices $\Fq$ determining \acp{FCNC} couplings are $4\times 4$ square matrices, with 
\begin{equation} \label{eq:vnc1}
    \Fq= {\VqR}^\dagger \diag(0,0,0,1)\, \VqR \,, \qquad (\text{for }\n=1)\,.
\end{equation}
The relations between the $\Fq$ and $\VR$ in~\cref{eq:vnc} follow from~\cref{eq:ABunit}. They directly connect the existence of such tree-level \acp{FCNC} to the non-unitarity of the \ac{RH} charged-current mixing matrix. 

\vskip 2mm

As for the interactions with the Higgs boson $\h$, the mass matrices are not proportional to the Yukawa matrices and flavour non-diagonal couplings are also generated. The relevant Lagrangian reads
\begin{equation} \label{eq:Lh}
\begin{aligned}
\mathcal{L}_\h
\,=\,
&-\frac{\h}{\sqrt{2}}
\left[
\overline{\U^0_L} 
\begin{pmatrix}
    \yu  &    0 \\[2mm]     
    \yU  &    0
\end{pmatrix}
\U^0_R
+
 \overline{\D^0_L}
\begin{pmatrix}
    \yd  &    0 \\[2mm]     
    \yD  &    0
\end{pmatrix}
\D^0_R
\right]
+ \text{h.c.}
\\[2mm]
\,=\,
&- \frac{\h}{\sqrt{2}}
\left[
\overline{\U_L}\,\,
\frac{\Du}{\vw} \left(\id - \Fu \right)
\,\U_R
+
\overline{\D_L}\,\,
\frac{\Dd}{\vw} \left(\id - \Fd \right)
\,\D_R
\right]
+ \text{h.c.}
\,,
\end{aligned}
\end{equation}
where we have used the fact that
\begin{equation} \label{eq:A1mF}
    {\AqR}^\dagger \AqR
    \,=\,
    \id -  \Fq
    \,=\,
   {\VqR}^\dagger\, \diag(1,1,1,0,\ldots,0)\, \VqR \,.
\end{equation}
Hence, Higgs-mediated \acp{FCNC} are controlled by the same quantities as $Z$-mediated \acp{FCNC}, namely $\Fu$ and $\Fd$.
Given the presence of the matrices $\Dq$ in~\cref{eq:Lh}, one sees that, as in the case of singlet \acp{VLQ}, Higgs-mediated transitions between the lighter quarks are suppressed by ratios $\msq_i/\vw$ with respect to $Z$-mediated transitions. 
On the other hand, there emerge non-negligible couplings between right-handed light states and left-handed heavy states, while
heavy-light transitions of the opposite chirality are generically suppressed by small \ac{SM} Yukawas.%
\footnote{In the isosinglet \ac{VLQ} case, the products $\Fq\Dq $ appear in $\mathcal{L}_\h$ instead of $\Dq \Fq$, so that the reverse happens, i.e.~the non-negligible transitions are those between 
\ac{LH} light states and \ac{RH} heavy states.
}

\vskip 2mm

It is worth noting that the last $\n$ columns of the mass matrices are shared between $\Mu$ and $\Md$, i.e.~the
$(3+\n)\times \n$ bare mass matrix $\Mb$ is common to both sectors. Keeping in mind that
\begin{equation}
    {\BqR}^\dagger = {\VqR}^\dagger
    \begin{pmatrix}
        0_{3\times \n} \\
        \id_{\n\times \n}
    \end{pmatrix}\,,
\end{equation}
and recalling~\mbox{\cref{eq:mqM}}, which implies
\begin{equation} \label{eq:LRrelation1}
\VuL\,\Du\,{\BuR}^\dagger \,=\,
\VdL\,\Dd\,{\BdR}^\dagger \,,
\end{equation}
one finds the relations
\begin{equation} \label{eq:LRrelation2}
\Du\, \VR \,=\, \VL\, \Dd\, \Fd\,,
\qquad
\Dd\, \VR^\dagger \,=\, \VL^\dagger\, \Du\, \Fu\,,
\end{equation}
and
\begin{equation} \label{eq:LRrelation3}
\Du\,\Fu\,\Du\,\VL \,=\, \VL\, \Dd\, \Fd\, \Dd
\end{equation}
between \ac{LH} and \ac{RH} couplings in charged and neutral currents.

\vskip 2mm

The discussion above applies to a general number $\n$ of \ac{SM}-like \acp{VLQ}. In the simplest case of $\n = 1$, $\mB$ is a number, $\mb$, $\yU$ and $\yD$ are 3-dimensional vectors, and the matrices $\Du = \diag(m_u,m_c,m_t,\MT)$, $\Dd = \diag(m_d,m_s,m_b,\MB)$, $\VL$ (unitary), $\VR$ (not unitary), $\Fu$ and $\Fd$ (Hermitian) are $4\times 4$ square matrices. The counting of independent physical parameters contained in these matrices, for both this case and the most general one, is carried out in the next section.

\clearpage

\section{Weak bases and physical parameters}
\label{sec:WB}

Not all of the $18 + 9 \n + \n^2$ complex Yukawa couplings and bare mass terms in the Lagrangian of~\cref{eq:LY2} represent independent physical parameters. Indeed, owing to the invariance of the S-matrix under field redefinitions and the underlying symmetries of QFT, many choices are equivalent. Each choice corresponds to a so-called \ac{WB} in flavour space. These are connected by \acp{WBT} which keep the kinetic terms invariant, i.e.~do not modify the form of the \ac{EW} gauge interactions of~\cref{sec:gauge}.
Any two sets of Yukawa and mass matrices related by a \ac{WBT} represent the same physical system.

In the class of models under consideration,  it is straightforward to identify the following transformations as the most general \acp{WBT},  describing the redundancy in the determination of the number of independent parameters:
\begin{equation} \label{eq:WB}
\begin{array}{l@{\qquad}l}
    \U^0_{L}   \,\to\, \WL \,   \U^0_{L}
  \,,
  &
    \D^0_{L}   \,\to\, \WL \,   \D^0_{L}
  \,,
  \\[3mm]
  u^0_R \,\to\, \WRu  \, u^0_R
  \,,
  &d^0_R \,\to\, \WRd  \, d^0_R
  \,,\\[3mm]
  \T^0_R \,\to\, \WR  \, \T^0_R
  \,,
  &\B^0_R \,\to\, \WR  \, \B^0_R
\,,
\end{array}
\end{equation}
where the matrices $\WRq$ ($q=u,d$) and  $\WR$  are respectively $3\times 3$ and $\n\times\n$ unitary matrices. Note that the $3+\n$ species of \ac{LH} doublets have identical quantum numbers, and can thus be connected via a larger $(3+\n) \times (3+\n)$ unitary matrix, $\WL$. In the case $\n=1$, $\WR=e^{i\WRp}$ is simply a phase.

Under a \ac{WBT}, as given in~\cref{eq:WB}, the Yukawa and mass matrices are not left invariant, but instead transform as
\begin{equation} \label{eq:MWBT}
\Yq \, \to \, \WL^\dagger\,\Yq \, \WRq\,, 
\qquad
\Mb \, \to \, \WL^\dagger\,\Mb\,\WR \,,
\end{equation}
where $\vw\,\Yq $ is the $(3+\n)\times 3$ submatrix of $\Mq$ arising from Yukawa couplings to the Higgs doublet and $\Mb$ is the $(3+\n)\times \n$ matrix corresponding to the last $\n$ columns of the mass matrices $\Mq$, see~\cref{eq:defsM}.
It follows that the Hermitian combinations 
\begin{equation} \label{eq:hq-H}
\hq \,\equiv \, \mq \mq^\dagger
= \vw^2 \Yq \Yq^\dagger
\,, \qquad
\C \,\equiv \, \Mb \Mb^\dagger
\end{equation}
transform as
\begin{equation} \label{eq:WL-hq-H}
\hq \,\to\, \WL^\dagger \,\hq\, \WL \,,\qquad
\C \,\to\,\WL^\dagger \,\C\, \WL \,,
\end{equation}
where $\mq=\vw\,\Yq $, c.f.~\cref{eq:defsM}.
Since any product involving these matrices transforms in the same way, they are useful to build \acp{WBI}, i.e.~physical quantities which do not depend on the \ac{WB} choice, via traces or determinants.%
\footnote{Notice that also the determinant of a matrix can always be rewritten as a simple polynomial of traces of powers of that matrix and thus traces alone can be used to form complete bases of \acp{WBI}.}
\Acp{WBI} allow one to extract the actual physical content of a set of Yukawa and mass matrices (see~\cref{sec:CPeven,sec:CPodd}).

\subsection{Convenient weak bases}
\label{sec:WBs}

\subsubsection{A minimal weak basis}
\label{sec:minimalweak}

The freedom present in the transformations of~\cref{eq:WB} allows one to shape the mass matrices $\Mq$ ($q=u,d$) without affecting their physical content, via~\cref{eq:MWBT}. For instance, one can always move to a \ac{WB} where the mixed bare mass terms vanish, $\mb = 0$, by an appropriate choice of $\WL$, using the fact that the $3+\n$ species of \ac{LH} doublets have identical quantum numbers.
Further \acp{WBT} can be used to make the $\mB$ block become diagonal and non-negative, $\mB \to \mBd$.
Thus, here and in what follows, without loss of generality, we take as a starting point the
pre-\ac{EWSB} Lagrangian 
\begin{equation} \label{eq:L1}
\begin{aligned}
-\LY \,=\,\,
 & \left(\yu\right)_{ij} \, \overline{q^0_{L}}_i \, \tilde{\Phi} \, u^0_{Rj} 
 \,+\, \left(\yd\right)_{ij} \,  \overline{q^0_{L}}_i \, \Phi \, d^0_{Rj} 
 \,+\, \text{h.c.}
\\[2mm]
\,+\, &\left(\yU\right)_{\alpha j} \, \overline{Q^0_{L}}_\alpha \, \tilde{\Phi} \, u^0_{Rj}
\,+\, \left(\yD\right)_{\alpha j} \, \overline{Q^0_{L}}_\alpha \,\Phi\,  d^0_{Rj} \,+\, \text{h.c.}
\\[2mm]
\,+\, &\,\left(\mBd\right)_{\alpha\beta} \,\, \overline{Q^0_{L}}_\alpha \,Q^0_{R\beta} \,+\, \text{h.c.}
\,,\,
\end{aligned}
\end{equation}
and the corresponding mass matrices
\begin{equation} \label{eq:LM}
\Mu \,=\, 
\begin{pmatrix}
    \vw\, \yu  &     0 \\[2mm]     
    \vw\, \yU  &    \mBd
\end{pmatrix}\,,
\qquad
\Md \,=\,
\begin{pmatrix}
    \vw\,\yd  &     0 \\[2mm]     
    \vw\,\yD  &    \mBd
\end{pmatrix}
\,,
\end{equation}
following \ac{EWSB}. At this stage, one has $18+7 \n$ complex parameters, which are still not independent and physical.

The $3\times 3$ block containing $\yq$ can be made diagonal (non-negative) via bidiagonalization in one of the two sectors, say the up sector.
Explicitly, one may write the singular value decompositions
\begin{equation} \label{eq:Uhat}
    \yu = \VUW\, \Ydu\, \WUW^\dagger\,, \qquad
    \yd = \VDW\, \Ydd\, \WDW^\dagger\,, 
\end{equation}
where the $\Ydq$ are diagonal and non-negative, 
\begin{equation} \label{eq:Yhat}
\Ydu  \equiv \diag(\hat{y}_u,\hat{y}_c,\hat{y}_t), \qquad \Ydd \equiv \diag(\hat{y}_d,\hat{y}_s,\hat{y}_b)\,,
\end{equation}
while $\VQW$ and $\WQW$ are $3\times 3$ unitary matrices.
One is free to choose the order of the non-zero elements in any of the diagonal matrices.
Then, $\WL$ and $\WRu$ can be used to cancel $\VUW$ and $\WUW$ in the matrix $\yu$, leading to $\yu \to \Ydu$, while $\yd \to \ydp = \VUW^\dagger \yd$.

One is also free to use 
the matrix $\WRd$
to remove the rightmost unitary rotation from the (general) matrix $\ydp = \VUW^\dagger \VDW \Ydd\WDW^\dagger$, which becomes $\ydp \,\to\, \VWB\, \Ydd$, 
where we have defined
\begin{equation} \label{eq:Vhat}
    \VWB  \equiv  \VUW^\dagger \VDW  \, .
\end{equation}
After these \acp{WBT}, the matrices $\yU$ and $\yD$ now read
\begin{equation} \label{eq:zqs}
    \yU \,\to\, \yup = \yU\, \WUW \,, \qquad
    \yD \,\to\, \ydb = \yD\, \WDW  \,. 
\end{equation}
As in the \ac{SM}, 5 phases can be removed from $\VWB$ by rephasing standard quark fields.%
\footnote{The sixth phase is not removable due to the unbroken baryon number. In this procedure, the diagonal matrix $\Ydu$ is unchanged by an appropriate rephasing with $\WRu$. The columns of both $\ydb$ and $\yup$ are rephased.}
Such rephasings can be included from the start in the definition of the matrices $\VQW$, $\WQW$.
Then, $\VWB$ can be parameterized as the standard \ac{CKM}, with three mixing angles and a phase. Thanks to the possibility of removing these phases, the angles can be chosen to lie in the first quadrant, while the range of the (sole) physical phase is a priori unrestricted.
With these choices, $\VWB$ becomes the \ac{SM} $\CKM$ in the limit of \ac{VLQ} doublet decoupling ($\yup, \ydb \to 0$, $\mBd\to \infty\, \id$).

Note that 
$\yup$ and $\ydb$ 
have remained general complex, up to this point. 
One can finally use the $\n\times\n$ lower-right block of $\WL$, which we will denote $\PWB$, to remove $\n$ of the 6$\n$ phases contained in $\yup$ and $\ydb$.%
\footnote{The diagonal matrix $\mBd$ is unchanged by an appropriate rephasing with $\WR$ affecting both sectors.}
As a concrete possibility, we consider removing one phase in each row of $\yup$, making its first column real.%
\footnote{One could just as well have removed $\n$ phases from the first column of $\ydb \to \ydbh$, as considered in~\cref{sec:mapping,sec:SMEFT}.}
As a result of the above sequence of transformations, the Yukawa and mass matrices now take the form
\begin{equation} \label{eq:minimalY}
\Yu \,=\,  
\begin{pmatrix}
    \Ydu \\[2mm]     
    \yub 
\end{pmatrix}\,,
\qquad
\Yd \,=\,
\begin{pmatrix}
    \VWB\, \Ydd\\[2mm]     
    \ydb
\end{pmatrix}\,,
\qquad
\Mb \,=\,
\begin{pmatrix}
      0 \\[2mm]     
   \mBd
\end{pmatrix}
\end{equation}
({\it minimal \ac{WB}}),
which after \ac{EWSB} correspond to the mass matrices
\begin{equation} \label{eq:minimal}
\Mu \,=\, 
\begin{pmatrix}
    \vw\,\Ydu &     0 \\[2mm]     
    \vw\,\yub &    \mBd
\end{pmatrix}\,,
\qquad
\Md \,=\,
\begin{pmatrix}
    \vw\,\VWB\,\Ydd  &     0 \\[2mm]     
    \vw\,\ydb &    \mBd
\end{pmatrix}\,.
\end{equation}
Here, $\yub = P\, \yup$ has $\n$ phases less than the general complex $\yup$.

\newcolumntype{x}[1]{>{\centering\arraybackslash\hspace{0pt}}p{#1}}
\begin{table}[t!]
  \centering
  \begin{tabular}{lx{1cm}x{1cm}x{1cm}x{1cm}x{1cm}x{1cm}x{2cm}}
    \toprule
    & $\Ydu$ & $\yub$ & $\VWB$ & $\Ydd$ & $\ydb$ & $\mBd$  & Total\\
    \midrule
    Moduli & $3$ & $3\n$ & $3$ & $3$ & $3\n$ & $\n$  & $9+7\n$  \\
    Phases & $0$ & $2\n$ & $1$ & $0$ & $3\n$ & $0$ & $1+5\n$  \\
    \midrule
    Total & $3$ & $5\n$ & $4$ & $3$ & $6\n$ & $\n$ & $10+12\n$ \\
    \bottomrule
  \end{tabular}
  \caption{Physical parameter count in the minimal \ac{WB} of~\cref{eq:minimalY,eq:minimal}, in the presence of $\n$ isodoublet \acp{VLQ}. }
  \label{tab:minimal}
\end{table}

\newcolumntype{y}[1]{>{\centering\arraybackslash\hspace{0pt}}p{#1}}
\begin{table}[t!]
  \centering
  \begin{tabular}{ly{1.5cm}y{1.5cm}y{1.5cm}y{1.5cm}y{1.5cm}}
    \toprule
    & \multirow{2}{*}{\ac{SM}} & New ($\n=1$) & Total ($\n=1$) & New ($\n \geq 1$) & Total ($\n \geq 1$) \\
    \midrule
    Quark masses & 6 & 2 & 8 & $2\n$ & $6+2\n$  \\
    Mixing angles & 3 & 5 & 8 & $5\n$ & $3+5\n$  \\
    Total moduli & 9 & 7 & 16 & $7\n$ & $9+7\n$ \\
    \midrule
    Phases & 1 & 5 & 6 & $5\n$ &$1+5\n$  \\
    Total angles\,+\,phases & 4 & 10 & 14 & $10\n$ & $4+10\n$ \\
    \midrule
    Total (parameters) & 10 & 12 & 22 & $12\n$ & $10+12\n$ \\
    \bottomrule
  \end{tabular}
  \caption{Summary of the number of physical parameters in \ac{SM} extensions  with one ($\n = 1$) and several ($\n > 1$)
   \ac{VLQ} doublets with hypercharge $1/6$.}
  \label{tab:countsummary}
\end{table}

Having exhausted the \ac{WB} freedom, the minimal \ac{WB} of~\cref{eq:minimalY,eq:minimal} allows one to count the number of physical parameters. One thus finds a total of $10 + 12\n$ parameters, with the corresponding breakdown being shown in~\Cref{tab:minimal}. Out of this total, $9+7\n$ parameters correspond to physical moduli, while $1+5\n$ correspond to physical phases. 
Anticipating the correspondence of the moduli to physical masses and mixing angles, the parameter count in the scenarios of interest is summarized in~\Cref{tab:countsummary}.
 One should keep in mind that, depending on the choice of minimal \acp{WB} one might obtain distinct countings with distinct numbers of phases and angles. This happens because some phases can be traded for the same number of mixing angles via \acp{WBT}~\cite{Bastos:2024afz,Silva-Marcos:2002upu}. However the total number of mixing parameters (angles\,+\,phases) is of course the same in all minimal \acp{WB}.
 Independent ways of counting physical parameters are presented in~\cref{sec:counting}.

\subsubsection{A ``stepladder'' weak basis (\texorpdfstring{$\n=1$}{\n=1})}
\label{sec:stepladderdefs}

In the case $\n=1$, one can move to a special ``stepladder'' \ac{WB}
which allows for a sequential and straightforward determination of the Yukawa couplings from a set of \acp{WBI}.

We take the minimal \ac{WB} of~\cref{eq:minimal} as a starting point. Then, one can move all 5 physical phases in $\yub$ and $\ydb$ to $\VWB$ by rephasing standard quarks,
undoing the aforementioned rephasing and
restoring $\VWB$ in this context to a general unitary matrix $\VWB \to \VWT$ (not \ac{CKM}-like), while $\yup$ and $\ydb$ are now real 3-dimensional row vectors. The mass matrices will have the structure
\begin{equation} \label{eq:struct1}
\Mu \,\sim\, 
\begin{pmatrix}
    \mathbb{R}  &  &  &      \\ 
     & \mathbb{R}  &   &      \\
    & & \mathbb{R}  &      \\
    \mathbb{R}  & \mathbb{R} & \mathbb{R} & \mathbb{R} 
\end{pmatrix}\,,
\qquad
\Md \,\sim\,
\begin{pmatrix}
    \VWT & \\
    & 1
\end{pmatrix}
\begin{pmatrix}
    \mathbb{R}  &  &  &      \\
     & \mathbb{R}  &   &      \\
    & & \mathbb{R}  &      \\
    \mathbb{R}  & \mathbb{R} & \mathbb{R} & \mathbb{R} 
\end{pmatrix}\,.
\end{equation}
One can use the matrices $\WRq$ to perform an orthogonal rotation in each sector, mixing the first two columns of $\Mu$ and of (the rightmost matrix of) $\Md$ in order to cancel the $(4,1)$ element in each of these matrices. Then, via a similar orthogonal rotation of the second and third columns in each sector, one may cancel the $(4,2)$ entries, leading to a \ac{WB} where, generically,
\begin{equation}
\Mu \,\sim\, 
\begin{pmatrix}
    \mathbb{R}  & \mathbb{R} & \mathbb{R} &      \\ 
    \mathbb{R} & \mathbb{R}  & \mathbb{R}  &      \\
     & \mathbb{R} & \mathbb{R}  &      \\
     &  & \mathbb{R} & \mathbb{R} 
\end{pmatrix}\,,
\qquad
\Md \,\sim\,
\begin{pmatrix}
    \VWT & \\
    & 1
\end{pmatrix}
\begin{pmatrix}
    \mathbb{R}  & \mathbb{R} & \mathbb{R} &      \\
    \mathbb{R} & \mathbb{R}  &  \mathbb{R} &      \\
      & \mathbb{R} & \mathbb{R}  &      \\
      &  & \mathbb{R} & \mathbb{R} 
\end{pmatrix}\,.
\end{equation}
Now, the upper-left block of $\WL$ can be used as an orthogonal rotation between the first two lines, enforcing a zero in the $(1,3)$ position of $\Mu$. A zero in the same position in the rightmost matrix of $\Md$ can be obtained by absorbing orthogonal rotations (including the one from $\WL$) in the general $\VWT$.
Similarly, one can enforce zeros in the $(2,3)$ positions by mixing the second and third rows and reach a \ac{WB} where 
\begin{equation}
\Mu \,\sim\, 
\begin{pmatrix}
    \mathbb{R}  & \mathbb{R} &  &      \\ 
    \mathbb{R} & \mathbb{R}  &  &      \\
    \mathbb{R} & \mathbb{R} & \mathbb{R}  &      \\
     &  & \mathbb{R} & \mathbb{R} 
\end{pmatrix}\,,
\qquad
\Md \,\sim\,
\begin{pmatrix}
    \VWT & \\
    & 1
\end{pmatrix}
\begin{pmatrix}
    \mathbb{R}  & \mathbb{R} &  &      \\
    \mathbb{R} & \mathbb{R}  &   &      \\
    \mathbb{R}  & \mathbb{R} & \mathbb{R}  &      \\
      &  & \mathbb{R} & \mathbb{R} 
\end{pmatrix}\,.
\end{equation}
Finally, one may use \ac{RH} rotations between the first two columns to cancel the $(3,1)$ elements and \ac{LH} rotations between the first two rows to cancel the $(1,2)$ elements, arriving at the desired {\it stepladder \ac{WB}},
\begin{equation} \label{eq:stepladder}
\Mu \,\sim\, 
\begin{pmatrix}
    \mathbb{R}  &  &  &      \\ 
    \mathbb{R} & \mathbb{R}  &   &      \\
     & \mathbb{R} & \mathbb{R}  &      \\
     &  & \mathbb{R} & \mathbb{R} 
\end{pmatrix}\,,
\qquad
\Md \,\sim\,
\begin{pmatrix}
    \VWT & \\
    & 1
\end{pmatrix}
\begin{pmatrix}
    \mathbb{R}  &  &  &      \\
    \mathbb{R} & \mathbb{R}  &   &      \\
      & \mathbb{R} & \mathbb{R}  &      \\
      &  & \mathbb{R} & \mathbb{R} 
\end{pmatrix}\,.
\end{equation}
Note that all intermediate basis are valid \ac{WB} choices.
Details on the extraction of couplings in the \ac{WB} of~\cref{eq:stepladder} from a set of invariants is postponed to~\cref{sec:stepladderwb}.
In what follows, we will consider the following explicit notation for the entries of the mass matrices in the stepladder \ac{WB}:
\begin{equation} \label{eq:stepladder_2}
\Mu = \mB
\begin{pmatrix}
    r_{u5}  & 0 & 0 &  0    \\ 
    r_{u4} & r_{u3}  & 0  & 0     \\
    0 & r_{u2} & r_{u1}  &   0   \\
    0 & 0 & r_{u0} & 1
\end{pmatrix}\,,
\quad
\Md = \mB
\begin{pmatrix}
    \VWT & \\
    & 1
\end{pmatrix} 
\begin{pmatrix}
    r_{d5}  & 0 & 0 &  0    \\ 
    r_{d4} & r_{d3}  & 0  & 0     \\
    0 & r_{d2} & r_{d1}  &   0   \\
    0 & 0 & r_{d0} & 1
\end{pmatrix}\,,
\end{equation}
where $r_{qi}$ ($i=0,\ldots,5$) are dimensionless (note the overall $\mB$ factors). 
Given these definitions, one expects $r_{qi} \lesssim \vw/\mB$.

\vskip 2mm
We emphasize that, in the $\n=1$ case, one can always switch between this \ac{WB} and that of~\cref{eq:minimal} via \acp{WBT}, while keeping the number of parameters at a minimum. In this sense, both \acp{WB} can be said to be minimal.
This remains true even when some of the parameters characterizing these \acp{WB} vanish. In fact, one can prove that, for a given sector, when any number of off-diagonal real parameters vanishes in the stepladder \ac{WB}, the same number of real couplings between \ac{SM} quarks and the \ac{VLQ} doublets must vanish in the minimal \ac{WB}.
Namely,
it is straightforward to show that:
\begin{itemize}
    \item If $r_{q4}=0$, then when transforming into the minimal \ac{WB} one finds that the Yukawa coupling vector $\yqp$ defined in~\cref{sec:minimalweak} has one vanishing entry.
    \item If $r_{q2}=0$, then $r_{q4}$ can be eliminated using \acp{WBT} and when switching into the minimal \ac{WB} one finds that $\yqp$ has two vanishing entries.
    \item If $r_{q0}=0$, then both $r_{q4}$ and $r_{q2}$ can be eliminated using \acp{WBT}, resulting in $\yqp=0$ when considering the minimal \ac{WB}. In that case the \ac{VLQ} is decoupled from that sector.
    \item $r_{q5,3,1}\neq 0$ correspond to non-zero Yukawa couplings $\hat{y}_{\alpha}$ or $\hat{y}_{i}$.
\end{itemize}
This link between both \acp{WB} will be relevant later on to establish a connection between the results of~\cref{sec:stepladderwb} and those of~\cref{sec:CPodd}.

\subsection{Rephasing invariants}
\label{sec:reph-inv}
Observables can only depend on
phase-convention-independent quantities.
In the \ac{SM}, one can construct rephasing-invariant quantities in terms of the mixing using as building blocks the moduli $|V_{\alpha i}|^2$ of \ac{CKM} entries or quartets of the form $Q_{\alpha i \beta j}=V_{\alpha i}V_{\beta j}V^*_{\alpha j}V^*_{\beta i}$, involving four different quarks. When $\alpha\neq \beta$ and $i\neq j$, these
are the simplest generically-complex rephasing invariants. The non-vanishing of the imaginary parts of the quartets is then directly connected to the existence of CP violation in the \ac{SM}.
Namely, given the unitarity of $\CKM$, all nine quantities $\im Q_{\alpha i \beta j}$ ($\alpha\neq \beta$, $i\neq j$) are equal to the invariant $J$~\cite{Jarlskog:1985ht,Jarlskog:1985cw,Bernabeu:1986fc} up to a sign,
\begin{equation} \label{eq:Jar}
    J \, \sum_{\gamma,k}\epsilon_{\alpha\beta\gamma}\,\epsilon_{ijk}= \im(V_{\alpha i}V_{\beta j}V^*_{\alpha j}V^*_{\beta i})
    \,,    
\end{equation}
and all CP-violating effects in the \ac{SM} are proportional to $J$.

When introducing \ac{VLQ} doublets, apart from the mixing of \ac{LH} quarks in the charged currents, we will also have mixing of \ac{RH} quarks in both charged and neutral currents. Hence, one can expect new types of rephasing-invariant quantities to be relevant and potentially new sources of CP violation to arise. In this subsection, we briefly present some of the simplest examples of such quantities in the context of these extensions.

We turn our attention to the gauge interactions in~\cref{eq:LZ,eq:LW}, from which we can conclude that under general rephasings of the quark mass eigenstates $\U_\alpha\rightarrow \U_\alpha \, e^{i \varphi_\alpha}$ and $\D_i\rightarrow \D_i  \, e^{i \varphi_i}$, the mixing matrices transform as 
\begin{equation} \label{eq:rephasing}
\begin{aligned}
        \VLa_{\alpha i} &\,\rightarrow\, e^{-i(\varphi_\alpha-\varphi_i)}\, \VLa_{\alpha i}\,,\\
        \VRa_{\alpha i} &\,\rightarrow\, e^{-i(\varphi_\alpha-\varphi_i)}\, \VRa_{\alpha i}\,,\\
        \Fua_{\alpha \beta} &\,\rightarrow\, e^{-i(\varphi_\alpha-\varphi_\beta)} \,\Fua_{\alpha \beta}\,,\\ 
        \Fda_{ij} &\,\rightarrow\, e^{-i(\varphi_i-\varphi_j)}\, \Fda_{ij}\,,
\end{aligned}
\end{equation}
where chirality and quark-type subscripts were changed to superscripts to avoid index cluttering.
Now, taking these transformations into account, we provide some examples of rephasing invariants in the general case of $\n\geq 1$ doublets, and then focus on the particular case of a single doublet ($\n=1$), for which the number of independent rephasing invariants is lower.

\subsubsection{The general case (\texorpdfstring{$\n\geq 1$}{\n>=1})}

Since $\VL$ and $\VR$ transform in the same manner, any bilinear of the form
\begin{equation} \label{eq:bilinears}
    \bi^{\chi \chi'}_{\alpha i}
    \,\equiv\,
    \Vr^\chi_{\alpha i}\, \Vr^{\chi'*}_{\alpha i}
    \,=\,
    \left(\bi^{\chi'\chi}_{\alpha i}\right)^*
\end{equation}
is a rephasing-invariant quantity. Here $\chi,\chi'=L,R$ are chirality indices. For $\chi=\chi'$, these quantities reduce to the squared moduli $|\VLa_{\alpha i}|^2$ or $|\VRa_{\alpha i}|^2$. When $\chi\neq\chi'$, we obtain bilinears 
$\bi^{LR}_{\alpha i} = \VLa_{\alpha i}\, \VRac_{\alpha i}$ which are complex in general and have no analogue in the \ac{SM} (or even in extensions with \ac{VLQ} singlets).
These are potentially related to new sources of CP violation.
In particular, this means one may expect CP-violating effects in charged currents which involve only two quarks. 
The next simplest rephasing invariants one can build
 are complex trilinears such as
\begin{equation} \label{eq:trilinears}
\begin{aligned}
 \tri^{\chi \chi'}_{i, \alpha \beta}
 \,&\equiv\,
 \Vr^{\chi*}_{\alpha i}\,  \Vr^{\chi'}_{\beta i} \, \Fua_{\alpha\beta}
 \,=\,
 \left(\tri^{\chi' \chi}_{i, \beta \alpha}\right)^*\,,
    \\[1mm]
\tri^{\chi \chi'}_{\alpha ,i j  }
 \,&\equiv\,
 \Vr^\chi_{\alpha i}\, \Vr^{\chi'*}_{\alpha j} \, \Fda_{ij}
 \,=\,
 \left(\tri^{\chi' \chi}_{\alpha ,j i }\right)^*\,.
\end{aligned}
\end{equation}
These invariants, which are associated with \acp{FCNC},
involve three quarks.
Other meaningful rephasing-invariant quantities that have analogues in the \ac{SM} are the quartets. Here, in general we have
\begin{equation} \label{eq:Q}
         \qua^{\chi\chi'\xi \xi'}_{\alpha i \beta j}
         \, \equiv\, 
         \Vr^\chi_{\alpha i}\, \Vr^{\chi'}_{\beta j}\, 
         \Vr^{\xi *}_{\alpha j} \, \Vr^{\xi' *}_{\beta i}
         \,=\,
         \left(\qua^{\xi' \xi \chi' \chi }_{ \beta i \alpha  j}\right)^* 
         = \left(\qua^{\xi\xi'\chi \chi'}_{\alpha j \beta i}\right)^* 
         = \qua^{\chi'\chi \xi' \xi}_{\beta j \alpha i}\,,
\end{equation}
where $\xi,\xi'=L,R$ are also chirality indices.

Note that trilinears can be written as sums of quartets. For instance,
\begin{equation}
\begin{aligned}
    \sum_i
    \qua^{R\chi\chi' R}_{\alpha i\beta j} 
    &=
    \Fua_{\alpha\beta} \,\Vr^{\chi}_{\beta j}\, \Vr^{\chi'*}_{\alpha j}
    = \tri^{\chi'\chi}_{j, \alpha \beta }\,,
    \\
    \sum_\alpha
    \qua^{R\chi R \chi'}_{\alpha i\beta j}
    &=
    \Fda_{ji} \,\Vr^{\chi}_{\beta j} \, \Vr^{\chi'*}_{\beta i}
    = \tri^{\chi\chi'}_{\beta , j i }\,.
\end{aligned}
\end{equation}
Additionally, one has
\begin{equation}
     \sum_{i,j}
    \qua^{RRRR}_{\alpha i\beta j}=|\Fua_{\alpha\beta}|^2
    \qquad \text{and}
    \qquad
    \sum_{\alpha,\beta}
    \qua^{RRRR}_{\alpha i\beta j}=|\Fda_{ij}|^2\,.
\end{equation}
In these equations, the sums go over all $3+\n$ quarks of a certain type (including the new states).
Replacing $R\to L$ in these relations and making use of the unitarity of $\VL$, we can relate sums of quartets to the bilinears of~\cref{eq:bilinears}, namely
\begin{equation}
\begin{aligned}
    \sum_i
    \qua^{L\chi\chi' L}_{\alpha i\beta j}
    &=
    \delta_{\alpha\beta}\, \Vr^{\chi}_{\beta j} \, \Vr^{\chi' *}_{\alpha j}
    = \delta_{\alpha\beta}\, \bi^{\chi\chi'}_{\alpha j}
    \,,
\\
    \sum_\alpha
    \qua^{L\chi L \chi' }_{\alpha i\beta j}
    &=
    \delta_{ji} \,\Vr^{\chi}_{\beta j} \, \Vr^{\chi'*}_{\beta i}
    = \delta_{ij}\, \bi^{\chi\chi'}_{\beta i}
    \,.
\end{aligned}
\end{equation}
We further obtain the constraints
\begin{equation}
    \sum_{i,j}
    \qua^{LLLL}_{\alpha i\beta j}=\delta_{\alpha\beta}
    \qquad \text{and} \qquad
    \sum_{\alpha,\beta}
    \qua^{LLLL}_{\alpha i\beta j}=\delta_{ij}\,,
\end{equation}
which follow directly from the above result and also hold in the absence of \acp{VLQ}.

In principle one can continue building other distinct rephasing invariants by appropriately assembling arbitrarily large products of the entries of $\VL$, $\VR$ and $\Fq$. However, here we choose to leave this discussion at the quartet level, as, at least in the case of one \ac{VLQ} doublet, after this point rephasing-invariant quantities are expected to be less relevant from a phenomenological standpoint.

\subsubsection{A single doublet (\texorpdfstring{$\n=1$}{\n=1})}

The case of a single doublet is special because the number of distinct invariants is considerably reduced when compared to cases with $\n>1$. This is connected to the fact that, in this specific case, we have 
\begin{equation}
        \VRa_{\alpha i}\, \VRa_{\beta j} \,=\,    \VRa_{\beta i} \, \VRa_{\alpha j} \quad \Rightarrow \quad
        \left\{\begin{array}{l}
        \VRa_{\beta i} \, \Fua_{\alpha \beta}
        \,=\, \VRa_{\beta i}\, \sum_j \VRa_{\alpha j}\VRac_{\beta j}
        \,=\, \VRa_{\alpha i}\,\sum_j |\VRa_{\beta j}|^2\,, \\[2mm]
        \VRa_{\alpha i} \, \Fda_{ij}
        \,=\, \VRa_{\alpha i}\, \sum_\beta \VRac_{\beta i}\VRa_{\beta j}
        \,=\, \VRa_{\alpha j}\,\sum_\beta |\VRa_{\beta i}|^2\,, 
        \end{array}\right.
\end{equation}
as a consequence of having $\VRa_{\alpha i} = (\BuR)^*_\alpha\, (\BdR)_i$, where, for $\n=1$, the $\BqR$ are vectors in flavour space. 
It then follows that 
 the only independent trilinears are those of the form
\begin{equation}
    \tri^{LL}_{i, \alpha \beta } =
    \left(\tri^{LL}_{i, \beta \alpha }\right)^*
    =
 \VLac_{\alpha i}\,  \VLa_{\beta i} \, \Fua_{\alpha\beta}
\qquad \text{and}
    \qquad
\tri^{LL}_{\alpha , i j }
=
\left(\tri^{LL}_{\alpha ,j i}\right)^*
=
 \VLa_{\alpha i}\, \VLac_{\alpha j} \, \Fda_{ij}
 \,,
\end{equation}
while the only generically-complex independent quartets are those of the type
\begin{equation}
\begin{aligned}
   \qua^{LLL\chi}_{\alpha i \beta j}
   =
   \qua^{LL\chi L}_{\beta j \alpha i}
   =
   \left(\qua^{L\chi LL}_{\alpha j \beta i}\right)^*
   =
   \left(\qua^{\chi LLL}_{\beta i \alpha j}\right)^*
   = \VLa_{\alpha i} \VLa_{\beta j} \VLac_{\alpha j} \Vr^{\chi *}_{\beta i}
   \,,
\end{aligned}
\end{equation}
with $\chi = L,R$.
Quartets like $\qua^{RRLL}_{\alpha i \beta j}$, for instance, can be written as the product of two bilinears
$\bi^{RL}_{\alpha j}\bi^{RL}_{\beta i}$ and do not contain new information on complex phases.

\vskip 2mm
Note that no complex trilinears depend on $\VR$ and that no complex quartets depend solely on $\VR$.
This can be traced to the fact that when $\n=1$
all phases in $\VR$ can be brought into $\VL$ via 
rephasings, while the reciprocal is not true. 
In fact, under the rephasing of the quark mass eigenstates 
$\U_\alpha\rightarrow \U_\alpha \, e^{i \varphi_\alpha}$ and $\D_i\rightarrow \D_i  \, e^{i \varphi_i}$
one has
\begin{equation}
     (\BuR)_\alpha \,\to\, (\BuR)_\alpha \,e^{i \varphi_\alpha}
     \qquad \text{and} \qquad
     (\BdR)_i \,\to\, (\BdR)_i \,e^{i \varphi_i}\,,
\end{equation}
where we again emphasize that the $\BqR$ are vectors in flavour space. Hence, with an appropriate choice of $\varphi_\alpha$ and $\varphi_i$,
all $\BqR$ can be made real, and consequently also $\VR = {\BuR}^\dagger \BdR$, $\Fu = {\BuR}^\dagger\BuR$ and $\Fd = {\BdR}^\dagger \BdR$ will be real.
One may thus eliminate all phases from the \ac{RH} currents in the $\n=1$ case, which then emerge in the \ac{LH} currents.

\vfill
\clearpage

\section{Rotating to the mass basis}
\label{sec:rotating}

As previously seen, the relevant parts of the Lagrangian in the mass basis read:
\begin{equation} \label{eq:lagr}
\begin{aligned}
\Lm \,&=\,
-\overline{\U}_L \,\Du\, \U_R \,-\, \overline{\D}_L \,\Dd\, \D_R 
+ \text{h.c.}\,,
\\
    \mathcal{L}_W \,&=\,
   - \frac{g}{\sqrt{2}} \,W_\mu^+ \left[
   \overline{\U_L}  \,\gamma^\mu\, \VL \, \D_L
   \,+\, \overline{\U_R}  \,\gamma^\mu\, \VR \,  \D_R
   \right]
+ \text{h.c.} \,,
\\
    \mathcal{L}_Z \,&=\,
    - \frac{g}{2 c_W} \,Z_\mu \left[
  \overline{\U_L} \, \gamma^\mu \, \U_L
- \overline{\D_L} \, \gamma^\mu \, \D_L
+\overline{\U_R} \, \gamma^\mu \, \Fu \, \U_R
-\overline{\D_R} \, \gamma^\mu \, \Fd \, \D_R
-2 s_W^2 J^\mu_\text{em}
\right]\,,
\\
\mathcal{L}_\h
\,&=\,
- \frac{\h}{\sqrt{2}}
\left[
\overline{\U_L}\,\,
\frac{\Du}{\vw} \left(\id - \Fu \right)
\,\U_R
+
\overline{\D_L}\,\,
\frac{\Dd}{\vw} \left(\id - \Fd \right)
\,\D_R
\right]
+ \text{h.c.}
\,,
\end{aligned}
\end{equation}
after \ac{EWSB}.
In what follows, we write the physical quark masses, the mixing matrices $\VL$ and $\VR$, and the \ac{FCNC} matrices $\Fu$ and $\Fd$ for $\n = 1$,
expanding in the small parameter $\vw/\mB$. This allows us to identify an alternative set of rephasing-invariant quantities.
In \cref{sec:param} we present exact parameterizations for $\VL$ and $\VR$ in these theories, for any $\n\geq 1$.

\subsection{Approximate masses and mixing (\texorpdfstring{$\n=1$}{\n=1})}
\label{sec:app1}

We consider for simplicity the case $\n = 1$.%
\footnote{This analysis can be extended to $\n > 1$. We make use of such a formalism for the cases with $\n = 2$ in~\cref{sec:CPodd,sec:pheno}.}
Let us start from the general Lagrangian of~\cref{eq:L1}, where mixed mass terms of the type $\mb \,\overline{q_L}\,  Q_R$ have been rotated away, so that the mass matrices of up-type and down-type quarks read
\begin{equation} 
\Mu \,=\, 
\begin{pmatrix}
    \vw\, \yu  &     0 \\[2mm]     
    \vw\, \yU  &    \mB
\end{pmatrix}\,,
\qquad
\Md \,=\,
\begin{pmatrix}
    \vw\,\yd  &     0 \\[2mm]     
    \vw\,\yD  &    \mB
\end{pmatrix}
\,.
\end{equation} 
These mass matrices can be diagonalized via bi-unitary transformations, as in~\cref{eq:diag}:
\begin{equation}
\begin{aligned}
{\VuL}^\dagger \Mu \VuR 
\, &= \, 
\Du
\, = \,  
\diag(y_u \vw,\, y_c \vw,\, y_t \vw,\, \MT)\,, \\
{\VdL}^\dagger \Md \VdR 
\, &= \, 
\Dd
\, = \,  
\diag(y_d \vw,\, y_s \vw,\, y_b \vw,\, \MB) \,,
\end{aligned}
\end{equation}
and the rotation matrices $\V$ can be obtained from the Hermitian products $\Mq\Mq^\dagger$ and $\Mq^\dagger\Mq$, see~\cref{eq:svd2}.
The initial states are related to the mass eigenstates via
\begin{equation} \label{eq:mass_rot} 
    \begin{pmatrix} u_1 \\ u_2 \\ u_3 \\ T \end{pmatrix}_{\!\!\!L,R}
    \!\!\!\!=\, \Vu_{L,R}\,
   \begin{pmatrix} u \\c \\ t \\ T' \end{pmatrix}_{\!\!\!L,R}\!\!\!\!,
\qquad\quad
    \begin{pmatrix} d_1 \\ d_2 \\ d_3 \\ B \end{pmatrix}_{\!\!\!L,R}
    \!\!\!\!=\, \Vd_{L,R}\,
   \begin{pmatrix} d \\ s \\ b \\ B' \end{pmatrix}_{\!\!\!L,R}\!\!\!\!,
\end{equation}
where we have expanded and simplified the notation, cf.~\cref{eq:rot}.
Finally, it is useful to keep in mind the singular value decompositions of $\yu$ and $\yd$
 introduced in~\cref{eq:Uhat,eq:Yhat}. Namely, recall that
\begin{equation} \label{eq:Yhat2}
    \VUW^\dagger \,\yu \,\WUW = \Ydu = \diag(\hat{y}_u,\hat{y}_c,\hat{y}_t)\,, \qquad
    \VDW^\dagger \,\yd \,\WDW = \Ydd = \diag(\hat{y}_d,\hat{y}_s,\hat{y}_b)\,,
\end{equation}
where the diagonal elements are not yet the physical masses in units of $\vw$ (only approximately so, see later~\cref{eq:mstd}). We also define
\begin{equation} \label{eq:z}
\yup \equiv \yU\, \WUW=(\hiu,\hic,\hit)\, , \qquad \ydb\equiv\yD\, \WDW =(\hid,\his,\hib)\,.
\end{equation}
Then, one can directly decompose the mass matrices as
\begin{equation}
    \Mq \,=\, \begin{pmatrix}
    \VQW &     0 \\[2mm]     
    0  &    1
\end{pmatrix}
\begin{pmatrix}
    \vw\, \Ydq  &     0 \\[2mm]     
    \vw\, \yqp  &    \mB
\end{pmatrix}
\begin{pmatrix}
    \WQW^\dagger  &     0 \\[2mm]     
    0  &    1
\end{pmatrix}
\qquad (q=u,d)
\,,
\end{equation}
and carry out their diagonalization. For both chiralities $\chi=L,R$, one finds that the rotation matrices $\V$ can be schematically parameterized as
\begin{equation}
\begin{aligned}
\Vq_{\chi} &= \begin{pmatrix}
\UW_{q\chi} & \\
 & 1 
\end{pmatrix} 
\begin{pmatrix}
\dU_{q\chi} & \\
 & 1 
\end{pmatrix}
\bigg(
\quad\Rf^{q\chi}\quad
\bigg)
\,,
\end{aligned}
\end{equation}
in each sector, where each $\Rf$ matrix is a product of $(1,4)$, $(2,4)$ and $(3,4)$ unitary rotations.

The present experimental bounds on \ac{VLQ} masses are 
$\MT, \MB \gtrsim 1.15$~TeV~\cite{ATLAS:2024zlo} for vector-like doublets coupling to light generations ($u$, $d$ and $s$ quarks) and
$\MT, \MB \gtrsim 1.3$--$1.5$~TeV~\cite{ATLAS:2022hnn,CMS:2022fck,CMS:2020ttz} for extra doublets coupling to the third generation
(however, the limits also depend on the couplings and branching ratios).
For perturbative couplings, the bare mass $\mB$ (here a number) is expected to source the \ac{VLQ} scale at the TeV or beyond, and thus the ratio $\vw/\mB$ is a small parameter.
Taking this into account, masses and mixings in both the \ac{RH} and \ac{LH} sectors can be directly related to Lagrangian parameters through power series in $\vw/\mB$ (see also Ref.~\cite{Belfatto:2021jhf}).
Note that, beyond agreeing with the \ac{EFT} expansion, which we revisit in~\cref{sec:EFT}, when computing tree-level amplitudes at relatively low energies, this ${\vw}/{\mB}$ expansion can be used at energies above the \ac{VLQ} threshold, unlike the \ac{EFT} one.

Before discussing the rotations bidiagonalizing the $\Mq$, we first obtain expressions for the eigenvalues of $\Mq\Mq^\dagger$, i.e.~for the squares of physical quark masses. One finds
\begin{equation} \label{eq:Msplit}
\begin{aligned}
    \MT^2&\,=\, \mB^2 \left[ 1+|\yup |^2 \tfrac{\vw^2}{\mB^2} \, + \mathcal{O}\left(\tfrac{ \vw^4}{\mB^4}\right)
    \right]
    \,,
  \\[1mm]
      \MB^2&\,=\, \mB^2 \left[ 1+|\ydb |^2 \tfrac{\vw^2}{\mB^2} \, + \mathcal{O}\left(\tfrac{ \vw^4}{\mB^4}\right)
    \right]
    \,,
\end{aligned}  
\end{equation}
for the masses of the new states,
implying $\MT \approx \MB$,
and
\begin{equation} \label{eq:mstd}
\begin{aligned}
    m_\alpha^2 =\, \vw^2 y_\alpha^2  =\,  \vw^2 \hat{y}_\alpha^2 \left[1  - |\hi_\alpha|^2\tfrac{\vw^2}{\mB^2} + \mathcal{O}\left(\tfrac{\vw^4}{\mB^4}\right) \right]&
    \qquad
    (\alpha = u,c,t)\,,
    \\[1mm]
    m_i^2 \,=\, \vw^2 y_i^2  \,=\,  \vw^2 \hat{y}_i^2 \left[1  - |\hi_i|^2\tfrac{\vw^2}{\mB^2} + \mathcal{O}\left(\tfrac{\vw^4}{\mB^4}\right) \right]&
    \qquad
    (i = d,s,b)\,,
\end{aligned}
\end{equation}
for the masses of standard quarks, where up- and down-type quarks are labelled by Greek and Latin indices respectively. Note that, to first order in $\vw/\mB$, the physical Yukawa couplings defined above as $y_{\alpha,i} = m_{\alpha,i} / \vw$ coincide with the elements $\hat{y}_{\alpha,i}$ of the diagonal matrices $\Ydq$ and with the couplings of standard quarks to the Higgs boson.

\subsubsection{The \acs{RH} sector}

An explicit parameterization for the down-sector \ac{RH} rotation reads
\begin{equation} \label{eq:angoli}
\begin{aligned}
\VdR
 \,\equiv\,&
\left(\begin{array}{cccc}
\V^R_{1d} & \V^R_{1s} & \V^R_{1b} & \V^R_{1B'} \\[1mm]
\V^R_{2d} & \V^R_{2s} & \V^R_{2b} & \V^R_{2B'} \\[1mm]
\V^R_{3d} & \V^R_{3s} & \V^R_{3b} & \V^R_{3B'} \\[1mm]
\V^R_{Bd} & \V^R_{Bs} & \V^R_{Bb} & \V^R_{BB'} 
\end{array}\right) 
\,=\,
\left(\begin{array}{c@{\hspace{0.5\tabcolsep}}c@{\hspace{0.5\tabcolsep}}c@{\hspace{1\tabcolsep}}c}
&  &  &  \\
 & \UW_{dR} &  &  \\
 & &  &  \\
 &  &  & 1
\end{array}\right)\!
\left(\begin{array}{c@{\hspace{0.5\tabcolsep}}c@{\hspace{0.5\tabcolsep}}c@{\hspace{1\tabcolsep}}c}
&  &  &  \\
 & \dU_{dR} &  &  \\
 & &  &  \\
 &  &  & 1
\end{array}\right)
\times
\\[2mm]
&\,\,\, \times
\underbrace{
\begin{pmatrix}
c^d_{R1} &  &  & -\tilde{s}^{d*}_{R1} \\
    & 1 &  &  \\
    &  & 1 &  \\
\tilde{s}^{d}_{R1} &  &  & c^d_{R1}
\end{pmatrix}
\begin{pmatrix}
1 &      &  & \\
  & c^d_{R2} &  & -\tilde{s}^{d*}_{R2} \\
  &     & 1 &  \\
 & \tilde{s}^{d}_{R2}  &  & c^d_{R2}
\end{pmatrix}
\begin{pmatrix}
1  &  &     &  \\
  & 1 &     &  \\
  &  & c^d_{R3} & -\tilde{s}^{d*}_{R3} \\
  &  & \tilde{s}^{d}_{R3} & c^d_{R3}
\end{pmatrix}
}_{\Rf^{dR}}
\,,
\end{aligned}
\end{equation}
resulting in
\begin{equation} \label{eq:angles}
\Rf^{dR} = 
\begin{pmatrix}
c^d_{R1} & -\tilde{s}^{d}_{R2} \tilde{s}^{d*}_{R1} & -c^d_{R2}\tilde{s}^{d}_{R3}\tilde{s}^{d*}_{R1} & -\tilde{s}^{d*}_{R1}c^d_{R2}c^d_{R3} \\[1mm]
 0   & c^d_{R2} & -\tilde{s}^{d}_{R3}\tilde{s}^{d*}_{R2} & -\tilde{s}^{d*}_{R2}c^d_{R3} \\[1mm]
 0   & 0 & c^d_{R3} & -\tilde{s}^{d*}_{R3} \\[1mm]
\tilde{s}^{d}_{R1} & \tilde{s}^{d}_{R2}c^d_{R1} & \tilde{s}^{d}_{R3}c^d_{R1}c^d_{R2} & c^d_{R1}c^d_{R2}c^d_{R3}
\end{pmatrix}\,,
\end{equation}
and a similar one is valid for the up-type quarks.
Here,
\begin{equation} \label{eq:tildes}
c^q_{Ri} \,\equiv\,  \cos\theta^{q}_{R\,i4}\,,
\qquad
\tilde{s}^q_{Ri} \,=\,  s^q_{Ri} \, e^{i \delta^q_{Ri}} \,\equiv\, \sin\theta^q_{R\,i4}\, e^{i \delta^q_{Ri}}
\end{equation}
are real cosines and complex sines of angles in the $(1,4)$, $(2,4)$, and $(3,4)$ family 
planes parameterizing the mixing of the first three families
with the \acp{VLQ}. We assume the analogous parameterization with real cosines and complex sines also for the other unitary matrices $\dU_{qR}$, $\UW_{qR}$.

As it is clear from~\cref{eq:VLVR1,eq:vnc1}, the \ac{RH} charged currents and flavour-changing couplings 
exclusively depend on the last rows
of the matrices $\VuR$, $\VdR$,
that is, on the mixings of \ac{SM} quarks with the vector-like doublet.
For couplings $\lesssim \mathcal{O}(1)$, these mixings
are determined by the small parameter $\vw/\mB$.
In fact, up to terms of order $\vw^2/\mB^2$, one can write $\VdR$ as
\begin{equation} \label{eq:UdR}
\VdR
\,=\,  
\begin{pmatrix}
 \UW_{dR}  &  \\
 & 1
\end{pmatrix}
\begin{pmatrix}
\id_{3\times 3} &   \ydb^\dagger\, \frac{\vw}{\mB}  \\[2mm]
  - \ydb\, \frac{\vw}{\mB} & 1
\end{pmatrix}
 + \mathcal{O}\left(\tfrac{\vw^2}{\mB^2}\right)
\,,
\end{equation}
where we have used the fact that $\dU_{dR} \approx \id$ at this order.
A similar expression holds for the rotation $\VuR$ of up-type quarks.
Thus, the mixing elements of interest, corresponding to the last row of $\VdR$, are given by
\begin{equation} \label{eq:4row}
\V^R_{Bd} = \tilde{s}_{R1}^{d} \approx -\hid \frac{\vw}{\mB} \,, \quad  
\V^R_{Bs} \approx \tilde{s}_{R2}^{d} \approx -\his \frac{\vw}{\mB} \,, \quad
\V^R_{Bb} \approx \tilde{s}_{R3}^{d} \approx -\hib \frac{\vw}{\mB} \,,
\end{equation}
in first approximation.
In particular, notice that the phases of these mixing elements exactly coincide with the phases of the complex sines. As is clear from~\cref{eq:4row}, these coincide with those of the extra Yukawa couplings $\ydb$,
\begin{equation} \label{eq:phases}
    \arg (\V^R_{Bi}) =
    \arg \tilde{s}_{Ri}^{d} \,(\equiv \delta^d_{Ri}) =
    \arg \hi_i
\,,
\end{equation}
at all orders (this result is also independent on the parameterization choice of the order of the product in $\Rf^{dR}$). In other words, the higher-order corrections to~\cref{eq:4row} share the same phase as the leading term.
The matrix $\dU_{dR} \approx \id$ contains tiny corrections given by small complex sines  
\begin{equation}
\tilde{s}_{i < j} \,\approx \, \hi_i \hi_j^*\, \frac{\hat{y}_i^2}{\hat{y}_j^2-\hat{y}_i^2}  \frac{\vw^2}{\mB^2}
\,\approx \, \hi_i \hi_j^*\, \frac{y_i^2}{y_j^2}  \frac{\vw^2}{\mB^2}
\qquad
(i,j = 1,2,3)
\,,
\end{equation}
at leading order, where we have taken into account~\cref{eq:mstd} and the hierarchy between \ac{SM} quark masses.%
\footnote{With a slight abuse of notation, it should be clear that $\hi_i = (\ydb)_i$ when $i=1,2,3$.}
A completely analogous discussion applies to up-type quarks.

Focusing on the last rows of $\VuR$ and $\VdR$, one can find explicit expressions for the mixings up to order $\vw^3/\mB^3$. In the up sector, for instance, we have
\begin{equation} \label{eq:4rowu}
\begin{aligned}
\V^R_{T\alpha} &\,\approx\, -\hi_\alpha \frac{\vw}{\mB}
+ \hi_\alpha\left(
\frac{1}{2} |\hi_\alpha|^2
- y_\alpha^2 + \sum_{\beta < \alpha} |\hi_\beta|^2
\right)
\frac{\vw^3}{\mB^3}
+
\mathcal{O}\left(\tfrac{\vw^5}{\mB^5} \right)
\,,  \\
\V^R_{TT'} &\,=\, 1-\frac12 |\yup|^2 \frac{\vw^2}{\mB^2}+\mathcal{O}\left(\tfrac{\vw^4}{\mB^4} \right)
\,,  
\end{aligned}
\end{equation}
with $\alpha=u,c,t$ and where we have once again taken into account the hierarchies between standard quark masses. Similar expressions hold for down-type quarks.
Given the above, one can construct an approximation for the non-unitary
mixing matrix in the \ac{RH} charged current. 
At leading order, it reads
\begin{equation} \label{eq:vckmR}
\begin{aligned}
\VR&= {\VuR}^\dagger \diag(0,0,0,1)\, \VdR
 =
\begin{pmatrix}
\VRa_{ud} & \VRa_{us} & \VRa_{ub} & \VRa_{uB'}  \\[1mm]
\VRa_{cd} & \VRa_{cs} & \VRa_{cb} & \VRa_{cB'}  \\[1mm]
\VRa_{td} & \VRa_{ts} & \VRa_{tb} & \VRa_{tB'}  \\[1mm]
\VRa_{T'd} & \VRa_{T's} & \VRa_{T'b} & \VRa_{T'B'} 
\end{pmatrix}
\\[2mm]
&= \left(\begin{array}{cccc}
\hiu^* \hid \, \frac{\vw^2}{\mB^2}  & \hiu^* \his\, \frac{\vw^2}{\mB^2}   &  \hiu^* \hib  \, \frac{\vw^2}{\mB^2} 
& -\hiu^* \frac{\vw}{\mB}   \\[2mm]
\hic^* \hid\, \frac{\vw^2}{\mB^2}  & \hic^* \his \, \frac{\vw^2}{\mB^2}   & \hic^* \hib \, \frac{\vw^2}{\mB^2} 
& - \hic^*  \frac{\vw}{\mB}  \\[2mm]
\hit^* \hid\, \frac{\vw^2}{\mB^2}  & \hit^* \his \, \frac{\vw^2}{\mB^2} &  \hit^* \hib\, \frac{\vw^2}{\mB^2} 
 & - \hit^*   \frac{\vw}{\mB}      \\[2mm]
-\hid   \frac{\vw}{\mB} & -\his    \frac{\vw}{\mB}  &  -\hib     \frac{\vw}{\mB} &  1
- \frac{1}{2}\left(|\yup|^2 + |\ydb|^2\right) \frac{\vw^2}{\mB^2} 
\end{array}\right) + \mathcal{O}\left(\tfrac{\vw^3}{\mB^3}\right)\,.
\end{aligned}
\end{equation}
Similarly, one can construct the matrices of \ac{RH} \ac{FCNC} couplings.
In the up sector,
$\Fu= {\VuR}^\dagger \diag(0,0,0,1)\, \VuR$,
explicitly reads
\begin{equation} \label{eq:Fzz}
\Fu
=
\left(\begin{array}{cccc}
|\hiu|^2 \, \frac{\vw^2}{\mB^2}  & \hiu^* \hic\, \frac{\vw^2}{\mB^2}   &  \hiu^* \hit  \, \frac{\vw^2}{\mB^2} 
& -\hiu^* \frac{\vw}{\mB}   \\[2mm]
\hic^* \hiu\, \frac{\vw^2}{\mB^2}  & |\hic|^2 \, \frac{\vw^2}{\mB^2}   & \hic^* \hit \, \frac{\vw^2}{\mB^2} 
& - \hic^*  \frac{\vw}{\mB}  \\[2mm]
\hit^* \hiu\, \frac{\vw^2}{\mB^2}  & \hit^* \hic \, \frac{\vw^2}{\mB^2} &  |\hit|^2 \, \frac{\vw^2}{\mB^2} 
 & - \hit^*   \frac{\vw}{\mB}      \\[2mm]
-\hiu   \frac{\vw}{\mB} & -\hic   \frac{\vw}{\mB}  &  -\hit     \frac{\vw}{\mB} &  1
- |\yup|^2 \frac{\vw^2}{\mB^2} 
\end{array}\right) 
+ \mathcal{O}\left(\tfrac{\vw^3}{\mB^3}\right)
\,,
\end{equation}
while the down sector
$\Fd= {\VdR}^\dagger \diag(0,0,0,1)\, \VdR$
has an analogous structure.

\subsubsection{The \acs{LH} sector}
\label{sec:LH}

Regarding the \ac{LH} sector,  recall that,
since the four quark species make up $SU(2)_L$ doublets, the mixing matrix 
$\VL = {\VuL}^\dagger \VdL$
is a unitary matrix.
The extra elements in the $\VqL$ describe the mixing of \ac{SM} quarks with the \ac{VLQ} doublets.
These mixings can be parameterized analogously to what was done in~\cref{eq:angles}, with the substitution $R\rightarrow L$. 
The corresponding mixing angles are suppressed by the ratio $\vw^2/\mB^2$, and are associated to complex sines which at leading order read $\tilde{s}_{Li} \approx -y_i \hi_i\, \vw^2/\mB^2$ in the down sector ($i\to \alpha$ in the up sector).

Focusing on the last rows of $\VuL$ and $\VdL$, one can once again find explicit expressions for the mixing matrix elements. In the up sector, for instance, we have
\begin{equation} \label{eq:4rowL}
\begin{aligned}
\V^L_{T\alpha} &\,\approx\, -\hi_\alpha {y}_\alpha \frac{\vw^2}{\mB^2}
-\hi_\alpha y_\alpha\left(
y_\alpha^2 - \frac{1}{2} |\hi_\alpha|^2 - \sum_{\beta < \alpha} |\hi_\beta|^2
\right)
\frac{\vw^4}{\mB^4}
+
\mathcal{O}\left(\tfrac{\vw^6}{\mB^6} \right)
\,,  \\
\V^L_{TT'} &\,=\, 1-\frac{1}{2}\sum_{\beta} |\hi_\beta|^2 y_\beta^2\, \frac{\vw^4}{\mB^4}+\mathcal{O}\left(\tfrac{\vw^6}{\mB^6} \right)
\,,  
\end{aligned}
\end{equation}
with $\alpha=u,c,t$ and where we took into account the hierarchy between \ac{SM} quark masses in writing terms of order $\vw^4/\mB^4$ in the first line. At this order, the last row and last columns of the matrix $\Rf^{uL}$ in the parameterization of $\VuL$ are related by $(\Rf^{uL})_{4i} \approx -(\Rf^{uL})_{i4}^*$ to a good approximation, as one can anticipate from~\cref{eq:angles}. 
Using the fact that
$\dU_{uL}$ deviates from $\id$
by small rotations with complex sines given by $\tilde{s}_{i < j} \approx \hi_i \hi_j^*\, \frac{y_i}{y_j}  \frac{\vw^2}{\mB^2}$ ($i,j = 1,2,3$)
at leading order, one finds
\begin{equation}
\V^L_{kT'} \,\approx\, 
\sum_{\alpha=u,c,t}
(\UW_{uL})_{k\alpha} \left[
\hi_\alpha^* {y}_\alpha \frac{\vw^2}{\mB^2}
+\hi_\alpha^* y_\alpha\left(
y_\alpha^2 + \frac{1}{2} |\hi_\alpha|^2 - |\yup|^2
\right)
\frac{\vw^4}{\mB^4}
\right]
+
\mathcal{O}\left(\tfrac{\vw^6}{\mB^6} \right)
\,,
\end{equation}
where $k=1,2,3$. Similar expressions hold for down-type quarks. 
Keeping only terms up to order $\vw^2/\mB^2$, the mixing matrix in the \ac{LH} charged currents 
reads
\begin{equation} \label{eq:VL}
\begin{aligned}
\VL &=
\begin{psmallmatrix}
c^u_{L1} & 0 &0 & \tilde{s}^{u*}_{L1} \\[1mm]
-\tilde{s}^{u*}_{L2} \tilde{s}^{u}_{L1} & c^u_{L2} &0 &\tilde{s}^{u*}_{L2}c^u_{L1} \\[1mm]
-c^u_{L2}\tilde{s}^{u*}_{L3}\tilde{s}^{u}_{L1} &  -\tilde{s}^{u*}_{L3}\tilde{s}^{u}_{L2} & c^u_{L3} &\tilde{s}^{u*}_{L3}c^u_{L1}c^u_{L2} \\[1mm]
-\tilde{s}^{u}_{L1}c^u_{L2}c^u_{L3} & -\tilde{s}^{u}_{L2}c^u_{L3} &  -\tilde{s}^{u}_{L3} & c^u_{L1}c^u_{L2}c^u_{L3}
\end{psmallmatrix}
\begin{pmatrix}
 \VLh  &  \\
 & 1
\end{pmatrix}
\begin{psmallmatrix}
c^d_{L1} & -\tilde{s}^{d}_{L2} \tilde{s}^{d*}_{L1} & -c^d_{L2}\tilde{s}^{d}_{L3}\tilde{s}^{d*}_{L1} & -\tilde{s}^{d*}_{L1}c^d_{L2}c^d_{L3} \\[1mm]
 0   & c^d_{L2} & -\tilde{s}^{d}_{L3}\tilde{s}^{d*}_{L2} & -\tilde{s}^{d*}_{L2}c^d_{L3} \\[1mm]
 0   & 0 & c^d_{L3} & -\tilde{s}^{d*}_{L3} \\[1mm]
\tilde{s}^{d}_{L1} & \tilde{s}^{d}_{L2}c^d_{L1} & \tilde{s}^{d}_{L3}c^d_{L1}c^d_{L2} & c^d_{L1}c^d_{L2}c^d_{L3}
\end{psmallmatrix}
\\[1mm]
&\approx
\begin{pmatrix}
\id_{3\times 3} &   -\Ydu\yup^\dagger\, \frac{\vw^2}{\mB^2}  \\[2mm]
  \yup\Ydu\, \frac{\vw^2}{\mB^2} & 1
\end{pmatrix}
\begin{pmatrix}
 \VLh  &  \\
 & 1
\end{pmatrix}
\begin{pmatrix}
\id_{3\times 3} &   \Ydd\ydb^\dagger\, \frac{\vw^2}{\mB^2}  \\[2mm]
  - \ydb \Ydd\, \frac{\vw^2}{\mB^2} & 1
\end{pmatrix}
 + \mathcal{O}\left(\tfrac{\vw^4}{\mB^4}\right)\,,
\end{aligned}
\end{equation}
where $\VLh \equiv \dU_{uL}^\dagger\UW_{uL}^\dagger\UW_{dL}\dU_{dL}$. Here, $c^{q}_{Li}$ and $\tilde{s}^{q}_{Li}$ are the cosines and complex sines which make up the $\Rf^{qL}$ matrices, defined as in the \ac{RH} sector.
For the last row and column of $\VL$, one has
\begin{equation} \label{eq:LH4mix}
\begin{aligned}
     \VLa_{T'i}\,&=\,\left( {\textstyle\sum_{\alpha}}\,\hi_\alpha y_\alpha\, \Vha_{\alpha i}  -\hi_i y_i \right)  \frac{\vw^2}{\mB^2}+\mathcal{O}\left(\tfrac{\vw^4}{\mB^4} \right)\,,  \\[1mm]
     \VLa_{\alpha B'} \,&=\, \left( {\textstyle\sum_{i}}\,   \Vha_{\alpha i} \,\hi_i^* y_i -\hi_\alpha^* y_\alpha \right) \frac{\vw^2}{\mB^2}+\mathcal{O}\left(\tfrac{\vw^4}{\mB^4} \right)\,,
\end{aligned}    
\end{equation}
where $i = d,s,b$ and $\alpha = u,c,t$. At this order, $\VLa_{T'B'} \approx 1$.

The matrix $\VLh$ introduced in~\cref{eq:VL} can be approximately obtained from the rotations defined in~\cref{eq:Yhat2}, namely
$\VLh \approx \UW_{uL}^\dagger\UW_{dL}$,
up to small extra rotations:
\begin{equation} \label{eq:tiny-unitary-corr}
\Vha_{\alpha i} \,\approx\,  
  \left[\UW_{uL}^\dagger\UW_{dL}\right]_{\alpha i}
\! +
  \left(
   \sum_{j\neq i} \hi_i\hi_j^* \frac{y_i\, y_j}{y_j^2-y_i^2}
   \left[\UW_{uL}^\dagger\UW_{dL}\right]_{\alpha j}
   + 
   \sum_{\beta\neq \alpha} \hi_\alpha^*\hi_\beta \frac{y_\alpha\, y_\beta}{y_\beta^2-y_\alpha^2}
   \left[\UW_{uL}^\dagger\UW_{dL}\right]_{\beta i}
   \right) \!
   \frac{\vw^2}{\mB^2} \,,
\end{equation}
with $\alpha, \beta =u,c,t$ and $i,j =d,s,b$.
Indeed, aside from the $\vw^2/\mB^2$ suppression, one can see that these corrections are always additionally suppressed by ratios of quark masses of different generations and \ac{FCNC} constraints (note that, for instance, $\Fda_{ij} \approx \hi_i^* \hi_j\, \vw^2/\mB^2$).
Recall that $\UW_{uL}^\dagger\UW_{dL}$ is actually the $3\times 3$ \ac{CKM} matrix in the limit of a decoupled \ac{VLQ} doublet.
Moreover, $\VLh$ is a unitary $3\times 3$ matrix by definition --- the corrections in~\cref{eq:tiny-unitary-corr} are induced by the unitary transformations $\dU_{uL}$ and $\dU_{dL}$ between the three \ac{SM} quarks.
One can also see that, to a very good approximation, $\VLh$
corresponds to the $3\times 3$ upper-left submatrix of $\VL$, i.e.
\begin{equation} \label{eq:deltaVL}
\VLa_{\alpha i}  
\,=\, 
 \Vha_{\alpha i}  \,+\,   \mathcal{O}\left(y^2\hi^2\tfrac{\vw^4}{\mB^4}\right)
\,,     
\end{equation}
for $\alpha=u,c,t$ and $i=d,s,b$.  
In summary, we have schematically
\begin{equation} \label{eq:vckmL}
\VL=
\begin{pmatrix}
\VLa_{ud} & \VLa_{us} & \VLa_{ub} & \VLa_{uB'}  \\[1mm]
\VLa_{cd} & \VLa_{cs} & \VLa_{cb} & \VLa_{cB'}  \\[1mm]
\VLa_{td} & \VLa_{ts} & \VLa_{tb} & \VLa_{tB'}  \\[1mm]
\VLa_{T'd} & \VLa_{T's} & \VLa_{T'b} & \VLa_{T'B'} 
\end{pmatrix}
\approx \left(\begin{array}{c:c}
&  \\[2mm]
  \UW_{uL}^\dagger\UW_{dL} \approx \CKM 
&  \mathcal{O}\left(yz\tfrac{\vw^2}{\mB^2}\right) \!\!\!  \\[2mm]
 &  \\[2mm] \hdashline[4pt/2pt]
 & { }\\[-2mm]
\mathcal{O}\left(yz\tfrac{\vw^2}{\mB^2}\right)
&  1
\end{array}\right) + \mathcal{O}\left(\tfrac{\vw^4}{\mB^4}, \tfrac{m}{m'}\tfrac{\vw^2}{\mB^2}\right)\,,
\end{equation}
at leading order, where $m/m'$ indicates mass ratios of different-generation quarks. The explicit form of the $\mathcal{O}(y \hi \vw^2/\mB^2)$ off-diagonal blocks can be found in~\cref{eq:LH4mix}.

\subsubsection{Effective rephasing invariants}
\label{sec:reph-inv-eff}

Let us note that the \ac{SM} quark fields can be rephased in order to absorb 5 phases in $\VLh$, e.g.~one row and one column of $\VLh$ can always be made real. Hence, $\VLh$ can be parameterized by three angles and one phase, as in the usual \ac{CKM} parameterization.
When rephasing the \ac{LH} quark fields,
the corresponding \ac{RH} fields must go through the same phase transformation in order to keep the diagonal Yukawa terms real.
The effect of these transformations
(e.g.~$u_{L,R}\rightarrow u_{L,R}\, e^{i\delta_u}$)
can be absorbed in the $\Rf$ matrices, and corresponds to changing the phases of the complex sines, i.e.~of the Yukawa couplings $\hi_\alpha$ and $\hi_i$ (for instance as $s_{R,L1}^{u} \to s_{R,L1}^{u}\, e^{i\delta_u}$ and $\hi_u \to \hi_u\, e^{i\delta_u}$),
where we have used the fact that relations analogous to~\cref{eq:phases} hold for the \ac{LH} sector.
One can then readily identify the following rephasing-invariant quantities:
\begin{equation} \label{eq:reph-hat} 
    \hi_\alpha \hi_i^* \Vha_{\alpha i}\,, \quad
    \hi_\alpha \hi_\beta^*  \Vha_{\alpha i} \Vhac_{\beta i} \,, \quad
    \hi_i^* \hi_j \Vha_{\alpha i} \Vhac_{\alpha j} \,, \quad
    \hi_\alpha \hi_j^* \Vhac_{\beta i} \Vha_{\alpha i} \Vha_{\beta j} \,,
\end{equation}
with $\alpha=u,c,t$ and $i=d,s,b$.
These quantities are related to the bilinears, trilinears and quartets described in~\cref{sec:reph-inv}. In fact, let us define%
\footnote{\label{foot:hatdefs}
The quantities $\VRha_{\alpha i} \equiv \frac{\vw^2}{\mB^2}\hi_\alpha^* \hi_i$,
$\Fuha_{\alpha\beta} \equiv \frac{\vw^2}{\mB^2}\hi_\alpha^* \hi_\beta$, and
$\Fdha_{ij} \equiv \frac{\vw^2}{\mB^2}\hi_i^* \hi_j$, defined here for $\n=1$, are useful shorthands used in the following sections (and generalized in~\cref{sec:EFT,sec:pheno} for $\n> 1$). To leading order in $\vw^2/\mB^2$, they approximate the $3\times 3$ upper-left submatrices of $\VR$, $\Fu$, and $\Fd$, respectively.
}
\begin{equation} \label{eq:reph-inv-eff}
\begin{aligned}
    \hbi_{\alpha i} &\,\equiv\, \frac{\vw^2}{\mB^2}\hi_\alpha^* \hi_i \Vhac_{\alpha i}
    \,\equiv\, \VRha_{\alpha i} \Vhac_{\alpha i}
    \,, \\
    \htr_{i,\alpha\beta} &\,\equiv\, \frac{\vw^2}{\mB^2}\hi_\alpha^* \hi_\beta  \Vhac_{\alpha i} \Vha_{\beta i}
    \,\equiv\, \Fuha_{\alpha\beta}   \Vhac_{\alpha i} \Vha_{\beta i}
    \,, \\
    \htr_{\alpha,ij} &\,\equiv\, \frac{\vw^2}{\mB^2}\hi_i^* \hi_j \Vha_{\alpha i} \Vhac_{\alpha j}
    \,\equiv\, \Fdha_{ij} \Vha_{\alpha i} \Vhac_{\alpha j}
    \,, \\
    \hqu_{\alpha i\beta j}&\,\equiv\, \frac{\vw^2}{\mB^2}\hi_\alpha^* \hi_i \Vha_{\beta j} \Vhac_{\alpha j} \Vhac_{\beta i} 
     \,\equiv\, \VRha_{\alpha i}\Vha_{\beta j} \Vhac_{\alpha j} \Vhac_{\beta i} 
    \,.
\end{aligned}
\end{equation}
While these do not exactly coincide with the previously-defined unhatted versions of~\cref{sec:reph-inv} --- consider~\cref{eq:4rowu,eq:vckmR,eq:deltaVL} --- they differ from them by rephasing-invariant quantities. 
The indices are now limited to those of light quarks,
so that these are essentially \emph{effective rephasing invariants}.
Moreover, they are approximately equal at leading order, i.e.
\begin{equation} \label{eq:hatvsunh}
\begin{aligned}
    \hbi_{\alpha i} &\,\approx\, \VRa_{\alpha i} \VLac_{\alpha i} =  \bi^{RL}_{\alpha i} \,, \\
    \htr_{i,\alpha\beta} &\,\approx\,  \Fua_{\alpha\beta}\, \VLac_{\alpha i} \VLa_{\beta i}
    = \tri^{LL}_{\alpha \beta, i}
    \,, \\
    \htr_{\alpha,ij} &\,\approx\, \Fda_{ij}\, \VLa_{\alpha i} \VLac_{\alpha j}
    = \tri^{LL}_{i j, \alpha}
    \,, \\
    \hqu_{\alpha i\beta j} &\,\approx\, \VRa_{\alpha i}  \VLa_{\beta j} \VLac_{\alpha j} \VLac_{\beta i}
    = \qua^{RLLL}_{\alpha i \beta j}
    \,, 
\end{aligned}
\end{equation}
meaning that for the purposes of phenomenology (see~\cref{sec:pheno}) one may focus on the hatted rephasing invariants.
We also define the non-calligraphic hatted invariants
\begin{equation} \label{eq:Qhat}
    \Qh_{\alpha i \beta j} \,\equiv\, \Vha_{\alpha i} \Vha_{\beta j} \Vhac_{\alpha j} \Vhac_{\beta i}
    \,\approx\, \VLa_{\alpha i} \VLa_{\beta j} \VLac_{\alpha j} \VLac_{\beta i}
    \,=\, \qua^{LLLL}_{\alpha i \beta j}\,,
\end{equation}
which are the practical generalization of the \ac{SM} quartets ($\Vha_{\alpha i} = \VLa_{\alpha i}$ in the \ac{SM} limit).

\vskip 2mm
Let us also notice that in presence of only one doublet, 
if $\hi_\alpha\neq 0$ and $\hi_i\neq 0$ the trilinears and quartets can be written as
\begin{equation} \label{eq:f-bilinears}
\begin{aligned}
  \htr_{i,\alpha\beta} &= \frac{\hbi_{\alpha i} \hbi_{\beta i}^*}{\frac{\vw^2}{\mB^2}|\hi_i|^2}
    \approx \frac{\hbi_{\alpha i} \hbi_{\beta i}^*}{\Fda_{ii}} \,,
    \\
   \htr_{\alpha, ij}  &= \frac{\hbi_{\alpha i}^* \hbi_{\alpha j}}{\frac{\vw^2}{\mB^2}|\hi_\alpha|^2}
   \approx  \frac{\hbi_{\alpha i}^* \hbi_{\alpha j}}{\Fua_{\alpha\alpha}} \,,
 \\
  \hqu_{\alpha i\beta j}&= \frac{\htr_{\beta, ij}^*\hbi_{\alpha j} }{\frac{\vw^2}{\mB^2}|\hi_j|^2}= \frac{\htr_{j,\alpha\beta} \hbi_{\beta i} }{\frac{\vw^2}{\mB^2}|\hi_\beta|^2}=\frac{\hbi_{\alpha j}\hbi_{\beta j}^*\hbi_{\beta i} }{\frac{\vw^4}{\mB^4}|\hi_j|^2|\hi_\beta|^2}
  \approx 
  \frac{\hbi_{\alpha j}\hbi_{\beta j}^*\hbi_{\beta i} }{\Fda_{jj}\Fua_{\beta\beta}}  \,,
\\
    \Qh_{\alpha i \beta j}&=
    \frac{\hqu_{\alpha i\beta j} \hbi_{\alpha i}^*}{\frac{\vw^4}{\mB^4}|\hi_\alpha|^2|\hi_i|^2}
    =\frac{\hbi_{\alpha j}\hbi_{\beta j}^*\hbi_{\beta i} \hbi_{\alpha i}^*}{\frac{\vw^8}{\mB^8}|\hi_j|^2|\hi_\beta|^2|\hi_\alpha|^2|\hi_i|^2}
    \approx
    \frac{\hbi_{\alpha j}\hbi_{\beta j}^*\hbi_{\beta i} \hbi_{\alpha i}^*}{|\Fda_{ij}|^2|\Fua_{\alpha\beta}|^2}
    \,.
\end{aligned}
\end{equation}
Thus, the physical phases may be identified as the six independent phases of the bilinears.
These relations do not hold in scenarios with 
vanishing couplings,
e.g.~when the vector-like doublet is coupling only to one sector (up or down). 
In these cases, the phases of the trilinears and quartets are independent CP-violating phases%
\footnote{It is interesting to note that the \ac{SM} limit can be regarded as such a singular case, where only the quartet phase is relevant for CP violation.}
(see~\cref{sec:CPodd} for further details). This correspondence is strictly related to the 
type of \acp{WBI} needed to describe CP violation.

We also have the rephasing-invariant elements
\begin{equation}
 \Fuha_{\alpha\alpha}=\frac{\vw^2}{\mB^2}|\hi_{\alpha}|^2\approx \Fua_{\alpha\alpha}  \,, \qquad \Fdha_{ii}=\frac{\vw^2}{\mB^2}|\hi_{i}|^2\approx \Fda_{ii} \,,
\end{equation}
together with the other CP-even quantities which are explicitly moduli ($|\Vha_{\alpha i}|$, $|\VRha_{\alpha i}|$, etc.).

\subsection{Parameterizing the mixing exactly}
\label{sec:param}

In this section, we derive exact parameterizations for the mixings in these extensions, namely for the matrices $\VL$, $\VR$ and consequently for the $\Fq$, for an arbitrary number of \ac{VLQ} doublets $\n$ with hypercharge $1/6$.
These are given explicitly in terms of (physical) angles and phases, in a way that is reminiscent of the PDG parameterization for the standard \ac{CKM} matrix~\cite{Chau:1984fp,ParticleDataGroup:2022pth} or the Botella-Chau parameterization~\cite{Botella:1985gb} for the mixing in \ac{SM} extensions with a fourth generation.
While this is done for completeness and may be useful for comprehensive scans of the parameter space of the model(s), an expansion like the one considered in~\cref{sec:rotating} should be enough for most phenomenological applications.

Throughout this section we will make use of the following definitions for the $(3+\n)\times (3+\n)$ orthogonal rotation matrices $\0_{ij}$,
\begin{equation} \label{eq:param_orthon}
    \left(\0_{ij}\right)_{\alpha\beta} \equiv \begin{cases}
       \sin\theta_{ij}\,, & \text{for } \alpha=i,\,\beta=j\,,\\
       -\sin\theta_{ij}\,, & \text{for } \alpha=j,\,\beta=i\,,\\
       \cos\theta_{ij}\,, & \text{for } \alpha=\beta=i \text{ or } j\,,\\
       \delta_{\alpha\beta}, & \text{otherwise}\,,
    \end{cases}
\end{equation}
with $\theta_{ij}\in [0,\pi/2]$,
and for the $(3+\n)\times (3+\n)$ diagonal phase matrices $\K_{ij}$,
\begin{equation} \label{eq:param_phasesn}
    \left(\K_{ij}\right)_{\alpha\beta} \equiv \begin{cases}
       e^{i \varphi_{ij}}\,, & \text{for } \alpha=\beta=i \,,\\
       \delta_{\alpha\beta}\,, & \text{otherwise}\,,\\
    \end{cases}
\end{equation}
with $\varphi_{ij}\in[0,2\pi]$,
implying e.g.~that $\K_{31} =\diag\left(1,1,e^{i\varphi_{31}},1\right)$
and that $\K_{24}\,\K_{14} = \diag\left(e^{i\varphi_{14}},e^{i\varphi_{24}},1,1\right)$, for $\n = 1$.

\subsubsection{A single doublet (\texorpdfstring{$\n=1$}{\n=1})}
\label{sec:1}

For the case of the \ac{SM} extended with a single doublet \ac{VLQ} we can parameterize the unitary mixing matrix of the charged \ac{LH} currents as a general $4\times 4$ unitary matrix
\begin{equation} \label{eq:VL_param1}
    \VL
    \,=\,
    \K\, \0_{34}\0_{24}\0_{14}\K_{24}\K_{14}\,
    \0_{23}\K_{31}\0_{13}\0_{12}\,\K'\,,
\end{equation}
where we have used the $4\times 4$ matrices $\0_{ij}$ and $\K_{ij}$ defined above, as well as the diagonal phase matrices given by
\begin{equation} \label{eq:param_phases1}
\begin{aligned}
        \K& =\diag\left(e^{i\varphi_1},e^{i\varphi_2},e^{i\varphi_3},e^{i\varphi_4}\right)\,,\\[2mm]
        \K'&  =\diag\left(1,e^{i\varphi'_2},e^{i\varphi'_3},e^{i\varphi'_4}\right)
        \,.
\end{aligned}
\end{equation}
Obtaining a parameterization for $\VR$ is less immediate due to its non-unitary. Nonetheless, it is helpful to recall that $\VR$ is given by the product
\begin{equation}
    \VR= {\VuR}^\dagger \diag\left(0,0,0,1\right)\VdR\,,
\end{equation}
where the $\VqR$ \emph{are} $4\times 4$ unitary matrices. Then, using again the general parameterization for a unitary matrix in~\cref{eq:VL_param1} to express both $\VuR$ and $\VdR$,%
\footnote{Without loss of generality, we reverse the order of the product in~\cref{eq:VL_param1} to parameterize the $\VqR$.}
one can write
\begin{equation} \label{eq:VR_param1}
    \VR\,=\,\K_u^* \0^{uT}_{34}\0^{uT}_{24}\0^{uT}_{14} \, \diag\left(0,0,0,1\right)\, \0^d_{14}\0^d_{24}\0^d_{34}\K_d \,,
\end{equation}
so that the matrices controlling the \acp{FCNC} are given by
\begin{equation} \label{eq:Fq_param1}
    \Fq \,=\, \K_q^*\0^{qT}_{34}\0^{qT}_{24}\0^{qT}_{14}\ \diag\left(0,0,0,1\right)\0^q_{14}\0^q_{24}\0^q_{34}\K_q\,,
\end{equation}
for $q=u,d$. Here, the $\0^q_{ij}$ matrices are $4\times 4$ orthogonal matrices as defined in~\cref{eq:param_orthon} and the $\K_u$ and $\K_d$ matrices are diagonal phase matrices similar to $\K$ in~\cref{eq:param_phases1}.

Now, by rephasing the \ac{LH} fields as $\U_L\rightarrow \K\, \U_L$ and $\D_L\rightarrow \K'^* \, \D_L$, followed by an identical transformation of the \ac{RH} fields, i.e.~$\U_R\rightarrow \K\, \U_R$ and $\D_R\rightarrow \K'^* \, \D_R$ which maintains the mass terms in~\cref{eq:LM2} invariant, one can simply write
\begin{equation} \label{eq:VLparam}
    \VL
    \,=\,
    \0_{34}\0_{24}\0_{14}\K_{24}\K_{14}\,
    \0_{23}\K_{31}\0_{13}\0_{12}\,,
\end{equation}
which has a form identical to the Botella-Chau parameterization of the quark mixing matrix for the case of the \ac{SM} extended with one \ac{VLQ} isosinglet. On the other hand, $\VR$ maintains the form in~\cref{eq:VR_param1}, of course with $\K_u$ and $\K_d$ being redefined to absorb the effects of the rephasing
\begin{equation}
        \K_u\,\K \,\rightarrow\, \K_u \,, \qquad
        \K_d\,\K'^* \,\rightarrow\, \K_d
        \,,        
\end{equation}
so that all factorizable phases are brought to the \ac{RH} charged currents (absorbed in $\VR$).

\vskip 2mm
Note that, right now, we have parameterizations for $\VL$ and $\VR$ that appear to depend on a number of parameters that largely exceeds the 14 mixing parameters we counted in~\cref{sec:WB}, see~\Cref{tab:countsummary}. This issue is resolved when enforcing the constraint of~\cref{eq:LRrelation1}, which allowed us to derive~\cref{eq:LRrelation2,eq:LRrelation3}.
For instance by writing
\begin{equation} \label{eq:Bparam_1}
    \BqR
    \,\,=\,
    \begin{pmatrix} 0 &0 &0 &1 \end{pmatrix}\,
    \VqR  \,\,=\,
    \begin{pmatrix} 0 &0 &0 &1 \end{pmatrix}\,    \0^q_{14}\0^q_{24}\0^q_{34}\K_q
\end{equation}
and using~\cref{eq:MQ1} which directly follows from~\cref{eq:LRrelation1} for $\n=1$, one finds
\begin{equation} \label{eq:LR_param_rel}
    \Du \,\K_u^*\0^{uT}_{34}\0^{uT}_{24}\0^{uT}_{14}
    \begin{pmatrix}0 \\ 0 \\ 0 \\ 1\end{pmatrix}
    \,=\,
    \VL\,\Dd\, \K_d^*\0^{dT}_{34}\0^{dT}_{24}\0^{dT}_{14}
\begin{pmatrix}0 \\ 0 \\ 0 \\ 1\end{pmatrix}
\,,
\end{equation}
that demonstrates the existence of relations between the (interdependent) parameters.

One can choose to keep the 6 angles and 3 phases of $\VL$ as given in~\cref{eq:VLparam} as independent parameters --- different choices are possible, c.f.~\cref{sec:massbasis1}.
In this case, one can plug~\cref{eq:Fq_param1} into the leftmost part of~\cref{eq:LRrelation2} to obtain
\begin{equation} \label{eq:duckmr}
     \VR
     \,=\,
     \Du^{-1}\VL \Dd \,
     \K_d^*
     \0^{dT}_{34}\0^{dT}_{24}\0^{dT}_{14} \diag\left(0,0,0,1\right)
     \0^d_{14}\0^d_{24}\0^d_{34}\K_d
     \,,
\end{equation}
after multiplying by $\Du^{-1}$ from the left.%
\footnote{We do not consider cases where lighter masses vanish, which would have to be treated differently.}
Here, one can parameterize the phase matrix $\K_d$ as
\begin{equation}
    \K_d=\diag\left(e^{i \varphi^d_1},e^{i \varphi^d_2},e^{i \varphi^d_3},1\right)\,,
\end{equation}
with the fourth phase being factorized out as a global phase, which cancels with that from $\K^*_d$.
Note that any mention of the $\0^u_{i4}$ ($i=1,2,3$) and $\K_u$ matrices has disappeared in this parameterization of $\VR$, and consequently from $\Fu=\VR^{} \VR^\dagger$ and $\Fd= \VR^\dagger \VR^{}$.
Finally, using the relation $\BuR {\BuR}^\dagger = 1$, see~\cref{eq:ABunit}, and recalling that $ {\BuR}^\dagger = \Du^{-1}\VL \Dd \,{\BdR}^\dagger $,
one finds an extra constraint on $\BdR$ and $\VL$. 
Taking into account~\cref{eq:Bparam_1}, this constraint can be expressed as
\begin{equation} \label{eq:constraint}
    \left[\0^d_{14}\0^d_{24}\0^d_{34}\K_d\, \Dd \VL^\dagger \Du^{-2} \VL \Dd \, \K^*_d\0^{dT}_{34}\0^{dT}_{24}\0^{dT}_{14}\right]_{44} = 1 \,.
\end{equation}
Since the right-hand side of this equation is trivially real, this relation will correspond to one single condition that has to be met by the mixing parameters. This allows us to eliminate, for instance, the dependence on the mixing angle in $\0^d_{14}$.
In summary, the parameters of the theory can be chosen as:
\begin{itemize}
    \item the 8 quark masses,
    \item the 6 mixing angles ($\theta_{12},\theta_{13},\theta_{23},\theta_{14},\theta_{24}$ and $\theta_{34}$) and 3 non-factorizable phases ($\delta_{13},\delta_{14}$ and $\delta_{24}$) in $\VL$,
    \item the 3 independent phases in $\K_d$ and the 2 independent mixing angles present in the product $\0^d_{14}\0^d_{24}\0^d_{34}$.
\end{itemize}
Hence all the mixing is parameterized by 8 mixing angles and 6 phases, for a total of 14 physical mixing parameters, as anticipated in~\Cref{tab:countsummary}.

\subsubsection{More than one doublet (\texorpdfstring{$\n>1$}{\n>1})}
\label{sec:n}
In a more general extension of the \ac{SM} with $\n$ doublets, one can parameterize the $(3+\n)\times (3+\n)$ unitary mixing matrix of the charged \ac{LH} currents in a way analogous to that of~\cref{eq:VL_param1} for the $\n=1$ case (see also e.g.~\cite{Branco:1986my}). We have 
\begin{equation} \label{eq:VLn}
\VL = \uu_{3+\n} \ldots \uu_{5}  \ \uu_{4} \ \uu_3\,,
\end{equation}
where the upper-left  $3\times 3$ block of the unitary matrix $\uu_3 \equiv \0_{23} \K_{31} \0_{13} \0_{12}$ corresponds to the \ac{SM} \ac{CKM} matrix in the limit of decoupled \ac{VLQ} doublets.
The remaining $\uu_i$ ($i>3$) matrices are also unitary and are given by
\begin{equation} \label{eq:VLn1}
    \uu_{i} \,\equiv\,
    \Bigg(\prod^{i-1}_{j=1} \0_{i-j,\, i} \Bigg) \Bigg(\prod^{i-1}_{k=2} \K_{i-k,\, i}\Bigg)
    \qquad
(i = 4,\,\ldots,\,3+\n)
\,,
\end{equation}
where the matrices $\0_{ij}$ and $\K_{ij}$
have been defined in~\cref{eq:param_orthon,eq:param_phasesn}.

For $\n > 1$,~\cref{eq:LRrelation2} still holds, implying $\VR$ can be obtained from $\VL$ and $\Fd$ via
\begin{equation} \label{eq:LR_relation}
    \VR  = \Du^{-1} \VL \Dd \Fd \,,
\end{equation}
where
\begin{equation}
    \Fd \,=\,
    {\BdR}^\dagger \BdR
    \,=\,
     {\VdR}^\dagger\, \diag(0,0,0,1,\ldots,1)\, \VdR\,. 
\end{equation}
The matrix ${\BdR}^\dagger$ can be written as%
\footnote{Note that any square matrix can be written as the product of an upper-triangular matrix and a unitary matrix (RQ decomposition).
By decomposing the last $\n$ rows of ${\BdR}^\dagger$ in this way, one can extract the unitary $\uu_{\n\times \n}$ to the right while keeping the 3 first rows generically complex.
The diagonal elements can be made real by making an appropriate rephasing of each column, which is absorbed by the unitary matrix.}
\begin{equation} \label{eq:BdR}
    {\BdR}^\dagger \,=\, {\VdR}^\dagger
    \begin{pmatrix}
        0_{3\times \n} \\
        \id_{\n\times \n}
    \end{pmatrix}
    \,=\,\begin{pmatrix}
        c & c & \cdots & c & c\\
        c & c & \cdots & c & c\\
        c & c & \cdots & c & c\\
        r & c & \cdots & c & c\\
        0 & r & \cdots & c & c\\
        0 & 0 & \ddots & \vdots & \vdots \\
        \vdots & \vdots &  \ddots & r & c \\
        0 & 0 &  \cdots & 0 & r
    \end{pmatrix}\,\uu_{\n\times \n} 
    \,\equiv\,
    {\overline{\BdR}}^\dagger \ \uu_{\n\times \n}
    \,,
\end{equation}
with $\uu_{\n\times \n}$ being an $\n\times \n$ unitary matrix and ${\overline{\BdR}}^\dagger$ a $(3+\n) \times \n$ rectangular matrix.
Here, each $r$ represents a real entry while each $c$ represents a complex entry.
Note that, just like for ${\BdR}^{\dagger}$, the columns of ${\overline{\BdR}}^{\dagger}$ are orthonormal. This is because
\begin{equation} \label{eq:Bbar_unit}
    \overline{\BdR}\, \overline{\BdR}^{\dagger} \,=\,
    \uu_{\n\times \n} \BdR \, {\BdR}^{\dagger} \uu^\dagger_{\n\times \n}=\id_{\n\times \n}
    \,.
\end{equation}
At the same time, we also have
\begin{equation} \label{eq:Fdbars}
    \Fd \,=\, \overline{\BdR}^{\dagger} \overline{\BdR}\,.
\end{equation}
One can now decompose ${\overline{\BdR}}^\dagger$ as
\begin{equation}
    \overline{\BdR}^{\dagger} \,=\,\K^{d*}_{34}\K^{d*}_{24}\K^{d*}_{14}
    \,\,\,
    \0^{dT}_{34}\0^{dT}_{24}\0^{dT}_{14} 
    \,
    \begin{pmatrix}
        0 & c & \cdots & c & c\\
        0 & c & \cdots & c & c\\
        0 & c & \cdots & c & c\\
        r_{41} & c_{42} & \cdots & c_{4,\n-1} & c_{4 \n}\\
        0 & r & \cdots & c & c\\
        0 & 0 & \ddots & \vdots & \vdots \\
        \vdots & \vdots &  \ddots & r & c \\
        0 & 0 &  \cdots & 0 & r\\
    \end{pmatrix}
    \,,
\end{equation}
where $\0^d_{ij}$ and $\K^d_{ij}$ are also of the forms described in~\cref{eq:param_orthon,eq:param_phasesn}, respectively.
From the orthonormality condition in~\cref{eq:Bbar_unit}, one must have $r_{41}=1$ and $c_{4i}=0$, so that
\begin{equation}
    \overline{\BdR}^{\dagger} \,=\,\K^{d*}_{34}\K^{d*}_{24}\K^{d*}_{14}
    \,\,\,
    \0^{dT}_{34}\0^{dT}_{24}\0^{dT}_{14} 
    \,
    \begin{pmatrix}
        0 & c & \cdots & c & c\\
        0 & c & \cdots & c & c\\
        0 & c & \cdots & c & c\\
        1 & 0 & \cdots & 0 & 0\\
        0 & r & \cdots & c & c\\
        0 & 0 & \ddots & \vdots & \vdots \\
        \vdots & \vdots &  \ddots & r & c \\
        0 & 0 &  \cdots & 0 & r\\
    \end{pmatrix}
    \,.
\end{equation}
Now, by repeating this process $\n-1$ more times, once for each remaining column of $\overline{\BdR}^{\dagger}$, one can arrive at the form
\begin{equation} \label{eq:BbarVbar}
    \overline{\BdR}^{\dagger} \,=\,
    \Bigg(\prod_{i=4}^{\n+3} \K^{d*}_{3i}\K^{d*}_{2i}\K^{d*}_{1i}
    \,\,
    \0^{dT}_{3i}\0^{dT}_{2i}\0^{dT}_{1i}\Bigg)
    \begin{pmatrix}
        0_{3\times \n} \\
        \id_{\n\times \n}
    \end{pmatrix}
    \,\equiv\,
    \overline{\VdR}^{\dagger}
    \begin{pmatrix}
        0_{3\times \n} \\
        \id_{\n\times \n}
    \end{pmatrix} \,.
\end{equation}
Note that each time we apply this procedure, only three phases and three orthogonal rotations can be factorised from each column $j=1,\ldots,\n$. 
After that,~\cref{eq:Bbar_unit} implies that
$r_{3+j, j}=1$ and $c_{3+j, k} = 0$ for all $k \neq j$.
In the end, since one has
\begin{equation}
    \Fd \,=\,
    \overline{\BdR}^\dagger \overline{\BdR}
    \,=\,
     \overline{\VdR}^\dagger\, \diag(0,0,0,1,\ldots,1)\, \overline{\VdR}\,,
\end{equation}
by plugging this expression for $\Fd$ in~\cref{eq:LR_relation}, we obtain a parameterization for $\VR$ in terms of a reduced number of parameters, which only grows linearly with $\n$.

Finally, using the relation $\BuR {\BuR}^\dagger = \id_{\n\times \n}$ and recalling that $ {\BuR}^\dagger = \Du^{-1}\VL \Dd \,{\BdR}^\dagger $
and $\BdR=\uu^\dagger_{\n\times \n}\overline{\BdR}$, we have
\begin{equation} \label{eq:n2_constraints}
    \overline{\BdR} \Dd \VL^\dagger \Du^{-2} \VL \Dd {\overline{\BdR}}^\dagger
    \,=\,
    \id_{\n \times \n}
    \,,
\end{equation}
or alternatively, given~\cref{eq:BbarVbar},
\begin{equation}
    \begin{pmatrix}
        0_{\n\times 3} & \id_{\n\times \n}
    \end{pmatrix} \,\hh\, \begin{pmatrix}
        0_{3\times \n} \\ \id_{\n\times \n}
    \end{pmatrix}
    \,=\,
    \id_{\n\times \n}
    \,,
\end{equation}
where
\begin{equation}
    \hh\,\equiv\, \overline{\VdR} \Dd \VL\Du^{-2} \VL^{\dagger}\Dd \overline{\VdR}^{\dagger}
    \,,
\end{equation}
which results in $\n^2$ independent constraints for the mixing parameters. Of these, $\n$ can be expressed as
\begin{equation} \label{eq:const_one}
    \hh_{ii}=1\,,
\end{equation}
with $i=4,\ldots,3+\n$, while the remaining $\n(\n-1)$ constraints can be written as
\begin{equation} \label{eq:const_zero}
   \hh_{ij}=0\,,
\end{equation}
with $i<j$. Note that since $\hh$ is Hermitian,~\cref{eq:const_one} corresponds to one constraint for each value of $i$, while~\cref{eq:const_zero} corresponds to two distinct constraints for each pair $(i,j)$ --- one on a modulus and one on a phase.

\vskip 2mm
A quick counting of parameters shows us that this method results in a minimal parameterization of the mixing in a general $1/6$-hypercharge \ac{VLQ} doublet model. 
For instance, to parameterize $\VL$ we have:
\begin{itemize}
    \item $(\n+3)(\n+2)/2$ mixing angles,
    \item $(\n+2)(\n+1)/2$ physical phases,
\end{itemize}
as can be inferred from~\cref{eq:VLn,eq:VLn1},
whereas to parameterize $\VR$ we are additionally using:
\begin{itemize}
    \item $3\n$ mixing angles,
    \item $3\n$ physical phases,
\end{itemize}
as can be seen from~\cref{eq:LR_relation,eq:Fdbars,eq:BbarVbar}.
Moreover, the naïve quadratic dependence of the number of mixing parameters on the number of \ac{VLQ} doublets is removed by taking into account the constraints in~\cref{eq:const_one,eq:const_zero}.
In summary, our parameterization depends on
\begin{equation}
    \frac{(\n+3)(\n+2)}{2} + \frac{(\n+2)(\n+1)}{2} + 6\n - \n^2 = 4+10 \n
\end{equation}
independent mixing parameters ($3+5\n$ angles and $1+5\n$ phases), in agreement with~\Cref{tab:countsummary}. Note that when one accounts for the $6+2\n$ quark masses of theory, one finds the total of $10+12\n$ physical parameters we obtained before.

\vfill
\clearpage

\section{Weak-basis invariants}
\label{sec:CPeven}

As mentioned in~\cref{sec:WB}, weak-basis-invariant quantities can be constructed in the quark sector from Hermitian combinations of the quark mass matrices. In this section we show how these quantities can be built in an explicit and systematic way and how they can be related to the physical quark masses and mixings. 
We then demonstrate how one can determine all parameters describing $\n \geq 1$ \ac{VLQ} doublet extensions in terms of \acfp{WBI}.

\subsection{Constructing \acsp{WBI}}
\label{sec:constr_WBI}

To construct \ac{WBI} quantities, it is useful to define the Hermitian squared mass matrices
\begin{equation} \label{eq:defsH}
\HqL \,\equiv \, \Mq \Mq^\dagger
\,, \qquad
\HqR \,\equiv \, \Mq^\dagger \Mq\,,
\end{equation}
with $\Mq = \left(\,\mq\,\,\text{\textbrokenbar}\,\,\Mb\, \right)$, see~\cref{eq:defsM}. 
Recall the Hermitian combinations defined in~\mbox{\cref{sec:WB}}, namely $\hq \,\equiv \, \mq \mq^\dagger$ and $\C \,\equiv \, \Mb \Mb^\dagger$,
which transform under \acfp{WBT} as
\begin{equation}
\hq \,\to\, \WL^\dagger \,\hq\, \WL \,,\qquad
\C \,\to\,\WL^\dagger \,\C\, \WL \,.
\end{equation}
Noting that $\HqL = \hq+\C$, one sees that also $\HqL$ transforms as
\begin{equation}
\HqL \,\to\, \WL^\dagger \,\HqL\, \WL
\end{equation}
under \acp{WBT}.
Since they transform in the same way under \acp{WBT}, combinations of the three matrices $\HqL$, $\hq$, $\C$ and of their powers can be used to construct \ac{WBI} quantities. This can be done, for instance, by taking traces or determinants, since e.g.~$\tr(\WL^\dagger \HqL \WL) = \tr\HqL$ and $\det(\WL^\dagger \HqL \WL) = \det\HqL$.
Given that the three building blocks are related, not all combinations necessarily lead to different \acp{WBI}. All independent \acp{WBI} can then be built from traces of products of matrices involving only two of the three matrices $\HqL$, $\hq$, and $\C$.

\vskip 2mm
Let us note that, in the case of a single \ac{VLQ} doublet ($\n=1$), some simplification is possible. Namely, since $\Mb$ is a column vector, $\Mb^\dagger \Mb$ is a number and $\C = \Mb \Mb^\dagger$ is a rank-1 matrix. Then, $\Mb^\dagger \Mb=\tr \left(\Mb \Mb^\dagger\right)$, resulting in
\begin{equation} \label{eq:tr1VLQ}
    \C^2=\left(\tr \C\right) \C\,, \qquad
    \text{for }\n=1    \,,
\end{equation}
and thus reducing the actual degrees of freedom when building independent \acp{WBI}. Indeed,~\cref{eq:tr1VLQ} directly implies
\begin{equation} \label{eq:tr1VLQ_2}
\tr(A \C^{r})=(\tr\C)^{r-1}\tr(A \C)\,, \qquad
    \text{for }\n=1    \,.
\end{equation}
Moreover, since $\Mb^\dagger A \Mb$ is a number, equaling its own trace, one also finds that
\begin{equation} \label{eq:tr1VLQ_3}
\tr\left( \C A \C B \right)= \tr\left( \C A \right) \tr\left( \C B \right)\,, \qquad
    \text{for }\n=1    \,.
\end{equation}

\vskip 2mm

The matrix $\HqR$ defined above can also be relevant when building \acp{WBI}. One has
\begin{equation}
    \HqR \,=\, \Mq^\dagger \Mq \,=\,\begin{pmatrix}
        \mq^\dagger \mq  & \mq^\dagger \Mb\\[2mm]
        \Mb^\dagger \mq  & \Mb^\dagger \Mb 
    \end{pmatrix}\,,
\end{equation}
which under a general \ac{WBT} transforms as
\begin{equation}
    \HqR
    \,\to\, 
     \begin{pmatrix}
        {\WRq}^\dagger & \\[2mm]
         & \WR^\dagger 
    \end{pmatrix}
    \HqR
     \begin{pmatrix}
        \WRq & \\[2mm]
         & \WR 
    \end{pmatrix}
    =
        \begin{pmatrix}
        {\WRq}^\dagger \mq^\dagger \mq \WRq & {\WRq}^\dagger \mq^\dagger \Mb \WR\\[2mm]
        \WR^\dagger \Mb^\dagger \mq \WRq & \WR^\dagger \Mb^\dagger \Mb \WR
    \end{pmatrix}
    \,.
\end{equation}
Inspection of the matrix blocks motivates
considering
the $3\times 3$ Hermitian matrices
$\hqR \equiv \mq^\dagger \mq$ and $\mq^\dagger \Mb \Mb^\dagger \mq $, 
and the $\n\times \n$ Hermitian matrices 
$\CR \equiv \Mb^\dagger \Mb$ and $\Mb^\dagger \mq \mq^\dagger \Mb $.
These are also viable building blocks of \acp{WBI}, directly connected to the transformations of \ac{RH} quark fields, which are physical in this \ac{BSM} scenario.
Within each pair, matrices share the same dimensions and transformation properties, so that their powers can be combined within traces to form two sets of independent \acp{WBI}.
Note, however, that \acp{WBI} built from these four \ac{RH} Hermitian matrices are not independent from the ones built from the \ac{LH} Hermitian matrices, $\HqL$, $\hq$, and $\C$, due to the cyclicity of the traces.

\vskip 2mm
It is useful to relate these building blocks to the quark masses and the unitary rotations connecting the flavour and mass bases.
Using~\cref{eq:svd2,eq:mqM}, one has
\begin{equation} \label{eq:Hermitian}
\begin{aligned}
    \HqL &\,=\, \VqL\,\Dq^2\,{\VqL}^\dagger \,, 
    \\
    \hq&\,=\, \VqL\,\Dq\,{\AqR}^\dagger\AqR\,\Dq\,{\VqL}^\dagger \,,
\end{aligned}
\qquad\quad
\begin{aligned}
    \HqR &\,=\, \VqR\,\Dq^2\,{\VqR}^\dagger \,, 
    \\
    \hqR  &\,=\, \AqR\,\Dq^2\,{\AqR}^\dagger\,.
\end{aligned}
\end{equation}
The remaining Hermitian matrices depend on $\Mb$, which can be written in two different ways, using matrices from either one sector or the other, recall~\cref{eq:mqM,eq:LRrelation1}.
Therefore, the remaining matrices can be connected to the masses and unitary rotations in more than one way. 
For instance, 
\begin{equation} \label{eq:H_diff_forms}
        \C 
        \,=\,\VuL \Du \VR \Dd {\VdL}^\dagger
        \,=\,\VqL \Dq \Fq \Dq {\VqL}^\dagger
        \qquad
        (q=u,d)\,.
\end{equation}
Additionally, using $\HqL=\hq+\C$, we have $\hq =\VqL \Dq^2 {\VqL}^\dagger-\C$, which could also be obtained from~\cref{eq:Hermitian} via~\cref{eq:A1mF}.

Now, we are in a position to construct \acp{WBI} and relate them to the quark masses and mixings. The masses of all $q$-sector quarks can be obtained from a set of $3+\n$ \acp{WBI} constructed solely from $\HqL$. These are given by 
\begin{equation} \label{eq:WBI_masses}
\chi_\sigma\left(\HqL\right)\,=\,\frac{1}{\sigma!}\left[\frac{\partial^\sigma}{\partial x^\sigma}\det\left(\HqL+x\,\id_{3+\n} \right)\right]_{x=0}
\,,
\end{equation}
where  $0\leq \sigma \leq 2+\n$.
They are closely related to the coefficients of a characteristic polynomial,
with e.g.~$\chi_0(\HqL)=\det \HqL = (\msq_1 \ldots \Msq_\n)^2$ and $\chi_{2+\n}(\HqL)=\tr\HqL = (\msq_1)^2 +  \ldots + (\Msq_\n)^2$.

\vskip 2mm
\acp{WBI} that are written in terms of the mixings involve matrices from both quark sectors. We have, for instance, the following set of CP-even \acp{WBI}
\begin{equation} \label{eq:modV}
 \tr\left(\HuL^r \HdL^s\right)=\tr\left(\VL^\dagger\,\Du^{2r}\, \VL\, \Dd^{2s}\right)=m^{2r}_{\alpha}m^{2s}_{i}\left|\VLa_{\alpha i}\right|^2\,,
\end{equation}
which relate to the mixing of \ac{LH} quarks. 
Here, the implied sums in $\alpha$ and $i$ go over all $3+\n$ quark masses in the up and down sectors, respectively.
Similarly, after straightforward manipulations, one finds
\begin{equation}
    \tr\left( \HuL^{r-1}\C \HdL^{s-1}\C \right)=\tr\left(\VR^\dagger\,\Du^{2r}\, \VR\, \Dd^{2s}\right)=m^{2r}_{\alpha}m^{2s}_{i}\left|\VRa_{\alpha i}\right|^2\,,
\end{equation}
which are related to the mixing of \ac{RH} quarks.

\vskip 2mm
As for CP-odd \acp{WBI}, i.e.~\acp{WBI} with non-vanishing imaginary parts, one has to look into combinations of at least three distinct Hermitian matrices and their powers. Some of these are of the form
\begin{equation} \label{eq:wbiodd-sec5}
\begin{aligned}
\tr\left([\HuL^r, \HdL^s]\C\right) &\,=\, 2i \, \im
\tr\left(\HuL^r \HdL^s \C\right)\\
&\,=\,
2i \, \im \tr\left(\Du^{2r+1}\,\VL\, \Dd^{2s+1} \,\VR^\dagger\right)\\
&\,=\,
 2i \, m^{2r+1}_{\alpha}m^{2s+1}_{i}\im \bi^{LR}_{\alpha i}\,,
\end{aligned}
\end{equation}
where we have recalled the definition of the bilinears $\bi^{LR}_{\alpha i} = \VLa_{\alpha i}\, \VRac_{\alpha i}$ from~\cref{eq:bilinears}.
As before, the implied sums in $\alpha$ and $i$ go over all $3+\n$ quark masses in the up and down sectors, respectively.
Note that, while in the \ac{SM} the CP-odd \ac{WBI} with the lowest mass dimension has dimension $\M = {12}$, here, for the case $r=s=1$, one finds a CP-odd \ac{WBI} of dimension $\M = 6$ which is non-zero in general. In that specific case we have
\begin{equation} \label{eq:lowest_CP_odd}
    \tr\left([\HuL,\HdL]\C\right)
    =\tr\left([ \hu,\hd]\C\right)
    =2 i \, m_\alpha^3 m_i^3 \im \bi^{LR}_{\alpha i}\,. 
\end{equation}
In principle, this could imply the presence of new and important sources of CP violation in the quark sector, as we will show in~\cref{sec:CPodd,sec:pheno}. In~\cref{sec:CPodd}, we also identify other types of CP-odd \acp{WBI} which can be relevant in enhancing CP violation.

\subsection{Reconstructing minimal parameterizations from \acsp{WBI}}
\label{sec:reco}

In the last section, we showed how one might construct \acp{WBI} and relate them to quark masses and mixings.
Here, we are instead interested in the possibility of reconstructing the
 quark mass matrices from a set of \acp{WBI}.
To do this in a comprehensive way, it is important to fix a suitable weak basis. In fact, this is a crucial ingredient; for certain \ac{WB} choices, the mass-matrix parameters present in the relations may be entangled in a convoluted way,
that obscures or even obstructs a viable solution.

In~\cref{sec:stepladderwb}, we will show how to characterize all 22 parameters of the stepladder \ac{WB}, in the $\n=1$ case, in terms of a relatively small number of \acp{WBI}.
In~\cref{sec:mapping}, we present an alternative procedure which considers the minimal \ac{WB}. Although slightly less direct, the latter can be applied to the case of $\n \geq 1$ doublets.
A numerical example is presented in~\cref{sec:numex}.

\subsubsection{Reconstructing the stepladder \acs{WB}}
\label{sec:stepladderwb}

Recall the stepladder weak basis 
introduced in~\cref{sec:WBs},~\cref{eq:stepladder_2},
for the case $\n = 1$: 
\begin{equation} \label{eq:specWB1}
\Mu = \begin{pmatrix}
    \tilde{m}_u & 0 \\ 
\mB\, r_u & \mB
\end{pmatrix}\,,
\qquad
\Md =\begin{pmatrix}
    \VWT & 0\\
    0 & 1
\end{pmatrix}\begin{pmatrix}
    \tilde{m}_{d} & 0 \\ 
\mB\, r_{d} & \mB
\end{pmatrix}\,.
\end{equation}
Here $\VWT$ is a general $3\times 3$ unitary matrix having three real angles
and six complex phases, one of which is an internal phase, while
the other five are factorizable phases. The real vectors $r_{d,u}$ and real matrices $\tilde{m}_{d,u}$ are, in both sectors,
of the special form 
\begin{equation} \label{eq:specWB2}
    r_q =  \begin{pmatrix}
    0 & 0 & r_{q0}
\end{pmatrix}, \hspace{5mm}
\tilde{m}_q = \mB \begin{pmatrix}
    r_{q5} & 0 & 0 \\ 
r_{q4} & r_{q3} & 0 \\ 
0 & r_{q2} & r_{q1}
\end{pmatrix}\,,
\end{equation}
while $\mB$ is a real number. In fact, we may choose $\mB$ and all the (dimensionless) $r_{qi}$ to be non-negative,
as any physical minus sign can be incorporated into $\VWT$ via suitable rephasings of the quark fields.
In this way, the matrices
\begin{equation} \label{eq:specWB3}
    \begin{pmatrix}
        \tilde{m}_q & 0 \\ 
\mB\,r_q & \mB
    \end{pmatrix}=\mB \begin{pmatrix}
        r_{q5} & 0 & 0 & 0 \\ 
r_{q4} & r_{q3} & 0 & 0 \\ 
0 & r_{q2} & r_{q1} & 0 \\ 
0 & 0 & r_{q0} & 1
    \end{pmatrix}
\end{equation}
have a ``stepladder'' shape. As we shall see, this
distinctive structure will be crucial to establish direct
relations between the mass-matrix parameters and \ac{WBI} quantities.
For starters, in terms of the former, the 
\ac{RH} Hermitian squared mass matrices
$\HqR \equiv  \Mq^\dagger \Mq$
are real even for $q=u$ (they do not depend on $\VWT$) and read
\begin{equation} \label{eq:specWB_HR}
\HqR
=
\begin{pmatrix}
    \tilde{m}_q^T \tilde{m}_q + \mB^2\,  r_q^T r_q & \mB^2\,  r_q^T\\
    \mB^2\, r_q & \mB^2
\end{pmatrix}\,.
\end{equation}

Before proceeding, recall
that this is also a ``minimal'' \ac{WB}.
Indeed, by counting the parameters that characterize this \ac{WB} --- the 12 real parameters $r_{qi}$, the mass scale $\mB$, and the 9 parameters that determine $\VWT$ --- one finds a total of 22 parameters, which coincides with the number of physical parameters of the model (cf.~\Cref{tab:countsummary} and the discussion at the end of~\cref{sec:stepladderdefs}).

\subsubsection*{A ``step-by-step'' solution} 

Now, as it is clear from the discussion in~\cref{sec:constr_WBI}, using $\HqR$ one may construct the following CP-even \acp{WBI},
\begin{equation}
\Iqp{k} \equiv \tr\left( \mathcal{K}_0 \, \HqR^{k+2}\right)\,, \quad 
\mathcal{K}_0 \equiv\diag\left( 0,0,0,1\right) \,, 
\end{equation}
where $k\in\{-1,0,1,2,\ldots\}$.

Note that these invariants differ in nature from those in the \ac{SM}.
We recall that in the \ac{SM}, with the $3\times 3$ \ac{RH} Hermitian squared mass matrix $h=m^\dagger m$, we can only obtain information about the masses. 
There are no $Z$-mediated currents involving the \ac{RH} fields.
The limited information regarding physical parameters that is contained in $h$ can be accessed by computing, e.g.~$\tr h$, $\tr (h^2)$, and $\tr (h^3)$ or $\det h$.
The same applies to the \ac{RH} Hermitian squared mass matrix of extensions of the \ac{SM} with \ac{VLQ} isosinglets, as in that case, there are also no $Z$-mediated currents involving the \ac{RH} fields.
Clearly this is not the case for extensions of the \ac{SM} with doublet \acp{VLQ}, where we must deal with extra \ac{NP} present in the mixing of \ac{RH} quarks. Thus, in the considered model with one doublet \ac{VLQ}, one expects more information to be contained in \acp{WBI} built from \ac{RH} Hermitian matrices. 
Indeed, and in contrast with the \ac{SM}, there are \ac{RH} weak-basis rotations which are necessarily shared between the up and down sectors, allowing for the construction of the new types of \acp{WBI} considered here.

The real parameters $r_{qi}$ and $\mB$, contained in~\cref{eq:specWB3},
can be determined in terms of the $\Iqp{k}$.
We find, for both sectors ($q=u,d$), 
\begin{equation} \label{eq:j41}
\Iqp{-1}= \mB^2 \,,   
\end{equation}
and then, making use of the rescaled invariants
\begin{equation}
 \Iqs{k} \equiv \frac{1}{\mB^{2(k+2)}}\Iqp{k}\,,\qquad k = 0,1,2,\ldots\,,
\end{equation}
we obtain 
\begin{equation} \label{eq:step_r}
\begin{aligned}
&\Is{0}=1+r_{0}^{2}\,,
\\[2mm]
&\Is{1}=\left( 1+r_{0}^{2}\right)^2 +r_{0}^{2}r_{1}^{2}\,,
\\[2mm]
&\Is{2}=\left[
\left(1+r_{0}^{2}\right)^3+2r_{0}^{2}r_{1}^{2}+r_{0}^{2}r_{1}^{4}+2r_{0}^{4}r_{1}^{2}\right] +r_{0}^{2}r_{1}^{2}r_{2}^{2}\,,
\\[2mm]
&\Is{3}=P_3(r^2_0,r^2_1,r^2_2)+r_{0}^{2}r_{1}^{2}r_{2}^{2}r_{3}^{2}\,,
\\[2mm]
&\Is{4}=P_4(r^2_0,r^2_1,r^2_2,r^2_3)+r_{0}^{2}r_{1}^{2}r_{2}^{2}r_{3}^{2}r_{4}^{2}\,,
\end{aligned}
\end{equation}
where $P_k$ are polynomial functions of $r^2_0,\ldots, r^2_{k-2}$ and $r^2_{k-1}$. Here, and for now, we omit the superscript $q$ when expressions apply to both sectors. It should, however, be stressed that one generically needs to compute two invariants of each kind, one for each sector.

Given the structure of the stepladder \ac{WB}, the relations of~\cref{eq:step_r} also follow a ``stepladder pattern''.
By this we mean the following:
the invariant $\Is{1}$ is the sum of a polynomial in $r_0$, 
which can be determined from $\Is{0}$,
and a single term depending on
the next parameter $r_{1}$, to wit, the term 
$r_{0}^{2}r_{1}^{2}$. 
This pattern repeats itself:
each invariant $\Is{k}$ can be written as the sum of a polynomial $P_k(r_0^2,\ldots,r_{k-1}^2)$ and a term proportional to $r_k^2$.
This allows one to iteratively extract each of the $r_i$.
Finally to obtain $r_{5}$, one may compute $\Is{5}$ or, alternatively, evaluate 
\begin{equation} \label{eq:j6}
\det \HR \,=\, \mB^{8}\, r_{1}^{2}r_{3}^{2}r_{5}^{2}\,, 
\end{equation}
for each sector. Thus, in both sectors, one can directly extract $\mB$ and all $r_k$ from a small set of \acp{WBI}.

However, it should be noted that this method appears to fail when some $r_k=0$. 
In fact, if a given $r_k=0$, then none of these invariants depends on any of the $r_{l > k}$. 
For example, if $r_2=0$ then $\Is{3}$ and $\Is{4}$ (as well as all $\Is{k>4}$) are independent of $r_3$ and $r_4$.
Still, in those cases%
\footnote{
We disregard the cases where $r_1$, $r_3$ or $r_5$ vanish, as those amount to vanishing masses, cf.~\cref{eq:j6}.}
one can always determine all non-vanishing $r_i$ parameters,
as there is a stepladder \ac{WB} where the off-diagonal $r_{l>k}$ also vanish (see the discussion at the end of~\cref{sec:stepladderdefs}).
In this stepladder \ac{WB}, the mass matrix is block-diagonal and
only the $r_i$ of the upper-left (diagonal) block remain undetermined. They can be obtained with the help of the determinant of $\HR$ or traces of its powers.
For instance, if $r_2 = 0$, then $r_4=0$ in this \ac{WB}, and one would extract $r_{3,5}$ by computing $\tr\HR=\mB^2(1+r_0^2+r_1^2+r_3^2+r_5^2)$ and $\det\HR$ given in~\cref{eq:j6}.

\vskip 2mm

To complete our characterization, we
need to express all parameters of the unitary matrix $\VWT$ in terms of
\acp{WBI}. Here we will assume that all $r_i\neq 0$. From the form of $\Md$ in~\cref{eq:specWB1}, it is clear that the \acp{WBI} that may ``capture'' the entries of $\VWT$ are those which relate to \acp{WBT} of \ac{LH} fields. Since $\VWT$ is in general complex (in fact, it contains all physical phases) one must also consider CP-odd \acp{WBI}.

There is an analogous iterative way to fully determine $\VWT$.
We start by defining
\begin{equation} \label{eq:Jrs0}
    \Js{rs}\,\equiv\,\frac{1}{\mB^{2(1+r+s)}}\tr\left(\HuL^r \HdL^s \C\right)\,.
\end{equation}
Then,
\begin{equation} \label{eq:J11}
\Js{11}=\frac{1}{\mB^{6}}\tr\left( \HuL\HdL\C\right)
\,=\,\left( 1+r_{u0}^{2}\right) \left( 1+r_{d0}^{2}\right) +r_{u0}r_{u1}\,r_{d0}r_{d1}\,\VWT_{33} \,,
\end{equation}
which is linear in $\VWT_{33}$, and depends on the already-determined real parameters $r_{qi}$ and $\mB$. Thus, by computing $\Js{11}$, which is complex in general, both $|\VWT_{33}|$ and $\arg \VWT_{33}$ can be determined without discrete ambiguities. 
Next, we evaluate 
\begin{equation} \label{eq:J12}
\Js{12}=\frac{1}{\mB^{8}}\tr\left(\HuL\HdL^2\C\right)
\,=\,\mathcal{P}_{12} +r_{u0}r_{u1}\,r_{d0}r_{d1}r_{d2}r_{d3}\,\VWT_{32}\,,
\end{equation}
where $\mathcal{P}_{12}$ is again a simple polynomial which depends on already-known parameters, including $\VWT_{33}$. Explicitly, we have
\begin{equation} \label{eq:P12}
\begin{aligned}
\mathcal{P}_{12} &=  \left(1+ r_{d0}^{2}\right) \left[ \left( 1+r_{d0}^{2}\right)
\left(1+ r_{u0}^{2}\right) +r_{d0}r_{d1}\,r_{u0}r_{u1}\,\VWT_{33}\right] \\ 
&\quad +r_{d0}r_{d1}\left[ r_{d0}r_{d1}\left(1+ r_{u0}^{2}\right)
+r_{u0}r_{u1}\left( r_{d1}^{2}+r_{d2}^{2}\right) \VWT_{33}\right]\,.
\end{aligned}
\end{equation}
Notice that $\Js{12}$ is linear in $\VWT_{32}$. Thus, like $\VWT_{33}$, this entry can be fully determined without ambiguity.
Further computing 
\begin{equation} \label{eq:J13}
\Js{13}=\frac{1}{\mB^{10}}\tr\left(\HuL\HdL^3\C \right) \,=\,
\mathcal{P}_{13}+r_{u0}r_{u1} \,r_{d0} r_{d1} r_{d2}r_{d3}r_{d4}r_{d5}\,\VWT_{31}
\end{equation}
allows one to find $\VWT_{31}$. Once again, $\mathcal{P}_{13}$ obeys the stepladder pattern: it is a linear combination of the previously-determined entries of $\VWT$, whose coefficients are polynomials in the $r_{qi}$. The new invariant $\Js{13}$ is again linear in $\VWT_{31}$, which is found directly.

To obtain all other $\VWT_{ij}$, one needs to repeat this scheme,
carefully choosing which invariants to compute. By computing, in the following order, the invariants
\begin{equation} \label{eq:J4}
\begin{aligned}
\Js{21}&=\frac{1}{\mB^{8}}\tr\left( \HuL^2\HdL\C\right)
=\mathcal{P}_{21}+r_{u0}r_{u1}r_{u2}r_{u3}\;r_{d0}r_{d1}\,\VWT_{23}\,, \\ 
\Js{31}&=\frac{1}{\mB^{10}}\tr\left( \HuL^3\HdL\C\right)=\mathcal{P}_{31}+r_{u0}r_{u1}r_{u2}r_{u3}r_{u4}r_{u5}\;r_{d0}r_{d1}\,\VWT_{13}\,,\\
\Js{22}&=\frac{1}{\mB^{10}}\tr\left( \HuL^2\HdL^2\C\right) =\mathcal{P}_{22}+r_{u0}r_{u1}r_{u2}r_{u3}\;r_{d0}r_{d1}r_{d2}r_{d3}\,\VWT_{22}\,,\\
\Js{32}&=\frac{1}{\mB^{12}}\tr\left( \HuL^3\HdL^2\C\right) =\mathcal{P}_{32}+r_{u0}r_{u1}r_{u2}r_{u3}r_{u4}r_{u5}\;r_{d0}r_{d1}r_{d2}r_{d3}\,\VWT_{12}\,,\\
\Js{23}&=\frac{1}{\mB^{12}}\tr\left( \HuL^2\HdL^3\C\right) =\mathcal{P}_{23}+r_{u0}r_{u1}r_{u2}r_{u3}r_{u4}r_{u5}\;r_{d0}r_{d1}r_{d2}r_{d3}\,\VWT_{21}\,,\\
\Js{33}&=\frac{1}{\mB^{14}}\tr\left( \HuL^3\HdL^3\C\right) =\mathcal{P}_{33}+r_{u0}r_{u1}r_{u2}r_{u3}r_{u4}r_{u5}\;r_{d0}r_{d1}r_{d2}r_{d3}r_{d4}r_{d5}\,\VWT_{11}\,,
\end{aligned}
\end{equation}
one fully determines all the remaining $\VWT_{ij}$ in succession, where again the $\mathcal{P}_{ij}$ follow the stepladder pattern, being linear combinations of the $\VWT_{ij}$ obtained in previous steps.
As a corollary, one sees that the vanishing of all nine $\Js{ij}$ is a sufficient condition for CP invariance in the case of non-vanishing $r_{qi}$, with $i=0,1,\ldots,5$. 
In~\cref{sec:CPconservation} we study the conditions for CP conservation in the various cases where $r_{q0}$, $r_{q2}$ or $r_{q4}$ vanish.

A direct connection is thus established between \acp{WBI} and 
\ac{WB} parameters.
In summary, we considered 13 CP-even \acp{WBI}, namely $\Ip{-1}$ and the 6 invariants $\Iqs{0,\ldots,5}$ for each sector $(q=u,d)$, where the $\Iqs{5}$ can be exchanged for $\det\HqR$;
as well as the 9 \acp{WBI} $\Js{rs}$, with $r,s = 1,2,3$. This set of 22 \acp{WBI} can be described collectively by
\begin{equation}
    \mathcal{I}_{rs} \equiv \frac{1}{\mB^{2(1+r+s)}}\tr\left(\HuL^r\HdL^s\C\right)\,, 
\end{equation}
for $(rs)= \{(00)$, $(01),\ldots,(06)$, $(10),\dots,(60)$, $(11)$, $(12)$, $(13)$, $(21)$, $(22)$, $(23)$, $(31)$, $(32)$, $(33)\}$,
with $\mathcal{I}_{0s} = \Is{s-1}^d$ and $\mathcal{I}_{r0} = \Is{r-1}^u$.
In general, the 31 real constraints described above are required to determine the relevant 22 parameters. In most cases, the unitarity of $\VWT$ can reduce the number of necessary constraints.

\vfill
\clearpage

\section{CP violation}
\label{sec:CPodd}

In this section, we examine the CP-violating properties of a model containing vector-like quark doublets in a weak-basis-independent way. 
\ac{CPV} must be connected to the non-vanishing imaginary parts of weak-basis-invariant quantities,
e.g.~imaginary parts of traces of mass matrix combinations~\cite{Bernabeu:1986fc,Gronau:1986xb,delAguila:1996pa,delAguila:1997vn,Albergaria:2022zaq}.
We will show that there is \emph{a crucial connection between \acfp{WBI} and effective rephasing invariants}, which provides a compelling tool to analyze CP-violating phenomena (see also~\cref{sec:pheno}).

We start by illustrating the construction of CP-odd \acp{WBI} in~\cref{sec:CPwbi}.
In~\cref{sec:CPwbi-eff}, we will show in the scenario with one \ac{VLQ} doublet ($\n = 1$) how 
the information carried by each \ac{WBI} becomes evident using the parameters defined in~\cref{sec:app1}. 
These parameters are directly connected
to observables.
We show that the \acp{WBI} can be expressed in terms of the masses and mixings of the three families of \ac{SM} quarks (providing an ``effective'' description of the \acp{WBI}). In particular, \acp{WBI} can be given in terms of the rephasing invariants introduced in~\cref{sec:reph-inv-eff}.
In~\cref{sec:CPconservation}, we derive a complete set of conditions for CP conservation in the case of $\n = 1$, under the assumption of non-vanishing and non-degenerate quark masses.

\subsection{CP-odd weak-basis invariants}
\label{sec:CPwbi}

In the \ac{SM}, the presence of \ac{CPV} is connected to the non-vanishing of one weak-basis invariant~\cite{Jarlskog:1985ht,Jarlskog:1985cw,Bernabeu:1986fc}:
\begin{equation} \label{eq:cp-sm}
\begin{aligned}
& \tr[\HuL,\HdL]^3 \,=\, 3 \det[\HuL,\HdL]  \\  
&\quad=\, - 6 i \,(m_t^2-m_c^2)(m_t^2-m_u^2) (m_c^2-m_u^2) (m_b^2-m_s^2)(m_b^2-m_d^2) (m_s^2-m_d^2) \,J \,,
\end{aligned}
\end{equation}
where $\HqL \equiv \Mq \Mq^\dagger$ as defined in~\mbox{\cref{sec:constr_WBI}}, while
$J=\im(V_{ud}V_{cs}V_{us}^*V_{cd}^*)$ is the rephasing invariant of~\cref{eq:Jar}. 
The vanishing of the trace in~\cref{eq:cp-sm} is a necessary and sufficient condition for CP conservation in the \ac{SM}.
This trace is the non-real \ac{WBI} with lowest mass dimension which can appear in the context of the \ac{SM}, as $J$ is the simplest complex rephasing invariant that can be built.
In a scenario involving \acp{VLQ}, the extra couplings can act as new sources of \ac{CPV}, since new physical phases emerge. 
Also, different kinds of complex \acp{WBI} can be built with lower mass matrix powers with respect to the \ac{SM}, which can be 
traced to the presence of the new rephasing-invariant quantities, involving two or more quarks (recall~\cref{sec:reph-inv}).
In the following, we analyze the structure of different invariants of increasing mass dimension.

\subsubsection{With one or more \acs{VLQ} doublets}
\label{sec:WBI_one_doublet}

The ingredients to build \acp{WBI} were 
provided in~\cref{sec:constr_WBI}.
To construct CP-odd \acp{WBI}, one has to look
into combinations of at least three distinct 
Hermitian \ac{WBI} ``building blocks'', i.e. powers of $\HuL$, $\HdL$ and $\C$.
Then, the simplest \acp{WBI} are of the form (see also~\cref{eq:wbiodd-sec5}):
\begin{equation} \label{eq:CPo3blocks}
\begin{aligned}
    & \im\tr\Big[\HuL^n\, \HdL^k \, \C\Big] = \im\tr\left[\left(\Mu\Mu^\dagger\right)^n\left(\Md\Md^\dagger\right)^k \Mb \Mb^\dagger\right] \,  =  
      \\ &
    =\im\tr\Big[\, \Du^{2n+1} \VL \Dd^{2k+1}\VR^\dagger \,\Big]
   = \sum_{\alpha, i=1}^{3+\n} m_\alpha^{2n+1}m_i^{2k+1}\, \im \left(\VLa_{\alpha i}\, \VRac_{\alpha i}\right)\,,
\end{aligned}
\end{equation}
where, as before, the Greek (Latin) indices refer to
the up (down) sector
and $\C \equiv \Mb \Mb^\dagger$ as defined in~\cref{sec:WB}.
Ordering them
by mass dimension, the first CP-odd weak-basis invariant appears at mass dimension $\M=6$,
using three Hermitian blocks:
\begin{equation} \label{eq:CPo6}    
\im\tr\Big[\HuL\,\HdL \, \C \Big] =\, 
 \im\tr\Big[\, \Du^3 \VL \Dd^3\VR^\dagger \,\Big]
  = \sum_{\alpha, i=1}^{3+\n} m_\alpha^3 m_i^3\, \im \left(\VLa_{\alpha i}\, \VRac_{\alpha i}\right)\,.
\end{equation}
We also specified the expression of the invariant in the mass basis.
In the last term, the sum runs over all species, the three standard model families and the vector-like doublets.
Note that this lowest mass dimension CP-odd \ac{WBI} brings with it the rephasing-invariant
quantities $\bi^{LR}_{\alpha i}=\VLa_{\alpha i} \VRac_{\alpha i}$. 
In fact, 
The presence of these bilinears points to the fact that these extensions can induce CP-violating effects involving only two quarks.
For instance, contributions to electric dipole moments and 
direct CP violation in kaon decays $\epsilon'$ emerge at tree level from the imaginary part of 
$\VLa_{ud} \VRac_{ud}$
and $\VLa_{us} \VRac_{us}$ (see~\cref{sec:phenobilinears}).
Let us note that in the mass basis, using the relation in~\cref{eq:LRrelation1}, 
the \ac{WBI} in~\cref{eq:CPo6} can be written also in other ways involving more insertions of mixing matrices:
\begin{equation}\label{eq:wbi-ways}
\begin{aligned}
\tr\Big[\HuL\,\HdL\, \C \Big] 
& = \tr\big[\, \Du^3 \VL \Dd^3\VR^\dagger \,\big] = \sum_{\alpha, i=1}^{3+\n} m_\alpha^3 m_i^3 \; \VLa_{\alpha i}\, \VRac_{\alpha i}  \\
& = \tr\big[\Du^2 \VL \Dd^3 \Fd \Dd \VL^\dagger \big]=\sum_{\alpha, i=1}^{3+\n} m_\alpha^2 m_i^3m_j \; 
\VLa_{\alpha i}\, \VLac_{\alpha j} \, \Fda_{ ij}  \\
& = \tr\big[\Du^3 \VL \Dd^2 \VL^\dagger \Du \Fu \big]
=\sum_{\alpha, i=1}^{3+\n} m_\alpha^3 m_i^2  m_\beta \VLa_{\alpha i}\, \VLac_{\beta i} \Fuac_{ \alpha\beta}  \\
& = \tr\big[\Du^2 \VL \Dd^2 \VL^\dagger \Du \VR \Dd \VL^\dagger\big]
=\sum_{\alpha, i=1}^{3+\n} m_\alpha^2 m_i^2 m_\beta m_j 
\VLa_{\alpha i}\, \VLac_{\beta i}\, \VLac_{L\alpha j}\, \VRa_{\beta j}\,,
\end{aligned}
\end{equation}
that is, in terms of the other rephasing invariants, trilinears and quartets (see~\cref{sec:reph-inv}). 
We stress that the sum runs over all quarks, \ac{SM} and vector-like, involving the whole $(3+\n)\times (3+\n)$ mixing matrices $\VL$, $\VR$, $\Fud$.
In the next section, it will be shown in the case of one vector-like doublet 
(and in an example with two vector-like doublets)
that there exists a direct connection between the form of the weak-basis invariants and the effective rephasing invariants studied in~\cref{sec:reph-inv-eff}, which involve only the standard model quarks. 
In particular, it will be shown that the invariants built with three Hermitian matrices (of the type of~\cref{eq:CPo3blocks}) only involve bilinears $\hbi_{\alpha i}$, where $\alpha=u,c,t$ and $i=d,s,b$.

We may, of course, obtain higher-order \acp{WBI}, in general of mass dimension $\M=2\left(n+k+1\right)$, by choosing different values of $n$ and/or $k$. This means that, apart from the single 3-block \ac{WBI} for $\M=6$, we can obtain two \ac{WBI} for $\M=8$, three for $\M=10$, four for $\M=12$ and so on.
In each one of those cases we could also make use of~\cref{eq:LRrelation1} to write different but equivalent expressions.

Other distinct structures emerge when we consider \acp{WBI} with four Hermitian blocks of the form
\begin{equation} \label{eq:CPo-4blocks}
\begin{aligned}
& \im\tr\Big[\HuL^m \,
\HdL^\ell \,
\HuL^n \, \C \Big]=
\sum_{\alpha, i=1}^{3+\n} m_\alpha^{2m+1} m_i^{2\ell} m_\beta^{2n+1}  \, 
\im\left(\VLa_{\alpha i}\, \VLac_{\beta i} \, \Fua_{\beta\alpha} \right)\,,
 \\
& \im\tr\Big[\HdL^m \,
\HuL^\ell \,
\HdL^n \, \C \Big]= 
 \sum_{\alpha, i=1}^{3+\n} m_i^{2m+1} m_\alpha^{2\ell} m_j^{2n+1}  \, 
\im\left(\VLac_{\alpha i}\, \VLa_{\alpha j} \, \Fda_{ji} \right)\,,
\end{aligned}
\end{equation}
where $n\neq m$ in order to get a CP-odd invariant. The invariants of these type with lowest mass dimension are the two $\M=10$ \acp{WBI} obtained for $m=\ell=1$ and $n=2$. All these invariants contain now the imaginary part of the rephasing invariants $\VLa_{\alpha i} \VLac_{\alpha j} \Fda_{ij}$ and $\VLa_{\alpha i} \VLac_{\beta i} \,\Fua_{\beta\alpha} $ (trilinears) in their expression with the fewest mixing matrix insertions. 
The trilinears involve three quarks and are related to CP violation in flavour-changing neutral currents (see~\cref{sec:phenotrilinears}).
Using the relation in~\cref{eq:LRrelation1}, also 
these \acp{WBI} can be expressed with more insertions of mixing matrices
similarly to \cref{eq:wbi-ways},
so that
larger rephasing invariants appear, in particular quartets and quintuplets $\VLa_{\alpha i}\, \VLac_{\beta i} \,\VLa_{\beta j}\, \VLac_{\alpha k} \Fda_{jk}$ (and similarly interchanging up and down quarks).
The content of the weak-basis invariants will become clearer in the next section in terms of the effective rephasing invariants described in~\cref{sec:reph-inv-eff}.
We will show that in the case of one vector-like doublet 
the invariants built with four Hermitian blocks (of the type of~\cref{eq:CPo-4blocks}) involve bilinears $\hbi_{\alpha i}$ and trilinears
$\htr_{i,\alpha\beta}$ (or $\htr_{\alpha, ij}$),
where the indices only refer to \ac{SM} quarks $\alpha=u,c,t$, $i=d,s,b$.

A sixth invariant of mass dimension $10$ emerges in the form with five Hermitian blocks:
\begin{equation} \label{eq:CPo-5blocks}
  \im\tr\Big[\HuL^m\, \HdL^\ell\,
\HuL^n \,\HdL^k\, \C \Big] 
=\sum_{\alpha, i=1}^{3+\n} m_\alpha^{2m+1} m_i^{2\ell} m_\beta^{2n} m_j^{2k+1} \; 
\im\left(\VLa_{\alpha i}\, \VLac_{\beta i}\, \VLa_{\beta j}\, \VRac_{\alpha j}\right)\,,
\end{equation}
where $n=m=\ell=k=1$ at mass dimension $10$.
This invariant contains in its simplest expression the rephasing-invariant quartet
$\VLa_{\alpha i}\, \VLac_{\beta i}\VLa_{\beta j}\VRac_{\alpha j}$ involving four quarks. 
By doing the same exercise of expressing the matrix $\C$ in different ways in the mass basis with more insertions of mixing matrices, as in~\cref{eq:wbi-ways}, one can find again larger rephasing invariants,
including the sextet $\VLa_{\alpha i}\, \VLac_{\beta i}\, \VLa_{\beta j}\,\VLac_{\delta j}\, \VLac_{\alpha k}\, \VRa_{\delta k}$.
We will show that in the case of one vector-like doublet,
the invariants built with five Hermitian blocks are given by combinations of bilinears $\hbi_{\alpha i}$, trilinears
$\htr_{i,\alpha\beta}$, $\htr_{\alpha, ij}$
and the quartets $\hqu_{\alpha i \beta j}$ defined in~\cref{sec:reph-inv-eff}, involving only \ac{SM} quarks, $\alpha=u,c,t$, $i=d,s,b$.

At $\M=12$, apart from the \acp{WBI} of~\cref{eq:CPo3blocks,eq:CPo-4blocks,eq:CPo-5blocks}
containing the $\C$ block, we have
the \ac{SM}-like \ac{WBI} built with four Hermitian blocks which is not vanishing in the \ac{SM} limit:
\begin{equation} \label{eq:SMlikeCPinv}
    \im\tr\Big[\HuL^2\, \HdL^2 \, \HuL \, \HdL \Big] 
=\sum_{\alpha, i=1}^{3+\n}m_\alpha^4 m_i^4m_\beta^2m_j^2 \, \im\left( \VLa_{\alpha i} \VLac_{\beta i} \VLa_{\beta j}\VLac_{\alpha j} \right)\,,
\end{equation}
which contains the quartet of mixing elements of the left unitary matrix.

From the results found in this section we can already expect that the CP-violating content embedded in the invariants depends on both the mass dimension and the number of Hermitian blocks from which it is built. 
For instance, the \ac{WBI}
of mass dimension $6$ is built using three Hermitian matrices. The \acp{WBI} of mass dimension $8$, which are also built with three blocks, bring different powers of the quark masses but share the same structure of the invariant of mass dimension $6$,
whereas the invariants of mass dimension $10$ built from four or five blocks also bring new types of rephasing invariants.
We will uncover and analyze this connection in detail in~\cref{sec:CPwbi-eff}.
We will use the notation $\text{I}(\M,k)$ to identify each CP-odd weak-basis invariant, by the mass dimension $\M$ and by the number of blocks $k$ (e.g.~the 3-block \ac{WBI} with $\M=6$ is dubbed $\text{I}(6,3)$).

By proceeding to higher mass dimension, 
weak-basis invariants with the same number of blocks will manifest the same 
expressions of rephasing invariants and mass eigenvalues, differing only by the powers of the mass eigenvalues. We will show in~\cref{sec:CPconservation} that the invariants listed in this subsection are sufficient to describe CP violation in the presence of one doublet.

\subsubsection{For more than one \acs{VLQ} doublet}

New structures and new rephasing invariants will appear when the \ac{WBI} is built using a larger number of Hermitian blocks.
For instance, at mass dimension $\M = 14$, \ac{WBI}s can be built using six Hermitian blocks:
\begin{equation} \label{eq:CPo-6blocks}
\begin{aligned}
&  \im\tr\Big[\HuL^m\, \HdL^\ell\,
\HuL^n \,\HdL^p\,\HuL^r\, \C \Big] 
 =\sum_{\alpha, i=1}^{3+\n} m_\alpha^{2m+1} m_i^{2\ell} m_\beta^{2n} m_j^{2p} m_\delta^{2r+1} \; 
\im\left(\VLa_{\alpha i}\, \VLac_{\beta i}\, \VLa_{\beta j}\,\VLac_{\delta j}\, \Fua_{\delta\alpha }\right)\,,
\\
&  \im\tr\Big[\HdL^m\, \HuL^\ell\,
\HdL^n \,\HuL^p\,\HdL^r\, \C \Big] 
 =\sum_{\alpha, i=1}^{3+\n} m_i^{2m+1} m_\alpha^{2\ell} m_j^{2n} m_\beta^{2p} m_k^{2r+1} \; 
\im\left(\VLac_{\alpha i}\, \VLa_{\alpha j}\, \VLac_{\beta j}\,\VLa_{\beta k}\, \Fda_{ki}\right)\,,
\end{aligned}
\end{equation}
with $m\neq r$. Similarly other invariants can be constructed at higher mass dimension.
We can also mention that by adding the pair of blocks $\HuL^m\HdL^n$
to the \ac{SM}-like weak-basis invariant of~\cref{eq:SMlikeCPinv} one finds the rephasing invariants built only with the left-handed mixing matrix $\VL$. For instance, at mass dimension $\M=16$ one finds
\begin{equation} \label{eq:CPo-6blocksSM}
    \im\tr\Big[\HuL^2\, \HdL^2 \, \HuL \, \HdL
    \, \HuL \, \HdL\Big] 
=\sum_{\alpha, i=1}^{3+\n}m_\alpha^4 m_i^4m_\beta^2m_j^2 m_\delta^2 m_k^2 \, \im\left( \VLa_{\alpha i} \VLac_{\beta i} \VLa_{\beta j}\VLac_{\delta j} \VLa_{\delta k}\VLac_{\alpha k} \right)\,.
\end{equation}

In the presence of more than one vector-like quark doublet, at mass dimension $\M=8$ one also encounters other types of \acp{WBI}, such as:
\begin{equation}\label{eq:more-trilinear}
\begin{aligned}
     \im\tr\Big[\HuL\, \HdL \, \C^2 \Big] & =
        \im\tr\Big[\, \Du^{3} \VL \Dd^{3}\VR^\dagger \Du^{2} \Fu \,\Big]
   = \sum_{\alpha,\beta , i=1}^{3+\n} m_\alpha^{3}m_i^{3} m_\beta^2 \, \im \left(\VLa_{\alpha i}\, \VRac_{\beta i} \Fua_{\beta\alpha}\right)
    \\ 
    &=\im\tr\Big[\, \Du^{3} \VL \Dd^{3} \Fd \Dd^{2} \VR^\dagger \,\Big]
   = \sum_{\alpha, i, j=1}^{3+\n} m_\alpha^{3}m_i^{3} m_j^2 \, \im \left(\VLa_{\alpha i}\, \VRac_{\alpha j} \Fda_{ij}\right)\,,
\end{aligned}
\end{equation}
where other trilinear rephasing invariants appear. 

Starting from mass dimension $\M=10$, one can also construct \acp{WBI} using the product
$\HuL \C \HdL$, which results in rephasing invariants featuring the entries of $\VR$ instead of $\VL$:
\begin{equation} \label{eq:wbi-rh}
\begin{aligned}
    \im\tr\Big[\HuL \C \HdL \, \C^2 \Big] 
    & =\im\tr\Big[\, \Du^{4} \VR \Dd^{4}\VR^\dagger \Du^{2} \Fu \,\Big]
   = \sum_{\alpha,\beta , i=1}^{3+\n} m_\alpha^{4}m_i^{4} m_\beta^2 \, \im \left(\VRa_{\alpha i}\, \VRac_{\beta i} \Fua_{\beta\alpha}\right)
    \\ &  =\im\tr\Big[\, \Du^{4} \VR \Dd^{4} \Fd \Dd^{2} \VR^\dagger \,\Big]
   = \sum_{\alpha, i, j=1}^{3+\n} m_\alpha^{4}m_i^{4} m_j^2 \, \im \left(\VRa_{\alpha i}\, \VRac_{\alpha j} \Fda_{ij}\right)\,.
\end{aligned}
\end{equation}
Larger rephasing invariants containing the mixings of the right-handed sector emerge by building \acp{WBI} of higher mass dimension with the Hermitian matrix $\C$:
\begin{equation}
\begin{aligned}
    &  \im\tr\Big[\HuL \HdL \, \C^3\Big] 
      \\ &
    =\im\tr\Big[\, \Du^{3} \VR \Dd^{3}\VR^\dagger \Du^{2} \Fu \Du^{2} \Fu \,\Big]
   = \sum_{\alpha,\beta,\delta , i=1}^{3+\n} m_\alpha^{3}m_i^{3} m_\beta^2 m_\delta^2 \, \im \left(\VLa_{\alpha i}\, \VRac_{\beta i} \Fua_{\beta\delta} \Fua_{\delta\alpha}\right)
    \\ &
    =\im\tr\Big[\, \Du^{3} \VL \Dd^{3} \Fd \Dd^{2} \Fd \Dd^{2} \VR^\dagger \,\Big]
   = \sum_{\alpha, i, j,k=1}^{3+\n} m_\alpha^{3}m_i^{3} m_j^2 m_k^2 \, \im \left(\VLa_{\alpha i}\,  \Fda_{ij} \Fda_{jk} \VRac_{\alpha k} \right)
       \\ &
    =\im\tr\Big[\, \Du^{3} \VL \Dd^{3} \Fd \Dd^{2} \VR^\dagger \Du^{2} \Fu   \,\Big]
   = \sum_{\alpha,\beta, i, j=1}^{3+\n} m_\alpha^{3}m_i^{3} m_j^2 m_\beta^2 \, 
   \im \left(\VLa_{\alpha i}\, \Fda_{ij}\, \VRac_{\beta j}  \Fua_{\beta\alpha}\right)
         \\ &
    =\im\tr\Big[\, \Du^{3} \VL \Dd^{3} \VR^\dagger \Du^{2} \VR \Dd^{2}   \VR^\dagger   \,\Big]
   = \sum_{\alpha,\beta, i, j=1}^{3+\n} m_\alpha^{3}m_i^{3} m_j^2 m_\beta^2 \, 
   \im \left(\VLa_{\alpha i}\,  \VRac_{\beta i} \VRa_{\beta j}  \VRac_{\alpha j} \right)\,.
\end{aligned}
\end{equation}
Note that the different expressions in this equation all appear with the same minimal number of insertion of mixing matrices.

Let us remark that in the case of only one vector-like doublet, $\n=1$, the rephasing invariants (and weak-basis invariants) containing only matrix elements related to the right-handed sector as in~\cref{eq:wbi-rh} are real (e.g.~$\VRa_{\alpha i}\, \VRac_{\beta i} \Fua_{\beta\alpha}=\Fua_{\alpha\alpha}\Fua_{\beta\beta}\Fda_{ii}$ and $ \VRa_{\alpha i}\, \VRac_{\alpha j} \Fda_{ij}=\Fda_{ii}\Fda_{jj}\Fua_{\alpha\alpha}$ in~\cref{eq:wbi-rh}), whereas weak-basis invariants built with more than one $\C$ block can be written in terms of weak-basis invariants with lower mass dimension
containing only one power of $\C$ as shown in~\cref{eq:tr1VLQ,eq:tr1VLQ_2,eq:tr1VLQ_3} (e.g.~the trilinears in~\cref{eq:more-trilinear} reduce to the bilinear form times a modulus: $\im(\VLa_{\alpha i}\VRac_{\alpha i}) \Fua_{\beta\beta}$ and $\im(\VLa_{\alpha i}\VRac_{\alpha i}) \Fda_{jj}$ respectively).
Similar situations may happen in specific scenarios in presence of more vector-like doublets, e.g. if the row vectors $\hi_{n\alpha(i)}$
are orthogonal among themselves.

\subsection{Effective CP-odd \acsp{WBI}}
\label{sec:CPwbi-eff}

In this section we provide an ``effective'' description of CP-violating weak-basis invariants, in particular for $\n=1$.
Namely, we express the invariants in terms of the couplings involved in \ac{SM} quark interactions, 
which are connected to the $3\times 3$
submatrices of $\VL$, $\VR$, $\Fd$ and $\Fu$
describing the coupling of \ac{SM} quarks to $W$, $Z$ and Higgs bosons.
We show that in this way the information carried by weak-basis invariants can be more easily understood and connected to observables.
Moreover it is possible to find a set of CP-odd weak-basis invariants which can give a complete description of the CP-violating effects in presence of one vector-like doublet mixing with the \ac{SM} quarks.

This analysis is possible by taking into account that the ratio $\vw/\mB$ is a small parameter.
As described in~\cref{sec:app1}, 
masses and mixings can be expressed in terms
of Lagrangian parameters, more precisely the parameters in~\cref{eq:z}, by an expansion in $\vw/\mB$.
Then, these relations can be used to write 
expressions for the weak-basis invariants,
which are simply related to both the minimal set of Lagrangian parameters and mass-basis quantities, and ultimately, to observables.
In particular, for CP-odd weak-basis invariants
this expansion also allows to clarify the content 
of weak-basis invariants
in terms of rephasing-invariant quantities involving only the three \ac{SM} quarks. 
We illustrate the results and leave the details in~\cref{app:CP}.

Let us consider first CP-odd invariants of the form:
\begin{equation}
\text{I}(\M,3)\equiv \im\tr[\HuL^n\HdL^m \C]\,,
\end{equation}
where we used $\text{I}(\M,3)$ to denote the family of invariants with the mass dimension $\M=2(n+m+1)$ and built from 3 Hermitian blocks.
Using the expressions of mass eigenvalues and mixing
elements found in~\cref{sec:app1} we find (see~\cref{app:CP} for details):
\begin{equation} \label{eq:inv3bl}
\frac{1}{ \mB^{2(m+n+1)}}\im \tr\Big[\HuL^n \HdL^m \C\Big] 
=
\frac{\vw^4}{\mB^4}  \sum_{i,\alpha=1}^3
\im \left( \,\Vha_{\alpha i} \, \hi_\alpha \hi_i^* \right) \;   
y_\alpha \, y_i \:  
\left(1 +  k^{(m,n)}_{\alpha i}\, \right) \, ,
\end{equation}
where $k^{(m,n)}_{\alpha i}$ stand for real corrections of 
order $\vw^2/\mB^2$ or higher. 
CP violation is carried by the imaginary part of the rephasing-invariant quantities
\begin{equation}
   \frac{\vw^2}{\mB^2} \Vha_{\alpha i} \, \hi_\alpha \hi_i^* = \Vha_{\alpha i} \, \VRhac_{\alpha i} 
    =  \hbi_{\alpha i}^* 
    \approx
   \VLa_{\alpha i} \, \VRac_{\alpha i} \,. 
\end{equation}
In the general case, all couplings $\hi_\alpha,\hi_i$ are different from zero, and thus all physical phases would appear in this class of invariants (see~\cref{sec:reph-inv-eff}).
Even if the new physics only brings a real contribution, in the sense that only one phase survives in the model (e.g.~if all $\hi_\alpha,\hi_i$ were real in this parameterization), the CP-violating effect of the remaining phase would appear in these invariants.
If the condition 
$\im \big(\Vha_{\alpha i} \, \hi_\alpha \hi_i^* \big)=0$
held for each $\alpha,i$ (or for six independent combinations)
then there would be no CP violation.
These conditions change when some couplings are vanishing.
In this case, not all physical phases emerge in bilinears and
rephasing-invariant trilinears and quartets become essential in the description of the CP-violating effects.
In fact, in a situation in which one or more $\hi_{\alpha(i)}$ are zero, 
the rephasing-invariant trilinears and quartets $\htr_{i,\alpha\beta}$, $\htr_{\alpha,ij}$, $\hqu_{\alpha i\beta j}$ (see~\cref{eq:reph-inv-eff}) cannot be written in terms of bilinears $\hbi_{\alpha i}$
as shown in~\cref{eq:f-bilinears}
and they become non-redundant. 
As a consequence of this observation, we can already anticipate that in these scenarios the vanishing of the imaginary part of invariants containing only the bilinears will not suffice as conditions for CP conservation.

For completeness, let us notice that similar expansions hold for the real part of the same invariants. In first approximation, one has
\begin{equation}
\begin{aligned}
 \frac{1}{\mB^{2(n+m+1)}}\tr\Big[\HuL^n\,\HdL^m\, \C \Big] &\approx  1+\frac{\vw^2}{\mB^2}\sum_{\alpha i}(n|\hi_\alpha|^2+m|\hi_i|^2)\\
 & =1+n \sum_\alpha\Fuha_{\alpha\alpha}+m\sum_i\Fdha_{ii} \,.
 \end{aligned}
\end{equation}
If only two couplings are switched on (i.e.~only $\hiu,\his\neq 0$), this means that, at first order 
\begin{equation} \label{eq:effwbione}
\begin{aligned}
    \frac{1}{\mB^6}\tr\Big[\HuL\,\HdL\, \C \Big] & \approx
    1+\Fuha_{uu}+\Fdha_{ss} \, ,\\
    \frac{1}{\mB^6}\im\tr\Big[\HuL\,\HdL\, \C \Big] & \approx
    \frac{\vw^2}{\mB^2} \ y_u \,y_s \im\left(\Vha_{us}\,\VRhac_{us}\right)\,.
\end{aligned}     
\end{equation}

The next type of CP-odd invariants we can consider
are of the form:
\begin{equation}
\begin{aligned}
& \text{I}(\M,4)_d\equiv \im\tr[\HdL^m \HuL^\ell \HdL^n \C]\,,
\\[2mm]
& \text{I}(\M,4)_u\equiv \im\tr[\HuL^m \HdL^\ell \HuL^n \C]\,,
\end{aligned}
\end{equation}
with $\M=2\left(n+m+\ell+1\right)$.
Taking $n>m$, we have 
\begin{equation} \label{eq:inv4bl}
\begin{aligned}
& \frac{1}{\mB^{2(m+n+\ell+1)}} \im \tr \left[\HdL^m
\HuL^\ell
\HdL^n \C\right]  \\
&= 
\sum_{i,j,\alpha,\beta=1}^3 \bigg[\,
  \frac{\vw^{4+2m}}{\mB^{4+2m}} \,
  \im \left( \, \Vha_{\alpha i} \, \hi_\alpha \hi_i^* \,\right)\,  y_\alpha\, y_i^{2m+1} \,
 \left( 1 + k^{(n,m,\ell)}_{i\alpha}
  \right) 
    \\ & \hspace{1.6cm} +
   \frac{\vw^{4+2m+2\ell}}{\mB^{4+2m+2\ell}} \,
   \im \left( \, \Vha_{\alpha i} \Vhac_{\alpha j} \, \hi_i^* \hi_j \,\right)\, y_i^{2m+1}\, y_j\, y_\alpha^{2\ell} \left( 1+k^{(n,m,\ell)}_{\alpha ij} 
   \right)  
       \\ & \hspace{1.6cm}
     +   \frac{\vw^{8+2m}}{\mB^{8+2m}} \,
     \im \left( \,
     \Vha_{\alpha i} \hi_i^* \hi_\alpha \Vhac_{\beta j}\,  \hi_j  \hi_\beta^*  \right) 
     \, y_i^{2m+1} \,y_j \,y_\alpha\, y_\beta
        \left(1+ k^{(n,m,\ell)}_{ij\alpha\beta}   
        \right) 
        \bigg]\,,
\end{aligned}
\end{equation}
and similarly for the second invariant with the exchange of the down and up sectors.
These invariants contain the rephasing-invariant quantities involving three quarks
associated with flavour-changing neutral currents:
\begin{equation}
\begin{aligned}
    \frac{\vw^2}{\mB^2}\,\hi_i^* \hi_j \Vha_{\alpha i} \Vhac_{\alpha j} 
    &=\Fdha_{ij} \Vha_{\alpha i} \Vhac_{\alpha j}
    =\htr_{\alpha,ij} \approx 
  \VLa_{\alpha i} \VLac_{\alpha j} \Fda_{\, ij}
  \, ,  \\ 
   \frac{\vw^2}{\mB^2} \, \hi_\alpha \hi_\beta^*  \Vha_{\alpha i} \Vhac_{\beta i} &= \Fuhac_{\alpha\beta}\Vha_{\alpha i} \Vhac_{\beta i} = \htr_{i,\alpha\beta}^* \approx
 \VLa_{\alpha i} \VLac_{\beta i} \, \Fua_{\, \beta\alpha}\,.
\end{aligned}
\end{equation} 
The last term of~\mbox{\cref{eq:inv4bl}} involves the quantities of the type $\Vha_{\alpha i}\Vhac_{\beta j} \hi_i^* \hi_j \hi_\alpha \hi_\beta^*$ which can be expressed in terms of the rephasing invariants $\Vha_{\alpha i}\hi_\alpha \hi_i^*$.

Lastly, we can consider the 5-block weak-basis invariants
\begin{equation}
    \text{I}(\M,5)\equiv     \im \tr[\HuL^n \HdL^k \HuL^\ell \HdL^m \C] \,,
\end{equation}
with $\M=2\left(n+k+m+\ell+1\right)$.
Besides the leading-order terms proportional to $\Vha_{\alpha i} \, \hi_\alpha \hi_i^*$
and those associated with flavour-changing neutral currents $\Vha_{\alpha i} \Vhac_{\beta i} \, \hi_\alpha \hi_\beta^*$, 
$ \Vhac_{\alpha i} \Vha_{\alpha j} \, \hi_i \hi_j^*$, these invariants
contain the rephasing-invariant quantities 
involving four quarks
\begin{equation}
 \frac{\vw^2}{\mB^2} \, \hi_\beta \hi_j^*\Vhac_{\alpha i} \Vha_{\beta i} \Vha_{\alpha j} =\VRhac_{\beta j}\Vhac_{\alpha i} \Vha_{\beta i} \Vha_{\alpha j}=
    \hqu_{\beta j \alpha i}^* \approx 
 \VRac_{\beta j} \VLac_{\alpha i} \VLa_{\beta i} \VLa_{\alpha j}   \, .
\end{equation}
For instance, for the \ac{WBI} of mass dimension $\M = 10$,
we can write (see~\cref{app:CP} for details):
\begin{equation} \label{eq:5blockintext}
\begin{aligned}
&\frac{1}{\mB^{10}}\im \tr\left[\HuL \HdL
\HuL \HdL \C\right]
 \\ & \hspace{1cm} = 
     \frac{\vw^4}{\mB^4} \sum_{i,\alpha=1}^3 \im \Big( \Vha_{\alpha i} \, \hi_\alpha \hi_i^* \, \Big)  \, y_\alpha \, y_i   
 \left( 1 + k^{(10\text{a})}_{\alpha i}  
     \right)
          \\ & \hspace{1cm}
   +  \frac{\vw^8}{\mB^8} \sum_{i,j,\alpha,\beta=1}^3 
\im \Big( \Vha_{\alpha j} \Vha_{\beta i}  \Vhac_{\beta j}  \, \hi_\alpha \hi_i^* \, \Big) \, y_\beta^2 \,  y_\alpha\,  y_j^2\,  y_i \,  
   \left(1+k^{(10\text{b})}_{\alpha\beta i j} \right)  +
\mathcal{O}\left( \frac{\vw^{12}}{\mB^{12}} \right)\,.
\end{aligned}
\end{equation}
Finally we can consider the \ac{SM}-like weak-basis invariant 
shown in~\cref{eq:SMlikeCPinv} which
is built with four Hermitian building blocks without the insertion of $\C=\Mb \Mb^\dagger$ in~\cref{eq:SMlikeCPinv}.
This invariant contains, besides bilinears, trilinears and quartets, the SM-like quartet 
\begin{equation}
    \Qh_{\alpha i \beta j} = \Vha_{\alpha i} \Vha_{\beta j} \Vhac_{\alpha j} \Vhac_{\beta i}\,,
\end{equation}
recall~\cref{eq:Qhat}. Note that if we consider
the
rephasing invariant extracted solely from the LH mixing,
$J=\im(\VLa_{ud}\VLa_{cs}\VLac_{us}\VLac_{cd})$
(see~\cref{eq:Jar}), then we have 
\begin{equation}
    |\im \Qh_{\alpha i \beta j}|\approx|\im \VLa_{\alpha i}\VLac_{\beta i}\VLa_{\beta j}\VLac_{\alpha j}| \approx |J|\,.
\end{equation}

We can conclude from the result of this section that there is a unique connection between weak-basis invariants and rephasing invariants. Weak-basis invariants built with three Hermitian blocks only contain bilinears, while trilinears appear in 4-block invariants. The non-standard quartets appear in 5-block invariants. The \ac{SM}-like quartet explicitly emerges in the \ac{SM}-like \ac{WBI} of~\cref{eq:SMlikeCPinv}.

\subsection{Conditions for CP invariance}
\label{sec:CPconservation}

We can now study the weak-basis-independent conditions for \ac{CPI} in the presence of one vector-like doublet.
If the imaginary part of any weak-basis invariant is different from zero, then at least one physical phase is present in the model, inducing CP violation. The conditions for \ac{CPI} can be obtained by imposing that all CP-odd weak-basis invariants vanish, i.e.
\begin{equation}
\text{I}(\M,k)=0\,.
\end{equation}
More precisely, we want to find a set of invariants which may completely describe the CP properties of the model. Then, 
the vanishing of these invariants would also imply CP conservation.
We assume non-vanishing and non-degeneracy of quark masses.
The results obtained in~\cref{sec:CPwbi-eff} provide a simple guide in the identification of the set of invariants relevant in different scenarios, depending on the mixing of the \ac{VLQ} doublet with the standard quarks.
In the following, we illustrate the choice of weak-basis invariants as based on these results.
A more rigorous proof is given in~\cref{app:CPI}.

We summarize our results in~\Cref{tab:cp-odd-inv}. In the first two columns of this table the number of couplings between \ac{SM}-quarks and the extra \ac{VLQ} doublet which are switched on are indicated. In the third column we write the number of physical phases in each case. For each scenario, we indicate the type and number of weak-basis invariants which can provide a complete characterization of the CP properties of the model and whose vanishing would correspond to CP conservation. 
The entries with zeroes indicate that the imaginary part of the corresponding invariant vanishes in that scenario.
Note that if the \ac{VLQ} doublet is decoupled, or exhibits only a non-vanishing coupling with one generation in only one sector, then the sole non-zero CP-odd invariant is the usual \ac{SM}-like
invariant for $3+\n$ generations, with mass dimension $\M=12$ and built from 4 blocks (see~\cref{eq:SMlikeCPinv}).
The blank space indicates that the CP-odd \acp{WBI} is non-zero, but the condition is redundant.
In fact, the set of invariants indicated in the table is not minimal, as we use the most convenient set of \acp{WBI} in each scenario.

\begin{table}[t]
\renewcommand{\arraystretch}{1.4}
\centering
\begin{tabular}{ c c  cccccc}
    \toprule
\belowrulesepcolor{light-gray}
\rowcolor{light-gray}
\#$\hi_i$ & \#$\hi_{\alpha}$ & \#$\delta$ & \#$\text{I}(\M\geq 6,3)$ & \#$\text{I}(\M\geq 8,4)_{d}$  & \#$\text{I}(\M\geq 8,4)_{u}$ & $\text{I}(10,5)$ & $\text{I}^{\text{\tiny{SM-like}}}_{(12,4)}$  \\
\aboverulesepcolor{light-gray}
    \midrule
  $3$ & $3$ & $6$ & $9$ &  &  &  &   \\
\rowcolor{lighter-gray}
$3$ & $2$ & $5$ & $7$ &  &  & &   \\
\rowcolor{lighter-gray}
$2$ & $3$ & $5$ & $7$ &  &  & &   \\
  $2$ & $2$ & $4$ & $4$ &  &  &  &  \\
\rowcolor{lighter-gray}
  $3$ & $1$ & $4$ & 3 & 1 &  &  &  \\
\rowcolor{lighter-gray}
  $1$ & $3$ & $4$ & 3 &  & 1 &  &  \\
  $2$ & $1$ & $3$ & 2 & 1 &  &  &   \\
  $1$ & $2$ & $3$ & 2 &  & 1 &  &   \\
\rowcolor{lighter-gray}
  $3$ & $0$ & $3$ & $=0$ & 6 & $=0$ &  $=0$ & \\
\rowcolor{lighter-gray}
  $0$ & $3$ & $3$ & $=0$ & $=0$ & 6 &  $=0$ & \\
  $2$ & $0$ & $2$ & $=0$ & 2 & $=0$ &  $=0$ & \\
  $0$ & $2$ & $2$ & $=0$ & $=0$ & 2 &  $=0$ & \\
\rowcolor{lighter-gray}
  $1$ & $1$ & $2$ & 1 &  &  & 1  &   \\
  $1$ & $0$ & $1$ & $=0$ & $=0$ & $=0$ & $=0$  & 1  \\
  $0$ & $1$ & $1$ & $=0$ & $=0$ & $=0$ & $=0$  & 1  \\
\aboverulesepcolor{lighter-gray}
    \bottomrule
\end{tabular}
\caption{CP-odd weak-basis invariants needed for CP conservation in different scenarios, in the presence of one vector-like doublet, assuming non-vanishing quark Yukawas with a SM-like hierarchy (see the text for more details).}
\label{tab:cp-odd-inv}
\end{table}
\renewcommand{\arraystretch}{1}

\paragraph{All $\boldsymbol{\hi_i\neq 0}$, $\boldsymbol{\hi_\alpha \neq 0}$.}
Let us start by the most generic case in which all couplings are allowed to be different from zero.
In this scenario, all the elements of the matrices $\VR$, $\Fud$ are non-vanishing and right-handed charged and neutral currents, both flavour-changing and flavour-conserving, are generated between \ac{SM} quarks.
In this case, the \acp{WBI} with three blocks in~\mbox{\cref{eq:inv3bl}}
are linear combinations of the 9 bilinears $\im\big( \Vha_{\alpha i} \hi_\alpha \hi_i^* \big)$.
The 9 phases $\arg\big( \Vha_{\alpha i} \hi_\alpha \hi_i^* \big)$ are not independent,
but in this scenario all the 6 physical phases emerge in these combinations.
Moreover, the vanishing of all the $\im\big( \Vha_{\alpha i} \hi_\alpha \hi_i^* \big)$ for each $\alpha,i$
implies the vanishing of all physical phases, that is, CP invariance.
In fact, as we have found in~\cref{sec:stepladderwb} using the ``stepladder'' weak basis,
the sufficient and necessary conditions for \ac{CPI} in this scenario
correspond to the vanishing of the imaginary part of 9 \acp{WBI} built from three Hermitian blocks.

\paragraph{One $\boldsymbol{\hi_{\alpha(i)}=0}$.}
In this scenario, one row or one column of $\VR$ is zero, as well as one row and one column of $\Fu$ or $\Fd$.
There are 5 physical phases. 
Since there are 6 bilinears $\im\big( \Vha_{\alpha i} \hi_\alpha \hi_i^* \big)$ 
the 5 independent physical phases can be chosen among the 6 phases of the bilinears.
Hence, the situation is similar to the previous case, with 3-block \acp{WBI} containing the information on CP violation.
It can be shown using the stepladder \ac{WB} that the vanishing of the imaginary part of 7 \acp{WBI} with three blocks are the necessary and sufficient conditions for CP conservation in this scenario (see~\cref{app:CPI}).

\paragraph{One $\boldsymbol{\hi_{i}=0}$ and one $\boldsymbol{\hi_{\alpha}=0}$.}
In this scenario, one row and one column of $\VR$ are vanishing, as well as one row and one column of $\Fu$ and $\Fd$.
There are 4 physical phases and 4 combinations $\im\big( \Vha_{\alpha i} \hi_\alpha \hi_i^* \big)$.
Also in this case, the conditions $\im\big( \Vha_{\alpha i} \hi_\alpha \hi_i^* \big)=0$ for each $\alpha,i$
imply that all physical phases are zero.
This can be seen e.g.~by choosing the elements $ \Vha_{\beta j}$ of a row $\beta\neq \alpha$ and a row $j \neq i$ to be real
and also $\hi_\beta$ real 
(one column and one row of $\VLh$ can always be made real, together with one of the couplings $\hi_{\alpha(i)}$)  
and assigning three phases to the other three non-vanishing $\hi$-couplings and the remaining one to the $2\times 2$ submatrix of $\VLh$.
Then these $4$ phases are directly related to the phases of the bilinears in a trivial relation.
Then, also in this scenario, 3-block \acp{WBI} can characterize all the CP properties.
 It can be shown that the vanishing of 4 \acp{WBI} of the type
 $\text{I}(\M,3)=0$
 are indeed necessary and sufficient conditions for \ac{CPI} (see~\cref{app:CPI}).

\paragraph{Two $\boldsymbol{\hi_{\alpha(i)}=0}$ in the same sector.}
In this scenario, $2$ rows or $2$ columns of $\VR$ are zero, as well as $2$ rows and $2$ columns of $\Fu$ or $\Fd$, implying no flavour-changing couplings at tree level between the \ac{SM} quarks in one sector.
There are $4$ physical phases.
However, there are only 3 non-vanishing bilinears $\im\big( \Vha_{\alpha i} \hi_\alpha \hi_i^* \big)$.
Then,  the 3-block \acp{WBI} can be written as a linear combination of these 3 bilinears but
one physical phase is not captured by these invariants (neither by the quartet $\hi_\beta \hi_j^*\Vhac_{\alpha i} \Vha_{\beta i} \Vha_{\alpha j} $).
The remaining phase emerges in the trilinears of the type $\hi_i^* \hi_j \Vha_{\alpha i} \Vhac_{\alpha j}$ 
(for 2 couplings $\hi_{\alpha}=0$ in the up-sector)
or $\hi_\alpha \hi_\beta^*  \Vha_{\alpha i} \Vhac_{\beta i}$ (for 2 couplings $\hi_{i}=0$ in the down-sector).
These trilinears appear in weak-basis invariants with at least 4 Hermitian blocks.
In~\cref{app:CPI} we show that 3 conditions of the type $\text{I}(\M,3)=0$ and one of the type $\text{I}(\M,4)=0$ are necessary and sufficient for \ac{CPI}.

\paragraph{Two $\boldsymbol{\hi_{\alpha(i)}=0}$ in one sector and one $\boldsymbol{\hi_{i(\alpha)}=0}$ in the other sector.}
This scenario is analogous to the previous one.
There are $3$ physical phases and
there are $2$ non-vanishing bilinears $\im\big( \Vha_{\alpha i} \hi_\alpha \hi_i^* \big)$. 
The condition $\text{I}(\M,3)=0$ for two or more different $3$-block invariants would imply the vanishing of the imaginary parts of the $2$ bilinears $\im\big( \Vha_{\alpha i} \hi_\alpha \hi_i^* \big)=0$. 
The remaining phase emerges in the trilinears of the type $\hi_i^* \hi_j \Vha_{\alpha i} \Vhac_{\alpha j}$ 
(or $\hi_\alpha \hi_\beta^*  \Vha_{\alpha i} \Vhac_{\beta i}$).

For instance, by setting $\hic=\hit=0$,
we can choose the entries $\Vha_{ui}$ to be real. We can further choose the first column and $z_u$ to be real. By further setting $\hib=0$, we have, for the $4$-block \ac{WBI},
\begin{equation} 
\im \tr[\HdL
\HuL
\HdL^2 \C]  \propto
\im\left(\Vha_{td}\Vhac_{ts}\right) \,  \hid \his \,y_dy_s\left(y_t^2-y_c^2 \right)     
\left(y_d^2-y_s^2 \right) \,.
\end{equation}
Thus, the additional condition 
$\text{I}(\M,4)_d=0$ 
would imply CP conservation.

\paragraph{Three $\boldsymbol{\hi_{\alpha(i)}=0}$ in one sector.}
In this scenario, the $3\times 3$ submatrix of $\VR$ involving \ac{SM} quarks is vanishing, only its last column or last row survives, as well as one of the two matrices $\Fu$ or $\Fd$ depending on which sector is decoupled.
All bilinears disappear, so that every 3-block \ac{WBI} vanishes.
Still, there are 3 physical phases and these phases appear in the trilinear terms of the form
$\hi_i^* \hi_j \Vha_{\alpha i} \Vhac_{\alpha j} $ or $ \hi_\alpha \hi_\beta^* \Vha_{\alpha i} \Vhac_{\beta i}$.
Assuming for instance the case $\hi_\alpha=0$, $\alpha=u,c,t$ (the down-sector case being analogous), 
for the $4$-block invariants we have ($n>m$):
\begin{equation} \label{eq:inv3bl0}
\im \tr[\HdL^m
\HuL^\ell
\HdL^n \C] \propto
\sum_{i,j,\alpha=1}^3 
   \im \left( \, \Vha_{\alpha i} \Vhac_{\alpha j} \, \hi_i^* \hi_j \,\right)\, y_i^{2m+1} y_j y_\alpha^{2\ell} \left( 1+k^{(n,m,\ell)}_{\alpha ij} 
   \right)\,.  
\end{equation}
It can be shown that the vanishing of $6$ different invariants of this type would
provide the necessary and sufficient conditions for CP conservation in this scenario
(see~\cref{app:CPI}).

\paragraph{Three $\boldsymbol{\hi_{\alpha(i)}=0}$ in one sector and one $\boldsymbol{\hi_{i(\alpha)}=0}$ in the other sector.}
Let us take for illustration the case $\hi_\alpha=0$, $\alpha=u,c,t$ and $\hib=0$.
Then, for instance
\begin{equation}
\begin{aligned}
& \frac{1}{\mB^{10}}  \im \tr[\HdL
\HuL
\HdL^2 \C] 
\\ & 
 \approx \frac{\vw^{8}}{\mB^{8}} 
y_dy_s (y_d^2-y_s^2)\left[ (y_u^2-y_t^2) \im \left( \, \Vha_{ud} \Vhac_{us} \, \hid^* \his \,\right)
+(y_c^2-y_t^2) \im \left( \, \Vha_{cd} \Vhac_{cs} \, \hid^* \his \,\right)
\right]\,.
\end{aligned}
\end{equation}
The vanishing of the 2 physical phases would be signalled by the disappearance of invariants with at least $4$ blocks. In particular, the vanishing of two 4-block invariants (of the relevant type) implies CP conservation in this scenario (see~\cref{app:CPI}).

\paragraph{Two $\boldsymbol{\hi_{\alpha(i)}=0}$ in both sectors.}
In this scenario, only one coupling in each sector survives. There are no flavour-changing neutral currents at tree level between \ac{SM} quarks and only one non-vanishing entry in the $3\times 3 $ submatrix of $\VR$ involving the \ac{SM} quarks.
There are 2 physical phases. One phase appears in the surviving bilinear, and consequently in the 3-block invariants.
Let us assume as an illustration that only $\hiu\neq 0$, $\hid\neq 0$. Then we have
\begin{equation}
\frac{1}{\mB^6}
\im \tr[\HuL \HdL \C]
\propto
\frac{\vw^4}{\mB^4}\im \left( \,\Vha_{ud} \, \hiu \hid^* \right) \,  
y_u \, y_d \,.
\end{equation}
However, the second phase cannot manifest in either 3-block or 4-block invariants. The last phase only emerge in 
quartets, or, in terms of \acp{WBI}, in the $\M = 10$ invariant made up of 5 blocks:
\begin{equation}
\begin{aligned}
&\frac{1}{\mB^{10}}\im \tr[\HuL \HdL
\HuL \HdL \C] 
\approx
     \frac{\vw^4}{\mB^4}  \im \left( \Vha_{ud} \, \hiu \hid^* \, \right)   y_u \,y_d   
     \\ & \hspace{2cm} 
   +  \frac{\vw^8}{\mB^8} 
\im \left( \Vha_{u s} \Vha_{c d}  \Vhac_{c s}  \, \hiu \hid^* \, \right) \, y_u\,y_d\, \left(y_c^2-y_t^2 \right) \left(y_s^2-y_b^2 \right)\,.
\end{aligned}
\end{equation}
When the invariant of mass dimension $\M=6$ is zero, the invariant of mass dimension $10$ is proportional to the imaginary part of the quartet. The vanishing of the $2$ invariants implies CP conservation (in the realistic case of non-degeneracy). Again, we provide an equivalent proof in~\cref{app:CPI}.

\subsection{Motivated scenarios}
We now analyze four specific scenarios. The first two scenarios are relevant from the phenomenological point of view and they will be motivated in~\cref{sec:pheno}.
The last two address the 
extreme chiral limit, i.e.~the limit of extremely high energies, where \ac{SM} quark masses become negligible and CP violation can emerge only from \ac{NP} sources.

\subsubsection{With \texorpdfstring{$\n=1$}{\n=1} doublet}

In some scenarios, the first type of invariants built with three blocks may not be enough to describe the CP violation. As an illustration, let us consider the case in which there are no flavour-changing neutral currents at tree level in one sector (up or down) while in the other sector one vector-like species couples only to two families.
Let us consider a scenario which can be relevant for the Cabibbo angle anomaly related to kaon decays, with $\hiu,\his\neq 0$, in which the mass matrices can be written in the form
\begin{equation} \label{eq:lightmass1}
 {\cal M}_u = \left(\begin{array}{cccc}
\hat{y}_u \vw & 0 & 0 &  0  \\
0 & \hat{y}_u \vw & 0 &  0  \\
0 & 0 &  \hat{y}_t \vw &  0  \\
\hiu \vw & 0 & 0 &  \mB
\end{array}\right)\,,   \qquad 
{\cal M}_d = \left(\begin{array}{cccc}
 &  &  & 0   \\
 & Y_d \vw &  &  0  \\
 &  &   &  0  \\
0 & \his \vw & \hib \vw &  \mB
\end{array}\right)
\end{equation}
(we could consider in an analogous way $\hib= 0$, $\hit\neq 0$). The mixing matrices can be easily inferred from~\cref{eq:vckmR,eq:Fzz,eq:VL}.
In this scenario there still remain three physical phases.
We have in this case
\begin{equation}
\begin{aligned}
& \frac{1}{ \mB^{2(m+n+1)}}
\im \tr\Big[\HuL^n \HdL^m \C\Big] 
\\ & \qquad 
 = \frac{\vw^4}{\mB^4}  
\im \left( \his^* \hiu \Vha_{us} \right) \,
y_uy_s
\left(1    + k_{mn} \right)+
  \frac{\vw^4}{\mB^4}  
\im \left( \hib^* \hiu \Vha_{ub} \right) \,
y_uy_s
\left(1    + k'_{mn} \right)
\\[2mm] & \qquad 
\approx \frac{m_um_s}{\mB^2}\, \im \left( \VRac_{us} \VLa_{us} \right)
+\frac{m_um_b}{\mB^2} \, \im \left( \VRac_{ub} \VLa_{ub} \right)\,.
\end{aligned}
\end{equation}
Let us assume that in this parameterization the rephasing-invariant quantities $\Vha_{us}\VRhac_{us}$ and
$\Vha_{ub}\VRhac_{ub}$ are real. 
For instance, we can write $\VLh$ in the Kobayashi--Maskawa
parameterization (first row and first column real) and consider that $\Vha_{us}$, $\Vha_{ub}$, $\hiu$, $\his$ and $\hib$ are all real parameters.
Then, all the invariants in the above 3-block category would vanish. However, there still is one physical phase
which cannot be caught in these expressions.
The physical phase of the model would show in the other categories of invariants, in particular the $\M=10$ invariant of~\cref{eq:CPo-5blocks}.
The same effect manifests also in the
scenario
in which one doublet couples only to one generation,
e.g.~taking also $\hib=0$ in the previous example. 
In this case there are no flavour-changing neutral currents and all the CP-odd invariants with 3 and 4 building blocks vanish. The one physical phase would show in
the $\M=10$ invariant with 5 blocks and the non-zero imaginary part would result from the rephasing invariants involving four quarks,
\begin{equation}
  \frac{\vw^2}{\mB^2}\, \im\left( \Vhac_{\alpha i} \Vha_{u i} \Vha_{\alpha s} \, \hiu \his^* \right)
  \approx \im\left( \VLac_{\alpha i} \VLa_{u i} \VLa_{\alpha s} \VRac_{us} \right)\,.
\end{equation}

We can also imagine a situation in which the extra doublet only couples to one sector, up or down, that is, 
the $3\times 3$ submatrix of $\VR$ involving \ac{SM} quarks vanishes (only the last column or the last row survives) as well as one of the two matrices, $\Fu$ or $\Fd$, depending on which sector is decoupled.
Then, the 3-block invariants~\eqref{eq:inv3bl}
would vanish.
The three physical CP phases
would be contained in the rephasing-invariant quantities associated with \acp{FCNC},
\begin{equation} \label{eq:FCNCphase}
    \frac{\vw^2}{\mB^2}\im\left(\Vha_{\alpha s} \Vhac_{\alpha b}\, \his^* \hib\right) \approx 
 \im \left(  \VLa_{\alpha s} \VLac_{\alpha b}\, \Fda_{sb}\right) \,,
\end{equation}
appearing in the 4-block \acp{WBI}~\eqref{eq:inv4bl}.
In particular, these non-vanishing quantities would indicate a contribution to CP violation, for instance, in neutral meson mixing in interference with the \ac{SM}.

\subsubsection{With \texorpdfstring{$\n=2$}{\n=2} doublets}

Let us consider a scenario in which two vector-like doublets of quarks couple only with the light generations.
More specifically, let us imagine a specific pattern of couplings, with the doublets coupling only to the first generation in the up-sector, while coupling either with the first or the second in the down-sector.
This texture will be phenomenologically motivated in~\cref{sec:pheno} in the context of Cabibbo angle anomalies.
The mass matrices can be written as:
\begin{equation} \label{eq:mass-2doubl}
\begin{aligned} 
\Mu = \left(\! \begin{array}{c@{\hspace{-0.1\tabcolsep}} c@{\hspace{-0.1\tabcolsep}} cc@{\hspace{1\tabcolsep}}c}
 &  &  & 0  & 0 \\
 & \VUW\Ydu \vw &  &  0 & 0 \\
 &  &   &  0 & 0 \\
\hi_{1u} \vw & 0 & 0 &  \mB & 0 \\ 
\hi_{2u} \vw & 0 & 0 &  0 & a \mB
\end{array}\! \right), 
\quad
\Md = \left(\!\begin{array}{c@{\hspace{-0.1\tabcolsep}} c@{\hspace{-0.1\tabcolsep}} cc@{\hspace{1\tabcolsep}}c}
 &  &  & 0  & 0 \\
 & \VDW\Ydd \vw &  &  0 & 0 \\
 &  &   &  0 & 0 \\
\hi_{1d} \vw & 0 & 0 &  \mB & 0 \\ 
0 & \hi_{2s} \vw & 0 &  0 & a \mB
\end{array}\!\right)\,, 
\end{aligned}
\end{equation}
where $\VQW$ are the unitary matrices diagonalizing $Y_q$ from the left, $\VQW^\dagger \yq\yq^\dagger\VQW=\Ydq^2$, and we are
in the weak basis where the rotations $\WUW$, $\WDW$ have already been applied (as in the rest of this section).
Here, we defined $\mBi{_1} = \mB \, $ and $a =\mBi{_2}/\mBi{_1}\,.$

However, for practical purposes (as in  \cref{sec:CAAN2}), when $a<1$ ($\mBi{_1}>\mBi{_2}$) it can be more convenient to define $a_2=\mBi{_1}/\mBi{_2}$
and obtain all the expressions in terms of $a_2$ and $\mBi{_2}$ by making the substitution $a=1/a_2$,
$a\mB=\mBi{_2}$.

The couplings in charged and neutral \ac{RH} currents are determined by the last two rows
of the matrices $\VuR$, $\VdR$ diagonalizing the mass matrices in~\cref{eq:mass-2doubl}, by generalizing the procedure in~\cref{sec:app1}.
The mixing matrix of the \ac{RH} charged currents reads, at 
leading order in an expansion in powers of $\vw/\mBi{_{1,2}}$:
\begin{equation} \label{eq:vr-2doublets}
\VR\approx
    \left(\!\begin{array}{ccccc}
        \hi_{1u}^*\hi_{1d}\frac{\vw^2	}{\mB^2} & 
         \hi_{2u}^*\hi_{2s}\frac{\vw^2}{a^2\mB^2} & 0 &  
         -\hi_{1u}^*\frac{\vw}{\mB}  &  -\hi_{2u}^*\frac{\vw}{a\mB} \\
        0 & 0 & 0 & 0 & 0 \\
        0 & 0 & 0 & 0 & 0 \\
        - c_{45} \hi_{1d}\,\frac{\vw}{\mB} &  
       - \tilde{s}_{45}^* \hi_{2s}\,\frac{\vw}{a \mB} & 0 
        & c_{45} &  \tilde{s}_{45}^* \\
  \hi_{1d}\left( \tilde{s}_{45} +c_{45}\hi_{1u}^*\hi_{2u}\frac{\vw^2}{a\mB^2}\right) \frac{\vw}{\mB} & 
 -c_{45} \hi_{2s}\,\frac{\vw}{a\mB}  & 0 
 & -\tilde{s}_{45} -c_{45}\hi_{1u}^*\hi_{2u}\frac{\vw^2}{a\mB^2}
 & c_{45}
    \end{array}\!\right)\, ,
\end{equation}
where $c_{45}$ is the real cosine and $\tilde{s}_{45}$ is the complex sine of the angle,
\begin{equation}
\begin{aligned}
    \tilde{s}_{45}= \sin\theta_{45}\, \frac{\hi_{1u}^*\hi_{2u}}{|\hi_{1u}\hi_{2u}|} 
   \, , & &
    \frac{1}{2}\tan(2\,\theta_{45})\approx \frac{a\, |\hi_{1u}\hi_{2u}|\frac{\vw^2}{\mB^2}}{1-a^2+(|\hi_{1u}|^2-|\hi_{2u}|^2)\frac{\vw^2}{\mB^2}}\,. \\
    \end{aligned}
\end{equation}
For simplicity,
in the lower-right $2\times2$ submatrix in~\cref{eq:vr-2doublets}
we have omitted relative corrections of order $O(\vw^2/\mB^2)$ i.e.~$[1+O(\vw^2/\mB^2)]\tilde{s}_{45}\approx \tilde{s}_{45}$ and
$[1+O(\vw^2/\mB^2)]c_{45}\approx c_{45}$.

As for the neutral currents, we have
\begin{equation} \label{eq:Fu-2doubl}
\Fu\approx
    \left(\!\begin{array}{ccccc}
       \left( |\hi_{1u}|^2+\frac{|\hi_{2u}|^2}{a^2} \right)\frac{\vw^2}{\mB^2} & 
         0 & 0 &  
         -\left(\hi_{1u}^* c_{45} + \frac{\hi_{2u}^*}{a} \tilde{s}_{45} \right)\frac{\vw}{\mB}  & \left(\hi_{1u}^*\tilde{s}_{45}^*-\frac{\hi_{2u}^*}{a} c_{45} \right)\frac{\vw}{\mB} \\
        0 & 0 & 0 & 0 & 0 \\
        0 & 0 & 0 & 0 & 0 \\
         -\left(\hi_{1u}c_{45} +\frac{\hi_{2u}}{a}\tilde{s}_{45}^* \right)\frac{\vw}{\mB} &  
     0 & 0 & 1 &  O\left(\frac{\vw^2}{\mB^2}\right) \\
   \left(\hi_{1u}\tilde{s}_{45} -\frac{\hi_{2u}}{a}c_{45} \right)\frac{\vw}{\mB} & 
 0 & 0 
 & O\left(\frac{\vw^2}{\mB^2}\right)
 & 1
    \end{array}\!\right)
\end{equation}
and
\begin{equation} \label{eq:Fd-2doubl}
\Fd\approx
    \left(\!\begin{array}{ccccc}
        |\hi_{1d}|^2\frac{\vw^2}{\mB^2} & 
         0 & 0 &  
         - \hi_{1d}^*\frac{\vw}{\mB}  &  0 \\
        0 & |\hi_{2s}|^2\frac{\vw^2}{a^2\mB^2} & 0 & 0 & - \hi_{2s}^*\frac{\vw}{a\mB} \\
        0 & 0 & 0 & 0 & 0 \\
          - \hi_{1d}\frac{\vw}{\mB} &  
     0 & 0 & 1-|\hi_{1d}|^2\frac{\vw^2}{\mB^2} &  0 \\
   0 & - \hi_{2s}\frac{\vw}{a\mB} & 0 
 & 0 & 1-|\hi_{2s}|^2\frac{\vw^2}{a^2\mB^2}
    \end{array}\!\right)\, .
\end{equation}

For $a\neq 1$, in the limit
$|a-1|\gg |\hi_{1(2)u}|\vw/\mB$
(or $|a_2-1|\gg |\hi_{1(2)u}|\vw/\mB$)
the angle
$\theta_{45}$ is small and we have
\begin{equation} \label{eq:s45}
\begin{aligned}
&   \tilde{s}_{45} \approx -\hi_{1u}^*\hi_{2u}\, \frac{a}{a^2-1}\frac{\vw^2}{\mB^2} \, , \qquad  
\text{or}
\qquad  \tilde{s}_{45} \approx \hi_{1u}^*\hi_{2u}\, \frac{a_2^3}{1-a_2^2}\frac{\vw^2}{\mBi{_2}^2} \, ,
 \end{aligned}
\end{equation}
and $c_{45}\approx 1$. Then the right-handed currents can be written as
\begin{equation} \label{eq:vr-2doubl}
\VR\approx
    \left(\begin{array}{ccccc}
        \hi_{1u}^*\hi_{1d}\frac{\vw^2	}{\mB^2} & 
         \hi_{2u}^*\hi_{2s}\frac{\vw^2}{a^2\mB^2} & 0 &  
         -\hi_{1u}^*\frac{\vw}{\mB}  &  -\hi_{2u}^*\frac{\vw}{a\mB} \\
        0 & 0 & 0 & 0 & 0 \\
        0 & 0 & 0 & 0 & 0 \\
        -\hi_{1d}\frac{\vw}{\mB} &  
        \frac{\hi_{2s}\hi_{1u}\hi_{2u}^* \vw^3}{\mB^3  (a^2-1)} & 0 
        & 1-\frac{1}{2}\frac{(|\hi_{1d}|^2+|\hi_{1u}|^2)\vw^2}{\mB^2} & 
        -\frac{\hi_{1u}\hi_{2u}^*a\vw^2}{\mB^2(a^2-1)} \\
 -   \frac{\hi_{1d}\hi_{1u}^*\hi_{2u} \vw^3}{\mB^3 a (a^2-1)} & 
 -\hi_{2s}\frac{\vw}{a\mB} & 0 
 & \frac{\hi_{1u}^*\hi_{2u}\vw^2}{\mB^2a(a^2-1)} 
 & 1-\frac{1}{2}\frac{(|\hi_{2s}|^2+|\hi_{2u}|^2 )\vw^2}{a^2\mB^2}
    \end{array}\right)\,,
\end{equation}
at leading order.
The \acp{CAA} can be explained in the presence of the couplings
\begin{equation}
\begin{aligned}
  \VRa_{ud} & \approx \hi_{1d}^* \hi_{1u}\frac{\vw^2}{\mB^2}
   \approx - 0.78(27) \times 10^{-3}
   \, , \\  
   \VRa_{us} & \approx  \hi_{2s}^* \hi_{2u} \frac{\vw^2}{a^2\mB^2}
   \approx - 1.26(38) \times 10^{-3}\,,
\end{aligned}
\end{equation}
while flavour-changing neutral currents are suppressed since only diagonal elements are present in the $3\times 3$ submatrix of $\Fu$ and $\Fd$ (see Refs.~\cite{Belfatto:2021jhf,Belfatto:2023tbv}).
In what concerns the left-handed rotations,
the mixing matrix $\VL$ can be written analogously to \cref{eq:LH4mix}
with $\VLh = \VUW^\dagger\VDW$ (an equality in this scenario, with $\dU_{uL}=\dU_{dL}=\id$).

As for the $\M=6$ CP-odd weak-basis invariant we have
\begin{equation}
\begin{aligned}
\frac{1}{ \mB^6}
\im \tr\Big[\HuL \HdL \C\Big] 
 & = \frac{\vw^4}{\mB^4}  
\im \left( \hi_{1d}^* \hi_{1u} \Vha_{ud} \right) \,
\hat{y}_u\hat{y}_d+
  \frac{\vw^4}{\mB^4}  
\im \left( \hi_{2s}^* \hi_{2u} \Vha_{us} \right) \,
a^2\hat{y}_u\hat{y}_s
 \\[2mm] &  
\approx \frac{m_u m_d}{\mB^2}\, \im \left( \VRac_{ud} \VLa_{ud} \right)
+\frac{m_um_s}{\mB^2}a^4 \, \im \left( \VRac_{us} \VLa_{us} \right),
\end{aligned}
\end{equation}
which is valid even when $a=1$. Note that the vanishing of this category of CP-odd weak-basis invariants would not necessarily imply the absence of CP violation.
The missing physical phase is contained in other rephasing-invariant quantities, like
$\Vhac_{\alpha i} \Vha_{\beta i} \Vha_{\alpha j} \, \hi_\beta \hi_j^*$, which is contained
in the CP-odd invariant of mass dimension $\M=10$ of~\cref{eq:CPo-5blocks}.

\subsubsection{The extreme chiral limit with \texorpdfstring{$\n=1$}{\n=1} doublet}

Let us consider the limit in which only the third generation has non-zero mass. This so-called \ac{ECL} is relevant when considering scenarios with extremely high energies, where to good approximation the masses of the first two generations of quarks can be taken as zero. Notably, in \ac{VLQ} models \ac{CPV} is possible in this limit, in contrast to what happens in the \ac{SM}. For instance, the \ac{VLQ} singlet scenario has been studied in the \ac{ECL} \cite{delAguila:1997vn,Albergaria:2022zaq}, and it was found that one independent CP-odd \ac{WBI} still survives.
In the \ac{ECL}, for the $\n=1$ \ac{VLQ} doublet case, one can move to a \ac{WB} where
the mass matrices take the forms
\begin{equation} \label{eq:ex-ch}
{\cal M}_u = \left(\begin{array}{cccc}
0&0&0 &  0 \\
0&0&0 &  0 \\
0&0&  \hat{y}_t \vw &  0 \\
0 & \hic \vw & \hit \vw &  \mB  
\end{array}\right)\,,   \qquad
{\cal M}_d = \left(\begin{array}{cccc}
0 & 0 & 0 &  0 \\
0 & 0 & y_{23} \vw &  0 \\
0 & 0 &  y_{33} \vw &  0 \\
0 & \his \vw & \hib \vw &  \mB  
\end{array}\right)\,.
\end{equation} 
Note that all phases except one
can be absorbed by rephasing the quarks fields, so that there is a single physical phase which cannot be eliminated,
and we can consider all parameters to be real except e.g.~$\hib$. This single surviving phase induces \ac{CPV}, which means all CP properties of the model are determined by a single CP-odd \ac{WBI} built from 3 Hermitian blocks.

The mass matrices can be diagonalized by bi-unitary transformations as in~\cref{eq:diag}.
The right-handed mixing elements are given by the last rows
of the matrices $\VuR$ and $\VdR$,
which appear as in~\cref{eq:4rowu}, while
the mass eigenvalues read as in~\cref{eq:Msplit,eq:mstd}.
Regarding the left-handed rotations,
the unitary matrices can be written as
\begin{equation}
\begin{aligned}
    \VuL&=
    \begin{pmatrix}
        1 & 0 & 0 & 0 \\
        0 & 1 & 0 & 0 \\
        0 & 0 & 1 & \hit^*\hat{y}_t\frac{\vw^2}{\mB^2} \\
        0 & 0 & -\hit \hat{y}_t\frac{\vw^2}{\mB^2} & 1
    \end{pmatrix}    
    +\mathcal{O}\left(\frac{\vw^4}{\mB^4} \right)\, ,
 \\[2mm]
    \VdL&=
     \begin{pmatrix}
        1 & 0 & 0 & 0 \\
        0 & \Vha_{qq} & \Vha_{qb} & 0 \\
        0 & \Vha_{tq} & \Vha_{tb} & 0 \\
        0 & 0 & 0 & 1
        \end{pmatrix}
        \begin{pmatrix}
        1 & 0 & 0 & 0 \\
        0 & 1 & 0 & 0 \\
        0 & 0 & 1 &\hib^*\hat{y}_b\frac{\vw^2}{\mB^2} \\
        0 & 0 & -\hib \hat{y}_b\frac{\vw^2}{\mB^2} & 1
    \end{pmatrix}
    +\mathcal{O}\left(\frac{\vw^4}{\mB^4} \right)\, ,   
\end{aligned}
\end{equation}
where $q$ here refers to the combination of the massless quarks coupling to the top quark, 
and $\VL  = {\VuL}^\dagger\VdL$. We then have
\begin{equation}
\Vha_{tb}= 
  \frac{y_{33}}{\sqrt{y_{33}^2+y_{23}^2}} \, , \qquad
\Vha_{qb}= 
\frac{y_{23}}{\sqrt{y_{33}^2+y_{23}^2}}=-\Vha_{qb} 
\, ,
\qquad \hat{y}_b^2= y_{33}^2+y_{23}^2 \,.
\end{equation}
The relevant CP-odd invariant thus takes the simple form
\begin{equation}
\begin{aligned}
    \frac{1}{M^6_Q}\im\tr\Big[\HuL \HdL H\Big]
    &=\frac{v^4}{M^4_Q}\im\left(\hat{y}_t y_{33} \hit  \hib^* \right) 
 =\frac{\vw^2}{\mB^2}\hat{y}_t\hat{y}_b\im\left( \Vha_{tb}\VRhac_{tb}  \right) 
 \\
    &\approx \frac{m_t m_b}{M^2_Q}\im\left(\VRac_{tb}\VLa_{tb}\right)\,.
\end{aligned}
\end{equation}

\subsubsection{The extreme chiral limit with \texorpdfstring{$\n=2$}{\n=2} doublets}
\label{sec:ECL_2}

We now briefly
consider an even more extreme high-energy limit,
where the top quark is the only standard quark that has non-zero mass. In that case, 
one can move to a \ac{WB} where
the mass matrices take the minimal forms 
\begin{equation} \label{eq:2doublets_ECL}
    \Mu=\begin{pmatrix}
        0 & 0 & 0 & 0 & 0\\
        0 & 0 & 0 & 0 & 0\\
        0 & 0 & \hat{y}_t \vw & 0 & 0\\
       0 & \hi_{1c} \vw & \hi_{1t} \vw & \mB & 0\\
       \hi_{2u} \vw  & \hi_{2c} \vw & \hi_{2t} \vw & 0 & a \mB\\
    \end{pmatrix}\,, \quad
    \Md=\begin{pmatrix}
        0 & 0 & 0 & 0 & 0\\
        0 & 0 & 0 & 0 & 0\\
        0 & 0 & 0 & 0 & 0\\
       0 & 0 & \hi_{1b} \vw &  \mB & 0\\
       0 & \hi_{2s} \vw & \hi_{2b} \vw & 0 &  a \mB\\
    \end{pmatrix}\,,
\end{equation}
and all phases except two (e.g.~those of $\hi_{2t}$ and $\hi_{2b}$) can be eliminated via rephasings of the quark fields. Hence, we can write
\begin{equation}
   \frac{1}{M^6_Q}\im\tr\Big[\HuL \HdL \C\Big]= \frac{\vw^4}{M^4_Q}(a^2-1) \,\im\Big[ \hi_{1b} \hi^*_{2b} \left(\hi^*_{1c} \hi_{2c} + \hi^*_{1t} \hi_{2t}\right) \Big]\,,
\end{equation}
so that in general there is still \ac{CPV}, even when $\hat{y}_t=0$ (i.e.~the limit of vanishing top mass). Note that, although this invariant vanishes for $a=1$, in that case we still have $\im\tr\left[\HuL \HdL \HuL^2\C\right]\neq 0$, meaning that \ac{CPV} is still possible as far as $\hat{y}_t\neq 0$.

If, instead, both $\hat{y}_t=0$ and $a=1$, the \acp{WBI} vanish as a consequence of CP conservation. Indeed, one can check that the matrix $\Mu$ can, in this case, be brought to the same form as $\Md$ in~\cref{eq:2doublets_ECL} (since $\hat{y}_t=0$). Moreover, any unitary (left-handed) operation on the last two rows of the mass matrices can now be undone for the last two columns (via an extra right-handed notation). This is true since $a=1$, and thus the lower-right $2\times 2$ block is proportional to the identity. Then, all the phases in the blocks
$A_u = \begin{bsmallmatrix} 0 & \hi_{1t}  \\ \hi_{2c}   & \hi_{2t}  \end{bsmallmatrix}  \vw$
and 
$A_d = \begin{bsmallmatrix} 0 & \hi_{1b}  \\ \hi_{2s} & \hi_{2b} 
\end{bsmallmatrix} \vw$
are removable. This can be seen e.g.~by i) undoing the singular value decomposition in one sector, say $A_u = U_1 D U_2 \to A_u' = D$, followed by ii) undoing a polar decomposition in the other, $A_d' = P U \to P$, and finally iii) by rephasing away the off-diagonal phase in the Hermitian matrix $P$.

\vfill
\clearpage

\section{Phenomenology}
\label{sec:pheno}

The phenomenological effects of the presence of vector-like quarks have been the object of study of several works 
(see~\cite{Branco:1986my,delAguila:1989rq,Nir:1990yq,delAguila:2000rc,Barenboim:2001fd,Cacciapaglia:2010vn,Botella:2012ju,Ishiwata:2015cga,Wang:2016mjr,Biekotter:2016kgi,Bobeth:2016llm,Botella:2016ibj,Cacciapaglia:2018lld,Belfatto:2019swo,Belfatto:2021jhf,Branco:2021vhs,Botella:2021uxz,Belfatto:2023tbv,Balaji:2021lpr}).
However, it is compelling to conduct a comprehensive phenomenological analysis of scenarios with \ac{VLQ} doublets  adopting a new perspective based on invariants.

Invariants under \acp{WBT} have been a main object of study in this work. 
Any observable, as any minimal set of parameters in a specific weak basis, can ultimately be written in terms of weak-basis invariants. 
The possibility of such description in terms of invariants is intriguing, as it would avoid any potential misinterpretations due to the choice of weak basis (minimal or not).
This feature appears particularly attractive
when describing CP-violating phenomena, since the results cannot be the consequence of any parameterization choices as well as of any unphysical complex phase.
Moreover, the different structures of invariants emerging in presence of \ac{VLQ} 
doublets compared to the \ac{SM}, as studied in detail in~\cref{sec:CPeven,sec:CPodd}, provide hints on the processes
which one can expect to be most affected, 
particularly when involving CP violation.

In this section, we show how different observables can be written in terms of the invariant quantities analyzed in the previous sections, 
and how one can gain a clear
insight into which observables are modified, as well as how CP-even and CP-odd probes are related.
This exercise also highlights the distinct imprints of \ac{VLQ} doublets compared to other models that generate similar invariant structures.

We start by summarising the low-energy effects of \ac{VLQ} doublets in~\cref{sec:phenopresummary}. In~\cref{sec:phenorepinv}, we identify effective rephasing invariants as the most relevant objects for low-energy phenomenology, which provide a glimpse on which kind of observables are going to be more sensitive to \acp{VLQ} prior to any computation. 
In~\cref{sec:phenobilinears,sec:phenotrilinears},
we study the most relevant phenomena in terms of these rephasing invariants, in particular bilinears and trilinears respectively. In this way, we can
provide unambiguous constraints on the model 
and a clear connection between different observables.

\subsection{Weak currents}
\label{sec:phenopresummary}

Let us first summarize the effects of the presence of vector-like quarks doublets mixing with \ac{SM} quarks (see also~\cref{sec:app1,sec:EFT}).
Weak charged and neutral currents are modified as
\begin{equation} \label{eq:effRHcurrent}
\begin{aligned}
    \mathcal{L} \,\supset\, &
    - \frac{g}{\sqrt{2}} \,W_\mu^+ \left(\overline{\nu}_{L}\gamma^{\mu} e_{L}+\overline{u_L}\gamma^{\mu} \, \VLh \, d_{L}+ \, \overline{u_R}\gamma^{\mu} \, \VRh  \, d_{R}\right)
+\text{h.c.} 
\\
& - \frac{g}{2c_W} \,Z_\mu  \left(J^{Z,\text{SM}}_{\mu}+\,\overline{u_{R}} \gamma^{\mu} \, \Fuh \, u_{R}-\,\overline{d_{R}} \gamma^{\mu} \, \Fdh \, d_{R} \right)\,.
\end{aligned}
\end{equation}
The whole effect of vector-like quarks 
is captured by the matrices 
$\VLh$, $\VRh$, $\Fuh$, $\Fdh$ in good approximation
(see~\cref{eq:vckmR,eq:Fzz,eq:VR-F-eff}).
In particular, it is evident from the description in~\cref{sec:app1,sec:EFT} that the set of Lagrangian parameters $\hi_\alpha$, $\hi_i$ is directly connected to couplings of charged and neutral currents, and hence to observables.
For $\n$ \ac{VLQ} doublets one has
\begin{equation} \label{eq:effmatrices}
\begin{aligned}
& \hspace{2.8cm} \VRha_{\alpha i}\equiv \sum_{n} \hi_{n \alpha}^{*}\; \hi_{n i}\, \frac{\vw^2}{[\mBd]^2_{nn}}\, , \\
& \Fuha_{\alpha\beta}= \sum_{n} \hi_{n\alpha}^{*}\; \hi_{n\beta}\,\frac{\vw^2}{[\mBd]^2_{nn}} \, , \qquad \Fdha_{ij}=\sum_{n} \hi_{ni}^{*} \; \hi_{nj} \, \frac{\vw^2}{[\mBd]^2_{nn}} \, ,
\end{aligned}
\end{equation}
where the mass elements $[\mBd]_{nn}$ are defined in~\cref{eq:L1}, the couplings $\hi_{n\alpha(i)}$ are defined analogously to~\cref{eq:z,eq:VR-F-eff}
and the index $n$ runs from $1$ to $\n$, while $\alpha,\beta=u,c,t$ and $i,j=d,s,b$.
Then:
\begin{itemize}
\item \acp{FCNC} are generated at tree level, in contrast to the \ac{SM}, where they only appear at loop level and are further suppressed by the \ac{GIM} mechanism~\cite{Glashow:1970gm,Glashow:1976nt,Paschos:1976ay}. The mixing of \ac{SM} quarks with the vector-like quarks is thus strongly constrained.
\item The new couplings introduce potentially
CP-violating \ac{RH} quark currents at low energies, only suppressed by $\vw^2/\mB^2$. 
\end{itemize} 
Let us emphasize that the \ac{RH} \emph{charged} currents are intrinsically linked to the \ac{RH} \emph{neutral} currents, as both depend on the same Yukawa parameters (and the same rows of the unitary matrices diagonalizing the mass matrices). This connection is particularly relevant when comparing the phenomenological consequences of \ac{VLQ} doublets with other models that also generate effective \ac{RH} currents in the light-quark sectors. The relationship becomes even stronger with a lower number of doublets. In fact, for the case of a single doublet $\n =1$, knowing all non-zero couplings of \ac{SM} quarks associated with \acp{FCNC} would fully determine the moduli of all new Yukawa parameters. 
Remarkably, the knowledge of the last row and column of the whole $4\times 4$ matrices of neutral-current couplings 
would provide the moduli of all right-handed charged-current couplings at all orders in $\vw/\mB$.%
\footnote{
Recall that $\Fr_{T'\alpha}=\V^{R*}_{TT'}\V^R_{T\alpha}$ ($\alpha = u,c,t,T'$) and $\Fr_{B'i}=\V^{R*}_{BB'}\V^R_{Bi}$ ($i = d,s,b,B'$). Knowing them yields all $|\V^R_{T\alpha}|$ and $|\V^R_{Bi}|$ (the entries of $|\BqL|$), allowing for the reconstruction of $|\VRa_{\alpha i}| = |\V^R_{T\alpha}||\V^R_{Bi}|$.
}

\subsection{Phenomenology of effective rephasing invariants}
\label{sec:phenorepinv}

Establishing which \acp{WBI} will be more or less relevant in low-energy observables is not trivial, due to the strong hierarchy of the \ac{SM} parameters and the possible presence of hierarchies in \ac{VLQ} couplings.
Nevertheless, in~\cref{sec:CPodd} 
we have seen
how in a scenario with one vector-like doublet (and some phenomenologically motivated scenarios with more doublets)
starting from the \ac{UV} description and expanding the mixing matrices in powers of $\vw/\mB$, 
\acp{WBI} are naturally related to rephasing invariants involving only the three \ac{SM} quarks (see~\cref{sec:reph-inv-eff}). 
This dependence provides a direct connection to low-energy observables, 
which can always be written in terms of rephasing invariants. 

In a scenario with one or more vector-like doublets,
we can combine the restrictions that rephasing invariants provide on the form that any observable may take with the power counting given by the effective description at the amplitude level. 
In this way, 
we can find some physical insight at the observable level prior to any computation, especially regarding CP-violating probes.%
\footnote{Let us note how this exercise can be generalized to other models generating effective \ac{RH} currents. This generalization goes beyond the reach of this work.}
Let us thus recall the different possible rephasing invariants appearing in the effective description. 

The mixing matrix $\VLh$ appears at dimension $D=4$ in the effective Lagrangian.
In principle one may think that up to $6$ phases can appear in $\VLh$.
However, only one (physical) of them can be obtained from rephasing invariants with 
vertex insertions appearing only at $D=4$, that is only with $\VLh$ insertions.
In fact, the lowest rephasing invariant which can be obtained from the elements of
$\VLh$ is the \ac{SM}-like one,
corresponding to the imaginary part of the left-handed rephasing-invariant quartet:
\begin{equation}
    \Qh_{\alpha i \beta j}=
    \Vha_{\alpha i}\, \Vha_{\beta j} \, \Vhac_{\alpha j} \, \Vhac_{\beta i}\,.
\end{equation}
In the limit $\mB \rightarrow \infty$ ($\mB$ the lowest mass scale of vector-like quark doublets), the imaginary part
$\im(\Qh_{\alpha i \beta j})$ reduces to the $J$ invariant of~\cref{eq:Jar}, and thus to the one physical CP phase of the \ac{SM}
with the same $4-$vertex/$4-$quark \ac{SM} suppression.

At $\mathcal{O}(1/\mB^2)$ 
(but with a different number of \ac{SM} mixing insertions)
one finds new CP-odd structures, originating from the imaginary part of the following rephasing invariants:
\begin{equation}
\begin{aligned} \label{eq:reph-inv}
    &\textbf{$\bullet$ bilinears:} && \hbi_{\alpha i}=\Vhac_{\alpha i} \, \VRha_{\alpha i}\, , 
     \\
    &\textbf{$\bullet$ trilinears:} && \htr_{i,\alpha\beta}=\Vhac_{\alpha i} \,  \Vha_{\beta i} \, \Fuha_{\alpha\beta} \, ,
    \qquad
    \htr_{\alpha, ij}=\Vha_{\alpha i} \, \Vhac_{\alpha j}\, \Fdha_{ij}  \, ,
    \\ &\textbf{$\bullet$ quartets:} &&
    \hqu_{\alpha i\beta j} =
     \Vha_{\beta j} \, \Vhac_{\alpha j} \, \Vhac_{\beta i} \, \VRha_{\alpha i} \, , 
\end{aligned}
\end{equation}
corresponding to the ones defined in~\cref{eq:reph-inv-eff} for one \ac{VLQ} doublet.
The following information can be gathered regarding CP-violating processes:
\begin{itemize}
    \item[--] The imaginary part of the bilinears $\im\hbi_{\alpha i}$ is directly linked to potential CP violation in the charged-current sector, produced by a phase mismatch between the \ac{RH} and \ac{LH} sectors. 
    \item[--] In the presence of \acp{VLQ},
    $\hat{\Fr}_{u(d)}$ induces tree-level \acp{FCNC}. CP violation is induced with two extra $\Vha_{\alpha i}$ insertions. Indeed,
    a phase difference between the \ac{SM} amplitude (which is always induced at loop level) and the new contributions can induce CP violation in the flavour-changing up or down sector driven respectively by $\im \htr_{i,\alpha\beta}$ and $\im\htr_{\alpha,ij}$. 
    \item[--] The last type of invariant comes with the same insertions as in the \ac{SM}, but with a right-handed vertex. In general, there will be stronger contributions to the same process generated by the phases of bilinears and trilinears.
\end{itemize}
Let us mention that one cannot
 get CP-odd rephasing invariants at order $\mathcal{O}(1/\mB^4)$ through the interference of two tree-level $\mathcal{O}(1/\mB^2)$ amplitudes in the right-handed sector alone. Namely, CP-odd rephasing invariants
 cannot be made by
 two right-handed vertices (with elements of the mixing matrices $\Fuh$, $\Fdh$, $\VRh$) without \ac{CKM} vertex insertions.
 Moreover, in presence of only one vector-like doublet,
 one cannot obtain new
CP-odd rephasing invariants with more than
one insertion of a right-handed vertex.

In the following, we analyze how the new CP-violating invariant types directly relate to 
physical processes, focusing on some key flavour observables involving light quarks.
These results are then summarized in~\Cref{tab:vlq_invariants} and~\Cref{fig:kaons-fcnc}.

 \subsection{Bilinears and the Cabibbo sector}
 \label{sec:phenobilinears}
 
The bilinears $\hbi_{\alpha i}=\Vhac_{\alpha i} \VRha_{\alpha i}$
could potentially induce the largest \ac{VLQ} imprints, 
both in their CP-conserving ($\re \hbi_{\alpha i}$)
and CP-violating ($\im \hbi_{\alpha i}$) parts,
since they can appear in the charged current sectors at $\mathcal{O}(1/\mB^2)$ at tree level.
The presence of these right-handed couplings acquires 
highly-motivated phenomenological importance 
given the presence of the \acfp{CAA},
that is, the tensions between three different determinations of the Cabibbo angle $\theta_C$. In fact, it was shown that the vector-like quark charged as a doublet of $SU(2)_L$ 
appears as the favoured candidate 
in explaining the \acp{CAA}~\cite{Belfatto:2021jhf,Crivellin:2022rhw,Belfatto:2023tbv,Cirigliano:2023nol}.

\subsubsection{Cabibbo angle anomalies}

Three types of independent determinations of the Cabibbo angle $\theta_C$ are extracted with high precision and show
tension between each other. 
In particular,
the determination of $\vert V_{ud} \vert=\cos\theta_C$ is obtained 
from $\beta$ decays. 
Recent calculations of short-distance radiative corrections
(see e.g.~\cite{Seng:2018qru,Seng:2018yzq})
in $\beta$ decays led to an improved determination of 
$\vert V_{ud} \vert$.
The most precise determination is obtained
from super-allowed $0^+$-- $0^+$ nuclear $\beta$ decays, which are
pure Fermi transitions \cite{Hardy:2020qwl}
(even though
in this last survey the uncertainty is increased by a factor of $2.6$ due to 
new contributions in nuclear-structure corrections $\delta_{\text{NS}}$ \cite{Seng:2018qru,Gorchtein:2018fxl}, 
which now dominate the uncertainty).
One can determine $\vert V_{ud} \vert$ also from free neutron $\beta$ decay, using 
the average of the neutron lifetime from eight bottle experiments
\cite{UCNt:2021pcg,Ezhov:2014tna,Pattie:2017vsj,Serebrov:2017bzo,Arzumanov:2015tea,Steyerl:2012zz,Pichlmaier:2010zz,Serebrov:2004zf} (with
rescaled uncertainty) and the axial-vector coupling from the latest experiments measuring the parity-violating $\beta$-asymmetry parameter $A$ from polarized  
neutrons~\cite{Mund:2012fq,UCNA:2017obv,Markisch:2018ndu}.
Present lattice computations~\cite{FlavourLatticeAveragingGroupFLAG:2021npn,FlavourLatticeAveragingGroupFLAG:2024oxs} and analyses of electromagnetic radiative corrections~\cite{Seng:2021nar,Seng:2022wcw}
provide a precise determination of $\vert V_{us} \vert=\sin\theta_C$
from semileptonic $K_{\ell 3}$ kaon decays $K \to \pi \ell \nu$
($K_{L e3}$, $K_{S e3}$, $K^{\pm}_{e3}$, $K^{\pm}_{\mu3}$, $K_{L \mu3}$, $K_{S \mu3}$)\cite{Moulson:2017ive}.
The ratio  $\vert V_{us}/V_{ud}\vert =\tan\theta_C$ can be independently determined from
the ratio of the kaon and pion leptonic 
decay rates $K_{\mu 2}$ and $\pi_{\mu2}$, i.e.~$K\rightarrow \mu\nu(\gamma)$ and 
$\pi\rightarrow \mu\nu(\gamma)$~\cite{Marciano:2004uf}
using lattice QCD calculations~\cite{FlavourLatticeAveragingGroupFLAG:2021npn,FlavourLatticeAveragingGroupFLAG:2024oxs}
and including electroweak radiative corrections~\cite{Cirigliano:2011tm,DiCarlo:2019thl,Boyle:2022lsi}.
We get:%
\footnote{We follow Ref.~\cite{Belfatto:2023tbv} updating the result $f_{K}/f_{\pi}=1.1934(19)$ from Ref.~\cite{FlavourLatticeAveragingGroupFLAG:2024oxs}.
The obtained determinations are also in good agreement with the fit made in Ref.~\cite{Cirigliano:2023nol}, with a slightly different choice of inputs.} 
\begin{equation} \label{eq:vusvud-values}
 |V_{ud}|_\beta =  0.97372(26)~, \quad
    \vert V_{us} \vert_{K\ell 3} = 0.22308(55)~, \quad
      \left|\frac{V_{us}}{V_{ud}}\right|_{K\mu 2/\pi\mu 2} =0.23126(48)\,.
\end{equation}
After using these determinations,
there is approximately a $\sim 3 \sigma$ deficit in the \ac{CKM} first row unitarity (dubbed \ac{CAA}1 in~\cite{Belfatto:2023tbv})
when using the
value of $|V_{ud}|$ from $\beta$ decays with the value of $|V_{us}|$ from kaon decays. Additionally,
there is a tension of about $\sim 3 \sigma$ between
the determination of $\vert V_{us} \vert$ from $K_{\ell 3}$ decays
and the one obtained from leptonic $K_{\mu 2}$ and $\pi_{\mu 2}$ decay rates 
(\ac{CAA}2 in Ref.~\cite{Belfatto:2023tbv}). 

Various models have been suggested as possible solutions~\cite{Belfatto:2019swo,Grossman:2019bzp,Coutinho:2019aiy,Cheung:2020vqm,Crivellin:2020lzu,Endo:2020tkb,Belfatto:2021jhf,Branco:2021vhs,Capdevila:2020rrl,Crivellin:2020ebi,Kirk:2020wdk,Manzari:2020eum,Alok:2020jod,Crivellin:2020oup,Crivellin:2020klg,Crivellin:2021njn,Marzocca:2021azj,Botella:2021uxz,Crivellin:2022rhw,Dcruz:2022rjg} (see \cite{Fischer:2021sqw} for a review). 
A remarkable solution for the anomalies is given by
the vector-like quark charged as a doublet of $SU(2)_L$, as found in Refs.~\cite{Belfatto:2021jhf,Crivellin:2022rhw,Belfatto:2023tbv}, which can potentially explain all the tensions through right-handed charged currents~\cite{Belfatto:2021jhf,Grossman:2019bzp,Crivellin:2022rhw,Belfatto:2023tbv,Cirigliano:2023nol}.

\begin{figure}[t]
    \centering
\includegraphics[width=0.9\textwidth]{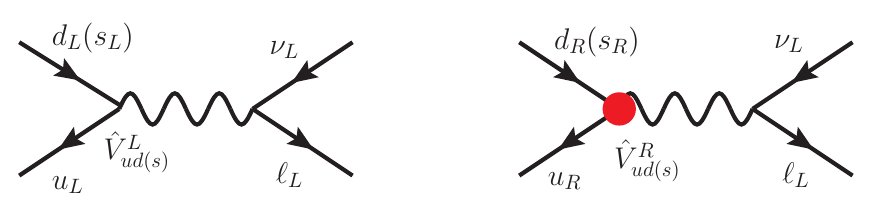}
    \caption{Diagrams contributing to a semileptonic process in the Cabibbo sector, leading to 
    $\Gamma_{H}^{\text{Cab.}}\sim \big|\Vha_{ui} \; (\pm)_H \; \VRha_{ui}\big|^2\approx \big|\Vha_{ui}\big|^2 \; (\pm)_H \; 2 \re \Vha_{ui}\VRhac_{ui}$.
    Here, $H$ refers to the specific hadronic process, typically mediated either by a vector current (plus sign) or by an axial-vector one (minus sign).} 
    \label{fig:cabbsketch}
\end{figure}

In fact, one must take into account that
the value of $V_{ud}$ which is obtained from $\beta$ decays
depends uniquely on the
vector part of the weak interaction
$G_V = G_F \vert V_{ud} \vert$.
Also semileptonic kaon decays $K_{\ell 3}$ determine the weak vector coupling, while leptonic decays $K_{\mu2}$ and $\pi_{\mu2}$ depend on the axial-vector current. In the \ac{SM} there is no difference between these
determinations since they are all given by left-handed charged currents and they determine the same mixing angle $\theta_C$.
In a model with right-handed currents instead, vector and axial-vector couplings are not equal and the three determinations correspond to different couplings.
In the scenario with vector-like doublets, 
right-handed currents are generated. Then,
the tensions in the different determinations can be
explained by the mixing of vector-like doublets with light standard model quarks. 
From~\cref{eq:effRHcurrent} we have
\begin{equation} \label{eq:lagr-udus}
\mathcal{L}_\text{CC} \supset 
- \frac{g}{\sqrt{2}} W_\mu^+  \Big[ 
\bar{u}_L \gamma^\mu \Vha_{ud} d_L + 
\bar{u}_L \gamma^\mu \Vha_{us} s_L + 
\bar{u}_R \gamma^\mu  \VRha_{ud} d_R +
\bar{u}_R \gamma^\mu  \VRha_{us} s_R  \Big]
+\text{h.c.}\,.
\end{equation}
Thus, in this scenario we have:
\begin{equation} \label{eq:caa-rh}
\begin{aligned}
&\Big|V_{ud}\Big|_\beta 
= \Big|\,\Vha_{ud}+\VRha_{ ud}\,\Big|   
\approx \big|\Vha_{ud}\big|+\,\frac{\re \left(\Vha_{ud}\,\VRhac_{ud} \right)}{\big|\Vha_{ud}\big|}
\,,  \\
&\Big|V_{us}\Big|_{K\ell 3}
=
\Big|\,\Vha_{us}+\VRha_{ us}\,\Big| \approx 
\big|\Vha_{us}\big|+\,\frac{\re \left(\Vha_{us}\,\VRhac_{us} \right)}{\big|\Vha_{us}\big|} \,, 
\\
& \left|\frac{V_{us}}{V_{ud}}\right|_{\frac{K\mu 2}{\pi\mu2}} 
=\frac{\Big|\,\Vha_{us}-\VRha_{ us}\,\Big|}{\Big|\,\Vha_{ud}-\VRha_{ ud}\,\Big|} 
\approx
\frac{\big|\Vha_{us}\big|}{\big|\Vha_{ud}\big|}\; \Bigg(1-
\,\frac{\re \left(\,\Vha_{us}\,\VRhac_{us}\,\right)}{\big|\Vha_{us}\big|^2}+
\,\frac{\re \left(\,\Vha_{ud}\,\VRhac_{ud}\,\right) }{\big|\Vha_{ud}\big|^2}\Bigg) \,.  
\end{aligned}
\end{equation}
where the approximations hold since
$|\VRha_{ud(s)}| \ll |\Vha_{ud(s)}|$.
Then, using the determinations in~\cref{eq:vusvud-values}, one obtains
the rephasing invariants:
\begin{equation} \label{eq:cabfit}
\begin{aligned}
&|\Vha_{ud}|=0.97450(8) \,, && 
\frac{\re \hbi_{ud}}{|\Vha_{ud}|}= \frac{\re\left(\,\Vhac_{ud} \VRha_{u d}\,\right)}{|\Vha_{ud}|}
=-\, 0.79(27) \times 10^{-3}   \,,
\\
& |\Vha_{us}|=0.22431(35)  \,, &&
\frac{\re \hbi_{us}}{|\Vha_{us}|}=\frac{\re\left(\,\Vhac_{us} \VRha_{us}\,\right)}{|\Vha_{us}|}=-\, 1.24(37) \times 10^{-3} \,.
\end{aligned}
\end{equation}
Therefore, the Cabibbo angle anomalies hint towards the presence of right-handed couplings, with a significant part of the mixing in the right-handed sector aligned in phase with the left-handed sector, with the opposite sign ($\pi$ phase difference).
The diagram of the processes is sketched in~\Cref{fig:cabbsketch}.

In principle, a single vector-like doublet mixing with the up, the down and the strange \ac{SM} quarks could explain all inconsistencies. However, 
as shown in Ref.~\cite{Belfatto:2021jhf}, 
the required couplings are excluded by limits on flavour-changing neutral currents.
Nevertheless, the presence of more than one doublet can be at the origin of the anomalies without contradicting experimental constraints
(see also~\cref{sec:CAAN2}).
In particular, 
a \ac{VLQ} doublet mixing with the up and the down \ac{SM} quarks 
($\hiu \neq 0$ and $\hid \neq 0$)
can resolve the tension between $V_{ud}$ and $V_{us}$ (\ac{CAA}1), 
while a \ac{VLQ} doublet coupling predominantly to the up and strange quarks $\hiu \neq 0$ and $\his \neq 0$
can be the cause of the tension between the $K_{\ell3}$ and $K_{\mu2}/\pi_{\mu2}$ determinations of $V_{us}$ (\ac{CAA}2)~\cite{Belfatto:2021jhf,Belfatto:2023tbv}.

\subsubsection{The neutral kaon system}
\label{sec:eps-prime}

The success of the \ac{SM} in high-precision measurements 
leads to very stringent bounds on \ac{NP} models. This is particularly true for observables that are highly suppressed in the \ac{SM}, such as CP-violating effects or \ac{FCNC} processes. 
One of the most relevant testing fields for 
the \ac{SM} is provided by kaon physics, e.g.~rare kaon decays,
kaon mixing both in the CP-conserving (mass splitting $\Delta m_K$) and CP-violating effects ($\epsilon$), direct CP violation, but also the decay $K\rightarrow\pi\pi$, into pions
with isospin $I=2$. 
The \ac{SM} prediction for the latter does not require loops or \acp{FCNC}, but the decay into pions with isospin $I=2$ is still suppressed, making it an additional interesting probe for \ac{BSM} physics.

In particular, when facing the Cabibbo angle anomalies, the same new couplings in~\cref{eq:lagr-udus} which can be responsible for the anomalies and contribute to leptonic and semileptonic kaon decays, also impact kaon hadronic decays, other leptonic and semileptonic decay channels, and kaon mixing.  
Strong constraints from kaon mixing and flavour-changing kaon decays prohibit large couplings with both the down and strange quarks of the same vector-like doublet. Then, 
as already mentioned,
the anomalies cannot be simultaneously explained by one doublet~\cite{Belfatto:2021jhf}. 

These \ac{FCNC} processes are determined by trilinears,
and we will discuss some of these constraints in~\cref{sec:phenotrilinears}.
We begin by addressing constraints from charged currents and flavour-conserving processes,
which are controlled by bilinears.
In the context of this work, we are interested in a rephasing-invariant description of physical processes. In the following we will focus on some processes in which a rephasing-invariant description can provide useful and unambiguous information on the constraints on \ac{NP} couplings. 

Quantities related to observables should be invariant under the rephasing of the states 
($|K^0\rangle$, $|\bar{K}^0\rangle$, $|K_{S,L}\rangle$, final states $|f\rangle$, etc.), rephasing of the quark fields,
and independent on phase choices in the CP transformations.
We are also interested in a description which is independent on the choice of parameterization of the mixing matrices.
Nevertheless, assuming CPT is a good symmetry, it is convenient, for simplicity, to fix the relative phase of $|K_S\rangle$ and $|K_L\rangle$ in such a way that the coefficients of $|K^0\rangle$ have the same phase and are therefore equal. This choice also implies that the coefficients of $|\bar{K}^0\rangle$ have equal phases (see e.g.~Ref.~\cite{Branco:1999fs}). This is the only phase convention that we assume.
Then, the Hamiltonian eigenstates can be written as
\begin{equation} \label{eq:KLKS}
|K_{L}\rangle=p_K\, |K^0\rangle+q_K\, |\bar{K}^0\rangle \, , \qquad |K_{S}\rangle=p_K\, |K^0\rangle-q_K\, |\bar{K}^0\rangle \, ,
\end{equation}
with $|p_K|^2+|q_K|^2=1$ and $\delta =|p_K|^2-|q_K|^2$.

\paragraph{\boldmath{$K\rightarrow \pi\pi$}.}
As noted in Ref.~\cite{Cirigliano:2023nol},
significant bounds on right-handed currents can come from the theoretical and 
experimental results on the decay $K \to \pi\pi$.
The decays to two pions $\pi^+\pi^-$ and $\pi^0\pi^0$ are dominant for $K_S$.
Because of the small CP violation in the kaon system, also $K_L$ can decay to two pions (CP conservation would prevent the decay into two pions for the CP-odd state). Additionally, the kaons predominantly decay to the state of two pions with isospin $I=0$, while the decay to two pions with isospin $I=2$ is suppressed.%
\footnote{Within the \ac{SM}, the leading contribution to the $\Delta I=3/2$ amplitude arises from the low-energy realization of the $(27_L,1_R)$-plet part of the product of two quark currents transforming as $(8_L,1_R) \times (8_L,1_R)$ under the $SU(3)_{L} \times SU(3)_{R}$ symmetry of the approximate massless QCD Lagrangian. The observed suppression of the corresponding non-perturbative coupling, leading to the $\Delta I=1/2$ rule, is not present for \ac{VLQ} doublets, whose corresponding $(8_L , 8_R)$ contributions come with an additional chiral enhancement. See for example Ref.~\cite{Pich:2021yll} for more detailed discussion.} Therefore, the decay of kaons to two pions 
is usually described by three useful parameters~\cite{Branco:1999fs}
\begin{equation} \label{eq:om-eps-epsp}
\begin{aligned}
\omega & = \frac{\bra{2}H_{\Delta S=1}\ket{K_S}}{\bra{0}H_{\Delta S=1}\ket{K_S}}\, , 
\quad\qquad
\epsilon =  \frac{\bra{0} H_{\Delta S=1}|K_L\rangle}{\bra{0}H_{\Delta S=1}\ket{K_S}}\, ,  \\[2mm]
\epsilon' &=  \frac{\bra{2}H_{\Delta S=1}\ket{K_L}\bra{0}H_{\Delta S=1}\ket{K_S} -\bra{0}H_{\Delta S=1}\ket{K_L}\bra{2}H_{\Delta S=1}\ket{K_S}}{\sqrt{2}\bra{0}H_{\Delta S=1}\ket{K_S}^2}\,,
\end{aligned}
\end{equation}
where $\bra{0}$ and $\bra{2}$ indicate the states $\bra{2\pi,I=0}$ and $\bra{2\pi,I=2}$ respectively.
With the choice of relative phase between $K_L$ and $K_S$ already mentioned in~\cref{eq:KLKS}, 
these three parameters are rephasing invariants.
Here,
$\omega$ is the parameter associated to $\Delta I=1/2$ rule violation, 
$\epsilon$ indicates mixing CP-violation, and
$\epsilon'$ direct CP violation. Bilinear rephasing invariants $\hbi_{\alpha i}$ affect $\omega$ and $\epsilon'$, while $\epsilon$ is affected by trilinears $\htr_{\alpha, ij}$. We will discuss the contribution to $\epsilon$ in the next section.

In presence of vector-like weak doublets of quarks mixing with \ac{SM} quarks 
at the weak scale we have from~\cref{eq:leftlag2} the effective Hamiltonian 
\begin{equation} \label{eq:Heff-Kpipi}
H_{\Delta S=1}^\text{NP} = 
\frac{4 G_F}{\sqrt{2}} \Big[ 
\Vhac_{us}\VRha_{ud}\, (\bar{s}_L\gamma^\mu u_L)(\bar{u}_R\gamma_\mu d_R)
+\VRhac_{us}\Vha_{ud}\, (\bar{s}_R\gamma^\mu u_R)(\bar{u}_L\gamma_\mu d_L)
\Big]+\text{h.c.}\,.
\end{equation}
Then, the amplitudes of the decays $K\rightarrow\pi\pi$ receive an additional \ac{NP} contribution:
\begin{equation} \label{eq:VLQAI}
 A_I e^{i\delta_I}= A_I^\text{SM} e^{i\delta_I}+A_I^\text{NP} e^{i\delta_I} \, ,
\end{equation}
where 
\begin{equation}
     \langle I |H_{\Delta S=1} | K^0 \rangle =A_I e^{i \delta_I}\,\sqrt{2} \, , \qquad 
    \langle I |H_{\Delta S=1} | \bar{K}^0 \rangle  = \bar{A}_I e^{i \delta_I}\,\sqrt{2} \, ,
\end{equation}
for $I=0$ and $I=2$ and the phases $\delta_I$ are the final-state-interaction phase shifts of the two pion states.%
\footnote{\label{foot:long}%
The factor $\sqrt{2}$ comes from the convention on the normalization of the amplitudes.
Depending on whether the charged pions are considered as identical particles or not, different normalizations are chosen for the isospin decomposition of the two pions states (see e.g.~Ref.~\cite{Branco:1999fs}).
If $\pi^+$ and $\pi^-$ are considered as distinguishable, the branching ratios are written as 
\begin{equation} \label{eq:norm-latt}
\begin{aligned}
\Gamma(K_S\rightarrow \pi^+\pi^-)&=
\frac{1}{16\pi m_K}\sqrt{1-\frac{4m_\pi^2}{m_K^2}}\left| \sqrt{\frac{1}{3}} \langle 2|H_{\Delta S=1}|K_S\rangle +\sqrt{\frac{2}{3}} \langle 0|H_{\Delta S=1}|K_S\rangle  \right|^2  \,, \\
\Gamma(K_S\rightarrow \pi^0\pi^0)&=
\frac{1}{32\pi m_K}\sqrt{1-\frac{4m_\pi^2}{m_K^2}}\left| \frac{2}{\sqrt{3}} \langle 2|H_{\Delta S=1}|K_S\rangle -\sqrt{\frac{2}{3}} \langle 0|H_{\Delta S=1}|K_S\rangle  \right|^2\,.
\end{aligned}
\end{equation}
This normalization is used in Refs.~\cite{Blum:2012uk,RBC:2015gro,RBC:2020kdj} 
as well as in Refs.~\cite{Bertolini:2013noa,Cirigliano:2016yhc} and gives 
$ |A_0|_\text{exp} = 3.3201(18)\times 10^{-7}\,\text{GeV}$~\cite{Blum:2012uk},
while in Ref.~\cite{Branco:1999fs} 
$\pi^+$ and $\pi^-$ are considered as identical particles and
the amplitudes $A_I$ are $\sqrt{2}$ times larger giving
$ |A_0|_\text{exp} = 4.6953(25)\times 10^{-7}\,\text{GeV}$.
However, in Refs.~\cite{Blum:2012uk,RBC:2015gro,Cirigliano:2016yhc} the amplitudes $A_I$ are defined as 
$\sqrt{2}A_I e^{i \delta_I}= \langle I |H_{\Delta S=1} | K^0 \rangle$ so that the matrix elements are the same as in Ref.~\cite{Branco:1999fs},
while in Refs.~\cite{Bertolini:2013noa,RBC:2020kdj} 
$A_I e^{i \delta_I}= \langle I |H_{\Delta S=1} | K^0 \rangle$ so that
the matrix elements are $\sqrt{2}$ times smaller than in Refs.~\cite{Branco:1999fs,Blum:2012uk,RBC:2015gro}.
We will use in the following the convention
$\sqrt{2}A_I e^{i \delta_I}= \langle I |H_{\Delta S=1} | K^0 \rangle$ with 
$ |A_0|_\text{exp} = 3.3201(18) \times 10^{-7}\,\text{GeV}$.}
In the \ac{SM} the main contribution to the amplitudes $A_I$ is at tree level:
\begin{equation} \label{eq:AI-SM-tree}
\sqrt{2}\, A_I^\text{SM}  e^{i \delta_I}\approx \frac{4 G_F}{\sqrt{2}}  \Vhac_{us}\Vha_{ud} 
\langle  I | (\bar{s}_L\gamma^\mu u_L)(\bar{u}_L\gamma_\mu d_L) | K^0 \rangle =
\frac{G_F}{\sqrt{2}}  \Vhac_{us}\Vha_{ud} \langle  I | Q_2| K^0 \rangle\,.
\end{equation}
The presence of vector-like doublets generates the additional contribution
\begin{equation} \label{eq:AI-new}
\sqrt{2} \, A_I^\text{NP} e^{i\delta_I}  =
\frac{G_F}{\sqrt{2}}  \left(\Vhac_{us}\VRha_{ud}-\VRhac_{us}\Vha_{ud} \right) \left(c_1
\langle  I | Q_1^{LR} | K^0 \rangle +c_2 \langle  I | Q_2^{LR} | K^0 \rangle  \right)\,,
\end{equation}
where $c_1(3\,\text{GeV})=0.9$ and $c_2(3\,\text{GeV})=0.4$ are the factors resulting 
from the QCD renormalization~\cite{Cirigliano:2016yhc} and
the operators are defined as~\cite{Bertolini:2013noa} 
(we rename $1 \leftrightarrow 2$):
\begin{equation}
\begin{aligned}
Q_1^{LR}&=\big( \bar{s} \gamma_\mu(1-\gamma_5)u \big)\big( \bar{u} \gamma_\mu(1+\gamma_5)d \big) \, ,\\[1mm]
Q_2^{LR}&=\big( \bar{s}_\alpha \gamma_\mu(1-\gamma_5)u_\beta \big)\big( \bar{u}_\beta \gamma_\mu(1+\gamma_5)d_\alpha \big) \,.
\end{aligned}
\end{equation}
The hadronic matrix elements of left-right operators
are chirally enhanced with respect to the tree-level \ac{SM} ones.\footnote{The matrix elements of left-right operators are $\mathcal{O}(p^0)$ in the chiral counting, i.e.~they do not vanish in the limit of zero quark/meson masses.}
Moreover, the corresponding amplitudes of the decay in the state with isospin $I=2$ do not undergo a suppression relative to the
$I=0$ final state. Therefore, the new contributions to the amplitudes may be significant.

In order to obtain both the amplitudes $A_0$ and $A_2$, it is useful to start with the computation of $A_2$ first.
In the decay of $K$ to two pions with isospin $I=2$ 
a relation exists between the matrix elements of the operators $Q_{1,2}^{LR}$ and the matrix elements of the electroweak penguin operators $Q_{7,8}$. In the isospin limit\cite{Chen:2008kt,Bertolini:2013noa},
\begin{equation}
\langle  2 | Q_1^{RL} | K^0 \rangle = \frac{2}{3} \langle  2 | Q_7 | K^0 \rangle \, , \qquad
\langle  2 | Q_2^{RL} | K^0 \rangle = \frac{2}{3} \langle  2 | Q_8 | K^0 \rangle\,.
\end{equation}
The final-state strong-interaction phase 
in~\cref{eq:AI-new} arises from the hadronic matrix element.
Then, we can use the lattice results in Ref.~\cite{Blum:2012uk} to get the hadronic matrix elements at the scale 
$\mu = 3$~GeV:
\begin{equation}
\frac{\re{\langle  2 | Q_1^{RL} | K^0 \rangle }}{\cos\delta_2} =  0.238(13) \, \text{GeV}^3    \,  , \qquad
\frac{\re{\langle  2 | Q_2^{RL} | K^0 \rangle }}{\cos\delta_2}= 1.078(61) \, \text{GeV}^3\,,
\end{equation}
in the suitable phase convention for $|K^0\rangle$.
However, let us remark that any rephasing of the initial or final states is irrelevant whenever considering ratios involving hadronic matrix elements with same initial and final states.%
\footnote{The \ac{SM} contribution to the amplitudes (also including the loop-level contribution) as well as the new-physics one present a structure (see~\cref{eq:AI-SM-tree,eq:AI-new}):
\begin{equation}
    \sqrt{2}\, \lambda_u^* A_I e^{i\delta_I} = \frac{G_F}{\sqrt{2}}\lambda_u^* \sum_i \lambda^{(i)} \langle I | Q_i | K^0 \rangle\,,
\end{equation}
where $\lambda^{(i)}$ include couplings and eventual Inami-Lim functions and we also explicitly multiply by
$\lambda_u^*=\Vha_{us}\Vhac_{ud}$ since in the expression of observables the amplitudes always emerge in the combination $\lambda_u^*A_I$
(see e.g.~\cref{eq:omega,eq:epsprime}).
The final-state strong-interaction phase 
in~\cref{eq:AI-new} arises from the hadronic matrix element.
Thus, one gets
\begin{equation}
\label{eq:realA}
\begin{aligned}
\sqrt{2}\, \re(\lambda_u^* A_I) &=
\frac{G_F}{\sqrt{2}}\sum_i \re(\lambda_u^* \lambda^{(i)}) \frac{\re \langle I | Q_i | K^0 \rangle}{\cos\delta_I}  \,, \\ 
\sqrt{2}\, \im(\lambda_u^*A_I ) &=
\frac{G_F}{\sqrt{2}}\sum_i \im(\lambda_u^*\lambda^{(i)}) \frac{\re \langle I | Q_i | K^0 \rangle}{\cos\delta_I}\,.
\end{aligned}
\end{equation}
and ratios with the same initial and final states can be written in a rephasing invariant way as
\begin{equation}
\frac{\text{Re}\big(A_I^{(k)} \Vha_{us} \Vhac_{ud}\big)}{\text{Re}(A_I^{(j)}  \Vha_{us} \Vhac_{ud})} =
\frac{ \re\big(\lambda_u^*\lambda^{(k)}\big) \,\re \langle I | Q_k | K^0\rangle }{  \re(\lambda_u^*\lambda^{(j)}) \,\re \langle I | Q_j | K^0 \rangle}=
\pm
\frac{ \text{Re}\big(\lambda_u^*\lambda^{(k)}\big) \,|\langle I | Q_k | K^0 \rangle|}{  \text{Re}(\lambda_u^*\lambda^{(j)}) \,|\langle I | Q_j | K^0 \rangle|}
\end{equation}
and similarly for the imaginary part.
} 
In the following analysis we will
consider rephasing-invariant ratios.
Hence, we can write:
\begin{equation}
\begin{aligned}
\sqrt{2}\, A_2^\text{NP}    &\approx
\frac{G_F}{\sqrt{2}}\, \left(\Vhac_{us}\VRha_{ud}-\VRhac_{us}\Vha_{ud} \right) \times 0.645(27)\,\text{GeV}^3 
  \\ & \approx 
\sqrt{2}\, \left(\Vhac_{us}\VRha_{ud}-\VRhac_{us}\Vha_{ud} \right) \times 3.76(16) \times 10^{-6}~\text{GeV} \, .
\end{aligned}
\end{equation}
As regards the amplitude of the decay to the $I=0$ state, we can use the relation 
$A_0^{\text{NP}}=-2\sqrt{2}A_2^{\text{NP}}$
in Ref.~\cite{Cirigliano:2016yhc}, valid in the chiral limit. Substituting in~\cref{eq:AI-new}, one finds
\begin{equation}
\begin{aligned}
\sqrt{2}\, A_0^\text{NP}  &\approx
-\frac{G_F}{\sqrt{2}}\, \left(\Vhac_{us}\VRha_{ud}-\VRhac_{us}\Vha_{ud} \right) \times 1.825(77)~\text{GeV}^3
 \\ & \approx
-\sqrt{2}\, \left(\Vhac_{us}\VRha_{ud}-\VRhac_{us}\Vha_{ud} \right) \times 1.064(45) \times 10^{-5}~\text{GeV}\,.
\end{aligned}
\end{equation}
These new contributions should be compared with the \ac{SM} expectation and experimental values.

\begin{table}[t!]
\renewcommand{\arraystretch}{1.2}
  \centering
  \begin{tabular}{cc}
    \toprule
    \textbf{Quantity} & \textbf{Experimental value} \\
    \midrule
$\omega_\text{exp} $ & $ 0.04454(12)$~\mbox{\cite{Blum:2012uk}} \\
$ |A_0|_\text{exp}  $ & $ 3.3201(18)\times 10^{-7}\,\text{GeV} $~\mbox{\cite{Blum:2012uk}} \\
$ |A_2|_\text{exp}    $ & $ 1.479(4)\times 10^{-8}\,\text{GeV}$~\mbox{\cite{Blum:2012uk}} \\
$ |A_2|_{\text{exp,} K_S}   $ & $1.570(53)\times 10^{-8}\,\text{GeV} $~\mbox{\cite{Blum:2012uk}} \\
$\epsilon_\text{exp} $ &  $ 2.228(11)\times 10^{-3}$~\mbox{\cite{ParticleDataGroup:2024cfk}} \\
$\re \left( {\epsilon'}/{\epsilon} \right)_\text{exp}$ &  $ 1.66(23)\times 10^{-3} $~\mbox{\cite{ParticleDataGroup:2024cfk}}\\
    \bottomrule
  \end{tabular}
  \caption{Values of the quantities used in the text, determined from experimental data.}
  \label{tab:values-exp}
\end{table}
  \renewcommand{\arraystretch}{1}

\paragraph{\boldmath{$K\rightarrow \pi\pi\, , I=0$}.}
 The experimental value of $|A_0|$ can be obtained from the decay rates in~\cref{eq:norm-latt},~\cref{foot:long}~\cite{Blum:2012uk}:
\begin{equation}
|A_0|_\text{exp} = 3.3201(18)\times 10^{-7}\,\text{GeV}\,.
\end{equation}
For reference, the experimental values of relevant quantities are summarized in~\Cref{tab:values-exp}.
In the \ac{SM}, the main contribution to the amplitude $A_0$ is at tree level:
\begin{equation} \label{eq:A0main}
\sqrt{2}\, A_0^\text{SM}  e^{i \delta_0} \approx \frac{4 G_F}{\sqrt{2}}  \Vhac_{us}\Vha_{ud} 
\langle  0 | (\bar{s}_L\gamma^\mu u_L)(\bar{u}_L\gamma_\mu d_L) | K^0 \rangle =
\frac{G_F}{\sqrt{2}}  \lambda_u \langle  0 | Q_2| K^0 \rangle\,,
\end{equation}
with $\re \langle  0 | Q_2| K^0 \rangle \,/\cos\delta_0  = 0.147(15)\,\text{GeV}^3$~\cite{RBC:2020kdj}, which accounts for 
$\sim 97\%$ of the \ac{SM} prediction  or $A_0$. We defined also
$\lambda_u=\Vhac_{us}\Vha_{ud}$ for convenience.
Therefore, the dominant contribution to the amplitude is proportional to $\Vhac_{us}\Vha_{ud}$
and in very good approximation 
\begin{equation} \label{eq:real}
\lambda_u^*\,A_0^\text{SM}\approx \re \Big[\lambda_u^*\,A_0^\text{SM}\Big]\,.
\end{equation}
 The \ac{SM} expectation was computed in Ref.~\cite{RBC:2020kdj}:
\begin{equation}
 \frac{\re \left(\lambda_u^*\,A_0^\text{SM}\right)}{\big|\lambda_u\big|}
    = 2.99(67)\times 10^{-7}\,\text{GeV} \,.
\end{equation}
For reference, we also summarize \ac{SM} predictions in~\Cref{tab:values-SM}.
Then, we can compare the \ac{SM} expectation with the experimental determination:
\begin{equation} \label{eq:A0sm-exp}
     \frac{\big|A_0\big|_\text{exp} - \big|A_0^\text{SM}\big|}{|A_0|_\text{exp} } = 1-
    \frac{\big|\re \left(\lambda_u^*\,A_0^\text{SM}\right)\big|}{|\lambda_u||A_0|_\text{exp} } =
0.10 \pm 0.20\,.
\end{equation}

\begin{table}[t!]
\renewcommand{\arraystretch}{1.2}
  \centering
  \begin{tabular}{cc}
    \toprule
    \textbf{Quantity} & \textbf{\ac{SM} value}  \\
    \midrule
$ \re A_0^\text{SM} $ & $ 2.99(67)\times 10^{-7}\,\text{GeV} $~\mbox{\cite{RBC:2020kdj}} \\
$  \im A_0^\text{SM} $ & $-6.98(1.57)\times 10^{-11}\,\text{GeV}  $~\mbox{\cite{RBC:2020kdj}} \\
$  \re A_2^\text{SM} $ & $ 1.50(15)\times 10^{-8}\,\text{GeV} $~\mbox{\cite{RBC:2015gro}}\\
$  \im A_2^\text{SM} $ & $ -8.34(1.03)\times 10^{-13}\,\text{GeV} $~\mbox{\cite{RBC:2020kdj}} \\
$ \re \left({\epsilon'}/{\epsilon}\right)_\text{SM} $ & $ 2.17(84)\times 10^{-3} $~\mbox{\cite{RBC:2020kdj}}  \\
    \bottomrule
  \end{tabular}
\caption{
Prediction of the \ac{SM} values. 
The values of the amplitudes 
are given in a particular phase convention, in which the tree-level contribution is real.
The amplitudes are given in the renormalization indicated in the text. }
  \label{tab:values-SM} 
\end{table}
\renewcommand{\arraystretch}{1}

When including extra quark doublets, we get an additional contribution to the amplitudes:%
\footnote{Since the dominant contributions to the \ac{SM} amplitudes are proportional to $\lambda_u$, we have $|\im \left( \lambda_u^{*}\,A_I^\text{SM} \right)|\ll |\re \left( \lambda_u^{*}\,A_I^\text{SM} \right)|$ and we can write:
$$ \frac{\re \left( A_I^{\text{SM}*}\,A_I^\text{NP} \right)}{\big|A_I^\text{SM}\big|^2} =
\frac{\re \left(\lambda_u^* \,A_I^\text{NP} \right)\re \left(\lambda_u\, A_I^{\text{SM}*} \right)-\im \left(\lambda_u^* \,A_I^\text{NP} \right)\im \left(\lambda_u\, A_I^{\text{SM}*} \right)}{\re \left(\lambda_u^* \,A_I^\text{SM} \right)^2 + \im \left(\lambda_u^*\, A_I^{\text{SM}} \right)^2}   
\approx \frac{\re \left( \lambda_u^{*}\,A_I^\text{NP} \right)}{\re \left( \lambda_u^{*}\,A_I^\text{SM} \right)}\,.
$$
}
\begin{equation} \label{eq:ASMplusANP}
    \big| A_0^\text{SM}+A_0^\text{NP} \big|\approx
   \big| A_0^\text{SM} \big|\left(1+
    \frac{\re \left( A_0^{\text{SM}*}\,A_0^\text{NP} \right)}{\big|A_0^\text{SM}\big|^2} \right)
     \approx  \big| A_0^\text{SM} \big|\left(1+
    \frac{\re \left( \lambda_u^{*}\,A_0^\text{NP} \right)}{\re \left( \lambda_u^{*}\,A_0^\text{SM} \right)} \right)\,.
\end{equation}
Then, it is useful to compare the new rephasing-invariant contribution
given by the extra quark doublet with the \ac{SM} contribution:
\begin{equation}
\begin{aligned}
& 
\frac{\re \left(\lambda_u^*\,A_0^\text{NP}\right)}{\re \left(\lambda_u^*\,A_0^\text{SM}\right)} 
\approx  \\ & 
\frac{-2\sqrt{2}\re \Big[\left(|\Vha_{us}|^2\Vhac_{ud}\VRha_{ud}-|\Vha_{ud}|^2\Vha_{us}\VRhac_{us} \right)\Big] \Big|0.9 \langle  0 | Q_1^{LR} | K^0 \rangle +0.4 \langle  0 | Q_2^{LR} | K^0 \rangle  \Big|}
{|\Vha_{us}|^2|\Vha_{ud}|^2 \Big|\langle  0 | Q_2| K^0 \rangle \Big|} \,.
\end{aligned}
\end{equation}
Let us underline that this ratio emerges in a rephasing-invariant form.
Using the values needed to explain the \acp{CAA} and the \ac{SM} estimate in Ref.~\cite{RBC:2020kdj}
in their central values, we obtain
\begin{equation}
\frac{\re \left[\left(\Vha_{us}\Vhac_{ud}\right)\,A_0^\text{NP}\right]}{\re \left[\left(\Vha_{us}\Vhac_{ud}\right)\,A_0^\text{SM}\right]} \approx -0.037\,,
\end{equation}
i.e.~the extra contribution would be $\sim27$ times smaller than the \ac{SM} one 
 (and, by comparing with~\cref{eq:A0sm-exp}, $\sim 30$ times smaller than the experimental determination) and also
6 times smaller than the $1\sigma$ uncertainty on the lattice \ac{SM} prediction. Therefore, we can neglect this effect.

\paragraph{$\Delta = \frac{1}{2} $ isospin rule.} 
The $\Delta = 1/2 $ rule violation and the CP-violating effects are suppressed in the \ac{SM} context. Then, they may provide stringent constraints on the mixings with the extra vector-like doublets.
Regarding the CP-conserving contribution of the isospin-suppressed part of the decay $K\rightarrow \pi\pi$, 
 it is conveniently described by the parameter $\omega$~\cite{Branco:1999fs}:
\begin{equation}
\omega = \frac{\langle 2|H_{\Delta S=1}|K_S\rangle}{\langle 0|H_{\Delta S=1}|K_S\rangle} 
\,\approx\, e^{i(\delta_2-\delta_0)}\re \frac{A_2}{A_0}
\,=\,e^{i(\delta_2-\delta_0)}\frac{\re \left(A_2 A_0^* \right)}{|A_0|^2} \,,
\end{equation}
where the approximation follows from taking into account the small value of $\epsilon$.
The main contribution to the CP-conserving part of the decay of the kaon into two pions with isospin $I=2$ in the \ac{SM}
is again given by the operator $Q_2$, which accounts for $\sim 93\%$ of the amplitude~\cite{Blum:2012uk}.
Using the fact that the dominant contribution to the amplitudes is proportional to $\Vhac_{us}\Vha_{ud}$
so that $|\re (\Vha_{us}\Vhac_{ud} A_I)|\gg|\im(\Vha_{us}\Vhac_{ud} A_I)|$, we can write~\cite{Branco:1999fs}
\begin{equation} \label{eq:omega}
| \omega |  \approx
\frac{\re \left(A_2 A_0^* \right)}{|A_0|^2}  
\approx  \frac{\re \left(\lambda_u^*\,A_2  \right)}{\re \left(\lambda_u^*\,A_0  \right)}\,.
\end{equation}
 The \ac{SM} lattice prediction for $A_2$ gives $\re(\lambda_u^* A_2)/|\lambda_u|= 1.50(15)\times 10^{-8}\,\text{GeV} $~\cite{RBC:2015gro}.
 The experimental value can be determined from the decay rate of $K^+\rightarrow\pi^+\pi^0$
 as reported in Ref.~\cite{Blum:2012uk}: 
\begin{equation} \label{eq:A2exp}
|A_2|_\text{exp} = 1.479(4)\times 10^{-8}\,\text{GeV} \, ,
\end{equation}
while from $K_S$ decays $ |A_2|_{\text{exp}, K_S}= 1.570(53)\times 10^{-8}$ GeV. 
The $\omega$ parameter represents the CP-conserving part of the ratio $A_2/A_0$,
while the parameter $\epsilon'$ represents the CP-violating part (direct CP violation).
Taking into account that the CP violation is very suppressed, the experimental value of the
amplitudes gives~\cite{Blum:2012uk}:
\begin{equation} \label{eq:omega-exp}
 |\omega|_\text{exp} = 0.04454(12)\,.
\end{equation}
We can compare the experimental result with the \ac{SM} expectation:
\begin{equation} \label{eq:A2sm-exp}
     \frac{\big|A_2\big|_\text{exp} - \big|A_2^\text{SM}\big|}{|A_2|_\text{exp} } = 1-
    \frac{\big|\re \big[\lambda_u^*\,A_2^\text{SM}\big]\big|}{|\lambda_u||A_2|_\text{exp} } =
-0.014\pm 0.098\,,
\end{equation}
and
\begin{equation}
\frac{|\omega|_\text{exp} - |\omega|_\text{SM}}{ |\omega|_\text{exp}}=
1- \frac{1}{| \omega|_\text{exp}} \,
\frac{\re \left(\lambda_u^*\,A_2  \right)}{\re \left(\lambda_u^*\, A_0  \right)}
= -0.13 \pm 0.28\,.
\end{equation}

In a scenario with \ac{VLQ} doublets, the amplitudes receive
additional contributions.
Following the same steps as in~\cref{eq:ASMplusANP} we can write
\begin{equation}
 \big| A_2^\text{SM}+A_2^\text{NP} \big|
     \approx  \big| A_2^\text{SM} \big|\left(1+
    \frac{\re \left( \lambda_u^{*}\,A_2^\text{NP} \right)}{\re \left( \lambda_u^{*}\,A_2^\text{SM} \right)} \right)\,,
    \label{eq:A2SMplusANP}
\end{equation}
where we have
\begin{equation}
\re \left(\lambda_u^*\,A_2^\text{NP}\right)  = 
\re \Big[ |\Vha_{us}|^2\Vhac_{ud}\VRha_{ud}-|\Vha_{ud}|^2\Vha_{us}\VRhac_{us} \Big] \times 3.76(16) \times 10^{-6}~\text{GeV} \,.
\label{eq:reA2NP}
\end{equation}
Regarding the parameter $\omega$, from~\cref{eq:omega} we have:
\begin{equation}
\omega_\text{SM}+\omega_\text{NP}=
   e^{i(\delta_2-\delta_0)}\frac{\re \Big[\lambda_u^* \left(A_2^\text{SM} +A_2^\text{NP}\right)\Big]}{\re \left(\lambda_u^*\,A_0^\text{SM}\right)}=
  \omega_\text{SM}\, \frac{\re \Big[\lambda_u^* \left(A_2^\text{SM} +A_2^\text{NP}\right)\Big]}{\re \left(\lambda_u^*\,A_2^\text{SM}\right)}\,.
\end{equation}
Using the values needed to explain the Cabibbo angle anomalies in~\cref{eq:cabfit}, we get:
\begin{equation}
 \frac{\re \left(\lambda_u^*\,A_2^\text{NP}\right)}{\re \left(\lambda_u^*\,A_2^\text{SM}\right) }  = \frac{\omega_\text{NP} }{\omega_\text{SM}}
  = 0.26\pm 0.10 \,.
\end{equation}
Let us emphasize again that this ratio emerges in a rephasing-invariant way.

Then, we can evaluate the impact of the new contribution 
against the experimental determinations. At $2\sigma$ we get
\begin{equation}
    -0.29 \,\,\lesssim\,\, \frac{\big|A_2\big|_\text{exp} - \big| A_2^\text{SM}+A_2^\text{NP} \big|}{\big|A_2\big|_\text{exp}} \,\,\lesssim\,\, 0.26\,,
\end{equation}
and from~\cref{eq:A2SMplusANP,eq:reA2NP}
\begin{equation}
    -1.1\times 10^{-3} \,\,\lesssim\,\, |\Vha_{us}|\frac{\re\big(\Vhac_{ud}\VRha_{ud}\big)}{|\Vha_{ud}|}
    -|\Vha_{ud}|\frac{\re\big(\Vha_{us}\VRhac_{us}\big)}{|\Vha_{us}|}
    \,\,\lesssim\,\, 1.0\times 10^{-3}\,.
\end{equation}
By inserting the values needed for Cabibbo angle anomalies,
\begin{equation}
|\Vha_{us}|\frac{\re\big(\Vhac_{ud}\VRha_{ud}\big)}{|\Vha_{ud}|}
    -|\Vha_{ud}|\frac{\re\big(\Vha_{us}\VRhac_{us}\big)}{|\Vha_{us}|}=
    1.05(37) \times 10^{-3}\,,
\end{equation}
we receive
\begin{equation}
\begin{aligned}
& \frac{\big|A_2\big|_\text{exp} - \big| A_2^\text{SM}+A_2^\text{NP} \big|}{\big|A_2\big|_\text{exp}} \approx 
1- \frac{ \big| A_2^\text{SM} \big|}{\big|A_2\big|_\text{exp}}\,\frac{\re \Big[\lambda_u^*\,\left(A_2^\text{SM}+A_2^\text{NP}\right)\Big]}{\re \left(\lambda_u^*\,A_2^\text{SM}\right) }  
= -0.28\pm 0.14\,,
\\
& 1- \frac{|\omega_\text{SM}+\omega_\text{NP}|}{ |\omega|_\text{exp}} = -0.42 \pm 0.35\,,
\end{aligned} 
\end{equation}
i.e., by comparison with~\cref{eq:A2exp,eq:omega-exp}, the \ac{CAA} values agree with the experimental results within $2\sigma$.

\paragraph{Direct CP violation.}
$\epsilon'\neq 0$ represents direct CP violation, i.e.~CP violation in the decay amplitudes:
\begin{equation}
\begin{aligned}
 \epsilon' &=  \frac{\langle 2|H_{\Delta S=1}|K_L\rangle\langle 0|H_{\Delta S=1}|K_S\rangle -\langle 0|H_{\Delta S=1}|K_L\rangle\langle 2|H_{\Delta S=1}|K_S\rangle}{\sqrt{2}\langle 0|H_{\Delta S=1}|K_S\rangle^2}
 \\
 & \approx  \frac{i}{\sqrt{2}} e^{i(\delta_2-\delta_0)}\im\frac{A_2}{A_0}\,,
\end{aligned}
\end{equation}
where the approximation makes use of the small value of $\epsilon$~\cite{Branco:1999fs}.
The ratio $\epsilon'/\epsilon$ can be experimentally determined from the relation
\begin{equation}
\left| \frac{\eta_{00}}{\eta_{+-}} \right| \approx 1-6 \re \left( \frac{\epsilon'}{\epsilon} \right)\,,
\end{equation}
where the parameters $\eta$ are defined as
$ |\eta_f| = \sqrt{(\Gamma_L/\Gamma_S)\br(K_L\rightarrow f)/\br(K_S \rightarrow f)}$.
Since $\epsilon$ and $\epsilon'$ exhibit the same phase, we also have
\begin{equation}
\left(\frac{\epsilon'}{\epsilon}\right)_\text{SM}\approx
\re \left(\frac{\epsilon'}{\epsilon}\right)_\text{SM}\,.
\end{equation}
Then, using the fact that the dominant contribution to the amplitudes is proportional to $\Vhac_{us}\Vha_{ud}$
and one expects $|\re (\lambda_u^*\,A_I)|\gg|\im(\lambda_u^*\,A_I)|$
we can write~\cite{Branco:1999fs}
\begin{equation} \label{eq:epsprime}
\begin{aligned}
\left(\frac{\epsilon'}{\epsilon}\right)_\text{SM}
   &   \approx  \frac{\im\left(A_2 A_0^*\right)}{\sqrt{2}\, |\epsilon|\, |A_0|^2}
      \approx    \frac{\im\left( \lambda_u^*\,A_2\right)\re\left( \lambda_u^*\,A_0\right) - \re\left( \lambda_u^*\,A_2\right) \im\left( \lambda_u^*\,A_0\right)}{\sqrt{2}\, |\epsilon|\, \big[\re \left( \lambda_u^*\,A_0\right)\big]^2}
      \\
    & 
   \approx \frac{\im\left(\lambda_u^*\,A_2\right) - |\omega| \im\left(\lambda_u^*\,A_0\right)}{\sqrt{2}\, \big|\epsilon \, \lambda_u^*\,A_0\big|}\,.
\end{aligned}
\end{equation}
Experimentally it is found that~\cite{ParticleDataGroup:2024cfk}
\begin{equation}
\re \left( \frac{\epsilon'}{\epsilon} \right)_\text{exp}=1.66(23)\times 10^{-3}\,.
\end{equation}
The lattice \ac{SM} prediction gives~\cite{RBC:2020kdj}
\begin{equation}
\left(\frac{\epsilon'}{\epsilon}\right)_\text{SM}\approx
\re \left(\frac{\epsilon'}{\epsilon}\right)_\text{SM}=2.17(84)\times 10^{-3}\,,
\end{equation}
in agreement with the estimate from chiral perturbation theory~\cite{Gisbert:2017vvj,Cirigliano:2019ani}, especially if one considers isospin-breaking corrections~\cite{Cirigliano:2019cpi}.

From~\cref{eq:epsprime} we get
\begin{equation} \label{eq:epsprime-tot}
\frac{\epsilon'}{\epsilon}
=  \left(\frac{\epsilon'}{\epsilon}\right)_\text{SM}+
\frac{\left(1+2\sqrt{2} |\omega|_\text{exp}\right)\im\left(\Vha_{us}\Vhac_{ud}\,A_2^\text{NP} \right)}{\sqrt{2}\, \big|\epsilon \,  \Vha_{us}\Vhac_{ud}\,A_0^\text{exp}\big|} \,.
\end{equation}
The comparison of the new contribution with the \ac{SM} one gives:
\begin{equation}
\begin{aligned}\label{eq:epsprime-np-sm-ratio}
    \frac{\left( \epsilon'/\epsilon \right)_\text{NP}}{\left( \epsilon'/\epsilon \right)_\text{SM}}
&\approx
    \frac{\left(1+2\sqrt{2} |\omega|_\text{exp}\right)\,\im\left(\Vha_{us}\Vhac_{ud}\,A_2^\text{NP}\right)}{\im\left(\Vha_{us}\Vhac_{ud}\, A_2^\text{SM}\right) - |\omega|_\text{exp} \im\left(\Vha_{us}\Vhac_{ud}\,A_0^\text{SM}\right)} 
     \\
    &\approx \frac{\im\Big[\Vha_{us}\Vhac_{ud} \left(\Vhac_{us}\VRha_{ud}-\VRhac_{us}\Vha_{ud} \right)\Big]\times  1.9\,(6)\times 10^{6}}{\big|\Vha_{us}\Vhac_{ud}\big|}\,.
\end{aligned}
\end{equation}
After inserting the values needed for the Cabibbo angle anomalies (see~\cref{eq:cabfit}),
this ratio indicates that, in order to get a contribution less than the \ac{SM} one, the 
CP-violating part of the bilinears $\hbi_{us}$, $\hbi_{ud}$ should be a factor $\sim 10^3$ smaller than the CP-conserving part.

By confronting the contribution of the model including vector-like quark doublets with the experimental result we can get a $95\%$ confidence level limit on the CP-violating part of the new-physics contribution:
\begin{equation} \label{eq:epsprime-constr}
- 5.7 \times 10^{-7} \,\,\lesssim \,\,
\frac{|\Vha_{us}|}{|\Vha_{ud}|}\;\frac{\im\Big[ \Vhac_{ud}\VRha_{ud}\Big]}{|\Vha_{ud}| } + \frac{\im\Big[\Vhac_{us}\VRha_{us} \Big] }{|\Vha_{us}| } \,\,
\lesssim \,\, 3.1 \times 10^{-7}\,.
\end{equation}
This constraint implies that the imaginary part of that combination
of bilinears should be at least $3$ orders of magnitude smaller than the CP conserving part 
requested by the Cabibbo angle anomalies:
\begin{equation}
\label{eq:epsprime-constr-caa}
  \left|\frac{ \im\Big[ |\Vha_{us}|^2\,\Vhac_{ud}\VRha_{ud}-|\Vha_{ud}|^2\,\Vha_{us}\VRhac_{us} \Big] }{\re \Big[ |\Vha_{us}|^2\,\Vhac_{ud}\VRha_{ud}-|\Vha_{ud}|^2\,\Vha_{us}\VRhac_{us} \Big]} \right|=
   \left|\frac{ \im\Big[ |\Vha_{us}|^2\,\hbi_{ud}+|\Vha_{ud}|^2\,\hbi_{us} \Big] }{\re \Big[ |\Vha_{us}|^2\,\hbi_{ud}-|\Vha_{ud}|^2\,\hbi_{us} \Big]}\right|\lesssim 10^{-3}\,.
\end{equation}
This description clarifies the constraint placed on new physics by the observed smallness of direct CP violation, namely,
the phases of the bilinears
$\hbi_{us}^*/|\Vha_{us}|$ and 
$ \sim 0.23 \, \hbi_{ud}/|\Vha_{ud}|$ should tend to align:
\begin{equation}
    \left|\arg\left(\frac{0.23 \, \Vhac_{ud} \VRha_{ud}}{|\Vha_{ud}|}- \frac{\Vha_{us}\VRhac_{us}}{|\Vha_{us}|}\right)\right| \lesssim 10^{-3}\,.
\end{equation}
The diagram of the process is sketched in~\Cref{fig:directcp}.

\subsubsection{Electric dipole moments}

A combination of the imaginary parts of the bilinears $\hbi_{ud}$, $\hbi_{us}$ is constrained by the electric dipole moments.
The experimental constraints for the neutron and proton electric dipole moments $d_n$ and $d_p$ at $90\%$ CL respectively are
\begin{equation}
\begin{aligned}
d_n &< \, 1.8\times 10^{-13}\: \text{e fm} \quad \text{\cite{Abel:2020pzs}} \, , \\
d_p &< \, 2.1\times 10^{-12}\: \text{e fm} \quad  \text{\cite{Sahoo:2016zvr}}\,.
\end{aligned}
\end{equation}
The contribution of right-handed couplings 
to electric dipole moments was computed in Ref.~\cite{Alioli:2017ces}
(neglecting subleading contributions from couplings other than to light generations ($u,d,s$)):
\begin{align} \label{eq:dnvlq}
 d_n &\approx \im\Big[(1.4\pm 0.7)\, \Vhac_{ud}\VRha_{ud} +(2.7\pm 1.3)\, \Vhac_{us}\VRha_{us}  \Big] 
\times 10^{-7} \text{ e fm} \,,\\*[2mm]
\label{eq:dpvlq}
d_p &\approx \im\Big[-(2.7\pm 1.3)\, \Vhac_{ud}\VRha_{ud} -(3.6\pm 1.5)\, \Vhac_{us}\VRha_{us}  \Big]
\times 10^{-7} \text{ e fm}\,. 
\end{align}
We can use the bounds on the neutron and proton electric dipole moment 
to obtain a limit on the combination of bilinears:
\begin{equation} \label{eq:edm-constr}
\begin{aligned}  
\left|\left(2.3\pm 1.1 \right)\,\frac{\im\Big[ \Vhac_{ud}\VRha_{ud}\Big]}{|\Vha_{ud}| } +\left(1.0\pm 0.5 \right)\,\frac{\im\Big[\Vhac_{us}\VRha_{us} \Big] }{|\Vha_{us}| } \right|
\,&\lesssim \, 3 \times 10^{-6} \, , \\[2mm]
\left|\left(3.3\pm 1.6 \right)\,\frac{\im\Big[ \Vhac_{ud}\VRha_{ud}\Big]}{|\Vha_{ud}| } +\left(1.0\pm 0.4 \right)\,\frac{\im\Big[\Vhac_{us}\VRha_{us} \Big] }{|\Vha_{us}| } \right|
\,&\lesssim \, 6 \times 10^{-6}\,,
\end{aligned}
\end{equation}
which can be compared to the constraint in~\cref{eq:epsprime-constr}. 
This condition requires the imaginary part of the combinations of
bilinears to be at least about $2$ orders of magnitude smaller than the CP-conserving part
indicated by the Cabibbo angle anomalies.

\subsubsection{Low-energy \acs{EW} flavour-conserving observables and \texorpdfstring{$Z$}{Z} decay}
\label{sec:Z-decay}

Cabibbo angle anomalies imply a large mixing of the vector-like species with the light quarks. 
In this context, $Z$-boson physics provides constraints on the magnitude of
the flavour-conserving couplings.
In fact,
mixing with the heavy doublet changes the prediction of the $Z$ decay rate into hadrons (and consequently the total decay rate) as 
\begin{equation} \label{eq:Zdo} 
\begin{aligned}
&\Gamma(Z\rightarrow\text{had})-\Gamma(Z\rightarrow\text{had})_\text{SM}=
 \Gamma(Z)-\Gamma(Z)_\text{SM}
 \\
&\qquad \approx\frac{G_FM^3_Z}{\sqrt{2}\pi}   \left[
-\frac{2}{3}\sin^2\theta_W\left(\hat{\Fr}_{uu}+\hat{\Fr}_{cc}\right)
-\frac{1}{3}\sin^2\theta_W \left(\hat{\Fr}_{dd}+\hat{\Fr}_{ss}+\hat{\Fr}_{bb}\right)\right]\,,
\end{aligned}
\end{equation}
where 
\begin{equation} \label{eq:toymodel11}
   \hat{\Fr}_{\alpha\alpha}=\frac{\vw^2}{\mB^2}|\hi_{\alpha}|^2\approx \Fua_{\alpha\alpha}  \, , \qquad \hat{\Fr}_{ii}=\frac{\vw^2}{\mB^2}|\hi_{i}|^2\approx \Fda_{ii}
\end{equation}
are rephasing-invariant moduli.
\Cref{eq:Zdo} implies that the predicted decay rate is lower than the \ac{SM} expectation.

First, we consider the total decay rate of the $Z$ boson as well as the partial decay rate into hadrons. The experimental measurements yield~\cite{ParticleDataGroup:2024cfk}
\begin{equation}
\begin{aligned}
& \Gamma(Z)_\text{exp}=2.4955\pm0.0023~\text{GeV}\,,  \\ 
&\Gamma(Z\rightarrow\text{hadr})_\text{exp}=1.7432\pm0.0019~\text{GeV}\,,
\end{aligned}
\end{equation}
while the corresponding \ac{SM} predictions are $\Gamma(Z)_\text{SM}=2.4940\pm0.0009\text{ GeV}$ and $\Gamma(Z\rightarrow\text{hadr})_\text{SM}=1.74088\pm0.00086\text{ GeV}$~\cite{ParticleDataGroup:2024cfk}.
At $2\sigma$ CL we obtain the limit 
\begin{equation} \label{eq:Zdecay-constr}
\hat{\Fr}_{uu}+\hat{\Fr}_{cc}+\frac{1}{2} \left(\hat{\Fr}_{dd}+\hat{\Fr}_{ss}+\hat{\Fr}_{bb}\right) 
\lesssim 5.7 \times 10^{-3}\,.
\end{equation}

Another set of flavour-conserving constraints originates from parity-violating effects at low-energy electron-hadron processes with $Z$-boson exchange, as well as
other low-energy electroweak observables as the oblique parameters
(see Refs.~\cite{Belfatto:2021jhf,Crivellin:2022rhw} for the effects of vector-like doublets; in the subsequent analysis of~\cref{sec:CAApheno} we use $g^{ep}_{AV,\text{exp}}=- 0.0356 \pm 0.0023$ for the weak charge of the proton and we update the constraint $55 g^{ep}_{AV}+78 g^{en}_{AV}=36.25\pm 0.21$ for atomic parity violation in Cesium, $Q^{55,78}_W(\text{Cs})_\text{exp}$~\cite{ParticleDataGroup:2024cfk}).

The importance of the \ac{VLQ} doublet in explaining the Cabibbo angle anomalies emerges particularly when considering these observables~\cite{Belfatto:2021jhf,Crivellin:2022rhw,Belfatto:2023tbv,Cirigliano:2023nol}.
In fact, in general, explanations for the \ac{CKM} unitarity deficit can be in tension with 
 electroweak low-energy observables, e.g.~the value of $m_W$.
For example, it was suggested in Ref.~\cite{Belfatto:2019swo} that a solution to \ac{CAA}1 which modifies the Fermi constant $G_F$ with respect to the muon decay constant $G_\mu$ also implies a deficit in $m_W$. Conversely,
in Ref.~\cite{Cirigliano:2022qdm}, it is shown that the models that predict a positive shift in the $W$ mass may also predict a huge violation of \ac{CKM} unitarity, much larger than the one indicated by the current anomaly.
The vector-like quark charged as a doublet of $SU(2)_L$ emerges as a favoured candidate for explaining the \acp{CAA} since the
mixing with the light generations of \ac{SM} quarks can resolve these anomalies
at tree level, while 
the contribution to $m_W$ is
at the one-loop level and can be relevant only in presence of a large coupling to the top quark.

\subsection{Trilinears and flavour-changing neutral currents}
\label{sec:phenotrilinears}

In the presence of \acp{VLQ}, flavour-changing neutral currents are induced at tree level as well as at loop level. 
These flavour-changing effects interfere with the \ac{SM}, both in CP-conserving and CP-violating processes. In this section, we illustrate the effects on relevant kaon decays and on kaon mixing, in particular concerning 
mechanisms of CP violation.
In fact, we have treated CP violation in decay amplitudes (direct CP violation) in~\cref{sec:eps-prime}. In what follows, we will illustrate the rephasing-invariant effects of the vector-like doublet in CP violation in neutral kaon mixing (indirect CP violation) and CP violation due to the phase mismatch between mixing parameters and decay amplitudes (`interference CP violation').

\subsubsection{Kaon mixing}

In the \ac{SM}, the short-distance contribution to the transition 
$K^0(d\bar{s})\leftrightarrow \bar{K}^0(\bar{d}s)$
arises from weak box diagrams.
The effective Lagrangian describing this contribution is given by:
\begin{equation} \label{eq:smkk}
\mathcal{L}^\text{SM}_{\Delta S=2}=-\frac{G^2_F m^2_W}{4\pi^2}\left(\lambda^{2}_c S_0(x_c)+\lambda^{2}_t S_0(x_t)+2\lambda_c\lambda_tS_0(x_c,x_t)\right)(\overline{s_L}\gamma^\mu d_L)^{2}
+\text{h.c.} \,,
\end{equation}
where $\lambda_a=V_{as}^*V_{ad}$, $x_a={m_a^2}/{m_W^2}$ and $S_0(x_i)$ are the Inami-Lim
functions~\cite{Inami:1980fz}.
The weak short-distance contribution to the mass splitting $\Delta m_K=m_{K_L}-m_{K_S}$
and the CP-violating effects 
are described by
the off-diagonal term $M_{12}$ of the mass matrix of neutral kaons
$M_{12}=- \langle K^0|\mathcal{L}_{\Delta S=2}|\bar{K}^0\rangle /( 2m_{K})$, which in the \ac{SM} is~\cite{Branco:1999fs}
\begin{equation} \label{eq:M12sm}
M_{12}^\text{SM}=
-\frac{G^2_F m^2_W f^2_Km_{K}B_K}{12\pi^2}
\left(\lambda^{*2}_c \, S(x_c)+\lambda^{*2}_t \, S(x_t)+2\,\lambda^*_c\lambda^*_t \, S(x_c,x_t)\right)\,,
\end{equation}
where $f_K$ is the kaon decay constant, which can be estimated in lattice QCD to be
$f_K=155.7(0.7)$ MeV~\cite{FlavourLatticeAveragingGroupFLAG:2021npn},
 $m_{K}=497.611\pm0.013$~MeV is the neutral kaon mass, 
the factor $B_{K}$ is the correction to the vacuum insertion approximation which is calculated in
lattice QCD,
$B_K=0.7533(91)$~\cite{FlavourLatticeAveragingGroupFLAG:2024oxs},
and $S(x_i)$ are the Inami-Lim
functions~\cite{Inami:1980fz} (corrected by short-distance QCD effects~\cite{Buchalla:1995vs}).
The modulus of the mixing mass $M_{12}$
describes short-distance contributions in the mass splitting.
The mass splitting is given by $\Delta m_K \approx 2|M_{12}|
+\Delta m_{K,\text{LD}} $~\cite{Buchalla:1995vs}, where
$\Delta m_{K,\text{LD}}$ is the long-distance contribution which is difficult to evaluate~\cite{Bai:2014cva,Bai:2018mdv}.
However, the short distance contribution 
gives the dominant contribution to
the experimental determination 
$\Delta m_{K,\text{exp}}=(3.484\pm 0.006) \times 10^{-15}$~GeV~\cite{ParticleDataGroup:2024cfk}.

From~\cref{eq:om-eps-epsp}, 
the CP-violating parameter $\epsilon$
can be written as~\cite{Branco:1999fs}:
\begin{equation} \label{eq:epsK}
\epsilon \,\approx\,  \frac{e^{i \pi/4}}{\sqrt{2}} \frac{2\,\im\left(M_{12}^*\Gamma_{12}\right)}{4\big|M_{12}\big|^2+\big|\Gamma_{12}\big|^2} 
\,\approx\, - e^{i \pi/4} \frac{\im \left( M_{12} A_0 \bar{A}_0^* \right)}{\sqrt{2}\,(\Delta m_K)\, \big|A_0\bar{A}_0\big|}\,
\end{equation}
(recall that $ \Delta m_K \approx 2 |M_{12}| \approx - \frac{1}{2} \Delta\Gamma \approx |\Gamma_{12}|\approx \frac{1}{2} \Gamma_S$).
The long-distance contribution to $M_{12}$ should have a phase 
$\arg\lambda_u^{*2}+\xi_s-\xi_d-\xi_K$ which does not modify
$\epsilon$~\cite{Branco:1999fs}, so that
the relevant contributions are the short-distance ones
(box diagrams).
The crucial approximation in the last formula 
is considering that the decay channel
$|\pi\pi,I=0\rangle$ is the dominant channel, $\Gamma_{12} = \sum_f A_f^*\bar{A}_f \approx A_0^*\bar{A}_0$
(where the sum is over all decay modes $f$).
In this approximation, the parameter $\epsilon$ represents the mixing CP violation in the neutral kaon system
and the phase $\arg\epsilon$ is given by the superweak phase $\phi_\epsilon=(43.52\pm 0.04)^\circ$.

In the \ac{SM}, at tree level,
\begin{equation} \label{eq:A0ratio}
\frac{A_0}{\bar{A}_0} =\frac{V_{us}^*V_{ud}}{V_{us}V_{ud}^*} e^{i (\xi_K +\xi_d-\xi_s)}\,.
\end{equation}
The phases $\xi_K$, $\xi_d$, $\xi_s$ are arbitrary phases in the definition of the CP transformations of $|K^0\rangle$ and the quark fields $d$ and $s$, respectively, which are cancelled by the 
same phases arising from the hadronic matrix element 
$\langle K^0 | (\overline{d_L}\gamma^\mu s_L)(\overline{d_L}\gamma_\mu s_L) |\bar{K}^0 \rangle \propto - e^{-i (\xi_K +\xi_d-\xi_s)}f^2_Km_K/3$
in $M_{12}$, and thus for brevity we will drop them in the formula for the parameter $\epsilon$, which is independent of it.
The relation in~\cref{eq:A0ratio} 
holds when $A_0$ and $\bar{A}_0$ are dominated by the tree-level diagram. 
Loop contributions 
are also proportional to $V_{ts}^*V_{td}$. However, because of the small experimental value of $\epsilon'/\epsilon$,
loop contributions to the phases of $A_0$ and $\bar{A}_0$ must be very small.
Then, from~\cref{eq:epsK,eq:A0ratio} we have
\begin{equation}
\epsilon \approx - e^{i \pi/4}\,\frac{\im\Big[ M_{12} \, \lambda_u^2 \Big]}{\sqrt{2}\,\Delta m_K \,\big| \lambda_u  \big|^2}\,.
\end{equation} 
From experimental data one obtains~\cite{ParticleDataGroup:2022pth}
\begin{equation}
 |\epsilon|_{\text{exp}}=(2.228 \pm 0.011) \times 10^{-3}\,.
\end{equation}
In the standard parameterization of the \ac{CKM} matrix, $\lambda_u$ is real and it is usually omitted.
However, we are interested in a rephasing-invariant analysis. 
Then, in the \ac{SM} we have:
\begin{equation} \label{eq:epsilonSM}
\begin{aligned}
   \epsilon &\,\approx\,
- \frac{e^{i \pi/4}}{\sqrt{2}\,\Delta m_K}
   \frac{\im\Big[ M^\text{SM}_{12} \, \, \lambda_u^2 \Big]}{ \big|  \lambda_u  \big|^2}
\,=\,
  \frac{e^{i \pi/4}f^2_Km_{K}B_K}{\sqrt{2}\,\Delta m_K}\frac{G^2_F m^2_W }{12\pi^2}
     \\ &\qquad \quad
\times \frac{1}{\big|  \lambda_u  \big|^2} \im\Big[
    \left( Q_{csud} \right)^2\, S(x_c)+
    \left( Q_{tsud} \right)^2\, S_0(x_t)+
    2 \, Q_{tsud}\, Q_{csud} \, S (x_c,x_t)\Big] \,.
\end{aligned}
\end{equation}
It is clear that the CP-violating effects in the \ac{SM} are given by the imaginary parts of quartets 
appearing in the weak-basis invariant of mass dimension $\M = 12$ in~\cref{eq:cp-sm}.
This is directly related to the fact that CP violation emerges at loop level, in particular, in the case of kaon mixing, in box diagrams with four quarks involved.
The last line of~\cref{eq:epsilonSM} can be explicitly written in terms of the 
$J$ invariant of~\cref{eq:Jar}:
\begin{equation}
   \frac{\im\Big[ M^\text{SM}_{12} \, \lambda_u^2\Big]}{\left|  \lambda_u  \right|^2}
   \propto
       \frac{2\, J}{\left|  \lambda_u  \right|^2}
\Big(\re \left( Q_{csud} \right) S(x_c)-
      \re \left( Q_{tsud} \right) S_0(x_t)+
\re\left(Q_{tsud} -Q_{csud} \right) S (x_c,x_t)  
        \Big)\,.
\end{equation}

\begin{figure}
\centering
\includegraphics[width=0.3\textwidth]{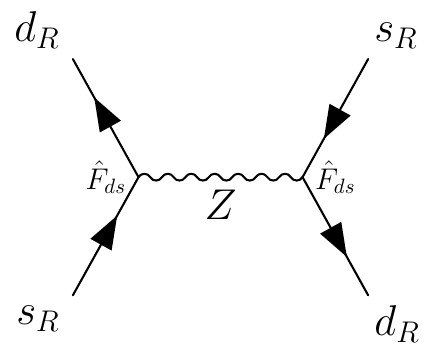}
\hspace{2cm}
\includegraphics[width=0.3\textwidth]{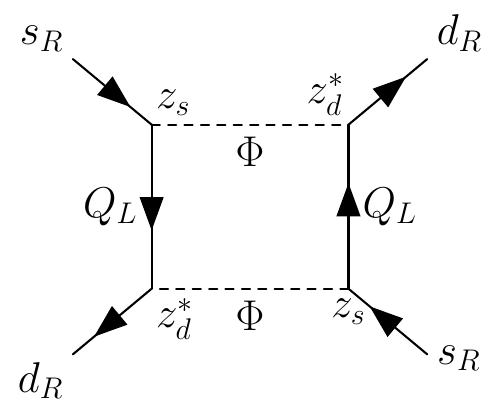}
\caption{\label{fig:kaons}
New diagrams relevant for neutral kaon mixing.}
\end{figure}
\begin{figure}[t]
    \centering
    \includegraphics[width=0.85\linewidth]{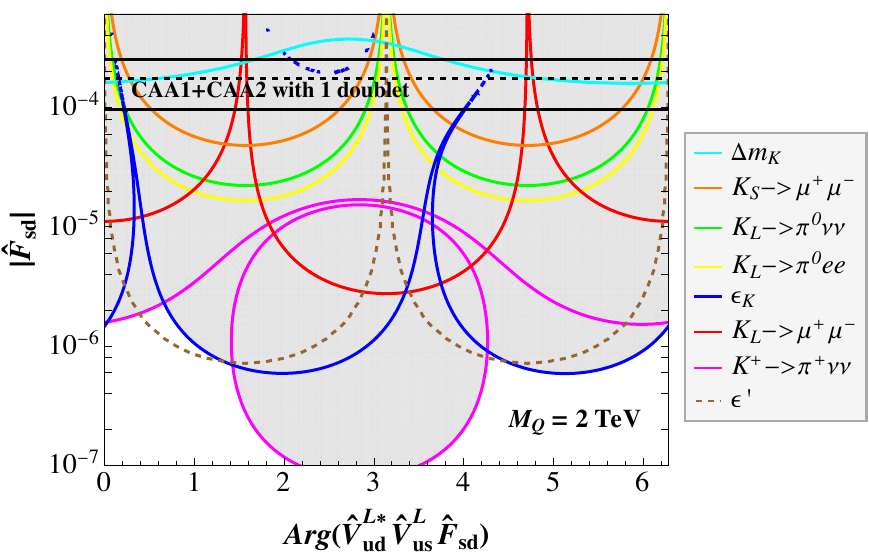}
    \caption{Upper limits obtained from kaon decays and neutral kaon mixing
on the flavour-changing coupling $|\htr_{u,ds}|/|\Vha_{ud}\Vhac_{us}|=|\hat{\Fr}_{ds}|=|\hat{\Fr}_{sd}|=|\hid\his|\vw^2/\mB^2$
as a function of the phase of the rephasing invariant 
$\htr_{u,ds}^*=\Vhac_{ud}\Vha_{us}\hat{\Fr}_{sd}$ (see text and Refs.~\cite{Belfatto:2021jhf,Belfatto:2023tbv} for details).
We also show the bound from $\epsilon'/\epsilon$ (dashed brown line) considering only the contribution of $\htr_{u,ds}$ which however would require $\hiu$ vanishingly small, which is not the case addressing the Cabibbo angle anomalies
(when $\hiu\neq 0$ the CP violation in $\epsilon'$ can receive contributions from
the bilinears and this bound disappears).
Recalling that $|\hat{\Fr}_{sd}|=|\VRhac_{ud}\VRha_{us}|/\hat{\Fr}_{uu}$, the black lines indicate the values of $|\hat{\Fr}_{sd}|$
obtained by considering the $1\sigma$ interval of 
the mixings $|\VRha_{ud}\VRhac_{us}|$ needed to explain the Cabibbo angle anomalies,
assuming $\hiu=0.8$ with $\mB=2\,\text{TeV}$, or equivalently $|\hiu|\vw/\mB=0.07$ (the maximum value allowed by the $Z$-decay constraint). 
}
\label{fig:kaons-fcnc}
\end{figure}

In the presence of \acp{VLQ}, a new contribution to the mixing mass term
is induced at tree level as well as at loop level. 
Then, $M_{12}=M^\text{SM}_{12}+M^{\text{NP}}_{12}$ with
\begin{equation}
\begin{aligned}
M^{\text{NP}}_{12} &\approx 
-\frac{1}{3}m_{K}f_{K}^{2} \, 0.43\, \bigg\{
\frac{G_F}{\sqrt{2}}\, \hat{\Fr}_{ds}^2 \,+ \frac{G_F^2}{4\pi^2}   \bigg[  
\,\frac{1}{2} \mB^2 \, \hat{\Fr}_{ds}^2   \\ &
\qquad\qquad-  3.1 \frac{m_{K}^2}{(m_d+m_s)^2}
 \,
\hat{\Fr}_{ds} \,
\Vhac_{td} \Vha_{ts} \, m_W^2 f(x_Q,x_t)
\bigg]\bigg\}\,,
\end{aligned}
\end{equation}
where for the last term
we used the results calculated in Refs.~\cite{Garron:2016mva,Buras:2001ra,Bobeth:2017xry}, $f(x_Q,x_t)\approx x_t \ln(x_Q)\,/\,4$. The first term represents the tree level contribution, the second term is the one loop right-right contribution (see~\Cref{fig:kaons}) 
and the last term is the mixed left-right contribution which emerges at loop level. The right-right loop-level contribution is relevant since it is parametrically stronger ($\propto 1/\mB^2$) than the tree-level contribution ($\propto 1/\mB^4$)
\cite{Ishiwata:2015cga,Belfatto:2021jhf}, so that 
the constraint on the mixing $\Fdha_{ij}$ becomes stronger with increasing mass of the extra doublet (see Ref.~\cite{Belfatto:2021jhf}).
We showed in~\cref{sec:eps-prime} that, in the context of Cabibbo angle anomalies,
the new-physics contribution to the amplitudes $A_0$, $\bar{A}_0$ is negligible (as expected).
Then, we can use~\cref{eq:epsK,eq:A0ratio} and write the new contribution to $\epsilon$:
\begin{equation}
\begin{aligned}
   \frac{\im\Big[ M^\text{NP}_{12} \, (\lambda_u)^2\Big]}{|\lambda_u|^2}
    &=
    \frac{1}{3}m_{K}f_{K}^{2} \, 0.43\, \im\bigg\{
\frac{G_F}{\sqrt{2}}\frac{\left(\htr_{u,ds}\right)^2}{|\lambda_u|^2} +
 \frac{G_F^2}{4\pi^2}   \bigg[  
\,\frac{1}{2} \mB^2 \frac{\left(\htr_{u,ds}\right)^2}{|\lambda_u|^2}  \\ &
\qquad\qquad \qquad -  3.1 \frac{m_{K}^2}{(m_d+m_s)^2} 
 \,
\frac{\htr_{u,ds} \, \Qh_{tsud}}{|\lambda_u|^2}  \, m_W^2 f(x_Q,x_t)
\bigg]\bigg\}\,,
\label{eq:epsNP}
\end{aligned}
\end{equation} 
with $\lambda_u=\Vhac_{us}\Vha_{ud}$.
Let us remark how the new contribution enters in $\epsilon$ only via
rephasing invariants. In particular, this flavour-changing and CP-violating contribution is determined by the trilinear $\htr_{u,ds} \equiv \Vha_{ud}\Vhac_{us}\hat{\Fr}_{ds} = \lambda_u \hat{\Fr}_{ds}$.
Constraints can be estimated as $|M^\text{NP}_{12}|<|M^\text{SM}_{12}|\,\Delta_{K}$, $|\im M^\text{NP}_{12}|<|\im M^\text{SM}_{12}| \, \Delta_{\epsilon_{K}} $.
Then, the bound on the new physics will be a function of 
the modulus $|\htr_{u,ds}|$ and phase $\arg\big(\htr_{u,ds}\big)$ of the rephasing invariant. 
Setting $\Delta_{K}=1$ and using the results in Ref.~\cite{Bona:2022zhn} at $95\%$ CL 
(which approximately corresponds to $\Delta_{\epsilon_{K}}= 0.3$) 
we obtain, for $\mB\approx 2$~TeV (the experimental limit on the mass of vector-like doublets coupling to light quarks is $\mB\gtrsim 1$~TeV~\cite{ATLAS:2024zlo}):
\begin{equation} \label{eq:vdshhkk}
 |\hat{\Fr}_{ds}|
 \,<\,  6 \times 10^{-7} \mbox{ --- } 2\times 10^{-4}\,,
\end{equation}
depending on the relative phase of the couplings.
We show the result in~\Cref{fig:kaons-fcnc} (blue line).
The limit becomes stronger with increasing mass $\mB$ because of the loop-level contribution (second term in~\cref{eq:epsNP}).

Notice that after considering flavour-changing processes together with the constraint from $Z$ decays in the one-doublet case we would have 
\begin{equation} \label{eq:1doub-fcvscaa}
|\VRha_{ud}|= \frac{\hat{\Fr}_{uu}
|\hat{\Fr}_{ds}|}{|\VRha_{us}|}\lesssim\frac{5\times 10^{-3}\times 2\times 10^{-6}}{10^{-3}}\approx 10^{-5}\ll 10^{-3}\,,
\end{equation}
after assuming the value of $\VRha_{us}$ needed for 
the Cabibbo angle anomaly \ac{CAA}2, that is, the \ac{CAA}1 tension
would not be explained with one doublet.

\subsubsection{\texorpdfstring{$K_L\rightarrow\pi^0\nu\nu$}{K\_L -> π0 ν ν}, 
\texorpdfstring{$K_{S,L}\rightarrow\mu\mu$}{K\_S,L -> μ μ} and other rare kaon decays}

The rephasing invariant $\htr_{u,ds}$ also generates 
contributions to
flavour-changing rare kaon decays, as $K\rightarrow\pi\nu\nu$, $K\rightarrow\mu\mu$, both CP-conserving and CP-violating.
Regarding CP violation, 
while $\epsilon'$ represents CP violation in the decay amplitudes (direct, $|\bar{A}/A|\neq 1$) and $\epsilon$ indicates CP violation in the mixing (indirect, $\arg(M_{12}^*\Gamma_{12})\neq 0$), 
CP violation in these decays is due
to the phase mismatch between the mixing parameters and the decay amplitudes.

The decay $K_L\rightarrow\pi^0\nu\nu$ is theoretically very clean. 
It can be shown that if lepton flavour is conserved ($\nu$ and $\bar{\nu}$ are each other's antiparticle and they are generated by the $\bar{\nu}_{\ell L}\gamma^\mu \nu_{\ell L}$ operator) the final state is  CP even~\cite{Littenberg:1989ix,Kayser:1996sv,Grossman:1997sk,Buchalla:1998ux}. Then the decay $K_L\rightarrow\pi^0\nu\nu$ is CP-violating.
It is short-distance dominated 
 and
the effective Lagrangian
in the \ac{SM} reads~\cite{Buchalla:1995vs} 
\begin{equation} \label{eq:LKpinunu-sm}
\mathcal{L}(K\rightarrow\pi\nu\nu)_\text{SM} = -\frac{g^4}{16 \pi^2 m_W^2}\Big[ \VLac_{cs} \VLa_{cd} X(x_c)+\VLac_{ts} \VLa_{td} X(x_t) \Big] \left( \bar{s}_L\gamma^\mu d_L \right) \sum_{e,\mu,\tau} \left( \bar{\nu}_{\ell L}\gamma^\mu \nu_{\ell L} \right)\,,
\end{equation}
where $x_{c,t}=m^2_{c,t}/M^2_W$ and $X(x_a)$ are the Inami-Lim functions~\cite{Inami:1980fz}.
The mixing with the vector-like quark doublets induces the new contribution
\begin{equation} \label{eq:LKpinunu-new}
\mathcal{L}(K\rightarrow\pi\nu\nu)_\text{NP} =
\frac{4G_F}{\sqrt{2}} \, \frac{1}{2}\, \hat{\Fr}_{sd} \,
(\bar{s}_R\gamma^\mu d_R) \sum_{e,\mu,\tau}(\bar{\nu}_{\ell{L}}\gamma_\mu\nu_{\ell{L}} )\,.
\end{equation}
The amplitude for the decay $K_L\rightarrow\pi^0\nu\nu$ is given by
\begin{equation}
A( K_L\rightarrow\pi^0\nu\nu ) = p_K A(K^0\rightarrow\pi^0\nu\nu) +q_K A(\bar{K}^0\rightarrow\pi^0\nu\nu)\,.
\end{equation}
The relevant CP-violating parameter in this decay is~\cite{Branco:1999fs}
\begin{equation}
\begin{aligned}
 \lambda_{\pi\nu\bar{\nu}}&= \frac{q_K}{p_K}\frac{A(\bar{K}^0\rightarrow\pi^0\nu\nu)}{A(K^0\rightarrow\pi^0\nu\nu) }
 \\ 
&\approx -\frac{\Vhac_{us}\Vha_{ud}}{\Vha_{us}\Vhac_{ud}} \;
\frac{\frac{\alpha}{2\pi \sin^2\theta_W}\left(\Vha_{cs} \Vhac_{cd} X(x_c)+\Vha_{ts} \Vhac_{td} X(x_t) \right)-\frac{1}{2} \hat{\Fr}_{sd}^*}{\frac{\alpha}{2\pi \sin^2\theta_W}\left(\Vhac_{cs} \Vha_{cd} X(x_c)+\Vhac_{ts} \Vha_{td} X(x_t) \right)-\frac{1}{2} \hat{\Fr}_{sd}}\,,
\end{aligned}
\end{equation}
and the decay amplitude can be written as
\begin{equation}
A( K_L\rightarrow\pi^0\nu\nu ) = p_K \left( 1+ \lambda_{\pi\nu\bar{\nu}}  \right) A(\bar{K}^0\rightarrow\pi^0\nu\nu)\,,
\end{equation}
where $\lambda_{\pi\nu\bar{\nu}}\neq - 1$ implies CP violation.
Since strong final state interaction phases are absent and diagrams with intermediate up quarks are suppressed by the \ac{GIM} mechanism, there is no direct CP violation ($|\bar{A}/A|=1$). Indirect CP violation is very small
($|q/p|$ can be considered as 1). The main reason why 
$|\lambda_{\pi\nu\bar{\nu}}|\neq 1$ in the \ac{SM} as well as in the new-physics scenario is because the phases of $q/p$
and $\bar{A}/A$ do not match, that is,
interference between mixing and decay~\cite{Littenberg:1989ix}.

Then, the total branching ratio results in
\begin{equation} \label{eq:brKLSMplusNP}
\frac{\br(K_L\rightarrow \pi^0\nu\bar{\nu})}{   \br(K_L\rightarrow \pi^0\nu\bar{\nu})_\text{SM}}
\approx  
  \left|1-\frac{\frac{1}{2}\im \Big[\Vha_{us}\Vhac_{ud}\,\hat{\Fr}_{sd}\Big]}{\im \Big[\Vha_{us}\Vhac_{ud}\,\frac{\alpha}{2\pi \sin^2\theta_W}\left(\Vhac_{cs} \Vha_{cd} X(x_c)+\Vhac_{ts} \Vha_{td} X(x_t) \right)\Big]}\right|^{2}\,.
\end{equation}
The experimental limit on this decay at $90\%$ CL is~\cite{ParticleDataGroup:2024cfk}
\begin{equation} \label{eq:kLexp}
\br(K_{L}\rightarrow \pi^0\nu\bar{\nu})_\text{exp}<3.0\times 10^{-9} \,,
\end{equation}
which is two orders of magnitude larger than the \ac{SM} expectation~\cite{Buras:2015qea}
\begin{equation}
\br(K_{L}\rightarrow \pi^0\nu\bar{\nu})_\text{SM}=(3.00\pm 0.30) \times 10^{-11} \,,
\end{equation}
so that the experimental limit can be applied 
to the \ac{NP} contribution.
Then, we obtain the rephasing-invariant bound:
\begin{equation}
  \frac{\frac{1}{2}\left|\im \Big[\left(\Vha_{us}\Vhac_{ud}\right)\,\hat{\Fr}_{sd}\Big]\right|}{\frac{\alpha}{2\pi \sin^2\theta_W}\left|\im \Big[\left(\Vha_{us}\Vhac_{ud}\right)\,\left(\Vhac_{cs} \Vha_{cd} X(x_c)+\Vhac_{ts} \Vha_{td} X(x_t) \right)\Big]\right|} <  9.3\,,
\end{equation}
that is, the new contribution can be $\sim 10$ times larger than the \ac{SM} contribution.

A similar analysis leads to rephasing-invariant constraints on the decays
$K_{L}\rightarrow\mu\mu$ and $K_{S}\rightarrow\mu\mu$,
 which involve the real part and the imaginary part of the invariant $\Vha_{us}\Vhac_{ud}\,\hat{\Fr}_{sd}$ respectively.

The same effective Lagrangian in~\cref{eq:LKpinunu-sm,eq:LKpinunu-new} 
leads to the decay $K^+\rightarrow\pi^+\nu\nu$.
Ref.~\cite{ParticleDataGroup:2024cfk} estimates the \ac{SM} contribution as $\br(K^+\rightarrow \pi^+\nu\bar{\nu})_\text{SM}=(0.81\pm 0.04)\times 10^{-10}$
(it also states that parametric uncertainty in the CKM angles can result in shifts of the central value differing from this one by up to $10\%$).
The experimental limit is $\br(K^+\rightarrow \pi^+\nu\bar{\nu})_\text{exp}=1.14^{+0.40}_{-0.33} \times  10^{-10}$ at $2\sigma$~\cite{ParticleDataGroup:2024cfk}.
One obtains 
\begin{equation} \label{eq:kpnncon0}
0.98\lesssim \left|-\frac{\frac{1}{2}\left(\Vha_{us}\Vhac_{ud}\right)\hat{\Fr}_{sd}}{\frac{\alpha}{2\pi \sin^2\theta_W}\left(\Vha_{us}\Vhac_{ud}\right)\,\left(\Vhac_{cs} \Vha_{cd} X(x_c)+\Vhac_{ts} \Vha_{td} X(x_t) \right)}+1\right|\,\lesssim\, 1.44\,,
\end{equation}
Notice that the \ac{SM} expectation is at the edge of the $2\sigma$
region determined by the experimental value. Then, 
a new contribution with 
modulus and phase
\begin{equation}
    |\hat{\Fr}_{ds}|\lesssim 3\times 10^{-6} \, ,
    \qquad
    0\leq \arg \big(\htr_{u,ds}^* \big)\lesssim 1.3 \,
    \quad \text{or} \quad
    4.3 \lesssim \arg\big(\htr_{u,ds}^* \big)\leq 2\pi
\end{equation}
would align the \ac{SM} prediction to the experimental result and it would always be allowed.
The result is shown in~\Cref{fig:kaons-fcnc}
(magenta line).

Let us note that CP-violating effects in the \ac{RH} sector are determined by the rephasing-invariant trilinears (see~\cref{eq:reph-inv-eff}), which appear in the weak-basis invariants of mass dimension $\M=10$. 
This is related to the fact that CP violation in neutral meson mixing due to the mixing with the vector-like doublet only involves two quarks of the same sector (down for $K$ and $B$, up for $D$ mesons).
As a consequence, the strength of CP violation can be expected to be stronger than in the \ac{SM}.

\subsubsection{CP violation in \texorpdfstring{$K\rightarrow\pi\pi$}{K -> π π} by trilinears}

\begin{figure}[t]
    \centering
\includegraphics[width=0.8\textwidth]{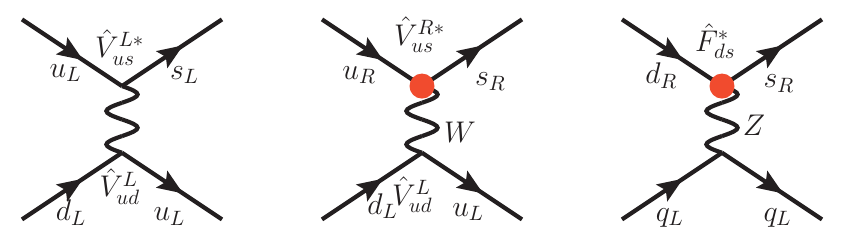}
    \caption{Simplified sketch of possible Feynman diagrams producing CP interference in $K\to\pi\pi$, probed in $\epsilon'$. On the left, the leading \ac{SM} contribution is shown, proportional to $\Vha_{ud}\Vhac_{us}$. In the middle, we show one of the contributions due to the mixing with the \ac{VLQ} doublet in charged currents, proportional to $\Vha_{ud}\VRhac_{ud}$ (see top panel of~\Cref{fig:integrate} and the operator in~\cref{eq:Heff-Kpipi}). On the right, we show the new contribution from neutral currents 
    proportional to $\hat{\Fr}_{ds}^*$ (see bottom panel of~\Cref{fig:integrate} and~\cref{eq:Leff-Kpipi-Fds}).} 
    \label{fig:directcp} 
\end{figure}

We observe how the new CP-odd rephasing invariant associated to \ac{RH} \acp{FCNC} appears and possible deviations from the \ac{SM} could be hinting for non-zero values of it.

Additional strong bounds to the same effective trilinear invariant $\htr_{u,ds}$ can be obtained from $\epsilon'$. The mechanism is sketched in~\Cref{fig:directcp}. The corresponding CP-violating $Z$-mediated \ac{RH} $s\rightarrow d$ transition, generates $\Delta S=1$ four-fermion operators below the \ac{EW} scale again potentially inducing a very large CP-violating $K\to(\pi\pi)_{I=2}$ amplitude. Indeed, using~\cref{eq:leftlag2} at the \ac{EW} scale and singling out the chirality-enhanced $8_L\times 8_R$ operators under $SU(3)_L\times SU(3)_R$ transformations ($q_{L(R)}^{l}\rightarrow L(R)q_{L(R)}^{l}$, $q^{l}\equiv (u,d,s)^T$), one finds 
\begin{equation} \label{eq:Leff-Kpipi-Fds}
\begin{aligned}
\mathcal{L}
&\supset \frac{4G_F}{\sqrt{2}}\left[\bar{u}_L\gamma^{\mu}u_L\left(\frac{1}{2}-\frac{2}{3}\sin^2\theta_W\right)-\left(\bar{d}_L\gamma^{\mu}d_L+\bar{s}_L \gamma^{\mu}s_L\right)\left(\frac{1}{2}-\frac{1}{3}\sin^2\theta_W\right)\right]\left(\bar{s}_R \hat{\Fr}_{sd} \gamma_\mu\, d_R\right)
\\ 
&=\frac{4G_F}{\sqrt{2}}  \left(-\frac{1}{3}\,\bar{q}^l_L\gamma^\mu q^l_L +\cos^2\theta_{W} \,\bar{q}_L^l \, Q \, \gamma^\mu q_L^l \right)\,\left( \bar{s}_R \hat{\Fr}_{sd} \gamma_{\mu} d_R \right) 
\\ &
\supset \frac{4G_F}{\sqrt{2}}\cos^2\theta_W \, \hat{\Fr}_{sd} \, \left( \bar{q}_L^l \, Q\, \gamma^{\mu} q_L^l \right)\,\left( \bar{s}_R \gamma_{\mu} d_R \right) \, .
\end{aligned}
\end{equation}
At leading order in the chiral counting, i.e.~in the limit of exact $SU(3)_V$ symmetry, the $K\to(\pi\pi)_{I=2}$ amplitudes generated by $\bar{q}_L^l Q \gamma_{\mu} q_L^l \bar{s}_R \gamma^{\mu} d_R$ and $\bar{u}_L \gamma_{\mu} d_L \bar{s}_R \gamma^{\mu} u_R$ entering in~\cref{eq:Heff-Kpipi} are identical.%
\footnote{One way of seeing this is using eqs.~(25) and~(29) of Ref.~\cite{Pich:2021yll}. The operator $\bar{q}_L^l Q q_L^l \bar{s}_R \gamma^{\mu} d_R$ induces $\tr[\frac{\lambda_{6}-i\lambda_{7}}{2}U Q U^{\dagger} ]$ with the same non-perturbative strength as $\bar{u}_L \gamma^{\mu} d_L \bar{s}_R \gamma_{\mu} u_R$ induces $\tr[\frac{\lambda_{4}-i\lambda_{5}}{2}U \frac{\lambda_{1}+i\lambda_{2}}{2}U^\dagger ]$. Using that the explicit $U$ convention is the one of eq.~(4) of Ref.~\cite{Pich:2018ltt}, one easily obtains the $K\to\pi\pi$ amplitudes.}
Thus the sensitivity of $A_2$ to $\frac{c_{W}^2}{\vw^2} \hat{\Fr}_{ds}$ is approximately the same as $-\frac{1}{\vw^2} (\VRha_{us}\Vhac_{ud})$, that is,
in this scenario the sensitivity to $c_W^2 \im \htr_{u,ds}$ is 
the same as the sensitivity to $-|\Vha_{ud}|^2 \im \hbi_{us}$ 
in~\cref{eq:epsprime-constr}. 
Thus, using~\cref{eq:epsprime-np-sm-ratio,eq:epsprime-constr}
we can write a bound to the non-standard contributions to $\epsilon'/\epsilon$ as:
\begin{equation} \label{eq:eps-prime-tri}
\begin{aligned}
\frac{\Big|\im \Big[ |\Vha_{us}|^2\hbi_{ud}+|\Vha_{ud}|^2\hbi_{us}-\htr_{u,ds}\cos^2\theta_W \Big] \Big|}{|\Vha_{us}\Vha_{ud}| }
\,\lesssim\,  5.6 \times 10^{-7}\,.
\end{aligned}
\end{equation}
Let us also notice that when $\hiu\neq 0$ this limit can be written as
\begin{equation}
\frac{\Big|\im \Big[ |\Vha_{us}|^2\hbi_{ud}+|\Vha_{ud}|^2\hbi_{us}-\hbi_{ud}^*\hbi_{us}\frac{\cos^2\theta_W}{\hat{\Fr}_{uu}^2} \Big] \Big|}{|\Vha_{us}\Vha_{ud}| }
\lesssim  5.6 \times 10^{-7}\,.
\end{equation}

\begin{table}[t]
\renewcommand{\arraystretch}{1.8} 
\centering
\begin{tabular}{y{3.4cm}y{4.cm}y{6cm}}
    \toprule
    \begin{tabular}{c}\bf Conventional \\[-4mm] \bf name\end{tabular}
    &     \begin{tabular}{c}\bf Rephasing-\\[-4mm] \bf invariant form\end{tabular} &   \textbf{Bounds on \ac{VLQ} invariants}  
\\ 
\midrule
$V_{ud}^{\beta}$ & $\big|\Vha_{ud}+\VRha_{ud}\big|$ & 
\multirow{3}{*}{$-\frac{\re \hbi_{ud}}{|\Vha_{ud}|}\approx 10^{-3}$,\,
$-\frac{\re \hbi_{us}}{|\Vha_{us}|}\approx 10^{-3}$} 
\\ 
$V_{us}^{K_{\ell 3}}$ & $\big|\Vha_{us}+\VRha_{us}\big|$ &  
\\ 
$\left(\frac{V_{us}}{V_{ud}}\right)_{\frac{K\mu 2}{\pi\mu2}}$ & $\frac{\big|\Vha_{us}-\VRha_{us}\big|}{\big|\Vha_{ud}-\VRha_{ud}\big|}$ &
\\
\midrule
$|\omega|$ & $\frac{\re (A_2 A_0^*)}{|A_0|^2}$ & 
$\frac{|\Vha_{us}|\re\mathcal{\hat{B}}_{ud}}{|\Vha_{ud}|}
    -\frac{|\Vha_{ud}|\re\mathcal{\hat{B}}_{us}}{|\Vha_{us}|}\lesssim
10^{-3}$
    \\
\midrule    
$\re\frac{\epsilon'}{\epsilon}$ & $\frac{\im(A_2 A_0^*)}{\sqrt{2}\, |\epsilon|\, |A_0|^2}$ & 
$
\Big| \frac{\left|\Vha_{us}\right|\im\mathcal{\hat{B}}_{ud}}{|\Vha_{ud}| } +\frac{\left|\Vha_{ud}\right|\im\mathcal{\hat{B}}_{us} }{|\Vha_{us}| }-\frac{\im\htr_{u,ds}\cos^2\theta_W }{|\Vha_{us}\Vha_{ud}| } \Big|
\lesssim \, 6 \times 10^{-7}$ \\ 
\midrule  
$d_{n,p}$ & \mbox{\cref{eq:dnvlq,eq:dpvlq}}  & $\frac{\im\mathcal{\hat{B}}_{ud}}{|\Vha_{ud}|}\, ,
    \frac{\im\mathcal{\hat{B}}_{us}}{|\Vha_{us}|}\lesssim
3 \mbox{ --- } 6 \times 10^{-6}$ \\ 
\midrule
$\Gamma(Z\rightarrow\text{had})$ & \mbox{\cref{eq:Zdo}}  & $\Fuha_{\alpha \alpha}, \Fdha_{ii} \lesssim 5\times 10^{-3}\,\, (\text{for } \alpha \neq t)$
\\
\midrule
$|\epsilon|$, $\Delta m_K$ & $\frac{|\im \left( M_{12} A_0 \bar{A}_0^* \right)|}{\sqrt{2}\,(\Delta m_K)\, \big|A_0\bar{A}_0\big|}$ , \small{$2|M_{12}|$} & 
$\frac{|\htr_{u,ds}|}{|\Vha_{ud}||\Vha_{us}|}
 \,<\,  6 \times 10^{-7} \mbox{ --- } 2\times 10^{-4}$
\\ 
\midrule  
$\br(K_{L}\rightarrow \pi^0\nu\bar{\nu})$ & 
$\lambda_{\pi\nu\bar{\nu}}$, \mbox{\cref{eq:brKLSMplusNP}} & $\frac{|\im\htr_{u,ds}|}{|\Vha_{ud}\Vha_{us}|}<2\times 10^{-5}$ \\
\midrule
$\br(K^+\rightarrow \pi^+\nu\bar{\nu})$ &  & $\frac{|\htr_{u,ds}|}{|\Vha_{ud}\Vha_{us}|}
 \,<\,(3 \mbox{ --- } 8 )\times 10^{-6}$ \\
\bottomrule
\end{tabular}
\caption{Summary table of constraints on rephasing-invariant mixing elements
in the presence of one \ac{VLQ} doublet of quarks
(assuming no cancellations).
}
\label{tab:vlq_invariants}
\end{table}
\renewcommand{\arraystretch}{1}

\subsection{\acs{CAA}-motivated scenarios}
\label{sec:CAApheno}
\subsubsection{One \acs{VLQ} doublet (\texorpdfstring{$\n=1$}{\n=1})}
\label{sec:CAAN1}

Here, we examine in more detail the scenarios with vector-like doublets relevant for the Cabibbo angle anomalies, using the rephasing-invariant formalism described in this work. As already recalled, it was
shown in Ref.~\cite{Belfatto:2021jhf} that
a single vector-like doublet mixing with the up, the down and the strange \ac{SM} quarks cannot explain all the tensions, since
the required couplings are excluded by limits on flavour-changing neutral currents (see also~\cref{eq:1doub-fcvscaa}).
We start by reviewing two scenarios with one extra doublet,
in one case exhibiting large mixing with the up and the down \ac{SM} quarks 
($\hiu \neq 0$ and $\hid \neq 0$), which can resolve the tension between $V_{ud}$ and $V_{us}$ (\ac{CAA}1), and in a second case
coupling predominantly to the up and strange quarks ($\hiu \neq 0$ and $\his \neq 0$), causing the tension between the $K\ell3$ and $K\mu2/\pi\mu2$ determinations of $V_{us}$ (\ac{CAA}2).
Then we study the scenario with the presence of two vector-like doublets
with the same pattern of couplings respectively.

\vskip 2mm

In the one-doublet scenarios,
the only non-zero matrix elements of the corresponding charged- and neutral-current matrices are either (\ac{CAA}1)
\begin{equation} \label{eq:model11}
\VRha_{ud}= \frac{\vw^2}{\mB^2}\hiu^{*}\hid\, ,\qquad \hat{\Fr}_{dd}=\frac{\vw^2}{\mB^2}|\hid|^2>0   \, , \qquad \hat{\Fr}_{uu}=\frac{\vw^2}{\mB^2}|\hiu|^2>0 \,,
\end{equation}
or (\ac{CAA}2)
\begin{equation} \label{eq:model12}
\VRha_{us}= \frac{\vw^2}{\mB^2}\hiu^{*}\his \, , \qquad\hat{\Fr}_{ss}=\frac{\vw^2}{\mB^2}|\his|^2>0   \, , \qquad \hat{\Fr}_{uu}=\frac{\vw^2}{\mB^2}|\hiu|^2>0 \,.
\end{equation}
By performing a $\chi^2$ fit of the determinations
of $|V_{ud}|_\beta$, $\vert V_{us} \vert_{K\ell 3}$, $ \left|V_{us}/V_{ud}\right|_{K\mu 2/\pi\mu 2} $
using the values in~\mbox{\cref{eq:vusvud-values}} and the parameters in~\mbox{\cref{eq:caa-rh}}
($\big|\Vha_{us}\big|$ and $\re \big(\Vhac_{ud}\,\VRha_{ud} \big)/\big|\Vha_{ud}\big|$ or $\re \big(\Vhac_{us}\,\VRha_{us} \big)/\big|\Vha_{us}\big|$ in the two cases, respectively)
one finds for \ac{CAA}1
\begin{equation} \label{eq:caa-values_ud}
\frac{\re \left(\Vhac_{ud}\,\VRha_{ud} \right)}{\big|\Vha_{ud}\big|}=-\,0.55\, (29) \times 10^{-3} \,,
\qquad  \big|\Vha_{us}\big|= 0.22451(38) \,,
\end{equation}
with $|\Vha_{ud}|=0.97447(9)$ obtained from unitarity of $\VLh$, or for \ac{CAA}2
\begin{equation} \label{eq:caa-values_us}
\frac{\re \left(\Vhac_{us}\,\VRha_{us} \right)}{\big|\Vha_{us}\big|}=-\, 1.09 \, (36)  \times 10^{-3} \,,
\qquad \big|\Vha_{us}\big|= 0.22453(34) \,,
\end{equation}
with $|\Vha_{ud}|=0.97446(8)$ from unitarity.

In these scenarios, very clean and distinct predictions can be made about non-standard flavour-conserving neutral currents, since the mixing matrices of the \ac{RH} charged
currents $\VRh$ and neutral currents $\Fqh$
originate from the same couplings (see e.g.~\cref{eq:model11,eq:model12}).
Hence, by assuming the presence of \ac{RH} charged currents, one is predicting a non-zero shift in \ac{RH} neutral currents, which can be tested for example in $Z$-pole observables. In particular, in these two cases we have
\begin{equation} \label{eq:Fud}
\hat{\Fr}_{dd} \hat{\Fr}_{uu}=\big|\,\VRha_{ud}\,\big|^2=\frac{\big[\re \hbi_{ud}\big]^2 + \big[\im \hbi_{ud}\big]^2}{\big|\Vha_{ud}\big|^2} \,,
\end{equation}
or
\begin{equation} \label{eq:Fus}
\hat{\Fr}_{ss} \hat{\Fr}_{uu}=\big|\,\VRha_{us}\,\big|^2=\frac{\big[\re \hbi_{us}\big]^2 + \big[\im \hbi_{us}\big]^2}{\big|\Vha_{us}\big|^2} \,.
\end{equation}
In order to explain the Cabibbo angle anomalies we can substitute the values of
$\re \hbi_{ud(s)}/|\Vha_{ud(s)}|$ 
in~\cref{eq:caa-values_ud,eq:caa-values_us}.
Then, one finds the relations
\begin{equation} \label{eq:geq_ud}
\hat{\Fr}_{dd} \hat{\Fr}_{uu}\,\geq \frac{\left(\,\re \hbi_{ud}\,\right)^2 }{|\Vha_{ud}|^2}=\left(0.55\, (29)\times 10^{-3}\right)^2 \,, 
\end{equation}
or 
\begin{equation} \label{eq:geq_us}
\hat{\Fr}_{ss} \hat{\Fr}_{uu}\,\geq \frac{\left(\,\re \hbi_{us}\,\right)^2}{|\Vha_{us}|^2}=\left(1.09 \, (36)\times 10^{-3}\right)^2 \,.
\end{equation}

\begin{figure}
\centering
\begin{subfigure}{0.37\textwidth}
\includegraphics[width=\textwidth]{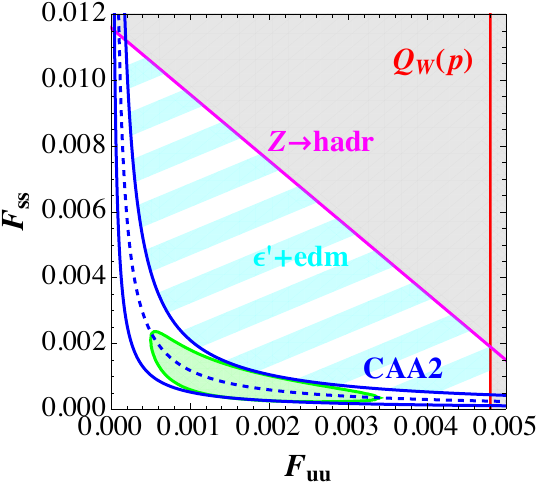}
\caption{\label{fig:1doub}}
 \end{subfigure}
\hspace{10pt}
\begin{subfigure}{0.367\textwidth}
\includegraphics[width=\textwidth]{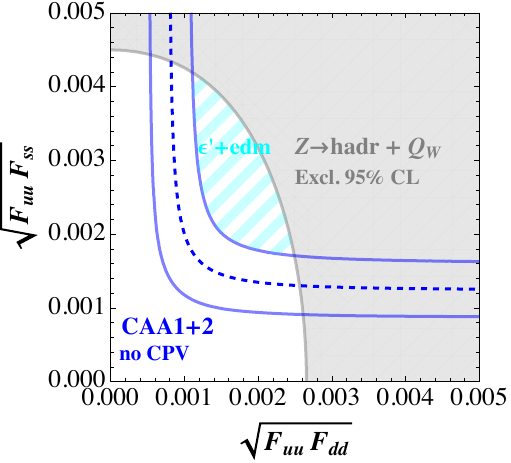}
\caption{\label{fig:hyperbola}}
 \end{subfigure}
\caption{\label{fig:hyperbola2}
Neutral current space corresponding (a) to~\cref{eq:geq_us}, one doublet case, and (b) to~\cref{eq:hyperbola}, two doublets case, together with the excluded region at $95\%$ CL considering $Z\to \mathrm{had}$ plus atomic parity violation (see text for details).}
\end{figure}

However,~\cref{eq:Fud,eq:Fus} should also be confronted 
with the strong experimental limits on CP-violating processes, which
constrain the CP-odd rephasing invariants $\im \hbi_{ud(s)}$ (see~\cref{eq:epsprime-constr,eq:edm-constr}).
After taking into account the CP-violating probes,
the product of couplings $\hat{\Fr}_{uu}\hat{\Fr}_{dd(ss)}$ can be sharply determined
and the relations in~\cref{eq:geq_ud,eq:geq_us} become equalities at the per mille level.

Moreover, these relations should also include the other experimental constraints, 
from electroweak and $Z$ boson observables (see e.g.~\cref{eq:Zdecay-constr} and Ref.~\cite{Belfatto:2023tbv}).
Then, after considering all phenomenological effects, these scenarios lead to precise predictions on each coupling $\hat{\Fr}_{uu}$,
$\hat{\Fr}_{dd}$ (or $\hat{\Fr}_{ss}$). 

By performing a $\chi^2$ fit for \ac{CAA}2 using the parameters $\hat{\Fr}_{uu}$, $\hat{\Fr}_{ss}$, 
$\Vha_{us}$,
including the Cabibbo angle determinations of~\cref{eq:vusvud-values} and the observables in~\cref{sec:Z-decay}  (i.e.~$\Gamma(Z \to \rm had)$, $Q_W(p)$, $Q_W(\text{Cs})$)
we obtain an improvement of $\chi^2_\text{SM}-\chi^2_\text{min}=10.1$ which is independent of the mass of the vector-like doublet.
The remaining discrepancy is due to the difference between the determinations of the Cabibbo angle from kaon and $\beta$ decays.
In~\Cref{fig:1doub} we show the $1\sigma$ ($\chi^2_\text{min}+1$) interval of the parameters (green region). 
The constraints at $2\sigma$ CL are also displayed, for the $Z$ boson decay to hadrons (magenta) and the weak charge of the proton (red). 
In addition,
we illustrate the relation~\eqref{eq:geq_us} in~\Cref{fig:1doub}.
In particular, we show the result of the fit of the anomalies alone,
namely~\cref{eq:vusvud-values}, as the blue curves, indicating the $1\sigma$ interval region ($\chi^2_\text{min}+1$). By assuming that the charged-current couplings
take the values
needed for the anomaly, that is~\cref{eq:caa-values_us}, 
CP-violation limits forbid the striped cyan region (which would be allowed by the CAA solution) and the 
neutral-current couplings should lie inside the blue band.

\subsubsection{Two \acs{VLQ} doublets (\texorpdfstring{$\n=2$}{\n=2})}
\label{sec:CAAN2}

\begin{figure}
\centering
\begin{subfigure}{\textwidth}
\centering
\includegraphics[width=0.37\textwidth]{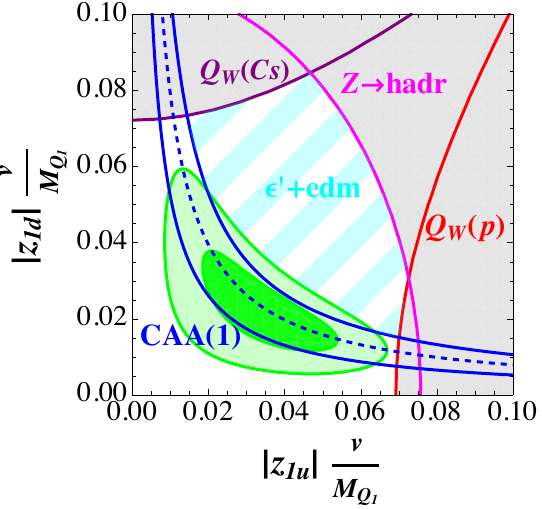}
\includegraphics[width=0.37\textwidth]{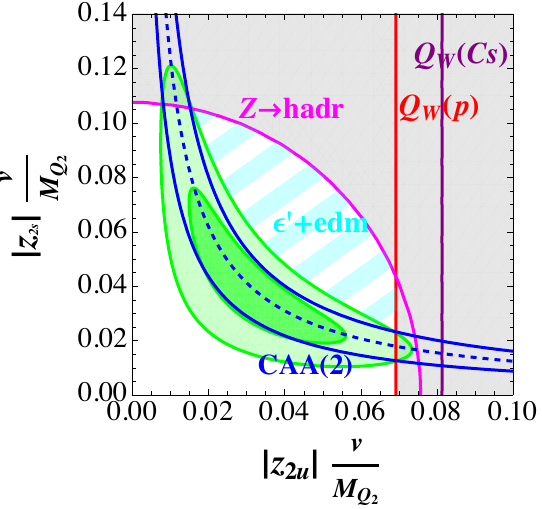}
\caption{\label{fig:vRuvRd}}
 \end{subfigure}
 \vskip 5mm
 \begin{subfigure}{\textwidth}
 \centering
\includegraphics[width=0.325\textwidth]{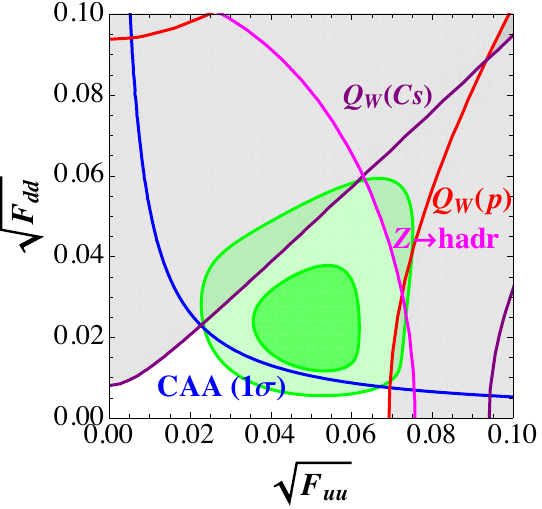}
\includegraphics[width=0.325\textwidth]{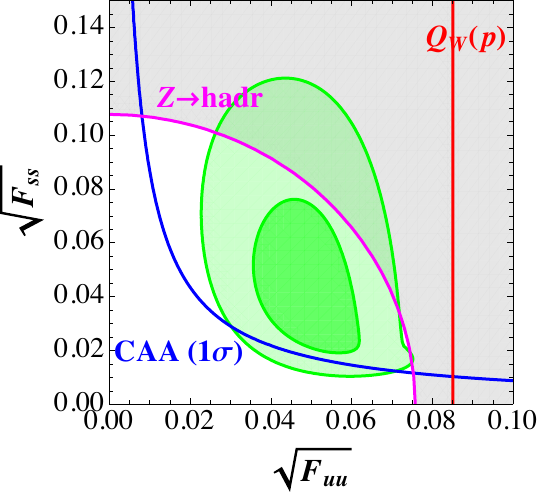}
\includegraphics[width=0.325\textwidth]{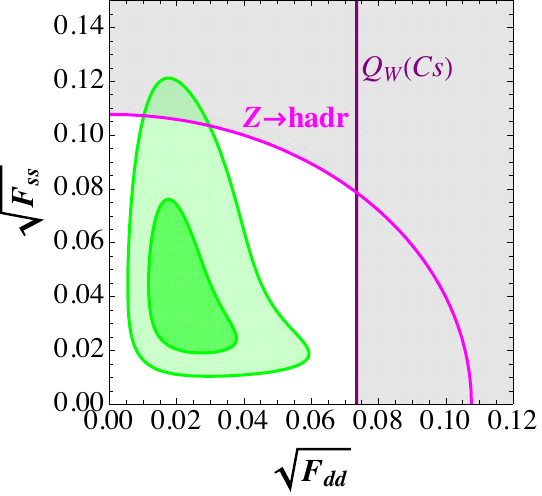}
\caption{\label{fig:FuFdFs}}
 \end{subfigure}
     \caption{Parameter space in the scenario with two
vector-like quark doublets coupling to up and down
and up and strange quarks, respectively, as predicted by resolving \ac{CAA}1 and \ac{CAA}2 (see text for details, $\mBi{_1}=\mB$, $\mBi{_2}=a\mB$).
}
\end{figure}

We now examine a scenario with two vector-like doublets,
which could be the source of all the tensions in the first row of the \ac{CKM} matrix while also not contradicting the stringent limits from flavour-changing neutral currents and other precision electroweak observables.
We assume that one doublet couples predominantly with the up quark and the down quark, while a second doublet has large couplings with the up quark and the strange quark.
The mass matrices related to this scenario are given in~\cref{eq:mass-2doubl},
with the mixing matrices $\VR$, $\Fu$, $\Fd$
given in~\cref{eq:vr-2doublets,eq:Fu-2doubl,eq:Fd-2doubl}.
Then, the relevant matrix elements of the corresponding charged and neutral current matrices read
\begin{equation} \label{eq:2doub-cc-nc}
\begin{aligned}
&\VRha_{ud}=\frac{\vw^2}{\mB^2} \, \hi_{1u}^{*}\hi_{1d} \,,  \qquad\VRha_{us}=\frac{\vw^2}{a^2\mB^2}\, \hi_{2u}^{*}\hi_{2s}  \,, 
\\
&\hat{\Fr}_{dd}=\frac{\vw^2}{\mB^2}\, |\hi_{1d}|^2 \,,   \qquad \hat{\Fr}_{ss}=\frac{\vw^2}{a^2\mB^2}\, |\hi_{2s}|^2  \,,
\qquad \hat{\Fr}_{uu}=\frac{\vw^2}{\mB^2}\left(|\hi_{1u}|^2+\frac{|\hi_{2u}|^2}{a^2}\right) \,,
\end{aligned}
\end{equation}
Let us also highlight that we have the following relations for the couplings $\hi_{ni(\alpha)}$:
\begin{equation}
\begin{aligned}
& \big| \hi_{1u} \big|\frac{\vw}{\mB}=\frac{\big| \VRha_{ud} \big|}{\sqrt{\hat{\Fr}_{dd}}} \,, \quad
   \big| \hi_{2u} \big|\frac{\vw}{a\mB}=\frac{\big| \VRha_{us} \big|}{\sqrt{\hat{\Fr}_{ss}}} \, , \quad \big|\hi_{1d}\big|\frac{\vw}{\mB} =\sqrt{\hat{\Fr}_{dd}}\, , \quad
\big|\hi_{2s}\big|\frac{\vw}{a\mB} =\sqrt{\hat{\Fr}_{ss}}\, .
\end{aligned}
\end{equation}
Again it is clear that there exists a direct connection between neutral and charged mixing elements. 
In particular, as right-handed charged currents are assumed in order to explain the anomalies, then modified neutral current couplings are predicted
with the peculiar relation
\begin{equation}
\hat{\Fr}_{uu}= \frac{\big|\VRha_{ud}\,\big|^2}{ \hat{\Fr}_{dd}}+\frac{\big|\VRha_{us}\,\big|^2}{ \hat{\Fr}_{ss}} \, .
\end{equation}
In order to explain the \acp{CAA}, the charged-current mixings
$\re \hbi_{ud} /|\Vha_{ud}|$ and $\re \hbi_{us}/|\Vha_{us}|$
should assume the values in~\cref{eq:cabfit}.
Then, one gets the relation between the charged-current couplings appearing in~\cref{eq:caa-rh} and the neutral-current couplings:
\begin{equation} \label{eq:diseq-2doubl}
     1=\frac{\big|\VRha_{ud}\,\big|^2}{\hat{\Fr}_{uu} \, \hat{\Fr}_{dd}}+\frac{\big|\VRha_{us}\,\big|^2}{\hat{\Fr}_{uu} \, \hat{\Fr}_{ss}}
    \,\geq\, 
    \frac{\big(\re \hbi_{ud}\big)^2}{|\Vha_{ud}|^2}
    \frac{1}{\hat{\Fr}_{uu} \, \hat{\Fr}_{dd}}+ \frac{\big(\re \hbi_{us}\big)^2}{|\Vha_{us}|^2}\frac{1}{\hat{\Fr}_{uu} \, \hat{\Fr}_{ss}}\,.
\end{equation}
Hence, the presence of right-handed currents generating the \acp{CAA}
would imply flavour-conserving neutral currents with couplings above the rectangular hyperbola of axes $\hat{\Fr}_{uu} \hat{\Fr}_{dd}$ and $\hat{\Fr}_{uu} \hat{\Fr}_{ss}$.
After applying the experimental bounds on the corresponding CP-violating
phases (see~\cref{eq:epsprime-constr-caa,eq:edm-constr}), we have that the bilinears should be real  at the per mille level when the real part assumes the values needed for the anomalies $\hbi_{ud(s)}/|\Vha_{ud(s)}|\approx \re \hbi_{ud(s)}/|\Vha_{ud(s)}|\approx 10^{-3}$. Then,
the product of flavour-conserving neutral-current couplings lie \emph{on} the hyperbola at the per mille level:
\begin{equation}\label{eq:hyperbola}
\hat{\Fr}_{uu}= \frac{\left( 0.79\,(27)\times 10^{-3}\right)^2}{\hat{\Fr}_{dd}}+\frac{\left(1.24\,(37)\times 10^{-3}\right)^2}{\hat{\Fr}_{ss}} \,,
\end{equation}
where we inserted in~\cref{eq:diseq-2doubl} the values in~\cref{eq:cabfit}.
This relation is shown in~\Cref{fig:hyperbola} (blue curves, corresponding to $\chi^2=1$ region). While the anomalies can be addressed in the region above the (blue) curves, CP-violation bounds force the couplings to remain 
within the blue band (excluding the striped cyan region) once the charged-current couplings acquire the values needed for \ac{CAA}1 and \ac{CAA}2.

The \ac{RH} charged currents explaining the anomalies
hence imply the presence of anomalous $Z$ couplings because of the connection between charged- and neutral-current couplings.
Neutral-current couplings are also subject to constraints 
from electroweak and $Z$-boson observables (see e.g.~\cref{eq:Zdecay-constr})
which should be included in the phenomenological analysis of the Cabibbo angle anomalies.
Then, after considering all phenomenological effects, the 
predictions for each coupling $\hat{\Fr}_{uu}$, $\hat{\Fr}_{dd}$, $\hat{\Fr}_{ss}$ become
quite accurate and testable, for instance, by improving the precision on $Z$-decay observables.

We show this connection and the corresponding parameter space in~\Cref{fig:vRuvRd,fig:FuFdFs}.
After considering the constraints on CP violation, we have
\begin{equation}\label{eq:z-2doublets}
\begin{aligned}
&\hi_{1u}^*\,\hi_{1d} \:\frac{\vw^2}{\mB^2}=\,
\frac{\hbi_{ud}}{\Vhac_{ud}}
\approx\,
\frac{\re \hbi_{ud}}{\Vhac_{ud}}=
-0.79\,(27)\times 10^{-3}\: \frac{\Vha_{ud}}{\big|\Vha_{ud}\big|} \, ,
 \\
& 
\hi_{2u}^*\,\hi_{2s} \:\frac{\vw^2}{a^2\mB^2}
\approx\,
\frac{\re \hbi_{us}}{\Vhac_{us}}=
-1.24\,(37)\times 10^{-3}\: \frac{\Vha_{us}}{\big|\Vha_{us}\big|} \, .
   \end{aligned}
\end{equation}
Taking into account these relations,
we perform a $\chi^2$ fit using the parameters $-\hi_{1u} \vw/\mBi{_{1}}$, $-\hi_{2u} \vw/\mBi{_{2}}$, $-\hi_{2s} \vw/\mBi{_{2}}$, $-\hi_{1d} \vw/\mBi{_{1}}$ and $\Vha_{us}$ ($\Vha_{ud}$ is determined by the unitarity of $\VLh$)
including the observables: determinations of the Cabibbo angle~\eqref{eq:vusvud-values}, $\Gamma(Z \to \rm had)$, $Q_W(p)$ and $Q_W(\text{Cs})$ (see~\cref{sec:Z-decay}).
We illustrate the result in~\Cref{fig:vRuvRd}.
The $1\sigma$ and $2\sigma$ interval regions
($\chi^2_\text{min}+1$, $\chi^2_\text{min}+4$) are displayed in green and light-green, marginalizing over the other variables and assuming real bilinears $\hbi_{ud}$ and $\hbi_{us}$. We receive 
an improvement over the \ac{SM}
of $\chi^2_\text{SM}-\chi^2_\text{min}=18.1$.
We also show the constraints at $2\sigma$ CL for the $Z$ boson decay to hadrons (magenta), the weak charge of the proton (red), and the atomic parity violation in $_{\hphantom{1}78}^{133}\text{Cs}$ (purple).
The fit including only the anomalies, 
namely~\cref{eq:vusvud-values} without the neutral-current flavour-conserving constraints and assuming real bilinears, is shown by the blue bands, which indicate the $1\sigma$ region ($\chi^2_\text{min}+1$) corresponding to eqs.~\eqref{eq:cabfit}. We also paint the cyan striped area indicating the region which is excluded by CP-violating phenomena once it is assumed that the real part of the couplings takes the values needed for the Cabibbo angle anomalies,~\cref{eq:cabfit,eq:z-2doublets}.

Under the assumption of real bilinears $\hbi_{ud(s)}$,
a $\chi^2$ fit can be performed
using the neutral-current couplings as parameters, i.e.~$\hat{\Fr}_{uu}$, $\hat{\Fr}_{ss}$, $\hat{\Fr}_{dd}$ (and $\Vha_{us}$) including the same observables in~\cref{eq:vusvud-values} and~\cref{sec:Z-decay}.
In~\Cref{fig:FuFdFs} we illustrate the results of the fit on the $\hat{\Fr}_{uu}$-$\hat{\Fr}_{ss}$, $\hat{\Fr}_{uu}$-$\hat{\Fr}_{dd}$ and $\hat{\Fr}_{dd}$-$\hat{\Fr}_{ss}$ planes, respectively, marginalizing over the other variables. The $1\sigma$ and $2\sigma$ interval regions 
($\chi^2_\text{min}+1$, $\chi^2_\text{min}+4$) are presented in green and light green. 
The relation in~\cref{eq:hyperbola} emerges
in the stand-alone fit of the anomalies, without including the flavour-conserving constraints, which at $1\sigma$ gives the blue curves in the plots: \ac{CAA}1 and \ac{CAA}2 can be resolved in the whole region above the curves, independently on the assumptions on CP violation.

Besides being the potential resolution of the Cabibbo angle anomalies, 
the Yukawa textures in~\cref{eq:mass-2doubl} are compatible
with all other experimental constraints, in particular the stringent limit
on flavour-changing neutral currents. In fact, as it can be inferred
from~\cref{eq:Fu-2doubl,eq:Fd-2doubl}, there are no
\acp{FCNC} at tree level. 
In principle, there is still a contribution at loop level e.g.~to neutral kaon mixing.
However,
in the scenario explaining the Cabibbo angle anomalies, i.e.
taking into account the values in~\cref{eq:cabfit}, we get
that the new contribution
is well below the experimental constraints
(at least two orders of magnitude smaller than the \ac{SM} contribution, see~\cref{app:kaon2}).

\vskip 2mm
Let us close this section with a couple of observations. If Cabibbo anomalies were to be confirmed and addressed by this scenario, two new puzzles would emerge: 
1) Why does each \ac{VLQ} (mostly) couple to one light \ac{SM} quark family in each sector (needed to evade \ac{FCNC} constraints)? 2) A priori, CP phases in nature are arbitrary. Why does one have $|\arg \hbi_{ud(s)}|=|\arg(\Vhac_{ud(s)} \VRha_{ud(s)})-\pi|\lesssim 10^{-3}$?
Perhaps a symmetry mechanism, beyond the reach of this work, could shed
some light on both issues.

Let us also note that, in an alternative scenario
in which e.g.~$\re \hbi_{ud} \sim \im \hbi_{ud}$, 
CP-violating probes on the invariant $|\im \hbi_{ud}|\lesssim 10^{-6}$ are naturally testing possible scenarios with \ac{VLQ} doublets at $\mB\sim 200\, \text{TeV}$, far beyond the direct reach of current and near-future high-energy colliders, thus becoming one of the highest-scale \ac{BSM} scenarios currently being tested and not violating any of the \ac{SM} symmetries (accidental or not).

\section{Summary and conclusions}
\label{sec:summary}

In this work, we studied models
of \ac{VLQ} doublets with standard charges, which are minimal extensions of the \ac{SM} and highly motivated from a model-building point of view. Notably, these models provide favoured solutions to the \acfp{CAA}~\cite{Belfatto:2021jhf,Crivellin:2022rhw,Belfatto:2023tbv,Cirigliano:2023nol}.
We emphasized that a prominent tool in the analysis of these models is obtained through a weak-basis invariant (\ac{WBI}) description.
In an extensive survey, we explained how to construct \acp{WBI} for \ac{VLQ} doublet models and how to relate these to physical parameters. The physical parameter count for an arbitrary number $\n$ of \ac{VLQ} doublets is given in~\Cref{tab:countsummary}.
 
An important and differentiating aspect of \acp{WBI} constructed for \ac{VLQ} doublet models --- and in contrast with the \ac{SM} or \ac{VLQ} singlet extensions --- is that, due to the presence of exotic right-handed currents new and simpler rephasing invariants and \acp{WBI} can be obtained, which can also be related to Lagrangian parameters and observables.
This extra character of doublet representations allows for lower-mass-dimension ($\M=6$) CP-odd \ac{WB} invariants. This could represent a significant enhancement of \ac{CPV} with respect to the \ac{SM} ($\M=12$) or even to the \ac{VLQ}-singlet extensions ($\M=8$).

For the one-doublet case, we found a special weak basis, referred to as the ``stepladder'' \ac{WB}, which in a simple and comprehensive way enables us to extract all physical parameters step-by-step, without having to solve complicated equations. We identified a set of \acp{WBI} totally characterizing the single \ac{VLQ} doublet scenario ($\n = 1$). An alternative procedure, using instead the minimal weak basis of~\cref{sec:minimalweak}, is also used to generalize this characterization to an arbitrary number of doublets ($\n > 1$).
Moreover, we presented useful parameterizations for the quark mixing in~\cref{sec:rotating}.

We then focused on studying rephasing invariants and CP-odd \acp{WBI}, whose imaginary parts signal the presence of \ac{CPV}. 
In the doublet scenario, new kinds of rephasing invariants arise, which result from the existence of right-handed charged and neutral currents mixing matrices $\VR$, $\Fu$ and $\Fd$. The interplay between the two types of charged currents, right-handed and left-handed, generates bilinears, of the form $\VLa_{\alpha i}\VRac_{\alpha i}$, involving only two quarks.
Flavour-changing neutral currents give rise to trilinear rephasing invariants of the form
$\Fua_{\alpha\beta} \VLac_{\alpha i}\VLa_{\beta i}$ and $\Fda_{ij} \VLa_{\alpha i}\VLac_{\alpha j}$, involving three quarks.

We found that the mass dimension and number of insertions of the mixing elements in \acp{WBI} play a crucial role. In fact, we uncover a direct connection between the type of \emph{effective} rephasing invariants contained in different \acp{WBI}
and their structure, i.e.~the number of Hermitian blocks from which the \acp{WBI} are built. 
We denote by \emph{effective} the rephasing invariants that involve only the three \ac{SM} quarks and that emerge after taking into account the hierarchy between the \ac{EW} and the \ac{VLQ} mass scale. 
These features then provide a deeper understanding of the links
between \acp{WBI}, rephasing invariants and observables.

We also find a complete set of \acp{WBI} which describe the CP properties of the one \ac{VLQ} doublet scenario and whose vanishing would ensure CP invariance. We present this set in~\Cref{tab:cp-odd-inv}. Moreover, we briefly studied the extreme chiral limit for the one- and two-\ac{VLQ} doublet scenarios. We show that in this regime of extremely high energies, even though the lighter quark masses can be taken as vanishing, \ac{CPV} can still be achieved, in contrast to what happens in the \ac{SM}.

We carried out a phenomenological analysis, using the \ac{WBI} description previously developed. 
Such description in terms of invariants provides unambiguous constraints on the model, which is particularly attractive
when describing CP-violating phenomena.
The crucial connection between \acp{WBI} and effective rephasing invariants 
hints at the strength of different CP-violating \ac{VLQ} contributions
---
the different structures of invariants emerging in the presence of \ac{VLQ} 
doublets suggest which phenomena can receive a large contribution,
particularly when involving CP violation.

The presence of vector-like doublets can affect several processes, both
CP-conserving and CP-violating, e.g.~hadron decays, low-energy \ac{EW} observables and $Z$ decay, kaon mixing, direct CP violation, rare kaon decays, and electric dipole moments.
We studied these phenomenological effects in their rephasing-invariant form. 
In particular, we analyzed the observables in terms of rephasing invariants, i.e.~moduli, bilinears, trilinears, quartets
and the interrelations between them. We imposed limits on the corresponding invariant quantities allowed by the model. These
results are summarized in~\Cref{tab:vlq_invariants}.

Finally, we analyzed specific scenarios addressing the Cabibbo Anomalies with \ac{VLQ} doublets, explicitly showing, using rephasing-invariant quantities, the remarkable relations between \ac{RH} neutral- and charged-current observables involving both CP-even and CP-odd probes.

\section*{Acknowledgements}

This work was partially supported by Fundação para a Ciência e a Tecnologia (FCT, Portugal) through the projects CERN/FIS-PAR/0002/2021, 2024.02004.CERN and CFTP-FCT Unit UIDB/00777/2020 and UIDP/00777/2020, which are partially funded through POCTI (FEDER), COMPETE, QREN and EU.
The work of B.B.~was supported by the Deutsche Forschungsgemeinschaft (DFG, German Research Foundation) under grant 396021762 - TRR 257.
A.R.S.~was supported by the Generalitat Valenciana (Spain) through the plan GenT program (CIDEIG/2023/12) and by the Spanish Government (Agencia Estatal de Investigación MCIN/AEI/10.13039/501100011033) Grants No. PID2023-146220NB-I00, No. PID2020-114473GB-I00, and No. CEX2023-001292-S.
J.F.B.~acknowledges funding from Fundação para a Ciência e a Tecnologia (FCT) through Grant PRT/BD/154581/2022 (https://doi.org/10.54499/PRT/BD/154581/2022).
F.A.~acknowledges funding from Fundação para a Ciência e a Tecnologia (FCT) through Grant UI/BD/153763/2022 (https://doi.org/10.54499/UI/BD/153763/2022).

\appendix

\section{Notation}
\label{app:notation}
\renewcommand{\arraystretch}{1.4}
  \begin{longtable}{llc}
    \caption{Notations introduced in the text and equations where they are defined.}
  \label{tab:notation} \\
    \toprule
    Notation & Description ($q=u,d$; $\chi=L,R$) & Definition \\
    \midrule
    \endfirsthead
    \caption{(continued)}\\
  \toprule
    Notation & Description ($q=u,d$; $\chi=L,R$) & Definition \\
\midrule \endhead
\bottomrule  \endfoot
    $\n$ & number of \ac{VLQ} doublets added to the \ac{SM} & \mbox{\cref{eq:doublets}} \\
    $\Mq$ & $(3+\n)$-dimensional mass matrix of $q$-type quarks & \mbox{\cref{eq:LM1}} \\
    $\yq$ & $3\times 3$ standard $q$-type Yukawa matrix & \mbox{\cref{eq:LY1}} \\
    $\yQ$ & $\n\times 3$ matrix of $q$-type Yukawa couplings to \acp{VLQ} & \mbox{\cref{eq:LY1}} \\
    $\mb$ & $3\times \n$ matrix of bare mass terms involving \ac{SM} quarks & \mbox{\cref{eq:LY1}} \\
    $\mB$ & $\n\times \n$ matrix of \ac{VLQ}-only bare mass terms & \mbox{\cref{eq:LY1}} \\
    $\mBd$ & $\mB$ in a basis in which it is diagonal and non-negative & \mbox{\cref{eq:L1}} \\
    $\Yq$ & $(3+\n)\times 3$ matrix of all $q$-type Yukawa couplings & \mbox{\cref{eq:defsM}} \\
    $\mq$ & the product $\vw\, \Yq$, i.e.~all masses terms or \ac{EW} origin & \mbox{\cref{eq:defsM}} \\
    $\Mb$ & $(3+\n)\times 3$ matrix of all bare masses terms  & \mbox{\cref{eq:defsM}} \\
    $\Vq_\chi$ & $(3+\n)$-dimensional unitary rotations bidiagonalizing $\Mq$ & \mbox{\cref{eq:rot}} \\
    $\Dq$ &  $(3+\n)$-dim.~diagonal matrix of physical $q$-type masses & \mbox{\cref{eq:diag}} \\
    $\Msq$ & new heavy physical masses ($\Msq_1,\ldots,\Msq_\n$) & \mbox{\cref{eq:diag}} \\
    $\Aq_\chi$ & matrix containing the first 3 rows of $\Vq_\chi$ & \mbox{\cref{eq:AB}} \\
    $\Bq_\chi$ &  matrix containing the last $\n$ rows of $\Vq_\chi$ & \mbox{\cref{eq:AB}} \\
    $\VL$ & $(3+\n)$-dim.~\ac{LH} charged-current mixing matrix (unitary) & \mbox{\cref{eq:VLVR}} \\
    $\VR$ &  $(3+\n)$-dim.~\ac{RH} charged-current mixing matrix (non-unit.) & \mbox{\cref{eq:VLVR}} \\
    $\CKM$ & $3\times 3$ non-unitary \ac{CKM} mixing matrix & \mbox{\cref{eq:CKM}} \\  
    $\Fq$ &  $(3+\n)$-dim.~Hermitian matrix controlling $q$-sector \acp{FCNC} & \mbox{\cref{eq:vnc}} \\
    $\VLh$ & 
    $\CKM$ at leading order, part of a parameterization of $\VL$
    & \mbox{\cref{eq:VL}} \\
    $\VRh$ & 
    $3\times 3$ upper-left block of $\VR$ at leading order, cf.~\mbox{\cref{eq:vckmR}}
    & \mbox{\cref{eq:reph-inv-eff}} \\
    $\Fqh$ & $3\times 3$ upper-left block of $\Fq$ at leading order, cf.~\mbox{\cref{eq:Fzz}} & \mbox{\cref{eq:reph-inv-eff}} \\
    $\WL$ & weak-basis rotation of \ac{LH} fields & \mbox{\cref{eq:WB}} \\
    $\WR$ & weak-basis rotation of \ac{RH} \ac{VLQ} fields & \mbox{\cref{eq:WB}} \\
    $\WRq$ & weak-basis rotation of \ac{RH} standard $q$-type quarks & \mbox{\cref{eq:WB}} \\
    $\Ydq$ & $\yq$ in a basis in which it is diagonal and non-negative & \mbox{\cref{eq:Uhat}} \\  
    $\UW_{q\chi}$ & $3\times 3$ unitary rotations bidiagonalizing $\yq$ & \mbox{\cref{eq:Uhat}} \\
    $\VWB$ & $3\times 3$ \ac{CKM}-like matrix in the definition of the minimal \ac{WB}
    & \mbox{\cref{eq:Vhat}} \\
    $\yqp$ & rotated $\yQ$ in the definition of the minimal \ac{WB} & \mbox{\cref{eq:zqs}} \\    
    $\yqh$ & rephased $\yqp$, with $\n$ less phases  & \mbox{\cref{eq:minimal}} \\
    $\VWT$ & $3\times 3$ unitary matrix in the definition of the stepladder \ac{WB}  & \mbox{\cref{eq:struct1}} \\    
    $\bi$ & rephasing-invariant bilinears, defined in terms of $\Vr_\chi$ and $\Fq$ & \mbox{\cref{eq:bilinears}} \\    
    $\tri$ & rephasing-invariant trilinears, defined in terms of $\Vr_\chi$ and $\Fq$ & \mbox{\cref{eq:trilinears}} \\    
    $\qua$ & rephasing-invariant quartets, defined in terms of $\Vr_\chi$ and $\Fq$ & \mbox{\cref{eq:Q}} \\
    $\hbi$& rephasing-invariant bilinears, defined in terms of $\Vch$ and $\Fqh$ & \mbox{\cref{eq:reph-inv-eff}} \\   
    $\htr$& rephasing-invariant trilinears, defined in terms of $\Vch$ and $\Fqh$ & \mbox{\cref{eq:reph-inv-eff}} \\   
    $\hqu$& rephasing-invariant quartets, defined in terms of $\Vch$ and $\Fqh$ & \mbox{\cref{eq:reph-inv-eff}} \\   
    $\Qh$ & rephasing-invariant quartets
    generalizing the \ac{SM} ones
    & \mbox{\cref{eq:Qhat}} \\    
    $\HqL$ & \ac{LH} Hermitian matrix $\Mq \Mq^\dagger = \hq + \C$ & \mbox{\cref{eq:defsH}} \\   
    $\hq$ & \ac{LH} Hermitian matrix $\mq \mq^\dagger$ & \mbox{\cref{eq:hq-H}} \\    
    $\C$ & \ac{LH} Hermitian matrix $\Mb \Mb^\dagger$ & \mbox{\cref{eq:hq-H}} \\   
    $\HqR$ & \ac{RH} Hermitian matrix $\Mq^\dagger \Mq $ & \mbox{\cref{eq:defsH}} \\   
    $a$ & ratio of \ac{VLQ} mass parameters in the $\n=2$ scenario & \mbox{\cref{eq:mass-2doubl}} \\
    $\lambda_u$ & shorthand for the product $\Vhac_{us}\Vha_{ud}$
    & \mbox{\cref{eq:A0main}} \\
\end{longtable}
\renewcommand{\arraystretch}{1}

\begin{acronym}[XXXXX] 
\acro{BAU}[BAU]{baryon asymmetry of the Universe}
\acro{BSM}[BSM]{beyond-the-Standard-Model}
\acro{CAA}[CAA]{Cabibbo angle anomaly}
\acroplural{CAA}[CAAs]{Cabibbo angle anomalies}
\acro{CKM}[CKM]{Cabibbo--Kobayashi--Maskawa}
\acro{CPI}[CPI]{CP invariance}
\acro{CPV}[CPV]{CP violation}
\acro{ECL}[ECL]{extreme chiral limit}
\acro{EFT}[EFT]{effective field theory}
\acro{EW}[EW]{electroweak}
\acro{EWSB}[EWSB]{electroweak symmetry breaking}
\acro{FCNC}[FCNC]{flavour-changing neutral current}
\acro{GIM}[GIM]{Glashow--Iliopoulos--Maiani}
\acro{LH}[LH]{left-handed}
\acro{NP}[NP]{New Physics}
\acro{RH}[RH]{right-handed}
\acro{SM}[SM]{Standard Model}
\acro{SMEFT}[SMEFT]{Standard Model effective field theory}
\acro{UV}[UV]{ultraviolet}
\acro{VEV}[VEV]{vacuum expectation value}
\acro{VLQ}[VLQ]{vector-like quark}
\acro{WB}[WB]{weak basis}
\acro{WBI}[WBI]{weak-basis invariant}
\acro{WBT}[WBT]{weak-basis transformation}
\end{acronym}

\vfill
\clearpage

\section{Alternative ways of counting physical parameters}
\label{sec:counting}

\subsection{Counting via a spurion analysis}
To independently check the (physical) parameter count of~\Cref{tab:minimal}, one may perform a spurion analysis~\cite{Santamaria:1993ah} (see also~\cite{Berger:2008zq,Alves:2023ufm}), which does not require fixing a~\ac{WB}. 
One starts by identifying the general flavour symmetry available in the absence of mass and Yukawa terms --- in direct correspondence with the \ac{WB} freedom of~\cref{eq:WB}, as mentioned above. Mass and Yukawa terms break the symmetry of the kinetic terms to the conserved baryon number symmetry. Schematically, one has
\begin{equation} \label{eq:chain}
    U(3+\n)_{\Q_L} \times U(3)_{u_R}\times U(3)_{d_R}\times U(\n)_{Q_R} \quad\to\quad U(1)_B\,.
\end{equation}
The number of physical parameters $N_\text{phys} = N_\text{phys}^\text{moduli} + N_\text{phys}^\text{phases}$ can be obtained from the differences
\begin{equation}
\begin{aligned}
    N_\text{phys}^\text{moduli} &= N_\text{general}^\text{moduli} - N_\text{broken}^\text{moduli}\,, \\[1mm]
    N_\text{phys}^\text{phases} &= N_\text{general}^\text{phases} - N_\text{broken}^\text{phases}\,,
\end{aligned}
\end{equation}
where $N_\text{general}$ refers to the number of parameters (even if unphysical) describing the $\Mq$ matrices in a generic \ac{WB}, i.e.
\begin{equation}
\begin{aligned}
&N_\text{general}^\text{moduli}  \,=\,
N_\text{general}^\text{phases}  \,=\,
\overbrace{2(9+ 3\n)}^{\yq,\,\yQ} + \overbrace{3\n}^{\mb} + \overbrace{\n^2}^{\mB} \,=\, 18+9\n+\n^2
\\[3mm]
\quad\Rightarrow \quad 
&N_\text{general} \,=\, N_\text{general}^\text{moduli} + N_\text{general}^\text{phases} \,=\, 36 + 18 \n + 2 \n^2
\,.
\end{aligned}
\end{equation}
As for $N_\text{broken}$, each $U(k)$ factor contributes with $\frac{1}{2}k(k-1)$ moduli and $\frac{1}{2}k(k+1)$ phases. The chain of~\cref{eq:chain} then implies
\begin{equation}
\begin{aligned}
&\left\{
\begin{array}{l}
    N_\text{broken}^\text{moduli} 
    \,=\, \dfrac{1}{2}(3+\n)(2+\n) + 3 + 3 + \dfrac{1}{2}\n(\n-1) - 0 
    \,=\, 9 + 2\n + \n^2
    \\[4mm]
    N_\text{broken}^\text{phases}
    \,=\, \dfrac{1}{2}(3+\n)(4+\n) + 6 + 6 + \dfrac{1}{2}\n(\n+1) - 1 
    \,=\, 17 + 4\n + \n^2
\end{array}\right.
\\[4mm]
&\quad\Rightarrow \quad 
N_\text{broken} \,=\, N_\text{broken}^\text{moduli} + N_\text{broken}^\text{phases} \,=\, 26+6\n+2\n^2
\,.
\end{aligned}
\end{equation}
As a result, we obtain
\begin{equation}
\left\{
\begin{array}{l}
    N_\text{phys}^\text{moduli} 
    \,=\, 9+7\n
    \\[2mm]
    N_\text{phys}^\text{phases}
    \,=\, 1+5\n
\end{array}\right.
\quad\Rightarrow \quad 
N_\text{phys} = 10 + 12 \n
\,,
\end{equation}
in agreement with~\Cref{tab:minimal}.

\vskip 2mm

The results above apply to a general number $\n$ of \ac{SM}-like \acp{VLQ}. In the simplest case of $\n = 1$, one counts 12 extra parameters in the quark sector (7 moduli and 5 phases), for a total of 22 physical parameters (16 moduli and 6 phases).

\subsection{Counting in the mass basis (\texorpdfstring{$\n=1$}{\n=1})}
\label{sec:massbasis1}

Consider the mass basis and the simplest case of $\n=1$. We have $8$ physical masses.
The mixing matrix in the \ac{LH} sector, $\VL$, is a unitary $4\times4$ matrix which can be parameterized using $6$ angles and $10$ phases. 
As regards the \ac{RH} sector, the matrices $\VR$, $\Fu$ and $\Fd$ are built from $\BuR$ and $\BdR$, i.e.~from the last rows of the unitary matrices $\VuR$ and $\VdR$, which contain a total of $6$ independent mixing angles (3 in each $\BqR$) and $7$ independent phases, as only differences of the $8$ phases (4 in each $\BqR$) appear.

However, not all of the $8+6+6 = 20$ moduli and $10+7 = 17$ phases identified above are physical. Recall that the extra species is vector-like. In the mass basis, this information is encoded in the relations of~\cref{eq:LRrelation1}, which in this case read
\begin{equation} \label{eq:MQ1}
\Du\,{\VuR}^\dagger 
\begin{pmatrix}0 \\ 0 \\ 0 \\ 1\end{pmatrix}
\,=\,
\VL\,\Dd\,{\VdR}^\dagger 
\begin{pmatrix}0 \\ 0 \\ 0 \\ 1\end{pmatrix}
\,.
\end{equation}
This equations represent $4$ real and $4$ imaginary conditions. 
Then, we have $8+6+6-4=16$ moduli describing the model. As regards the phases, note that, beyond the $4$ imaginary constraints set by~\cref{eq:MQ1}, $7$ phases can be eliminated (in total, between the \ac{RH} and \ac{LH} sectors) by rephasing the quark fields. 
Namely, one is free to rephase all 8 quark fields $u$, $c$, $t$, $T'$, and $d$, $s$, $b$, $B'$
in the mass basis with common phases for both chiralities (leaving the mass terms real), cf.~\cref{eq:rephasing}. Due to the conserved baryon number, this procedure can remove at most $7$ unphysical phases, as indicated above. 
Thus, we are left with $10+7-4-7=6$ physical phases describing the model.

As an example, out of the $10$ phases contained in $\VL$, one may choose to eliminate $5$ phases related to the \ac{LH} mixing of the three light quark species by rephasing the corresponding fields. Then, $4$ of the remaining $5$ phases are determined by the relations in~\cref{eq:MQ1}. This means that only one physical phase is contained in $\VL$.
By rephasing the $T'$ and $B'$ fields, $2$ out of $7$ phases can be eliminated from the \ac{RH} currents, leaving $5$ phases as free parameters in the \ac{RH} sector. This results in a total of $6$ physical phases, as anticipated.

This discussion can be extended to the case of more \ac{VLQ} doublets.
Counting in the context of a specific parameterization, for $\n \geq 1$, is carried out in~\cref{sec:param}.

\vfill
\clearpage


\section{Reconstructing the minimal \acs{WB}}
\label{sec:mapping}

In a less direct manner than what is described in \cref{sec:stepladderwb}, one may connect the minimal weak basis of~\cref{eq:minimal} to \acp{WBI} which are straightforward to compute, given the numerical values of the Yukawa matrices in any weak basis, for any value of $\n$.
Thus, the corresponding connection gives a general procedure to map any parameterization of the Yukawa and \ac{VLQ} (bare) mass matrices
to the minimal one studied above, also used to integrate out \acp{VLQ} (see~\cref{sec:EFT}). Finding this mapping is specially relevant in a top-down approach, where we may start from a different parameterization of the same model
(e.g.~from a specific flavour basis). 
In order to do so, we use the projection technique described below, 
which is valid up to singular directions, such as degenerate eigenvalues.

We define a projector $P$ as an object that under \acp{WBT} transforms as 
one of the relevant building blocks
and that satisfies $\left(P_{i}\right)_{ab}=\delta_{ai}\delta_{bi}$
in the considered \ac{WB}.
In our case, we are interested in the minimal \ac{WB} and require that $P \,\to\,\WL^\dagger \,P\, \WL$ under \acs{WBT}, just like the building blocks of~\cref{eq:WL-hq-H}.
In the minimal \ac{WB}, one has
\begin{equation} \label{eq:recallminWB}
\hu= 
\vw^2 \begin{pmatrix}
 \Ydu^{2}  & \Ydu \, \yup^{\dagger} \\[2mm]
  \yup \, \Ydu & \yup \, \yup^{\dagger}
\end{pmatrix} \,, 
\quad
\hd=  \vw^2
\begin{pmatrix}
  \VWB\Ydd^2 \, \VWB^{\dagger} & \VWB \, \Ydd \, \ydbh^{\dagger} \\[2mm]
  \ydbh \, \Ydd \, \VWB^{\dagger} & \ydbh \, \ydbh^{\dagger}
\end{pmatrix} \,,
\quad 
\C=\begin{pmatrix}
 0  & 0 \\[2mm]
  0 & \mBd^2
\end{pmatrix}\,.
\end{equation}
We assume, without loss of generality, that $\Ydq {_{i}}>\Ydq {_{j}}>0$ and $\mBd {_i}>\mBd {_j}>0$,
for $i>j$, with $i,j=1,2,3$ and $q=u,d$. 
Here and from now on, we use a single index when referring to matrix elements of the diagonal matrices $\Ydq$ and $\mBd$.
We also take the first column of $\ydb$ to be real and positive, $\ydb \to \ydbh$.
For the unitary matrix $\VWB$, we choose a (standard) parameterization in terms of three mixing angles $\theta_{13}\in [0, \pi/2]$, $\theta_{12}\in [0, \pi/2]$, $\theta_{23}\in [0, \pi/2]$ and a phase $\delta_{13}\in [-\pi, \pi]$.

For the \ac{WB} under consideration, one defines
\begin{equation}
P_{i\geq 4}\equiv \sum_{m=1}^{\n}  c^{(i)}_{m} \C^m \, .
\end{equation}
The corresponding projector condition, $\left(P_{i}\right)_{ab}=\delta_{ai}\delta_{bi}$ in a basis where $\C$ takes the form in~\cref{eq:recallminWB}, fixes the $c^{(i)}_{m}$ coefficients as functions of the (\ac{WBI}) non-zero eigenvalues of $\C$, the ${\mBd^2}_{a}$.
These coefficients
may be re-expressed in terms of $\{\tr \C,\tr \C^2,\dots, \tr \C^\n\}$, i.e.
\begin{equation}
   c^{(i)}_{m}\equiv c^{(i)}_{m}\left(\tr \C,\tr \C^2 ,\dots, \tr \C^\n \right)
\end{equation}
are \acp{WBI} themselves and can be computed in any basis.
Given that $P_{i\geq 4}$ is a polynomial in $\C$, 
it transforms as desired, and
any traces of combinations of the matrices $\hu,\hd,\C,P_i$ and their powers is also a \ac{WBI}. 

We may define now
\begin{equation}
P_{123}\equiv\id_\n-\sum_{i=1}^{\n}P_{i+3} \, .
\end{equation}
In the minimal weak basis,
\begin{equation}
\hu^{P}\equiv P_{123}\, \hu\, P_{123}\,=\,
\vw^2
\begin{pmatrix}
 \Ydu^{2}  & 0 \\
  0 & 0
\end{pmatrix} \,.
\end{equation}
Defining
\begin{equation}
P_{i<4}\equiv \sum_{m=1}^{3}  a^{(i)}_{m}\left(\tr \hu^{P},\tr \left(\hu^{P}\right)^2, \tr\left(\hu^{P}\right)^3 \right) \left(\hu^{P}\right)^m\, ,
\end{equation}
with $a^{(i)}_{m}$ such that these are projectors in the minimal weak basis, $\left(P_{i}\right)_{ab}=\delta_{ai}\delta_{bi}$, we now have all the $P_{i}$, 
with $i=1,\ldots,\n+3$, as (dimensionless) polynomials in $\hu$ and $\C$.
Hence, $P_i \,\to\,\WL^\dagger \,P_i\, \WL$ under \acp{WBT} and
they, together with $P_{123}$, can be used to build \acp{WBI}.

Now that we have all the projectors, we can trivially express all the moduli of some matrix $C=\hu, \hd, \C,\hu^2,\hu\hd,\ldots$, in this
minimal
basis as a function of \acp{WBI}, via
\begin{equation}
|C_{ij}|^2=\tr(P_i C P_j C) \,.
\end{equation}
Taking this into account, the mapping of the Yukawa
and bare mass
matrices in any basis to the minimal one can be achieved by following these steps:
\begin{enumerate}
\item Find the $\n +3$ projectors $P_i$ as explained above. This step
involves computing the non-zero eigenvalues of $\C$ and $\hu^{P}$, 
giving us the values of $\mBd$ and $\Ydu$.
\item $\Ydd^2$ are the non-zero eigenvalues of $P_{123} \hd P_{123}$. These weak-basis invariants can be trivially obtained, for example, by solving
\begin{equation}
\tr\left[(P_{123} \hd P_{123})^k\right]=\sum_{i=1}^{3} \left(\vw\, {\Ydd}{_i}\right)^{2k},
\end{equation}
where $k=1,2,3$.
\item The moduli of $\VWB$ can be reconstructed from
\begin{equation}
\begin{aligned}
\tr(P_{i} \hd) &=\sum_j |\VWB_{ij}|^2 \left(\vw{\Ydd}{}_{j}\right)^2 \,,\\  
\tr(P_{i} \hd P_{123} \hd) &= \sum_j |\VWB_{ij}|^2 \left(\vw{\Ydd}{}_{j}\right)^4 \,,
\end{aligned}
\end{equation}
where $i=1,2$ only. Taking into account the unitarity of $\VWB$, this gives $\theta_{12}$, $\theta_{13}$, $\theta_{23}$, and $|\delta_{13}|$ via $\cos \delta_{13}$. The sign of $\delta_{13}$ is the same as the one of $\im\left[\tr(P_{1}\hd P_{2} \hd P_{3} \hd)\right]$, since
\begin{equation}
\begin{aligned}
    \im\left[\tr(P_{1}\hd P_{2} \hd P_{3} \hd)\right] &=
    \vw^6 \left(({\Ydd}{_3})^2-({\Ydd}{_2})^2\right)\left(({\Ydd}{_3})^2-({\Ydd}{_1})^2\right)\left(({\Ydd}{_2})^2-({\Ydd}{_1})^2\right)
    \\
    &\,\, \times 
    \frac{1}{8} \sin2 \theta_{12}\sin2 \theta_{13}\sin2 \theta_{23} \cos \theta_{13} \sin \delta_{13}
    \,.
\end{aligned}
\end{equation}

\item For $\left|\yup\right|_{\alpha i}$ one simply has
\begin{equation}
\left|\yup\right|_{\alpha i}=\frac{ \sqrt{ \tr( P_{\alpha} \hu P_{i} \hu) } }{ \vw^2 {\Ydu}{_i} } \,,
\end{equation}
while $\left|\ydbh\right|_{\alpha i}$ can be obtained from
\begin{equation}
\tr\left[P_{\alpha}\hd(P_{123}\hd)^k\right]=\sum_{i}\left(\vw{\Ydd}{}_{i}\right)^{2k} \vw^2 \left|\ydbh\right|_{\alpha i}^{2}\,, 
\end{equation}
where $k=1,2,3$. 

\item The most involved step is finding the remaining phases. For each row $\alpha$, we first take
\begin{equation}
\tr\left[P_\alpha (\hd P_{123} \hd) P_i \hd\right]=\sum_{k,j}^{3}
\vw^6
\left(\ydbh\right)_{\alpha j}\left(\ydbh\right)_{\alpha k}^* ({\Ydd}{_j})^3 {\Ydd}{_k} \,\VWB_{ij}^* \VWB_{ik}\,,
\end{equation}
with $i=1,2$.
Since we have $\arg\left[\left(\ydbh\right)_{\alpha 1}\right]=0$, at this stage we can use those equations to fix the only remaining unknown parameters there, i.e., $\arg\left[\left(\ydbh\right)_{\alpha 2}\right]$ and $\arg\left[\left(\ydbh\right)_{\alpha 3}\right]$. This completes the determination of all parameters in $\Yd$ and thus $\hd$ in this weak basis, including the phases in $\hd$. The remaining $3\n$ phases in $\yup$  can be simply obtained from
\begin{equation}
\arg\left[\left(\yup\right)_{\alpha i}\right]=\arg\left[\left(\hu\right)_{\alpha i}\right]=\arg\left[\tr\left(P_{\alpha}\hu P_{i}\hd \right)\right]+\arg\left[\left(\hd\right)_{\alpha i}\right]\,,
\end{equation}
since $\tr\left(P_{\alpha}\hu P_{i}\hd \right) = \left(\hu\right)_{\alpha i}\left(\hd\right)_{i \alpha}$.
\end{enumerate}

\subsection{A numerical example}
\label{sec:numex}
For illustration, let us give a numerical example with one doublet. Starting from an arbitrarily convoluted set of values for the original Yukawa and mass matrices in~\cref{eq:LY1},%
\footnote{As it will become apparent below, it corresponds to a physically meaningful solution. In this specific basis, all Yukawa couplings are of the same order.}
\begin{equation}
\begin{aligned}
\Yu^0 &\approx \begin{pmatrix}
\phantom{+}0.200 - 0.242\,i &           -0.244 - 0.228\,i &           -0.157 + 0.101\,i\\
\phantom{+}0.209 + 0.217\,i & \phantom{+}0.244 - 0.207\,i &           -0.099 - 0.143\,i\\
\phantom{+}0.242 - 0.244\,i &           -0.240 - 0.244\,i & \phantom{+}0.089 - 0.180\,i\\
          -0.244 - 0.244\,i &           -0.243 + 0.244\,i &           -0.180 - 0.088\,i        
\end{pmatrix} \,,\\
\Yd^0 &\approx \begin{pmatrix}
\phantom{+}0.084 - 0.080\,i & \phantom{+}0.085 - 0.083\,i & \phantom{+}0.101 - 0.080\,i\\
\phantom{+}0.080 + 0.080\,i & \phantom{+}0.083 + 0.081\,i & \phantom{+}0.080 + 0.095\,i\\
          -0.082 + 0.080\,i &           -0.080 + 0.080\,i &           -0.080 + 0.087\,i\\
\phantom{+}0.080 + 0.082\,i & \phantom{+}0.080 + 0.080\,i & \phantom{+}0.088 + 0.080\,i
\end{pmatrix}\,,\\
\Mb^0&\approx\begin{pmatrix}
\phantom{+}4.21 - 4.22\,i \\
\phantom{+}4.21 + 3.95\,i \\
          -3.69 + 4.24\,i \\
\phantom{+}4.25 + 3.69\,i 
\end{pmatrix}\vw\,,
\end{aligned}
\end{equation}
the algorithm described in~\cref{sec:stepladderwb} returns the equivalent stepladder form
\begin{equation}
\vw\,\Yu =\begin{pmatrix}
     \tilde{m}_{u} \\
    \tilde{r}_{u}
\end{pmatrix} \,, \quad
\vw\,\Yd =\begin{pmatrix}
    \VWT & 0\\
    0 & 1
\end{pmatrix}\begin{pmatrix}
     \tilde{m}_{d} \\
    \tilde{r}_{d} 
\end{pmatrix} , \quad \Mb=\begin{pmatrix}
  0 \\
  \mB
\end{pmatrix} \, ,
\end{equation}
with $\mB\approx 11.49\, \vw$ ($\approx 2$~TeV) and 
\begin{equation}
\begin{aligned}
\begin{pmatrix}
\tilde{m}_{u} \\
\tilde{r}_{u}
\end{pmatrix} &\approx 
\begin{pmatrix}
0.00014	& 0 & 0 \\
0.0034 & 9.0 \times 10^{-4}	& 0 \\
0 & 0.92 & 0.18 \\
0 & 0 & 0.37
\end{pmatrix} \vw\,, \\
\begin{pmatrix}
\tilde{m}_{d} \\
\tilde{r}_{d}
\end{pmatrix} &\approx 
\begin{pmatrix}
1.4 \times 10^{-5} & 0 & 0 \\
3.7 \times 10^{-7} & 0.0035 & 0 \\
0 & 0.015 & 0.0013 \\
0 & 0 & 0.41
\end{pmatrix} \vw\,, \\
\VWT &\approx 
\begin{pmatrix}
0.98 + 5.6 \times 10^{-6}\,i & 0.18 +0.00040\,i & -0.043+0.0018\,i \\
0.18 + 0.00014\,i & -0.95 + 0.0081\,i & 0.24 + 0.036\,i \\
-0.0081 + 0.0033\,i & 0.15 + 0.19\,i & 0.48 + 0.84\,i
\end{pmatrix}\,.
\end{aligned}
\end{equation}
Here, the row vectors $\tilde{r}_q = (0\,\,\,0\,\,\,\mB r_{q0})$ have mass dimension one.

\vskip 2mm

Instead, the algorithm just described applied to any of the two equivalent Yukawa sets,%
\footnote{We have also checked that applying the stepladder algorithm to the equivalent minimal weak basis form also returns back the same stepladder result.}
leads to the equivalent minimal form
\begin{equation}
    \Yu= 
\begin{pmatrix}
 \Ydu   \\
  \yup 
\end{pmatrix}\, ,
\quad
\Yd=  
\begin{pmatrix}
  \VWB(\vec{\theta}_{\VWB})\,\Ydd  \\
  \ydbh  
\end{pmatrix} \, ,
\quad
\Mb=\begin{pmatrix}
  0 \\
  \mBd
\end{pmatrix} \,,
\end{equation}
with again $\mBd=\mB\approx 11.49\, \vw$, and
\begin{equation} \label{eq:numericalminWB}
\begin{aligned}
    \Ydu &\,=\, \diag(6.91 \times 10^{-6},\, 0.00344,\, 0.940) \,,\\ 
    \Ydd &\,=\, \diag(1.45 \times 10^{-5},\, 0.000292,\, 0.0158) \,,\\
    \yup  &\,=\, (0.365,\, 0.0180,\, -0.0702) \,,\\ 
    \ydbh &\,=\, (0.000501, \, -0.404, \,0.0334\, e^{2\pi i/3})\,,\\
    \vec{\theta}_{\VWB} &\,\equiv\, (\theta_{12}^{\VWB},\, \theta_{13}^{\VWB},\, \theta_{23}^{\VWB},\, \delta_{13}^{\VWB})\,=\,(0.227,\, 0.00369,\, 0.0418,\, 0.36 \pi)
    \,.
\end{aligned}
\end{equation}
This particular benchmark satisfies the constraints discussed in~\cref{sec:pheno}, while providing a solution to one of the  so-called \aclp{CAA} (in this case \ac{CAA}2, with $\hiu,\his \neq 0$; see also~\cref{sec:CAAN1}).

\vskip 2mm

A \texttt{Mathematica} implementation of the stepladder algorithm for $\n=1$ and of the minimal \ac{WB} algorithm for $\n=1,2,3$ can be found as supplementary material, mapping any set of (non-singular) $\Yu$, $\Yd$ and $\Mb$ to these minimal bases.

\section{CP-odd weak-basis invariants expanded in \texorpdfstring{$\vw/\mB$}{v/MQ}}
\label{app:CP}

Using the results found in~\cref{sec:app1},
we can write simplified expressions for weak-basis invariants in the case $\n=1$. 
For $3$-block invariants we have
\begin{equation} \label{eq:inv2C}
\begin{aligned}
\im \tr[\HuL^n \HdL^m \C] &= 
\im\tr\Big[\Du^{2n+1} \VL \Dd^{2m+1} \VR^\dagger\Big]
 \\[2mm]
&=\im\Big(\MT^{2n+1} \MB^{2m+1} \VLa_{T'B'}
\V^R_{TT'}\V^{R*}_{BB'}
+ m_\alpha^{2n+1} \MB^{2m+1} \VLa_{\alpha B'}
\V^R_{T\alpha}\V^{R*}_{BB'}
\\ 
&\qquad\quad\,
+ \MT^{2n+1} m^{2m+1}_i \VLa_{T'i}
\V^R_{TT'}\V^{R*}_{Bi}
\,+\,m_\alpha^{2n+1} m_i^{2m+1}\VLa_{\alpha i}\,
\V^R_{T\alpha}\V^{R*}_{Bi}
\Big)
 \\
&= \mB^{2m+2n+2} \, \frac{\vw^4}{\mB^4} 
 \sum_{i,\alpha=1}^3
\im \Bigg\{ \,\Vha_{\alpha i} \, \hi_\alpha \hi_i^*\;   
y_\alpha y_i \, 
\\ & \hspace{1.5cm}
\times
\bigg[\left( 1    + k_{Bm}\right)\left(1+  k_{Tn}\right)
\left( 1    + k_{RB'}\right)\left(1+  k_{RT'}\right)
\left(1+  k_{Li}\right)\left(1+  k_{L\alpha}\right)
 \\ & \hspace{1.5cm}
 -\frac{\vw^{2n}}{\mB^{2n}}y_\alpha^{2n} 
 \left( 1    + k_{Bm}\right)\left( 1    + k_{RB'}\right) \left(1 +k_{R\alpha} \right)
\left(1+  k_{Li}\right)
 \left(1+\frac{\dVhua_{\alpha i}}{\Vha_{\alpha i}}\right)
 \\ & \hspace{1.5cm}
-\frac{\vw^{2m}}{\mB^{2m}}y_i^{2m}\left(1+  k_{Tn}\right)\left(1+  k_{RT'}\right)\left(1+k_{Ri}\right)\left(1+  k_{L\alpha}\right)
\left(1+\frac{\dVhda_{\alpha i}}{\Vha_{\alpha i}}\right)
 \\ & \hspace{1.5cm}
 +\frac{\vw^{2n+2m}}{\mB^{2n+2m}}
 y_\alpha^{2n} y_i^{2m} \left(1+k_{R\alpha}\right)\left(1+k_{Ri}\right)
 \left(1+\frac{\delta\Vha_{\alpha i}}{\Vha_{\alpha i}}\right)
\bigg] \Bigg\}\,,
\end{aligned}
\end{equation}
where the Greek (Latin) indices refer to up-type (down-type) light quarks, $\alpha = u,c,t$ ($i=d,s,b$).
The quantities dubbed ``$k$''
account for the corrections to the eigenvalues and mixing elements of 
order at least $\vw^2/\mB^2$
which are \emph{real}.
In particular,
$k_{T(B)}$ includes the corrections in
$M_{T'(B')}=\mB(1+k_{T(B)})$, 
see~\cref{eq:Msplit}, with
\begin{equation}
\begin{aligned}
k_T &= \frac{1}{2} \sum_\alpha  |\hi_\alpha|^2 \frac{\vw^2}{\mB^2} 
+\frac{1}{2} \left[ \sum_\alpha  |\hi_\alpha|^2 y_\alpha^2 - \bigg(\frac{1}{2}\sum_\alpha  |\hi_\alpha|^2\bigg)^2 \right]\frac{\vw^4}{\mB^4}
+ \mathcal{O}\bigg(\frac{\vw^6}{\mB^6} \bigg)
\,,\\
k_B &= \frac{1}{2} \sum_i  |\hi_i|^2 \frac{\vw^2}{\mB^2} 
+\frac{1}{2} \left[ \sum_i  |\hi_i|^2 y_i^2 - \bigg(\frac{1}{2}\sum_i  |\hi_i|^2\bigg)^2 \right]\frac{\vw^4}{\mB^4}
+ \mathcal{O}\bigg(\frac{\vw^6}{\mB^6} \bigg)
\,,
\end{aligned}
\end{equation}
while $1+k_{Tn} \equiv (1+k_{T})^{2n+1}$ and $1+k_{Bn} \equiv (1+k_{B})^{2n+1}$. Recall also that $m_{\alpha(i)} = \vw\, y_{\alpha(i)}$.

To go from the second to the third equality in~\cref{eq:inv2C}, we additionally need to expand the \ac{RH}  elements $\V^R_{TT'}$ and $\V^R_{T\alpha}$ in the up sector, and $\V^R_{BB'}$ and $\V^R_{Bi}$ in the down sector.
The $k_{R\alpha(i)}$ are real quantities 
which account for
the corrections
of the elements $\V^R_{T\alpha}=-\hi_\alpha \vw/\mB(1+k_{R\alpha})$ and $\V^R_{Bi}=-\hi_i \vw/\mB(1+k_{Ri})$, see~\cref{eq:phases,eq:4rowu}, namely
\begin{equation}
\begin{aligned}
k_{R\alpha} &= - \bigg[\frac{1}{2} |\hi_\alpha|^2 - y_\alpha^2 + \sum_{\beta < \alpha} |\hi_\beta|^2 \bigg]\frac{\vw^2}{\mB^2} + \mathcal{O}\bigg(\frac{\vw^4}{\mB^4} \bigg)
\,, \\
k_{Ri} &= - \bigg[\frac{1}{2} |\hi_i|^2 - y_i^2 + \sum_{j < i} |\hi_j|^2 \bigg]\frac{\vw^2}{\mB^2} + \mathcal{O}\bigg(\frac{\vw^4}{\mB^4} \bigg)
\,,
\end{aligned}
\end{equation}
where
we also take into account the hierarchy of \ac{SM} Yukawa eigenvalues.
Note that both $\V^R_{TT'} = 1 +k_{RT'}$ and $\V^R_{BB'} = 1 + k_{RB'}$ are real,
with
\begin{equation}
\begin{aligned}
k_{RT'} &= - \frac{1}{2} \sum_\alpha|\hi_\alpha|^2 \frac{\vw^2}{\mB^2}
+ \left[\frac{3}{8}  \bigg(\sum_\alpha |\hi_\alpha|^2\bigg)^2 -\sum_\alpha|\hi_\alpha|^2 y_\alpha^2 \right]\frac{\vw^4}{\mB^4}
+ \mathcal{O}\bigg(\frac{\vw^6}{\mB^6} \bigg)
\,, \\
k_{RB'} &= - \frac{1}{2} \sum_i|\hi_i|^2 \frac{\vw^2}{\mB^2}
+ \left[\frac{3}{8} \bigg(\sum_i  |\hi_i|^2\bigg)^2 - \sum_i  |\hi_i|^2y_i^2  \right]\frac{\vw^4}{\mB^4}
+ \mathcal{O}\bigg(\frac{\vw^6}{\mB^6} \bigg)
\,,
\end{aligned}
\end{equation}
so that $k_{RT'} \approx -k_T$ and $k_{RB'} \approx -k_B$, up to $\mathcal{O}(\vw^4/\mB^4)$ differences.

In what concerns \ac{LH} mixing, recall that
\begin{equation}
    \VL = {\VuL}^\dagger \VdL = \Rf^{uL\dagger} \begin{pmatrix}
        \VLh & 0 \\ 0 & 1 
    \end{pmatrix} \Rf^{dL}\,,
\end{equation}
where the $\Rf$ matrices have the structure given in~\cref{eq:angles}. One then has
\begin{equation}
\begin{aligned}
    \VLa_{T'B'} &= \V^{L*}_{TT'} \V^L_{BB'} +
    \sum_{\alpha, i=1}^3  (\Rf^{uL})^*_{\alpha T'} \,\Vha_{\alpha i} \,(\Rf^{dL})_{i B'}
    \\
    &= 
    \V^L_{TT'} \V^L_{BB'} +
    \sum_{\alpha, i=1}^3  \V^{L}_{ T \alpha} \,\Vha_{\alpha i} \,\V^{L*}_{B i} \,(1+k'_{L\alpha})(1+k'_{Li})
    \,,
\end{aligned}
\end{equation}
where we have used $(\Rf^{uL})_{\alpha T'} = -(\Rf^{uL})^*_{T\alpha}(1+k'_{L\alpha})$ and $(\Rf^{dL})_{i B'} = -(\Rf^{dL})^*_{Bi}(1+k'_{Li})$, with real $k'_{L\alpha(i)}$ of at least $\mathcal{O}(\vw^4/\mB^4)$.
Note also that 
$(\Rf^{uL})_{T\alpha} = \V^L_{T \alpha}$, 
$(\Rf^{dL})_{Bi} = \V^L_{Bi}$, 
and that 
both $\V^L_{TT'} = 1 +k_{LT'}$ and $\V^L_{BB'} = 1 + k_{LB'}$ are real, with $k_{LT'(B')} = \mathcal{O}(\vw^4/\mB^4)$,
\begin{equation}
\begin{aligned}
k_{LT'} &= - \frac{1}{2} \sum_\alpha|\hi_\alpha|^2y_\alpha^2 \frac{\vw^4}{\mB^4} + \mathcal{O}\bigg(\frac{\vw^8}{\mB^8} \bigg)
\,, \\
k_{LB'} &= - \frac{1}{2} \sum_i|\hi_i|^2 y_i^2 \frac{\vw^4}{\mB^4} + \mathcal{O}\bigg(\frac{\vw^8}{\mB^8} \bigg)
\,.
\end{aligned}
\end{equation}
Recalling~\cref{eq:4rowL} and that a relation analogous to~\cref{eq:phases} holds for the \ac{LH} sector, we can write
\begin{equation}
\begin{aligned}
\V^L_{T\alpha}&=-\hi_\alpha y_\alpha \frac{\vw^2}{\mB^2} \frac{1+k_{L\alpha}}{1+k'_{L\alpha}}
\equiv -\hi_\alpha y_\alpha \frac{\vw^2}{\mB^2} (1+k''_{L\alpha})\,,\\
\V^L_{B i}&=-\hi_i y_i \frac{\vw^2}{\mB^2} \frac{1+k_{Li}}{1+k'_{Li}}
\equiv -\hi_i y_i \frac{\vw^2}{\mB^2} (1+k''_{Li})\,,
\end{aligned}
\end{equation}
leading to
\begin{equation}
    \im\VLa_{T'B'} = 
    \frac{\vw^4}{\mB^4}
    \sum_{\alpha, i=1}^3  \im\left[\hi_{\alpha} y_\alpha\, \Vha_{\alpha i} \,\hi^*_i y_i \right]\,(1+k_{L\alpha})(1+k_{Li})
    \,,
\end{equation}
which has been used in~\cref{eq:inv2C}. Here,
\begin{equation} \label{eq:kLs}
\begin{aligned}
k_{L\alpha} &= - \bigg[\frac{1}{2} |\hi_\alpha|^2 - y_\alpha^2 + \sum_{\beta < \alpha} |\hi_\beta|^2 \bigg]\frac{\vw^2}{\mB^2} + \mathcal{O}\bigg(\frac{\vw^4}{\mB^4} \bigg)
\,, \\
k_{Li} &= - \bigg[\frac{1}{2} |\hi_i|^2 - y_i^2 + \sum_{j < i} |\hi_j|^2 \bigg]\frac{\vw^2}{\mB^2} + \mathcal{O}\bigg(\frac{\vw^4}{\mB^4} \bigg)
\,,
\end{aligned}
\end{equation}
so that $k''_{L\alpha} \approx k_{L\alpha} \approx k_{R\alpha}$ and $k''_{Li} \approx k_{Li} \approx k_{Ri}$, up to $\mathcal{O}(\vw^4/\mB^4)$ differences, i.e.
\begin{equation}
k''_{L\alpha} -  k_{R\alpha} \,\equiv\, \Delta_\alpha\, \frac{\vw^{4}}{\mB^{4}}\,,\qquad
k''_{Lj} -  k_{Rj} \,\equiv\, \Delta_j\, \frac{\vw^{4}}{\mB^{4}}\,.
\end{equation}
In writing~\cref{eq:kLs}, we have once again taken into account the hierarchy of \ac{SM} Yukawas.

For the other elements of $\VL$ we find
\begin{equation}
\begin{aligned}
\VLa_{T'i}&=\frac{\vw^2}{\mB^2}\left[\sum_{\alpha=u,c,t} \hi_\alpha y_\alpha \left( \Vha_{\alpha i}+\dVhda_{\alpha i} \right) (1+k_{L\alpha})
-\hi_i y_i (1+k_{LT'})(1+k_{Li}'')\right]\,,
\\
\VLa_{\alpha B'}&= \frac{\vw^2}{\mB^2}\left[\sum_{i=d,s,b} 
\hi_i^* y_i \left( \Vha_{\alpha i}+\dVhua_{\alpha i} \right)(1+k_{Li})
- \hi_\alpha^* y_\alpha (1+k_{LB'})(1+k_{L\alpha}'')
\right]\,,
\end{aligned}
\end{equation}
expanding on~\cref{eq:LH4mix}.
Here,
$\dVhda_{\alpha i}$ and $\dVhua_{\alpha i}$ are small corrections 
to $\Vha_{\alpha i}$ due to the extra mixings
of order
$y^2\hi^2\vw^4/\mB^4$
as in~\cref{eq:deltaVL}:
\begin{equation} \label{eq:corrections}
\begin{aligned}
\Vha_{\alpha i}  +   \dVhda_{\alpha i} &=
    \Vha_{\alpha i} \left[
    1-\frac{1}{2}|\hi_i|^2y_i^2\frac{\vw^4}{\mB^4}
    -\sum_{j<i} \hi_j^* \hi_i  \frac{\Vha_{\alpha j}}{\Vha_{\alpha i}}y_j y_i \frac{\vw^4}{\mB^4} + \mathcal{O}\bigg(\frac{\vw^8}{\mB^8} \bigg)
    \right]
\,,  \\
\Vha_{\alpha i}  +   \dVhua_{\alpha i} &=
    \Vha_{\alpha i} \left[
    1-\frac{1}{2}|\hi_\alpha|^2y_\alpha^2\frac{\vw^4}{\mB^4}
    -\sum_{\beta<\alpha} \hi_\beta \hi_\alpha^*  \frac{\Vha_{\beta i}}{\Vha_{\alpha i}} y_\beta y_\alpha \frac{\vw^4}{\mB^4}+ \mathcal{O}\bigg(\frac{\vw^8}{\mB^8} \bigg)
    \right]
 \,.
\end{aligned}
\end{equation}
Note that, crucially, the $\mathcal{O}\big({\vw^8}/{\mB^8} \big)$ terms do not contain additional phases.
At first glance, the rightmost $\mathcal{O}(\vw^4/\mB^4)$ terms in each row of~\cref{eq:corrections} may seem to introduce new phases. In reality, when reinserted into~\cref{eq:inv2C} and after expanding the sums, one still finds that the complex structure reduces to a sum of terms proportional to $\im \big( \Vha_{\beta j} \, \hi_\beta \hi_j^* \big)$, with appropriate indices.
The remaining entries of the \ac{LH} mixing matrix read 
\begin{equation}
\VLa_{\alpha i} =\Vha_{\alpha i}  +   \delta\Vha_{\alpha i}
+ \frac{\vw^4}{\mB^4} \hi_\alpha^* \hi_i\, y_\alpha y_i (1+k_{L\alpha}'')(1+k_{Li}'')
\,,
\end{equation}
where the last term will vanish when taking the imaginary part in~\cref{eq:inv2C}. Here,
\begin{equation}
\begin{aligned}
&\Vha_{\alpha i}  +   \delta\Vha_{\alpha i} 
=
    \Vha_{\alpha i} \Bigg[
    1-\frac{1}{2}|\hi_i|^2y_i^2\frac{\vw^4}{\mB^4}
    -\frac{1}{2}|\hi_\alpha|^2y_\alpha^2\frac{\vw^4}{\mB^4}
    -\sum_{j<i} \hi_j^* \hi_i  \frac{\Vha_{\alpha j}}{\Vha_{\alpha i}}y_j y_i \frac{\vw^4}{\mB^4} 
   \\ \qquad \,\,& 
      -\sum_{\beta<\alpha} \hi_\beta \hi_\alpha^*   \frac{\Vha_{\beta i}}{\Vha_{\alpha i}}y_\beta y_\alpha\frac{\vw^4}{\mB^4}
+
    \sum_{j<i}\sum_{\beta<\alpha} \hi_\beta \hi_\alpha^*\hi_j^* \hi_i \frac{\Vha_{\beta j}}{\Vha_{\alpha i}} y_\beta y_\alpha y_j y_i\frac{\vw^8}{\mB^8} 
      +\mathcal{O}\bigg(\frac{\vw^8}{\mB^8} \bigg)
    \Bigg]\,,
\end{aligned}
\end{equation}
where, once again, the concealed $\mathcal{O}\big({\vw^8}/{\mB^8} \big)$ terms do not contain additional phases. 
Taking the above into account, we can write~\cref{eq:inv2C} in a more compact way,
\begin{equation}
\frac{1}{ \mB^{2(m+n+1)}}\im \tr[\HuL^n \HdL^m \C] 
= 
\frac{\vw^4}{\mB^4}  \sum_{i,\alpha=1}^3
\im \left( \,\Vha_{\alpha i} \, \hi_\alpha \hi_i^* \right) \;   
y_\alpha \, y_i \:  
\left(1 + k^{(m,n)}_{\alpha i}\, \right) \,,
\end{equation}
where the $k^{(m,n)}_{\alpha i}$ account for real corrections of 
order at least $\vw^2/\mB^2$ to the masses and mixing elements, matching~\cref{eq:inv3bl}.
We emphasize that no approximation has been made in deriving the complex structure of this result, revealing the exact phases at play.

As a further explicit example, consider \acp{WBI} with five ``blocks''. We find:
\begin{align} \label{eq:inv5bl}
&\frac{1}{\mB^{2n+2\ell+2m+2k+2}}\im \tr[\HuL^n \HdL^k
\HuL^\ell \HdL^m \C] 
\nonumber \\
&= \frac{1}{\mB^{2(n+\ell+m+k+1)}}
\im\tr\Big[\Du^{2n+1}\VL \Dd^{2k} \VL^\dagger \Du^{2\ell} \VL \Dd^{2m+1} \VR^\dagger\Big]
\nonumber \\
&= 
     \frac{\vw^4}{\mB^4} \sum_{i,\alpha=1}^3 \im \Big[ \Vha_{\alpha i} \, \hi_\alpha \hi_i^* \, \Big]  \, y_\alpha y_i   
 \left( 1 + k^{(\text{a})}_{\alpha i}  
     \right)
         \nonumber \\ & 
   +  \frac{\vw^{2k+2\ell+4}}{\mB^{2k+2\ell+4}} \sum_{i,j,\alpha,\beta=1}^3 
\im \Big[ \Vha_{\alpha j} \Vha_{\beta i}  \Vhac_{\beta j}  \, \hi_\alpha \hi_i^* \, \Big] \, 
y_\alpha y_i y_\beta^{2\ell} y_j^{2k}  \,  
   \left(1+k^{(\text{b})}_{\alpha\beta i j} \right)  
    \nonumber \\ & 
+ \frac{\vw^{8}}{\mB^{8}} \sum_{i,\alpha,\beta=1}^3 
 \im \Big[ \Vha_{\alpha i} \Vhac_{\beta i} \, \hi_\alpha \hi_\beta^* \,\Big]\, y_i^{2 k }  y_\alpha y_\beta  
 \Bigg[
\bigg(\frac{\vw}{\mB}\bigg)^{2(k+\ell-2)}\, y_\alpha^{2\ell} 
\,\,-\,
\bigg(\frac{\vw}{\mB}\bigg)^{2(k+n-2)}\, y_\alpha^{2n} 
  \Bigg]\left( 1 + k^{(\text{c})}_{\alpha\beta i}   \right)
                  \nonumber \\ & 
+ \frac{\vw^{8}}{\mB^{8}} \sum_{i,j,\alpha=1}^3 
\, \im \Big[ \Vhac_{\alpha i} \Vha_{\alpha j} \, \hi_i \hi_j^* \,\Big]\, y_\alpha^{2 \ell}  \,y_i\, y_j  
 \Bigg[
\bigg(\frac{\vw}{\mB}\bigg)^{2(m+\ell-2)}\, y_i^{2m} 
\,\,-\,
\bigg(\frac{\vw}{\mB}\bigg)^{2(k+\ell-2)}\, y_i^{2k} 
  \Bigg]\left( 1 + k^{(\text{d})}_{\alpha ij}   \right)
                  \nonumber \\ & 
 +  \frac{\vw^{12}}{\mB^{12}} \sum_{i,j,\alpha,\beta=1}^3 \im \Big[\Vha_{\alpha i} \, \hi_\alpha \hi_i^* \, \Vha_{\beta j} \, \hi_\beta \hi_j^*  \, \Big]  \,   y_\alpha y_i y_\beta  y_j 
 \nonumber \\
 & \qquad \times
\Bigg[
\bigg(\frac{\vw}{\mB}\bigg)^{2(k+n-2)}
  y_i^{2k}y_\alpha^{2n}
+\bigg(\frac{\vw}{\mB}\bigg)^{2(m+n-2)}
y_j^{2m}y_\alpha^{2n} 
 \nonumber \\
 & \qquad\,\,\,\,\,
+\bigg(\frac{\vw}{\mB}\bigg)^{2(k+\ell-2)}
y_i^{2k}y_\beta^{2\ell} 
+\bigg(\frac{\vw}{\mB}\bigg)^{2(m+\ell-2)}
y_j^{2m}y_\beta^{2\ell} 
\Bigg]
  \left( 1 + k^{(\text{e})}_{\alpha\beta i j}  \right)
\nonumber \\ & 
+ \frac{\vw^{12}}{\mB^{12}} \sum_{i,j,\alpha,\beta=1}^3 
\im \Big[ \Vha_{\alpha i} \, \hi_\alpha \hi_i^* \Vhac_{\beta j}    \, \hi_\beta^*  \hi_j \,\Big] \, y_i  y_j  y_\alpha y_\beta
 \nonumber \\
 & \qquad 
 \times
 \Bigg[
\bigg(\frac{\vw}{\mB}\bigg)^{2(\ell+m-2)} y_i^{2m}
\bigg\{y_\alpha^{2\ell} \delta_{\beta<\alpha}
+ y_\beta^{2\ell}\delta_{\alpha<\beta} \bigg\}
+
\bigg(\frac{\vw}{\mB}\bigg)^{2(k+n-2)} y_\alpha^{2n}
\bigg\{y_i^{2k}\delta_{j<i}
+ y_j^{2k} \delta_{i<j} \bigg\}
\nonumber \\
 & \qquad\,\,\,\,
+
\bigg(\frac{\vw}{\mB}\bigg)^{2(k+\ell-2)}  
\bigg\{
  y_j^{2k}  \left(y_\alpha^{2\ell} \delta_{\beta<\alpha} + y_\beta^{2\ell} \delta_{\alpha<\beta}\right)
+ y_\beta^{2\ell}  \left(y_i^{2k} \delta_{j<i} + y_j^{2k}\delta_{i<j}\right)
\bigg\}
  \Bigg]\left( 1 + k^{(\text{f})}_{\alpha\beta i j}   \right)
                  \nonumber \\ & 
- \frac{\vw^{12}}{\mB^{12}} \sum_{i,j,p,\alpha,\beta=1}^3 
 \im \Big[ \Vha_{\alpha i} \Vhac_{\alpha j} \, \hi_i^* \hi_j \, \Vha_{\beta p} \,  \hi_\beta  \hi_p^* \,\Big]\, y_\alpha^{2\ell}  y_i y_\beta y_j y_p 
 \nonumber \\
 & \qquad 
 \times  
 \Bigg[
\bigg(\frac{\vw}{\mB}\bigg)^{2(\ell+m-2)}\, y_i^{2m} 
+
\bigg(\frac{\vw}{\mB}\bigg)^{2(k+\ell-2)}  
\bigg\{y_p^{2k} \delta_{j<p}
+ y_j^{2k} \delta_{p<j} \bigg\}
  \Bigg]\left( 1 + k^{(\text{g})}_{\alpha\beta i j p}   \right)
                  \nonumber \\ & 
- \frac{\vw^{12}}{\mB^{12}} \sum_{i,j,\alpha,\beta,\delta=1}^3 
\im \Big[ \Vha_{\alpha i} \Vhac_{\beta i}  \hi_\beta^*  \hi_\alpha \,\Vha_{\delta j} \hi_\delta   \hi_j^* \, \Big] \,y_i^{2k}  y_j  y_\alpha y_\beta  y_\delta
 \nonumber \\
 & \qquad 
 \times   
 \Bigg[
\bigg(\frac{\vw}{\mB}\bigg)^{2(k+n-2)}\, y_\alpha^{2n} 
+
\bigg(\frac{\vw}{\mB}\bigg)^{2(k+\ell-2)}  
\bigg\{y_\delta^{2\ell} \delta_{\beta<\delta}
+ y_\beta^{2\ell} \delta_{\delta<\beta} \bigg\}
  \Bigg]\left( 1 + k^{(\text{h})}_{\alpha\beta\delta i j}   \right)
                  \nonumber \\ & 
+ \frac{\vw^{16}}{\mB^{16}} \sum_{i,j,p,\alpha,\beta,\delta=1}^3 
\im \Big[ \Vha_{\alpha i}  \hi_\alpha  \hi_i^* \, \Vha_{\beta j}  \hi_\beta \hi_j^* \, \Vhac_{\delta p}  \,   \hi_\delta^*   \hi_p \,\Big]
\,y_i  y_j y_p  y_\alpha y_\beta y_\delta
 \nonumber \\
 & \qquad 
 \times 
 \Bigg[
\bigg(\frac{\vw}{\mB}\bigg)^{2(k+n-2)} y_\alpha^{2n}
\bigg\{y_i^{2k} \delta_{p<i}
+ y_p^{2k} \delta_{i<p} \bigg\}
+
\bigg(\frac{\vw}{\mB}\bigg)^{2(\ell+m-2)} 
\bigg\{y_i^{2m} y_\alpha^{2\ell} \delta_{\delta<\beta}
+ y_j^{2m} y_\delta^{2\ell}  \delta_{\beta<\delta} \bigg\}
\nonumber \\
 & \qquad\,\,\,
+
\bigg(\frac{\vw}{\mB}\bigg)^{2(k+\ell-2)} 
\bigg\{y_i^{2k} y_\beta^{2\ell} \delta_{\delta<\beta}\delta_{p<i}
+ y_i^{2k} y_\delta^{2\ell} \delta_{\beta<\delta}\delta_{p<i}
+
y_p^{2k} y_\beta^{2\ell} \delta_{\delta<\beta}\delta_{i<p}
+ y_p^{2k} y_\delta^{2\ell} \delta_{\beta<\delta}\delta_{i<p}
\bigg\}
\nonumber \\
 & \qquad\,\,\,
+
\bigg(\frac{\vw}{\mB}\bigg)^{2(m+n-2)} y_j^{2m} y_\alpha^{2n}
  \Bigg]  \left( 1 + k^{(\text{i})}_{\alpha\beta\delta i j p} \right)
\,,
\end{align}
where the explicit forms of the higher-order $k$ factors in general depend on the exponents $n,k,\ell,m$. Here, $\delta_{i<j} =1$ for $i<j$ and is 0 otherwise.
Notice that the shown terms of type $(\text{c})$ or $(\text{d})$ survive only when $n\neq \ell$ or $k\neq m$, respectively. In the cases where both $n=\ell$ and $k=m$, also the shown type-$(\text{f})$ terms vanish. The leading terms of types $(\text{c})$, $(\text{d})$, and $(\text{f})$  are then of higher order in the $\vw/\mB$ expansion.
For instance, for the $\M = 10$ \ac{WBI}, which corresponds to taking $n=k=\ell=m=1$, we find:
\begin{align}
&\frac{1}{\mB^{10}}\im \tr\big[\HuL \HdL
\HuL \HdL \C\big] = \frac{1}{\mB^{10}}
\im\tr\Big[\Du^3\VL \Dd^2 \VL^\dagger \Du^2 \VL \Dd^3 \VR^\dagger\Big]
\nonumber \\
&= 
     \frac{\vw^4}{\mB^4} \sum_{i,\alpha=1}^3 \im \Big[ \Vha_{\alpha i} \, \hi_\alpha \hi_i^* \, \Big]  \, y_\alpha y_i   
 \left( 1 + k^{(10\text{a})}_{\alpha i}  
     \right)
         \nonumber \\ & 
   +  \frac{\vw^8}{\mB^8} \sum_{i,j,\alpha,\beta=1}^3 
\im \Big[ \Vha_{\alpha j} \Vha_{\beta i}  \Vhac_{\beta j}  \, \hi_\alpha \hi_i^* \, \Big] \,  y_\alpha   y_i y_\beta^2 y_j^2 \,  
   \left(1+k^{(10\text{b})}_{\alpha\beta i j} \right) 
           \nonumber \\ & 
         +  \frac{\vw^{12}}{\mB^{12}} \sum_{i,\alpha,\beta=1}^3 
  \im \Big[ \Vha_{\alpha i} \Vhac_{\beta i} \, \hi_\alpha \hi_\beta^* \, \Big] \,
  y_i^2   y_\alpha^3 y_\beta \,\Delta_\alpha    \left( 1 + k^{(10\text{c})}_{\alpha \beta i}  
     \right)
  \nonumber \\ &
 +  \frac{\vw^{12}}{\mB^{12}} \sum_{i,j,\alpha=1}^3 
  \im \Big[ \Vhac_{\alpha i} \Vha_{\alpha j} \, \hi_i \hi_j^* \, \Big] \, 
  y_\alpha^2  y_i y_j^3 \,\Delta_j   \left( 1 + k^{(10\text{d})}_{\alpha i j}  
     \right)
    \nonumber \\ & 
 +  \frac{\vw^{12}}{\mB^{12}} \sum_{i,j,\alpha,\beta=1}^3 \im \Big[\Vha_{\alpha i} \, \hi_\alpha \hi_i^* \, \Vha_{\beta j} \, \hi_\beta \hi_j^*  \, \Big]  \, y_\alpha y_i y_\beta y_j \left(y_i^2 + y_j^2\right)\left(y_\alpha^2 + y_\beta^2\right)  \,
       \left( 1 + k^{(10\text{e})}_{\alpha\beta i j}  
     \right)
                 \nonumber \\ & 
 - \frac{\vw^{12}}{\mB^{12}} \sum_{i,j,p,\alpha,\beta=1}^3 
 \im \Big[ \Vha_{\alpha i} \Vhac_{\alpha j} \, \hi_i^* \hi_j \, \Vha_{\beta p} \,  \hi_\beta  \hi_p^* \,\Big]
 \, y_\alpha^2 y_i y_\beta y_j y_k  
 \left(y_i^2 + y_p^2 \delta_{j<p}
+ y_j^2 \delta_{p<j} \right)
    \left( 1+ k^{(10\text{g})}_{\alpha\beta i j p}  \right)
                  \nonumber \\ & 
 - \frac{\vw^{12}}{\mB^{12}} \sum_{i,j,\alpha,\beta,\delta=1}^3 
\im \Big[ \Vha_{\alpha i} \Vhac_{\beta i}  \hi_\beta^*  \hi_\alpha \,\Vha_{\delta j} \hi_\delta   \hi_j^* \, \Big] \,  y_i^2 y_j y_\alpha y_\beta y_\delta 
 \left(y_\alpha^2 + y_\delta^2 \delta_{\beta<\delta}
+ y_\beta^2 \delta_{\delta<\beta} \right)
   \left( 1+  k^{(10\text{h})}_{\alpha\beta\delta i j }  \right)
                \nonumber \\ & 
 + \frac{\vw^{16}}{\mB^{16}} \sum_{i,j,\alpha,\beta=1}^3 
\im \Big[ \Vha_{\alpha i} \, \hi_\alpha \hi_i^* \Vhac_{\beta j}    \, \hi_\beta^*  \hi_j \,\Big] \,  
 y_i   y_j  y_\alpha y_\beta
\nonumber \\ & \quad
\times \frac{1}{4}\,
\bigg\{
\left(y_{i}^2 -y_{j}^2 \right)
\Big(\textstyle\sum_{p} |\hi_{p}|^2\Big)^2 y_{\beta}^2\, \delta_{\alpha<\beta}
+
\left(y_{\alpha}^2 -y_{\beta}^2 \right)
\Big(\textstyle\sum_{\delta} |\hi_{\delta}|^2\Big)^2 y_{i}^2\, \delta_{j<i}
   \bigg\}
   \left(1+k^{(10\text{f})}_{\alpha\beta i j}\right)
      \nonumber \\ & 
 + \frac{\vw^{16}}{\mB^{16}} \sum_{i,j,p,\alpha,\beta,\delta=1}^3 
\im \Big[ \Vha_{\alpha i}  \hi_\alpha  \hi_i^* \, \Vha_{\beta j}  \hi_\beta \hi_j^* \, \Vhac_{\delta p}  \,   \hi_\delta^*   \hi_p \,\Big] \,  y_i y_j y_p  y_\alpha y_\beta  y_\delta
\nonumber \\ & \quad
\times
\bigg\{
 \Big[y_{\delta }^2 \delta _{\beta <\delta }+y_{\alpha }^2 \Big]  \Big[y_i^2 \delta _{p<i}+y_p^2 \delta _{i<p}+y_j^2 \Big]+\delta_{\delta <\beta }  \Big[y_i^2 \left(y_{\beta }^2 \delta _{p<i}+y_{\alpha }^2\right)+y_p^2 y_{\beta }^2 \delta _{i<p} \Big]
\bigg\}
  \left(1 + k^{(10\text{i})}_{\alpha \beta\delta ijp}\right)\,,
\end{align}
which we have partially reported in the main text, see~\cref{eq:5blockintext}.

\vfill
\clearpage

\section{Conditions for CP Invariance}
\label{app:CPI}

In this appendix we illustrate the demonstration that the vanishing of the \acp{WBI} discussed in~\cref{sec:CPwbi} listed in~\Cref{tab:cp-odd-inv} would imply CP conservation in a scenario with one vector-like quark doublet.
The vanishing of all complex phases in a given parameterization gives sufficient conditions for \acf{CPI}. However, that is not strictly necessary. A less restrictive set of sufficient conditions consists in imposing that the imaginary parts of all rephasing invariants one can build vanish. For instance, in the case of the minimal \ac{WB} of~\cref{sec:minimalweak}, \ac{CPI} is achieved in a one-doublet scenario ($\n=1$) if%
\footnote{We emphasize that these are the rephasing invariants one can build directly from the minimal \ac{WB} of~\cref{sec:minimalweak} --- they are not exactly the same as the ones defined in~\cref{sec:reph-inv-eff}.}
\begin{equation} \label{eq:app_CPI}
\begin{aligned} 
&I_{\alpha i}\equiv \im \left(\hi_{\alpha} \VWB_{\alpha i} \hi_i^{*}\right)=\im \left(\yup P_{\alpha}\VWB P_i \ydb^{\dagger}\right)=0 \, ,\\
&I_{i\alpha j}\equiv \im \left(\hi_i \VWB^*_{\alpha i} \VWB_{\alpha j} \hi_j^{*}\right)=\im \left(\ydb P_{i} \VWB^\dagger P_\alpha \VWB P_j \ydb^{\dagger}\right)=0 \, ,\\
&I_{\alpha i \beta}\equiv \im \left(\hi_\alpha \VWB_{\alpha i} \VWB^*_{\beta i} \hi_\beta^{*}\right)=\im \left(\yup P_\alpha \VWB P_i \VWB^{\dagger} P_\beta \yup^\dagger\right)=0 \, ,\\
&I_{\alpha i \beta j}\equiv \im\left(\hi_\alpha \VWB_{\alpha i} \VWB^*_{\beta i} \VWB_{\beta j} \hi^{*}_j\right)=\im \left(\yup P_{\alpha}\VWB P_i \VWB^{\dagger} P_\beta \VWB P_j \ydb^{\dagger} \right)=0  \, ,\\
&I'_{\alpha i \beta j} \equiv \im\left(\VWB_{\alpha i}\VWB^*_{\beta i }\VWB_{\beta j}\VWB^*_{\alpha j}\right) =\im \tr\left[P_\alpha \VWB P_i \VWB^\dagger P_\beta \VWB P_j \VWB^\dagger \right]=0\,,
\end{aligned}
\end{equation}
where $\left(P_\alpha\right)_{ij}=\delta_{i\alpha}\delta_{j\alpha}$ and $\alpha,\beta,i,j=1,2,3$. We have used the fact that $\VWB$ is a $3\times3$ unitary matrix, so that any other rephasing invariant can be constructed by combining these 5 types of invariants.

In turn, assuming non-degeneracy, one can always write the projectors as
\begin{equation} \label{eq:app_projectors_odd}
    P_\alpha=\sum_{n=0,1,2} c^u_{n,\alpha} \hat{D}^{2n+1}_u\,, \quad \quad
    P_i=\sum_{n=0,1,2} c^d_{n,i} \hat{D}^{2n+1}_d\,,
\end{equation}
where $c^u_{n,\alpha}$ and $c^d_{n,i}$ are real non-zero coefficients which depend solely on the elements of $\hat{D}_u\equiv \vw\Ydu$ and $\hat{D}_d\equiv \vw \Ydd$, respectively. Thus, we have
\begin{equation} \label{eq:app_I_nm}
    \mathcal{I}_{nm}\equiv \im \tr\left[\yup \hat{D}_u^{2n+1} \VWB \hat{D}_d^{2m+1}\ydb^{\dagger}\right]=0 \quad\Rightarrow \quad I_{\alpha i} =0\,.
\end{equation}
Alternatively we can write the projectors in terms of different combinations such as
\begin{equation} \label{eq:app_projectors_even}
    P_\alpha=\sum_{n=1,2,3} c'^{u}_{n,\alpha} \hat{D}^{2n}_u\,, \quad \quad
    P_i=\sum_{n=1,2,3} c'^{d}_{n,i} \hat{D}^{2n}_d\,,
\end{equation}
so that using both~\cref{eq:app_projectors_odd} and~\cref{eq:app_projectors_even} we have
\begin{equation}
\begin{aligned}
&\mathcal{I}^d_{nmp}\equiv\im\tr\left[\ydb \hat{D}_{d}^{2n+1} \VWB^\dagger \hat{D}_{u}^{2m} \VWB \hat{D}_d^{2p+1} \ydb^{\dagger} \right]=0 \quad\Rightarrow\quad  I_{i\alpha j} =0\,,\\
&\mathcal{I}^u_{nmp}\equiv\im\tr\left[\yup \hat{D}_u^{2n+1} \VWB \hat{D}_d^{2m} \VWB^{\dagger} \hat{D}_u^{2p+1}\yup^\dagger \right]=0 \quad\Rightarrow\quad  I_{\alpha i \beta} =0\,, \\
&\mathcal{I}_{nmpq}\equiv\im\tr\left[\yup \hat{D}_u^{2n+1} \VWB \hat{D}_d^{2m} \VWB^{\dagger} \hat{D}_u^{2p} \VWB \hat{D}_d^{2q+1} \ydb^{\dagger}  \right]=0 \quad\Rightarrow\quad  I_{\alpha i \beta j}=0\,,\\
&\mathcal{I}'_{nmpq}\equiv\im\tr\left[\hat{D}^4_u \VWB \hat{D}^4_d \VWB^\dagger \hat{D}^2_u \VWB \hat{D}^2_d \VWB^\dagger \right]=0 \quad\Rightarrow\quad  I'_{\alpha i \beta j}=0\,.
\end{aligned}
\end{equation}

In order to connect this new set of conditions to the \acp{WBI} discussed in~\cref{sec:CPwbi} we note that
\begin{equation}
\HuL=  \left(\begin{array}{ccc}
\hat{D}_u^2  & \vw \hat{D}_u \yup^\dagger \\
\vw \yup \hat{D}_u  &  \vw^2\yup \yup^\dagger + \mB^2  
\end{array}\right)\,, \quad \quad \quad  \HdL= \left(\begin{array}{ccc}
\VWB \hat{D}_d^2 \VWB^\dagger  &  \vw \VWB \hat{D}_d \ydb^\dagger \\
\vw \ydb \hat{D}_d \VWB^\dagger  &   \vw^2\ydb \ydb^\dagger + \mB^2
\end{array}\right)\,,
\end{equation}
so that each one of the conditions in~\cref{eq:app_CPI} corresponds, in the same order, to the following conditions in terms of the Hermitian matrices $\HuL$ and $\HdL$,
\begin{equation} \label{eq:CPI_WBI}
\begin{aligned}
    & \mathcal{I}_{nm}=  \im \tr\left[P_4 \HuL\left[(\id-P_4)\HuL\right]^n\left[(\id-P_4)\HdL\right]^{m+1}\right]=0\,,\\
    & \mathcal{I}^d_{nmp} =\im \tr\left[P_4 \HdL\left[(\id-P_4)\HdL\right]^n\left[(\id-P_4)\HuL\right]^m \left[(\id-P_4)\HdL\right]^{p+1}\right]=0\,,\\
    & \mathcal{I}^u_{nmp}=\im \tr\left[P_4 \HuL\left[(\id-P_4)\HuL\right]^n\left[(\id-P_4)\HdL\right]^m \left[(\id-P_4)\HuL\right]^{p+1} \right]=0\,,\\
    & \mathcal{I}_{nmpq}=\im \tr\left[P_4 \HuL\left[(\id-P_4)\HuL\right]^n\left[(\id-P_4)\HdL\right]^m \left[(\id-P_4)\HuL\right]^p \left[(\id-P_4)\HdL\right]^{q+1} \right]=0\,,\\
    & \mathcal{I}'_{nmpq}=\im \tr\left[\left[(\id-P_4)\HuL\right]^2 \left[(\id-P_4)\HdL\right]^2 (\id-P_4)\HuL (\id-P_4)\HdL \right]=0\,,\\
\end{aligned}
\end{equation}
where $P_4=\id-P_1-P_2-P_3$ and $\id$ is the $4\times 4$ identity matrix.

To ultimately achieve a direct connection between these conditions and the \acp{WBI} studied in the main text, we must now remove the many $P_4$ that feature in these expressions, for which a useful property for $\n=1$ is
\begin{equation}\label{apptr_P4}
\tr\left( P_4 A P_4 B \right)= \tr\left( P_4 A \right) \tr\left( P_4 B \right) \, .
\end{equation}
Using this property, one can show that the form of $\mathcal{I}_{nm}$ in~\cref{eq:CPI_WBI} can be expanded into a convoluted combination of traces that are exclusively of three types:
\begin{equation} \label{eq:app_traces}
  \tr\left(P_4\HuL^a\right)\,, \quad \quad \tr\left(P_4\HdL^a\right)\,, \quad \quad \tr\left(P_4\HuL^a\HdL^b\right)\,.
\end{equation}
Thus, since only the last type of invariant is in general CP-odd, the vanishing of $\mathcal{I}_{nm}$ can be achieved if
\begin{equation} \label{eq:app_3block}
    \im\tr\left[P_4 \HuL^a \HdL^b\right]=0\,,
\end{equation}
for all $1\leq a\leq n+1$ and $1\leq b\leq m+1$.
Likewise, one can show that the expansion of $\mathcal{I}^d_{nmp}$ in~\cref{eq:CPI_WBI} depends solely on four types of traces: the ones in~\cref{eq:app_traces} plus
$\tr\left(P_4 \HdL^a\HuL^b \HdL^c\right)$, meaning that
\begin{equation} \label{eq:app_4blockd}
    \im\tr\left[P_4 \HdL^a \HuL^b \HdL^c\right]=0\,,
\end{equation}
for all $1\leq a\leq n+1$, $1\leq b\leq m$ and $1\leq c\leq p+1$, along with the condition of~\cref{eq:app_3block}, leads to the vanishing of $\mathcal{I}^d_{nmp}$.

In the same way, one can show that the vanishing of $\mathcal{I}^u_{nmp}$ can be achieved
if the conditions of~\cref{eq:app_3block} and
\begin{equation} \label{eq:app_4blocku}
 \im\tr\left[P_4 \HuL^a \HdL^b \HuL^c\right]=0
\end{equation}
are both true,
while the vanishing of $\mathcal{I}_{nmpq}$ is achieved if the conditions of~\cref{eq:app_3block,eq:app_4blockd,eq:app_4blocku} and
\begin{equation} \label{eq:app_5block}
   \im\tr\left[P_4 \HuL^a \HdL^b \HuL^c \HdL^d\right]=0
\end{equation}
are fulfilled. 
Finally, the last invariant vanishes if all conditions in~\cref{eq:app_3block,eq:app_4blockd,eq:app_4blocku,eq:app_5block} and
\begin{equation} \label{eq:app_5blocku}
   \im\tr\left[\HuL^2 \HdL^2 \HuL \HdL\right]=0
\end{equation}
are verified.
Hence, using the fact that $\C=\mB^2 P_4$ in the minimal \ac{WB}, 
and making use of the Cayley-Hamilton theorem to reduce the number of invariants to a finite set,
we can conclude that the conditions
\begin{equation} \label{eq:app_WBI_C-H}
\begin{aligned}
    & \im \tr\left[ \HuL^n \HdL^m \C\right]=0\,, \quad (n,m=1,2,3)\,,\\
    & \im \tr\left[\HdL^n \HuL^m \HdL^p \C\right]=0\,, \quad (n>p, \ n,p=1,2,3, \ m=1,2)\,,\\
    & \im \tr\left[\HuL^n \HdL^m \HuL^p \C\right]=0\,, \quad (n>p, \ n,p=1,2,3, \ m=1,2)\,,\\
    & \im \tr\left[\HuL^n \HdL^m \HuL^p \HdL^q \C\right]=0\,, \quad (n>p, \ n,q=1,2,3, \ m,p=1,2)\,,\\
    & \im \tr\left[\HuL^2 \HdL^2 \HuL \HdL\right]=0
\end{aligned}
\end{equation}
necessarily imply \ac{CPI} in a one-doublet model. These \acp{WBI} are the ones studied in~\cref{sec:WBI_one_doublet}. 

The set in~\cref{eq:app_WBI_C-H} still amounts to a very large number of conditions, and since the phase content of a one-\ac{VLQ}-doublet model corresponds to at most 6 phases, we expect the set of \ac{WBI} conditions that lead to \ac{CPI} to be much smaller. In fact, as we have shown in~\cref{sec:stepladderwb}, when no $\hi_\alpha$ or $\hi_i$ coupling vanishes and there exist 6 physical phases, we only need 9 conditions to ensure \ac{CPI}. In what follows we will analyze all of the remaining possible cases described in~\cref{sec:CPconservation} and presented in~\Cref{tab:cp-odd-inv} and, for each one of them, identify a reduced set of \ac{WBI} conditions that can lead to \ac{CPI}. 
A set of conditions which is also minimal (and valid also for general Yukawa matrices with mass degeneracies) could be achieved by a systematic application of the Cayley-Hamilton theorem, together with some Group Theory techniques as recently done in Ref.~\cite{deLima:2024vrn} for the scenario with one vector-like quark singlet case. We can anticipate that this set is about $\lesssim 10$ times larger for the doublet case.

\subsection{Case-by-case analysis}
Here we will demonstrate the results of~\Cref{tab:cp-odd-inv}. To that end, we make use of specific \acp{WB} for which these proofs are more straightforward. 
In particular, we start by using the minimal \ac{WB} of~\cref{sec:minimalweak}, while
at the end of this appendix we switch to the stepladder \ac{WB} of~\cref{sec:stepladderdefs} which is more convenient in scenarios where all couplings $\hi_{\alpha}$ or all $\hi_{i}$ are non-vanishing.
Let us recall that the goal is to achieve a set of \ac{WBI} conditions for \ac{CPI}, but the final result is independent of the particular choice of \ac{WB} exploited in the demonstration.

\subsubsection*{Using the minimal \acs{WB}:}

We start by using the minimal \ac{WB} of~\cref{sec:minimalweak}. Additionally, we recall that one column and one row of $\hat{V}_L$ can always be made real, together with one of the couplings $\hi_{i/\alpha}$, so that only one phase, which we denote by $\delta_0$, is contained in $\VWB$. Hence, $\VWB$ can take the form
\begin{equation}
    \VWB=\begin{pmatrix}
        \VWB^r_{ud} & \VWB^r_{us} & \VWB^r_{ub}\\
        \VWB_{cd} & \VWB_{cs} & \VWB^r_{cb}\\
        \VWB_{td} & \VWB_{ts} & \VWB^r_{tb}\\
    \end{pmatrix}
\end{equation}
in some version of the minimal \ac{WB}. Here the index $r$ identifies an entry as real. Note that the complex entries can be parameterized in the following way
\begin{equation}
    \VWB_{\alpha i}=a_{\alpha i}+b_{\alpha i}e^{i \delta_0},
\end{equation}
where $a_{\alpha i}$ and $b_{\alpha i}$ are real quantities.

\begin{itemize}
    \item \textbf{One $\hi_i= 0$ and one $\hi_\alpha =0$ :}
\end{itemize}
For the sake of illustration, we take $\hiu=\hid=0$. In this case, 4 physical phases are present, $\delta_0$ and the phases of $\hit,\his$ and $\hib$. The simplest type of invariant which is sensitive to all these phases is $\mathcal{I}_{nm}$ of~\cref{eq:app_I_nm} which takes the form
\begin{equation}\label{eq:app_zi_0_za_0}
\begin{aligned}
    \mathcal{I}_{nm}= & \ \hat{m}^{2n+1}_c\left[\hat{m}^{2m+1}_s \im\left(\VWB_{cs} \his^* \hic\right)+\hat{m}^{2m+1}_b \im\left(\VWB^r_{cb} \hib^* \hic\right)\right]\\
     + & \ \hat{m}^{2n+1}_t\left[\hat{m}^{2m+1}_s \im\left(\VWB_{ts} \his^* \hit\right)+\hat{m}^{2m+1}_b \im\left(\VWB^r_{tb} \hib^* \hit\right)\right]\,,
\end{aligned}
\end{equation}
where the 4 phases appear in 4 distinct rephasing-invariant combinations. Here, the $\hat{m}_i\equiv \vw\, \hat{y}_i $ denote the eigenvalues of the $\hat{D}_q$ matrices. 

Hence, \ac{CPI} can be achieved if we guarantee the simultaneous vanishing of four invariants of this type, such as
\begin{equation} \label{eq:app_4system}
    \mathcal{I}_{00}=\mathcal{I}_{10}=\mathcal{I}_{01}=\mathcal{I}_{11}=0\,,
\end{equation}
provided that these conditions form a non-singular system.

Now, note that
\begin{equation}
\mathcal{I}_{00}=\frac{1}{\mB^2}\im\tr\left[\HuL\HdL \C\right]
\end{equation}
vanishes when
\begin{equation}
    \im\tr\left[\HuL\HdL \C\right]=0\,.
\end{equation}
Assuming that is the case and using the property in~\cref{apptr_P4}, we then have
\begin{equation}
\begin{aligned}
        \mathcal{I}_{10} &=\frac{1}{\mB^2}\im\tr\left[\HuL^2 \HdL \C\right]\,,\\
        \mathcal{I}_{01} &=\frac{1}{\mB^2}\im\tr\left[\HuL \HdL^2 \C\right]\,,
\end{aligned}
\end{equation}
and their vanishing can be achieved via
\begin{equation}
    \im\tr\left[\HuL^2 \HdL \C\right]=\im\tr\left[\HuL \HdL^2 \C\right]=0\,.
\end{equation}
Then, in a similar way, this leads to
\begin{equation}
    \mathcal{I}_{11}=\frac{1}{\mB^2}\im\tr\left[\HuL^2\HdL^2 \C\right]\,.
\end{equation}
Therefore, the conditions in~\cref{eq:app_4system} can be achieved if
\begin{equation} \label{eq:app_4x3block}
    \im\tr\left[\HuL^{n+1} \HdL^{m+1} \C\right]=0\,,
\end{equation}
with $(n,m)=(0,0),(1,0),(0,1),(1,1)$. 

In turn, the system of linear equations in~\cref{eq:app_4system} implies
\begin{equation}
    \im\left(\VWB_{cs}\his^* \hic\right)=\im\left(\VWB^r_{cb} \hib^* \hic\right)=\im\left(\VWB_{ts}\his^* \hit\right)=\im\left(\VWB^r_{tb} \hib^* \hit\right)=0\,,
\end{equation}
assuming a non-singular system, i.e.
\begin{equation}
    \hat{m}_c^2 \hat{m}_t^2 \hat{m}_s^2 \hat{m}_b^2 (\hat{m}^2_b-\hat{m}^2_s)^2(\hat{m}^2_t-\hat{m}^2_c)^2\neq 0\,.
\end{equation}
Thus, in a non-degenerate scenario, we can eliminate four distinct rephasing-invariant combinations of the four physical phases and, therefore achieve \ac{CPI} if we require the vanishing of the four 4-block \ac{WBI} of the form of~\cref{eq:app_4x3block}.

From the form of~\cref{eq:app_zi_0_za_0} it may seem that the vanishing of some entries of $\hat{V
}$ could result in a great simplification of these invariants in a way that would impede any direct or useful connection between the physical phases and the set of \ac{WBI} in~\cref{eq:app_4system}. Still, we checked that in every case where any number of $\hat{V
}$ entries vanishes, there is a reduction in the number of physical phases such that, even if some (or all) of the invariants in~\cref{eq:app_4system} are trivially vanishing and therefore useless, imposing that the remaining subset of them vanishes will always guarantee the vanishing of all surviving phases and thus \ac{CPI}. We also checked that analogous considerations are valid for all the cases we shall consider next.

\begin{itemize}
     \item \textbf{Two $\hi_{\alpha(i)}= 0$ in one sector and one $\hi_{(i)\alpha}= 0$ in the other sector :}
\end{itemize}
We assume the case with $\hiu=\hid=\his=0$, where there are 3 phases present: $\delta_0$, the phase of $\hit$ and the phase of $\hib$. 

The simplest type of CP-odd invariants acquires the form
\begin{equation}
    \mathcal{I}_{nm} =\hat{m}^{2m+1}_b\left[ \hat{m}^{2n+1}_c\im\left(\VWB^r_{cb}\hib^* \hic\right)+\hat{m}^{2n+1}_t\im\left(\VWB^r_{tb} \hib^* \hit \right)\right]\,,
\end{equation}
which is sensitive to the phases of $\hit$ and $\hib$. The vanishing of all invariants of this type only requires
\begin{equation} \label{eq:app_2_3blocks}
     \im\tr \left[\HuL \HdL \C\right]=\im\tr \left[\HuL^2 \HdL \C\right]=0\,,
\end{equation}
since it would imply $\im \hit=\im \hib=0$, provided that the corresponding system of linear equations is non-singular, i.e.
\begin{equation}
     \hat{m}_t \hat{m}_c \hat{m}^2_b (\hat{m}^2_t-\hat{m}^2_c)\neq 0\,,
\end{equation}
which again amounts to having non-degeneracy.

 The next simplest and non-vanishing type of CP-odd invariant that is sensible to the remaining phase $\delta_0$ is now of the form\footnote{Here we make use of the unitarity condition $\VWB_{cs} \VWB^*_{ts}=-\VWB_{cd}\VWB^*_{td}-\VWB_{cb}\VWB^*_{tb}$ and the fact that $\VWB_{cb}$ and $\VWB_{tb}$ are real in the \ac{WB} we are using.}
\begin{equation} \label{eq:app_Iu_nmp}
\begin{aligned}
    \mathcal{I}^u_{nmp} & \ = A_{nmp} \im(\VWB_{cd} \VWB^*_{td}\hic \hit) +B_{nmp} \im(\VWB_{cs} \VWB^*_{ts}\hic \hit)\\
    & \ = A'_{nmp} \im(\VWB_{cd} \VWB^*_{td}\hic \hit) +B'_{nmp} \im(\VWB^r_{cb} \VWB^r_{tb}\hic \hit)\\
    & \ = A'_{nmp} \im(\VWB_{cd} \VWB^*_{td}\hic \hit) \\
    & \ = A'_{nmp} \, \frac{\hic \hit}{\VWB^r_{cb}\VWB^r_{tb}} \, \im(\VWB_{cd} \VWB^*_{td} \VWB^r_{cb}\VWB^r_{tb})  \,,
\end{aligned}
\end{equation}
where
\begin{equation}
    A'_{nmp}=\hat{m}^{2n+1}_c \hat{m}^{2n+1}_t \left(\hat{m}^{2(p-n)}_t-\hat{m}^{2(p-n)}_c\right)\left(\hat{m}^{2m}_d-\hat{m}^{2m}_s\right)\,,
\end{equation}
and assuming $\VWB^r_{cb},\VWB^r_{tb}\neq 0$.%
\footnote{Both these conditions are required to ensure that
the final step in~\cref{eq:app_Iu_nmp} is valid. Still, if either one of these conditions is not fulfilled, then phase $\delta_0$ is no longer an internal physical phase and can be rephased out of $\VWB$, meaning that $\mathcal{I}^u_{nmp}=0$, for all $n,m$ and $p$, and~\cref{eq:app_2_3blocks} is sufficient to achieve \ac{CPI} when there is no mass degeneracy.}

Given that at this point all $\hi_{\alpha(i)}$ couplings are real, we have $\mathcal{I}_{nm}=0$ for all $n$ and $m$, so that 
\begin{equation}
   \mathcal{I}^u_{011} =\frac{1}{\mB^2}\im\tr\left[\HuL \HdL \HuL^2 \C\right]\,,
\end{equation}
and the vanishing of this 4-block \ac{WBI}, i.e.
\begin{equation}
    \im\tr\left[\HuL\HdL \HuL^2 \C\right]=0\,,
\end{equation}
leads to $\mathcal{I}^u_{110}=0$. Assuming no degeneracy, we must have $A_{011}\neq 0$ so that the vanishing of this invariant ensures $\im\big( \VWB_{cd} \VWB^*_{td} \VWB^r_{cb}\VWB^r_{tb}\big)=0$ and similarly for all other quartets, meaning there is no \ac{CPV} induced by $ \delta_0$ in $\VWB$. 
Therefore, we need three conditions, such as
\begin{equation}
     \im \tr\left[ \HuL \HdL \C\right]=\im\tr\left[     \HuL^2 \HdL \C\right]=\im\tr\left[     \HuL \HdL \HuL^2 \C\right]=0\,,
\end{equation}
in order to achieve \ac{CPI}.

\pagebreak
\begin{itemize}
     \item \textbf{Three $\hi_{\alpha(i)}= 0$ in one sector and one $\hi_{(i)\alpha}= 0$ in the other sector :}
\end{itemize}
Here we take $\hiu=\hid=\his=\hib=0$ and only the phase $\delta_0$ and the phase of $\hit$ are physical. In this scenario we automatically have
\begin{equation} \label{eq:c.75}
    \mathcal{I}_{nm}=\mathcal{I}^d_{nmp}=\mathcal{I}_{nmpq}=0\,,
\end{equation}
while
\begin{equation} \label{eq:app_Inm}
    \mathcal{I}^u_{nmp}= A_{nmp} \im \left( \VWB^*_{cs} \VWB_{ts} \hic \hit^*\right)+B_{nmp}\im \left( \VWB^r_{cb} \VWB^r_{tb} \hic \hit^*\right)
\end{equation}
is sensitive to both physical phases and where
\begin{align}
    A_{nmp}&=\hat{m}^{2n+1}_c \hat{m}^{2n+1}_t\left(\hat{m}^{2(p-n)}_t-\hat{m}^{2(p-n)}_c\right)\left(\hat{m}^{2m}_s-\hat{m}^{2m}_d\right)\,,\\
     B_{nmp}&=\hat{m}^{2n+1}_c \hat{m}^{2n+1}_t\left(\hat{m}^{2(p-n)}_t-\hat{m}^{2(p-n)}_c\right)\left(\hat{m}^{2m}_b-\hat{m}^{2m}_d\right)\,.
\end{align}

The fact that all $\mathcal{I}_{nm}=0$ then allows us to write
\begin{equation}
     \mathcal{I}^u_{011}=\frac{1}{\mB^2}\im\tr\left[ \HuL \HdL \HuL^2 \C\right]\,,
\end{equation}
and the vanishing of this quantity can be achieved if
\begin{equation}
    \im\tr\left[ \HuL \HdL \HuL^2 \C\right]=0\,,
\end{equation}
which in turn leads to
\begin{equation}
    \mathcal{I}^u_{021}=\frac{1}{\mB^2}\im\tr\left[ \HuL \HdL^2 \HuL^2 \C\right]\,.
\end{equation}
Hence, the conditions 
$\mathcal{I}_{011}=\mathcal{I}_{021}=0$ are fulfilled by having
\begin{equation}
    \im\tr \left[\HuL \HdL \HuL^2 \C\right]=\im\tr \left[\HuL \HdL^2 \HuL^2 \C\right]=0\,.
\end{equation}
At the same time, from~\cref{eq:app_Inm}, these conditions correspond to the following linear system of equations
\begin{equation}
    \begin{pmatrix}
        A_{011} & B_{011} \\
        A_{021} & B_{021} \\
    \end{pmatrix} \begin{pmatrix}
        \im (\VWB^*_{cs}\VWB_{ts} \hic\hit^*)\\
        \im (\VWB^r_{cb}\VWB^r_{tb} \hic\hit^*)      
    \end{pmatrix}=0\,,
\end{equation}
which is non-singular if
\begin{equation}
    \hat{m}^2_c \hat{m}^2_t(\hat{m}^2_t-\hat{m}^2_c)^2(\hat{m}^2_s-\hat{m}^2_d)(\hat{m}^2_b-\hat{m}^2_d)(\hat{m}^2_b-\hat{m}^2_s)\neq 0\,,
\end{equation}
and since we assume no mass degeneracy, we conclude that the vanishing of these two invariants must imply 
\begin{equation}
    \im (\VWB^*_{cs}\VWB_{ts}\hic\hit^*)=\im (\VWB^r_{cb}\VWB^r_{tb}\hic\hit^*)=0\,,
\end{equation}
which leads to \ac{CPI}.

\begin{itemize}
     \item \textbf{Two $\hi_{\alpha(i)}= 0$ in one sector and two $\hi_{(i)\alpha}= 0$ in the other sector :}
\end{itemize}
Without loss of generality, we consider the case $\hid=\his=\hiu=\hic=0$. In this case we only have two physical phases: $\delta_0$ and the phase of $\hib$. Also one can check that
\begin{equation}
    \mathcal{I}^u_{nmp}=\mathcal{I}^d_{nmp}=0\,.
\end{equation}
As for invariants of the $\mathcal{I}_{nm}$ type one has
\begin{equation}
    \mathcal{I}_{nm} = \hat{m}^{2n+1}_t \hat{m}^{2m+1}_b \im\left( \hit\hib^* \VWB^r_{tb}\right)\,.
\end{equation}
Now, if one 3-block \ac{WBI} vanishes, then $\hib$ must be real. That is the case, for instance, if $\im\tr\left[ \HuL \HdL \C\right]=0$ because, with $\hit$ and $V^r_{33}$ being real, we have 
\begin{equation}
   \mathcal{I}_{00}=\im\tr\left[ \HuL \HdL \C\right]=0 \quad\Rightarrow\quad  \im \hib=0\,,
\end{equation}
and the only phase remaining is now $\delta_0$. 
 
 This phase is captured by invariants of the type
\begin{equation}
        \mathcal{I}_{nmpq}  = 
        \hat{m}^{2n+1}_t \hat{m}^{2q+1}_b\left(\hat{m}^{2p}_u-\hat{m}^{2p}_c\right)\left(\hat{m}^{2m}_d-\hat{m}^{2m}_s\right)\im\left(\VWB^r_{ud}\VWB^r_{ub}\VWB_{td}\hib^* \hit \right)\,\propto\,\sin\delta_0 
\end{equation}
which are proportional $\sin\delta_0$ due to now $\hib$ being real, along with $\hit$, $\hat{V}^r_{ud}$ and $\hat{V}^r_{ub}$.
At this point we have
\begin{equation}
   \mathcal{I}_{nm}=\mathcal{I}^u_{nmp}=\mathcal{I}^d_{nmp}=0\,,
\end{equation}
so that
\begin{equation}
    \mathcal{I}_{0110}= \frac{1}{\mB^2}\im\tr\left[ \HuL \HdL^{2} \HuL^{2} \HdL \C\right]\,.
\end{equation}
Thus, the vanishing of the remaining phase can be ensured if this 5-block \ac{WBI} vanishes, i.e.
\begin{equation}
   \mathcal{I}_{0110}= \frac{1}{\mB^2} \im\left[ \HuL \HdL \HuL \HdL \C\right]=0\,. 
\end{equation}
With no mass degeneracy, we then must have $\sin\delta_0=0$. 
Therefore, we can conclude that
\begin{equation}
    \im\tr\left[ \HuL \HdL \C\right]=\im\tr\left[ \HuL \HdL \HuL \HdL \C\right]=0
\end{equation}
ensures \ac{CPI}.

\pagebreak
\subsubsection*{Using the stepladder \acs{WB}:}

For the remaining three cases we drop the minimal \ac{WB} of~\cref{sec:minimalweak} and instead focus on the stepladder \ac{WB}. Recall from the discussion in~\cref{sec:stepladderdefs} that there is a direct connection between the number of $\hi_{\alpha(i)}$ couplings vanishing in the former \ac{WB} and the number of off-diagonal $r_q$ couplings vanishing in the latter. 

\begin{itemize}
    \item \textbf{One $\hi_{\alpha(i)}= 0$ :}
\end{itemize}
Consider now the case of $\hid=0$. To analyze this case, we switch to the stepladder \ac{WB} in the $r^d_4=0$ limit. Here we have 5 physical phases, which we 
include in the parameterization of
$\Tilde{V}$, making all $r_q$ couplings real,
just like we did in~\cref{sec:stepladderwb}.

Now, using~\cref{eq:J11,eq:J12,eq:P12,eq:J13,eq:J4}, we can compute seven of the CP-odd invariants $\im J_{ij}$, while the two remaining ones are identically zero ($J_{13},J_{33}\propto r^d_4=0$). In this way we can obtain the imaginary parts of all entries of $\tilde{V}$, except that of $\tilde{V}_{td}$ ($\tilde{V}_{ud}$ can be taken as real). However, using the fact that $\Tilde{V}$ is unitary, one can always extract the value of $\im \tilde{V}_{td}$ from simply having computed all the seven non-zero $\im J_{ij}$.

In fact, even in an extreme scenario where all seven \acp{WBI} vanish because $\tilde{V}$ has a form such as 
\begin{equation}
    \tilde{V}=\begin{pmatrix}
        0 & \tilde{V}^r_{us} & \tilde{V}^r_{ub}\\
        0 & \tilde{V}^r_{cs} & \tilde{V}^r_{cb}\\
        e^{i \alpha} & 0 & 0\\
    \end{pmatrix}\,,
\end{equation}
making it is impossible to determine the complex phase $\alpha$ via unitarity relations, there is still no CP violation due to $r^d_4=0$, since the phase $\alpha$ can be eliminated by rephasing the \ac{RH} down-type quark fields. 

Hence, \ac{CPI} is achieved if seven 3-block \acp{WBI} vanish. For instance,
\begin{equation}
    \im \tr\left[\HuL^n \HdL^m \C\right]=0\,,
\end{equation}
with $(n,m)=(1,1),(1,2),(2,1),(2,2),(2,3),(3,1),(3,2)$, ensures \ac{CPI}.

\begin{itemize}
     \item \textbf{Two $\hi_{\alpha(i)}= 0$ in the same sector :}
\end{itemize}
Consider the limit of $r^d_2=0$, which is analogous to having two $\hi_i=0$ in the down sector. In this scenario we can perform a rephasing of the quark fields that leads to the following form for the matrix $\Tilde{V}$:
\begin{equation}
    \Tilde{V}=\begin{pmatrix}
        \Tilde{V}^r_{ud} & \Tilde{V}^r_{us} & \Tilde{V}_{ub}\\
        \Tilde{V}_{cd} & \Tilde{V}_{cs} & \Tilde{V}_{cb}\\
        \Tilde{V}_{td} & \Tilde{V}_{ts} & \Tilde{V}_{tb}\\
    \end{pmatrix}\,,
\end{equation}
where the $r$-label identifies real entries. Additionally, we can identify $r^d_5=m_d$ and $r^d_3=m_s$.

In that case we can compute three of the CP-odd invariants in~\cref{eq:J11,eq:J12,eq:P12,eq:J13,eq:J4}, $\Js{11}$, $\Js{21}$ and $\Js{31}$, while the rest are trivially zero in this limit. These three will vanish if
\begin{equation} \label{eq:app_3blocks_SWB}
\im\tr\left[\HuL \HdL \C\right]=\im\tr\left[\HuL^2\HdL \C\right]=\im\tr\left[\HuL^3\HdL \C\right]=0\,,
\end{equation}
implying $\im\Tilde{V}_{\alpha b}=0$, for $\alpha=u,c,t$. At this point we have
\begin{equation}
    \Tilde{V}=\begin{pmatrix}
        \Tilde{V}^r_{ud} & \Tilde{V}^r_{us} & \Tilde{V}^r_{ub}\\
        \Tilde{V}_{cd} & \Tilde{V}_{cs} & \Tilde{V}^r_{cb}\\
        \Tilde{V}_{td} & \Tilde{V}_{ts} & \Tilde{V}^r_{tb}\\
    \end{pmatrix}\,,
\end{equation}
and the presence of \ac{CPV} hinges on having 
\begin{equation} \label{eq:app_CPV}
\im\left(\Tilde{V}_{cd}\Tilde{V}_{bs}\Tilde{V}^*_{cs}\Tilde{V}^*_{td}\right)=-\im\left(\Tilde{V}_{cd}\Tilde{V}^*_{td}\Tilde{V}^r_{cb}\Tilde{V}^r_{tb}\right)=-\im\left(\Tilde{V}_{cd}\Tilde{V}^*_{td}\right)\Tilde{V}^r_{cb}\Tilde{V}^r_{tb}\neq 0\,.    
\end{equation}
Assuming the conditions in~\cref{eq:app_3blocks_SWB} are verified, we then have
\begin{equation}
   \im\tr\left[\HuL\HdL\HuL^2 \C\right]=\left(r^u_0r^u_1\right)^2 r^u_2r^u_3\left(m^2_d-m^2_s\right)\im\left(\Tilde{V}_{td}\Tilde{V}^*_{cd}\right)\,.
\end{equation}
If there is no mass degeneracy, then having
\begin{equation}
    \im\tr\left[\HuL\HdL\HuL^2 \C\right]=0\,,
\end{equation}
leads to $\im\left(\Tilde{V}_{td}\Tilde{V}^*_{cd}\right)=0$ which from~\mbox{\cref{eq:app_CPV}} implies \ac{CPI}. 
Hence \ac{CPI} is achieved if
\begin{equation}
\im\tr\left[\HuL \HdL \C\right]=\im\tr\left[\HuL^2\HdL \C\right]=\im\tr\left[\HuL^3\HdL \C\right]=\im\tr\left[\HuL\HdL\HuL^2 \C\right]=0\,.
\end{equation}

\begin{itemize}
    \item \textbf{Three $\hi_{\alpha(i)}= 0$ in one sector :}
\end{itemize}
Now we choose to work the limit of $r^d_0=0$, which is analogous to having all down-type couplings $\hi_i=0$. 
Here we are interested in computing 4-block CP-odd \acp{WBI}, i.e.
\begin{equation}
    J_{nmp}\equiv\frac{1}{\mB^{2(1+n+m+p)}}\im\tr\left[\HuL^{n} \HdL^{m} \HuL^{p} \C\right]\,,
\end{equation}
with $n\neq p$, as all other types vanish. Note that in this limit we will have
\begin{equation}
    \HdL=\begin{pmatrix}
    \VWT & \\
    & 1
\end{pmatrix}\diag\left(m_d,m_s,m_b,
    \MB    \right)\begin{pmatrix}
    \VWT^\dagger & \\
    & 1
\end{pmatrix}\,,
\end{equation}
and in this scenario we have $\MB=\mB$.

Let us now consider the simplest invariants, given by
\begin{equation}
    J_{1m2}=\frac{\left(r^u_0 r^u_1\right)^2 r^u_2 r^u_3}{\mB^{2m}}\left\{\left(m^{2m}_{s}-m^{2m}_{d}\right)\im\left(\tilde{V}_{ts}\tilde{V}^*_{cs}\right)+\left(m^{2m}_{b}-m^{2m}_{d}\right)\im\left(\tilde{V}_{tb}\tilde{V}^*_{cb}\right)\right\}\,.
\end{equation}
If two of these vanish, let us say $J_{112}=J_{122}=0$, we will have the following system
\begin{equation}
    \begin{pmatrix}
        m^{2}_{s}-m^{2}_{d} & m^{2}_{b}-m^{2}_{d} \\
        m^{4}_{s}-m^{4}_{d} & m^{4}_{b}-m^{4}_{d}\\
    \end{pmatrix}\begin{pmatrix}
        \im\left(\tilde{V}_{ts}\tilde{V}^*_{cs}\right)\\
        \im\left(\tilde{V}_{tb}\tilde{V}^*_{cb}\right)\\
    \end{pmatrix}=0\,,
\end{equation}
which is non-singular for non-degenerate masses, in which case we must have
\begin{equation}
    \im\left(\tilde{V}_{ti}\tilde{V}^*_{ci}\right)=0
\end{equation}
and $J_{1m2}=0$ for all values of $m$.

Now consider the next simplest \ac{WBI} of the form
\begin{equation}
    J_{1m3}=\frac{\left(r^u_0r^u_1\right)^2 r^u_2r^u_3 r^u_4r^u_5}{\mB^{2m}}\left\{\left(m^{2m}_{s}-m^{2m}_{d}\right)\im\left(\tilde{V}_{ts}\tilde{V}^*_{us}\right)+\left(m^{2m}_{b}-m^{2m}_{d}\right)\im\left(\tilde{V}_{tb}\tilde{V}^*_{ub}\right)\right\}\,,
\end{equation}
where we assume not only that all $J_{1m2}$ computed above vanish, but also that no mass degeneracy is present. Thus, if for instance $J_{113}=J_{123}=0$, then we must have
\begin{equation}
\im\left(\tilde{V}_{ti}\tilde{V}^*_{ui}\right)=0\,.
\end{equation}
Finally, we may consider
\begin{equation}
    J_{2m3}=\frac{\left(r^u_0r^u_1 r^u_2r^u_3\right)^2 r^u_4r^u_5}{\mB^{2m}}\left\{\left(m^{2m}_{s}-m^{2m}_{d}\right)\im\left(\tilde{V}_{cs}\tilde{V}^*_{us}\right)+\left(m^{2m}_{b}-m^{2m}_{d}\right)\im\left(\tilde{V}_{cb}\tilde{V}^*_{ub}\right)\right\}\,,
\end{equation}
and having for instance $J_{213}=J_{223}=0$ will imply
\begin{equation}
    \im\left(\tilde{V}_{ci}\tilde{V}^*_{ui}\right) = 0\,.
\end{equation}

At this point, it is safe to say that all quartets vanish and there is no internal phase in $\VWT$. Any other phases present in $\VWT$ have to be common to all non-zero elements in a given column of this matrix, so that, given that $\Md$ is diagonal, they can be factored out via a rephasing of the \ac{RH} down-type quark fields.  
Hence, the vanishing of six 4-block \acp{WBI} is sufficient (although perhaps not minimal) to achieve \ac{CPI}. 

 As we did for the minimal \ac{WB} and the vanishing of $\hat{V}$ entries, we checked that the vanishing of any number of $\Tilde{V}$ entries does not invalidate these results. In those cases the set of invariants needed for CPI is just a subset of the set we arrived at in this appendix.

\clearpage

\newpage 

\section{Integrating out the \acsp{VLQ}}
\label{sec:EFT}

\subsection{\acs{SMEFT} description}
\label{sec:SMEFT}

Below the \ac{VLQ} mass thresholds, physical observables can be obtained from an \acf{EFT} that does not include the \ac{VLQ} fields as explicit degrees of freedom. The footprints of these fields, as well as those from any other heavy \ac{BSM} particle, are fully captured by $c$-numbers, known as Wilson coefficients, which are suppressed by powers of the heavy masses. Under general assumptions, this description is provided by the \ac{SMEFT}, which extends the \ac{SM} Lagrangian to incorporate operators of dimension greater than four ($D>4$):
\begin{equation}
\mathcal{L}_{\text{SMEFT}}\,=\,\mathcal{L}_{\text{SM}}+\sum_{i,\,D>4} C_{i,D}\,\mathcal{O}_{i,D} \, , 
\end{equation}
where $\mathcal{O}_{i,D}$ are operators of mass dimension $D$ constructed from \ac{SM} fields and $C_{i,D}$ are the corresponding Wilson coefficients. Their contributions to the amplitudes are suppressed by powers of the heavy masses, $1/\MBS^{D-4}$, and thus, by truncating at $D=6$, one can obtain the leading $\mathcal{O}(1/\MBS^2)$ corrections to the amplitudes. The resulting amplitudes in terms of the Wilson coefficients $C_{i,D}$ are identical for any heavy extension of the \ac{SM}, and one can consistently compute them at any loop order. Provided that these amplitudes and that the corresponding observables are known at a given order in terms of the Wilson coefficients, 
one needs only to match them to the parameters of the \ac{UV} model at the 
required level of approximation.
Apart from providing a clean comparison of the phenomenological consequences of different \ac{UV} models, 
this can simplify the low-energy phenomenology (see~\cref{sec:pheno}). 

The tree-level \ac{SMEFT} Wilson coefficients for the different \ac{UV} completions of the \ac{SM}, including \ac{VLQ} doublets, can be found for example in Ref.~\cite{deBlas:2017xtg}.%
\footnote{While for the \ac{EFT} applications of this work tree-level results suffice, it is worth mentioning that some partial results regarding the one-loop matching corrections can be found in the literature, for example in Refs.~\cite{Crivellin:2022fdf,Crivellin:2022rhw}. With the fast development of automated tools~\cite{Aebischer:2023nnv}, such as \texttt{MatchMakerEFT}~\cite{Carmona:2021xtq} or 
\texttt{Matchete}~\cite{Fuentes-Martin:2022jrf}, finding the \ac{SMEFT} expressions in the same (Warsaw) basis and implementing them consistently all the way to the observable level will soon become straightforward.}
Let us thus start from the Lagrangian in the form of~\cref{eq:L1},%
\footnote{One could write it in the more general form of~\cref{eq:LY1}
by simply undoing the set of unitary transformations that leads to this form.}
\begin{equation} \label{eq:LY1rest}
\begin{aligned}
-\LY \,=\,\,
 & \left(\yu\right)_{ij} \, \overline{q^0_{L}}_i \, \tilde{\Phi} \, u^0_{Rj} 
 \,+\, \left(\yd\right)_{ij} \,  \overline{q^0_{L}}_i \, \Phi \, d^0_{Rj} 
 \,+\, \text{h.c.}
\\[2mm]
\,+\, &\left(\yU\right)_{\alpha j} \, \overline{Q^0_{L}}_\alpha \, \tilde{\Phi} \, u^0_{Rj}
\,+\, \left(\yD\right)_{\alpha j} \, \overline{Q^0_{L}}_\alpha \,\Phi\,  d^0_{Rj} \,+\, \text{h.c.}
\\[2mm]
\,+\, &\,\left(\mBd\right)_{\alpha\beta} \,\, \overline{Q^0_{L}}_\alpha \,Q^0_{R\beta} \,+\, \text{h.c.}
\,.\,
\end{aligned}
\end{equation}
As shown in~\cref{sec:minimalweak}, a minimal parameterization is obtained taking $\yu$ diagonal and $\yd$ as the product of a \ac{CKM}-like unitary matrix times a diagonal matrix, plus setting to zero $\n$ phases from $\yD$ (e.g.~in its first column), thus avoiding redundancies that would translate into unphysical flat directions in the different fits. Finding the mapping of any weak basis to this minimal one was done in~\cref{sec:mapping}, by 
making use of \aclp{WBI}.
One can thus immediately obtain the \ac{EFT} parameters in terms of any set of values for the Yukawa matrices of the original \ac{UV} \ac{VLQ} completion.

After integrating them out, the low-energy tree-level effect of $\n$ \ac{VLQ} doublets, up to $D=6$, is fully encoded in
\begin{equation} \label{eq:smeft}
\begin{aligned}
    \mathcal{L}_{\text{SMEFT}}&\,\supset\, (C_{\Phi u})_{ij}\,(\Phi^{\dagger}i\overset\leftrightarrow{D}_{\mu}\Phi)\,\overline{u^0_{R}}_i\gamma^{\mu}u_{Rj}^0 \,+\,(C_{\Phi d})_{ij}\,(\Phi^{\dagger}i\overset\leftrightarrow{D}_{\mu}\Phi)\,\overline{d^0_{R}}_i\gamma^{\mu}d_{Rj}^0\\[2mm]
&\quad+[(C_{\Phi ud})_{ij}\,(\tilde{\Phi}^{\dagger}iD_{\mu}\Phi)\,\overline{u^0_{R}}_i\gamma^{\mu}d^0_{Rj}+\text{h.c.}] \\[2mm] 
&\quad+[(C_{ u \Phi})_{ij}\,(\Phi^{\dagger}\Phi)\,\overline{q^0_{L}}_i
\tilde{\Phi}\, u^0_{Rj}
+(C_{ d \Phi})_{ij}\,(\Phi^{\dagger}\Phi)\,\overline{q^0_{L}}_i \Phi\, d^0_{Rj}+\text{h.c.}] \,, 
\end{aligned}
\end{equation}
with
\begin{equation}
\begin{aligned}
C_{\Phi u} &= - \frac{1}{2\vw^2}\,  \yU^{\dagger} \frac{\vw^2}{\mBd^{2}} \yU \,,  \quad
C_{\Phi d} =\frac{1}{2\vw^2} \,  \yD^{\dagger} \frac{\vw^2}{\mBd^{2}} \yD \,,
\\ 
C_{\Phi ud}&= \,  \yU^{\dagger} \frac{1}{\mBd^{2}} \yD \, ,
 \\ 
C_{u \Phi} &=\frac{1}{2\vw^2}\yu \,  \yU^{\dagger} \frac{\vw^2}{\mBd^{2}} \yU \,, \quad
C_{d \Phi}=\frac{1}{2\vw^2}\yd \,  \yD^{\dagger} \frac{\vw^2}{\mBd^{2}} \yD
 \,.
\end{aligned}
\end{equation}

Notice how, regardless of the number of doublets one has, the full low-energy effect at tree level and of $\mathcal{O}(1/\mB^2)$, where $\mB = \MBS$, is encoded into three $3\times 3$ matrices, two of which are Hermitian, dramatically restricting the number of linear combinations of  physical parameters entering this regime.
Additionally, if the number of \ac{VLQ} doublets is small, one finds simple relations between the parameters in the charged-current sector and the ones in the neutral-current one.

\subsection{Electroweak symmetry breaking}
\label{sec:ESWB}

After \ac{EWSB}, the pure-\ac{VEV} terms in~\cref{eq:smeft} modify both the effective quark gauge currents and the mass matrices, 
\begin{equation}
    \mathcal{L} \,\supset\,
   \left[- \frac{g}{\sqrt{2}} \,W_\mu^+ J^{\mu}_W
+\text{h.c.}\right]- \left[\sum_{q=u,d} \overline{q^0_L} \,M^q_{\text{eff}} \,q_R^0  +\text{h.c.}\right]- \frac{g}{2c_W} \,Z_\mu J^{\mu}_Z \,.
\end{equation}
The \ac{SM} Higgs-fermion interactions are also modified,
\begin{equation}
\mathcal{L}\supset -\frac{\h}{\sqrt{2}}\sum_{q=u,d} \overline{q^0_L} \,Y^q_{\text{eff}}\, q^0_R \, .  
\end{equation}
One has%
\footnote{Note that in this work we are defining the mass (and Yukawa) terms using a left-right convention, i.e.~$\mathcal{L} \sim-\overline{\psi}_{L}\mathcal{M}\psi_{R}$, while in \ac{SMEFT} works (see e.g.~\cite{Jenkins:2017jig}) they are typically defined using a right-left convention, $\mathcal{L}\sim -\overline{\psi}_{R}\mathcal{M}\psi_{L}$. Notice also the different \ac{VEV} normalization.}
\begin{align} \label{eq:massmod}
M^q_{\text{eff}}&\,=\,\vw\, \yq \left[ \id-\frac{1}{2} \Fqh \right] \, ,\\*[2mm]
\label{eq:higgsfmod}
Y^{q}_{\text{eff}}&\,=\,\yq\left[ \id-\frac{3}{2} \Fqh \right]=\frac{M^q_{\text{eff}}}{\vw}\left[ \id- \Fqh\right] \, ,
\end{align}
and
\begin{equation} \label{eq:currentsEFT}
\begin{aligned}
J_{W}^{\mu}&\,=\, \overline{\nu}_{L}\gamma^{\mu} e_{L}+\overline{u^0_L}\gamma^{\mu} d^0_{L}+ \, \overline{u_R^0}\, \gamma^{\mu}  \VRh \,d^0_{R} \, ,\\[2mm]
J_{Z}^{\mu}&\,=\, J_{Z,\text{SM}}^{\mu}+\,\overline{u^0_{R}}\,  \gamma^{\mu} \Fuh\, u^0_{R}- \,\overline{d_{R}^0}\,  \gamma^{\mu} \Fdh \, d_{R}^0 \, ,
\end{aligned}    
\end{equation}
where
\begin{equation}
J^{\mu}_{Z,\text{SM}}\,\supset\, \overline{u^0_L}\gamma^{\mu} u^0_{L}-\overline{d^0_L}\gamma^{\mu} d^0_{L}-2s_W^2 \left(\frac{2}{3}\,\overline{u^0}\gamma^{\mu}u^0-\frac{1}{3}\,\overline{d^0}\gamma^{\mu}d^0\right)\,,
\end{equation}
with $u^0=u^0_{L}+u^0_{R}$ and $d^0=d^0_{L}+d^0_{R}$. 
We are omitting the well-known leptonic part of $J^{\mu}_{Z,\text{SM}}$, which  is not used in this work.

Without loss of generality, from now on we can work in the weak basis in which
$\yu=\Ydu$ is diagonal and $\yd=\VWB\, \Ydd$, i.e., a \ac{CKM}-like unitary matrix (free from external phases) times a diagonal matrix, so that $\yQ\to \yqp$ (corresponding to the trivial generalization of~\cref{eq:z} for $\n$ doublets). 
In~\cref{eq:massmod,eq:higgsfmod,eq:currentsEFT} we have used
\begin{equation} \label{eq:VR-F-eff}
\Fuh \,\equiv\, \yup^{\dagger} \frac{\vw^2}{\mBd^{2}} \yup \, , 
\qquad  \Fdh \,\equiv\, \ydb^{\dagger} \frac{\vw^2}{\mBd^{2}} \ydb \, ,
\qquad \VRh \,\equiv\, \yup^{\dagger} \frac{\vw^2}{\mBd^{2}} \ydb \, ,
\end{equation}
generalizing the definitions introduced in~\cref{eq:reph-inv-eff,foot:hatdefs} for the $\n=1$ case.

The resulting mass matrices are not proportional to $\yq$ and thus do not inherit the same form. Nevertheless we can perform the infinitesimal unitary field redefinitions
\begin{equation}
\begin{split}
    u^0_L &\,\rightarrow\, (\id+i\,\delta L_q)\,u^0_L \, ,\\
    u^0_R &\,\rightarrow\, (\id+i\,\delta R_u)\,u^0_R \, ,
\end{split}
\qquad
\begin{split}
    d^0_L &\,\rightarrow\, (\id+i\,\delta L_q)\,d^0_L \, ,\\
 d^0_R &\,\rightarrow\, (\id+i\,\delta R_d)\,d^0_R  \, ,
\end{split}
\end{equation}
where the infinitesimal matrices are $\mathcal{O}(1/\mB^2)$. Unitarity requires them to be Hermitian. Since $D \geq 8$ corrections of $\mathcal{O}(1/\mB^4)$ are neglected, any effect of this redefinition in the terms above that already contain an $\mathcal{O}(1/\mB^2)$ insertion can be neglected.
The only effect of the field redefinition at the working order is then in~\cref{eq:massmod,eq:higgsfmod},
\begin{align} \label{eq:massredefu}
M^u_{\text{eff}}&\,=\,{\vw}\,(\id-i\,\delta L_q)\,\Ydu\left[\id-\frac{1}{2} \Fuh \right] (\id+i\,\delta R_u) \, ,\\*[2mm] \label{eq:massredefd}
M^d_{\text{eff}}&\,=\,{\vw}\,(\id-i\,\delta L_q)\,\VWB \,\Ydd\left[\id-\frac{1}{2} \Fdh \right] (\id+i\,\delta R_d) \, .
\end{align}
Taking appropriate unitary matrices, one can always find a basis where
$M^u_{\text{eff}}=D^u \equiv \diag(m_u,m_c,m_t)$ and $M^d_{\text{eff}}=\VLh D^d \equiv\VLh \diag(m_d,m_s,m_b)$,   
up to (but not including) $\mathcal{O}(1/\mB^4)$ effects. 
Notice that $\VLh \neq \VWB$ in general.
Then, in the mass basis and at this order, the effective Lagrangian is described by
\begin{equation} \label{eq:eftmass}
M_{\text{eff}}^q=D^q \, ,\qquad
Y^{q}_{\text{eff}}=\frac{D^q}{\vw}\left[ \id- \Fqh \right] \, ,
\end{equation}
and
\begin{equation} \label{eq:eftckm}
\begin{aligned}
J_{W}^{\mu}&= \overline{\nu}_{L}\gamma^{\mu} e_{L}+\overline{u_L}\,\gamma^{\mu} \VLh \,d_{L}+ \, \overline{u_R}\,\gamma^{\mu}  \VRh \, d_{R} \, ,\\[2mm]
J_{Z}^{\mu}&= J_{Z,\text{SM}}^{\mu}+ \overline{u_{R}}\, \gamma^{\mu} \Fuh\, u_{R}- \overline{d_{R}}\, \gamma^{\mu} \Fdh\, d_{R} \,,
\end{aligned}
\end{equation}
consistently with the full-theory result, see e.g.~\cref{eq:vckmR,eq:Fzz}.
The matrix $\VLh$ can be parameterized as the usual \ac{CKM} matrix and thus, at order $\mathcal{O}(1/\mB^2)$, the tree-level \ac{LH} sector is left unmodified, cf.~\cref{eq:deltaVL}.

\begin{figure}[tb]
    \centering
\includegraphics[width=0.8\textwidth]{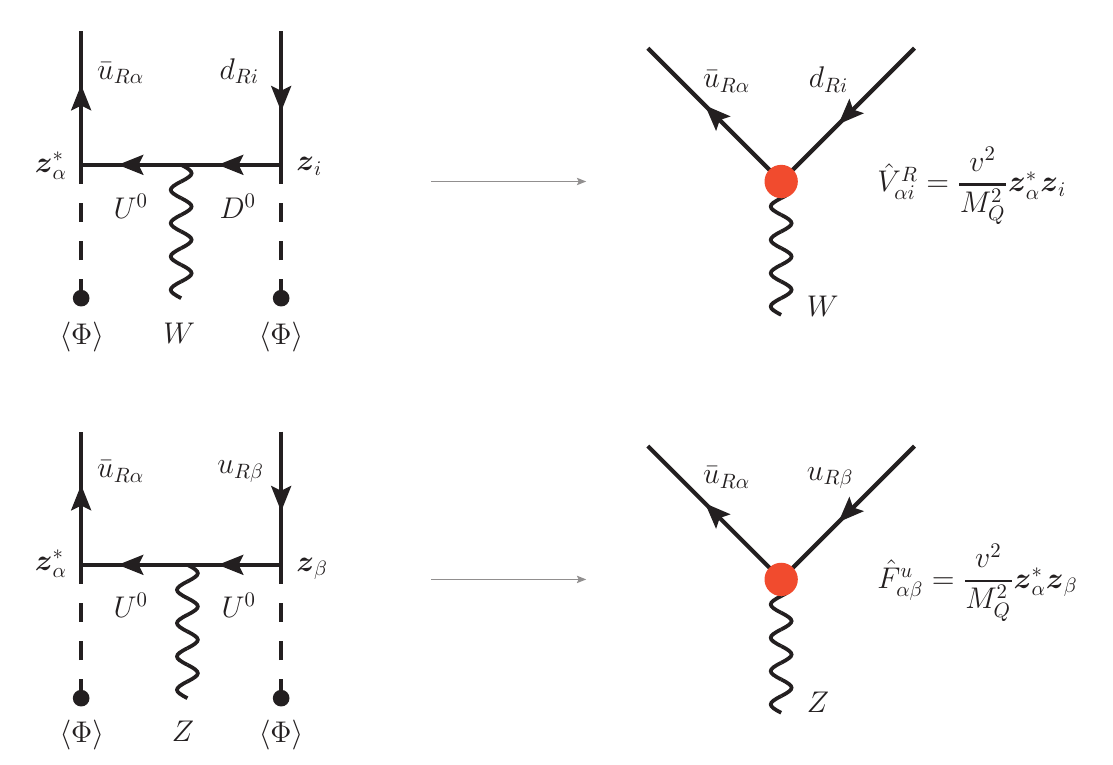}
    \caption{Graphical representation of the \ac{VLQ} diagrams generating the new effective gauge interactions. } 
    \label{fig:integrate}
\end{figure}

\vskip 2mm
A diagrammatic interpretation of the origin of the new vertices is depicted 
in~\Cref{fig:integrate}.
We relate
$\VLh$ and $D^{u,d}$ to the original Lagrangian parameters in~\cref{app:ckmeff}.

\subsection{Low-energy \acs{EFT}}
Below the \ac{EW} scale, the top quark and the electroweak bosons, as well as the Higgs, cannot be directly produced. Physical observables, both within and beyond the \ac{SM}, can, once again, be obtained from an \ac{EFT} free from the top quark, $W$ and $Z$ bosons or any other heavy \ac{BSM} particle.
In this case, the Wilson coefficients, now suppressed by $\mathcal{O}(1/\mB^2,\,1/M_{W,Z}^2,\,1/m_{t}^2,\,1/m_\h^2)$ factors, also contain pure-\ac{SM} contributions,
whose corresponding coefficients can be explicitly computed in terms of the heavy \ac{SM} masses,
\begin{equation} \label{eq:leftlag}
\mathcal{L}_{\text{LEFT}}\,=\,\mathcal{L}_{\text{QCD}+\text{QED}}\,+\,\sum_{i,\,D>4} \left(C_{i,D}^{\text{SM}}+C_{i, D}^{\text{BSM}}\right)\mathcal{O}_{i,D} \, .
\end{equation}
At tree level, the main operators induced both in the \ac{SM} and in our \ac{VLQ} scenario are the four-fermion operators generated by the interactions of two-fermion currents mediated by \ac{EW} bosons.%
\footnote{Higgs exchanges are also tree-level, but they are suppressed by $1/m_\h^2$  times  either light Yukawas or extra \ac{BSM} factors.}
In that approximation, one simply recovers a generalized version of the Fermi theory,
\begin{equation} \label{eq:leftlag2}
\mathcal{L}_{\text{LEFT}}\,\supset\,\mathcal{L}_{\text{QCD}+\text{QED}}-\frac{1}{\vw^2}J_W^\mu J_{W,\mu}^{\dagger}-\frac{1}{4\vw^2}J^\mu_{Z}J_{Z,\mu} \, .
\end{equation}
Any process mediated in the \ac{SM} by \ac{EW} currents is suppressed by $1/\vw^2$ and thus, even at very low energies, \ac{VLQ} interactions are suppressed with respect to the \ac{SM} only by $\mathcal{O}({\vw^2}/{\mB^2})$. In practice, we do not need to limit ourselves to the approximation of~\cref{eq:leftlag2} for the \ac{SM} computation, but
can use the results of \ac{SM} works at all known orders.
Instead,~\cref{eq:leftlag2} is often sufficient to compute \ac{BSM} effects,
since the neglected 
contributions are suppressed by both
loop factors and powers of $\vw^2/\mB^2$.%
\footnote{
One should however keep in mind that, when quark currents are involved, it is important to use renormalization group equations to, at least, re-sum large logarithms associated to gluonic corrections $\sum_n\alpha_s^n(\mu) \left(\ln {M_W^2}/{\mu^2}\right)^n$, since at the typical scale $\mu$ of the corresponding hadronic processes they can easily become $\mathcal{O}(1)$. The corresponding running equations can be found e.g.~in Ref.~\cite{Jenkins:2017jig}.}

\subsection{Effective \acs{CKM} angles vs.~Lagrangian parameters}
\label{app:ckmeff}

The effective matrices $D^q = \diag(\msq_1,\,\msq_2,\,\msq_3)$ and $\VLh$ in~\cref{eq:eftmass,eq:eftckm} are not directly $\Ydq$ and $\VWB$ of the minimal \ac{WB},~\cref{eq:minimal}. Let us explicitly relate them to the initial (minimal) parameters in $\Ydu$, $\Ydd$ and $\VWB$.
From~\cref{eq:massredefu,eq:massredefd} one has, 
up to $\mathcal{O}(1/\mB^4)$ corrections,
\begin{equation} \label{eq:DYF}
(D^q)_{ii}=\msq_i=\vw \,\Ydq{}_{i}\left(\id-\frac{1}{2} \Fqha_{ii}\right) \,,
\end{equation}
and
\begin{equation}
\begin{aligned}
i\,\alpha_i^{u}\equiv i\,(\delta L_q)_{ii}=i\,(\delta R_u)_{ii} \, ,
\\
i\,\alpha_i^{d}\equiv i\,(\delta L_d)_{ii}=i\,(\delta R_d)_{ii} \, ,
\end{aligned}
\end{equation}
where we have defined 
$i\, \delta L_{d} \equiv \id - \VLh^{\dagger}(\id-i\, \delta L_{q})\VWB$.
For $i\neq j$, one has instead
\begin{equation} \label{eq:VEFT}
\begin{split}
i\,(\delta L_q)_{ij}&\,\overset{i\neq j}{=}\,\frac{\msu_i\, \msu_j}{({\msu_i})^2-({\msu_j})^2}\,  \Fuha_{ij} \, , \\[2mm]
i\,(\delta R_{u})_{ij}&\,\overset{i\neq j}{=}\,\frac{1}{2} \,\frac{({\msu_i})^2+ ({\msu_j})^{2}}{({\msu_i})^2-({\msu_j})^2} \, \Fuha_{ij} \, ,
\end{split}
\qquad\,\,
\begin{split}
i\,(\delta L_d)_{ij} &\,\overset{i\neq j}{=}\,
\frac{\msd_i \, \msd_j}{({\msd_i})^2-({\msd_j})^2} \, \Fdha_{ij} \, , 
\\[2mm]
i\,(\delta R_{d})_{ij}&\,\overset{i\neq j}{=}\,\frac{1}{2}\, \frac{({\msd_i})^2+ ({\msd_j})^{2}}{({\msd_i})^2-({\msd_j})^2} \,\Fdha_{ij} \, .
\end{split}
\end{equation}

Recall that $\VLh$ and $\VWB$ have the same functional form, with $\VLh = V(\vec{\theta}_{\VLh}) \approx \CKM$ and $\VWB = V(\vec{\theta}_{\VWB})$. Thus,
$\vec{\theta}_{\VLh}=(\theta_{12},\theta_{13},\theta_{23},\delta_{13})$, while 
$\vec{\theta}_{\VWB}\equiv(\theta_{12}^{\VWB},\theta_{13}^{\VWB},\theta_{23}^{\VWB},\delta_{13}^{\VWB})$.
Expanding $V(\vec{\theta}\;\!)$ in the neighbourhood of $\vec{\theta}_{\VLh}$
and using the definition of $\delta L_d$, one finds
\begin{equation} \label{eq:expansion}
\left.\frac{\partial V}{\partial \vec{\theta}}\right|_{\vec{\theta}=\vec{\theta}_{\VLh}}\!\!\!\!\cdot (\vec{\theta}_{\VLh}-\vec{\theta}_{\VWB})\,=\,i\,\left[\VWB\,\delta L_d-\delta L_q\, \VWB \right] \, ,
\end{equation}
which can be used to extract the five unphysical external phase differences $\alpha_i^u-\alpha_j^d$ in the field redefinitions and, crucially, $(\theta_{12},\theta_{13},\theta_{23},\delta_{13})$.
The exact solution is relatively cumbersome and offers limited insight.
Replacing instead the \ac{CKM} angles and phase and quark masses by 
their experimental values 
(correct up to neglected second-order $\mathcal{O}(1/\mB^4)$ corrections) and keeping only the largest numerical coefficients, one finds
\begin{equation}
\begin{aligned}
\theta_{12} &\,\approx\, \theta_{12}^{\VWB} -0.050 \,  \re\Fdha_{12}\,, \\
\theta_{23} &\,\approx\, \theta_{23}^{\VWB} -0.018\, \re\Fdha_{23} \, , \\
\theta_{13} &\,\approx\, \theta_{13}^{\VWB} - 0.00033\, \re\Fdha_{13}-0.0015 \, \re\Fdha_{23}+0.00086\, \re\Fdha_{13} +0.0040\,\re\Fdha_{23} \, ,  \\
\delta_{13} &\,\approx\, \delta_{13}^{\VWB} +1.1 \, \re\Fdha_{23}\, .
\end{aligned}
\end{equation}
Notice how the leading prefactors in the corrections of the parameters happen to roughly scale as the values of the parameters themselves. 
Typically, the largest numerical prefactors in front of $\re\Fdha_{ij}$ for the $\theta_{ij}$ corrections scale approximately as $-{\msd_i}/{\msd_j}$, which accidentally happen to be of the same order as the $\theta_{ij}$ themselves.
More importantly, the rotation from the parameters in $\VWB$ to the ones in $\VLh$ is always suppressed by \ac{FCNC} factors. Thus, for phenomenological purposes ---  and unlike for the mass-like parameters $D^q\neq \vw\, \Ydq$, see~\cref{eq:DYF} --- the approximate identification of the angles and phase in $\vec{\theta}_{\VLh}$ with the ones of the original Lagrangian in the minimal weak basis, 
$\vec{\theta}_{\VLh} \approx \vec{\theta}_{\VWB}$, is legitimate.

\vfill
\clearpage

\vfill
\clearpage

\section{FCNCs in the two doublets scenario}
\label{app:kaon2}

\vskip 2mm
Besides being the potential resolution of the Cabibbo angle anomalies, 
the Yukawa textures in~\cref{eq:mass-2doubl} are compatible
with all other experimental constraints, in particular the stringent limit
on flavour-changing neutral currents. In fact, as it can be inferred
from~\cref{eq:Fu-2doubl,eq:Fd-2doubl}, there are no
\acp{FCNC} at tree level. There is still a contribution at loop level e.g.~to neutral kaon mixing:
\begin{equation} \label{eq:KK-2doubl}
\begin{aligned}
 &M_{12,\text{NP}}^K\frac{\lambda_u^2}{|\lambda_u|^2} \approx 
\frac{1}{3}m_{K}f_{K}^{2} \,  \frac{G_F^2m_W^2}{4\pi^2} 
\frac{\lambda_u^2}{|\lambda_u|^2}
\Bigg[
\Big(\tilde{s}^*_{45}c_{45}\hi_{1d}^*\hi_{2s}
\frac{\vw^2}{a\mB^2}\Big)^2 F\Big(\frac{M_{T^\prime}^2}{m_W^2}\Big)+
\\ & \qquad \qquad 
+\Big(\tilde{s}_{45}^*\hi_{1d}^*\frac{\vw}{\mB}
+c_{45}\hi_{1d}^*\hi_{1u}\hi_{2u}^*\frac{\vw^3}{a\mB^3}
\Big)^2\Big(c_{45}\hi_{2s} \frac{\vw}{a\mB}\Big)^2
F\Big(\frac{M_{T^{\prime\prime}}^2}{m_W^2} \Big) +
\\ & \qquad \qquad 
-2
\Big(\tilde{s}_{45}^*\hi_{1d}^*\frac{\vw}{\mB}
+c_{45}\hi_{1d}^*\hi_{1u}\hi_{2u}^*\frac{\vw^3}{a\mB^3}
\Big)\Big(\tilde{s}_{45}^*\hi_{1d}^*c_{45}^2\hi_{2s}^{2} \frac{\vw^3}{a^2\mB^3}\Big)
 F\Big(\frac{M_{T^{\prime\prime}}^2}{m_W^2},\frac{M_{T^\prime}^2}{m_W^2}  \Big)
\bigg]\,,
\end{aligned}
\end{equation}
where $F(x)$ are Inami-Lim functions~\cite{Inami:1980fz}.
Let us emphasize that the combination of mixing elements
is rephasing-invariant.
For large $x$ and $a\neq1$ approximately we have  $ F(x)\approx 0.25 x$ and 
$F(x,a^2 x)\approx 0.25 \,x\, a^2\ln(a^2)/(a^2-1)$.
In the limit $|a-1|\gg |\hi_{1(2)u}|\vw/\mB$,
$\tilde{s}_{45}$ is given by~\cref{eq:s45} and we get
\begin{equation} \label{eq:KK-2doubl2}
\begin{aligned}
 &M_{12,\text{NP}}^K\frac{\lambda_u^2}{|\lambda_u|^2} \approx 
\frac{1}{3}m_{K}f_{K}^{2} \,  \frac{G_F^2m_W^2}{4\pi^2} 
\frac{\lambda_u^2}{|\lambda_u|^2}
\Big(\hi_{1u} \hi_{1d}^* \, \hi_{2u}^*\hi_{2s}  \Big)^2 \frac{\vw^8}{a^4\mB^8} \,  \frac{\mB^2\,f(a) }{m_W^2}\,,
\end{aligned}
\end{equation}
where $f(a)$ is a function of $a$ which is smaller than $0.25$, $f(a)\approx 0.25\, a^4(1+1/a^2-4\log(a)/(a^2-1)) /(a^2-1)^2$.

We want to test this effect in the scenario explaining the Cabibbo angle anomalies.
Then, we should consider that there exists an upper limit on the vector-like doublet masses. In fact, given that $|\hiu^*\hi_{d(s)}|\vw^2/\mB^2\approx 10^{-3}$, assuming the couplings $\hi_{u,d,s}$ to be at most $1$ we have $\mB<6.2$~TeV, $a\,\mB<4.9$~TeV.
On the other hand, there is a lower bound on the mass given by the LHC limits,
$\mB,a\mB\gtrsim 1$~TeV~\cite{ATLAS:2024zlo}.
Then, the mass $\mB$ and parameter $a$ are constrained as
\begin{equation}
    1~\text{TeV} \lesssim \mB \lesssim 6.2~\text{TeV} \, , \qquad
    4.9>a > 0.16\,,
\end{equation}
with $a=0.16$ when $\mB=6.2$~TeV and $a=4.9$ when $\mB=1$~TeV. For the following computation, however, it is more convenient to dub $\mB=\min(\mBi{_1},\mBi{_2})$, $a>1$.

The new contribution should be compared to the \ac{SM} one in~\cref{eq:M12sm} and confronted with experimental constraints on $\Delta m_K$ and $\epsilon_K$.
Bounds on the \ac{NP} contribution can be estimated as $|M_{12,\text{NP}}^K|<|M_{12, \text{SM}}^K|\,\Delta_{K} $,
$ |\im M_{12,\text{NP}}^K|<|\im M_{12, \text{SM}}^K| \, \Delta_{\epsilon_{K}} $,
with $\Delta_{K}=1$ and, again using the results in Ref.~\cite{Bona:2022zhn}, $\Delta_{\epsilon_{K}}= 0.3$.
Then, by comparing the new contribution in~\cref{eq:KK-2doubl2} to the \ac{SM} one in~\cref{eq:M12sm} we have the constraints
\begin{equation} \label{eq:N2DeltaK}
\begin{aligned}
 \left|   \im\left[ \frac{\lambda_u^{*2}}{|\lambda_u|^2}\,
\left(\hi_{1u}^* \hi_{1d} \, \hi_{2u}\hi_{2s}^*  \right)^2 \right] \right|\frac{\vw^8}{a^4\mB^8}   
\frac{\mB^2\, f(a) }{m_W^2}
&\,\lesssim\, 1.6 \times 10^{-7} \, ,  \\[2mm]
 \left| \frac{\lambda_u^{*2}}{|\lambda_u|^2}\,
\left(\hi_{1u}^* \hi_{1d} \, \hi_{2u}\hi_{2s}^*  \right)^2 \right| \frac{\vw^8}{a^4\mB^8}   \frac{\mB^2\, f(a) }{m_W^2}
&\,\lesssim\, 2.3 \times 10^{-5} \, .
\end{aligned}
\end{equation}
Then, taking into account the values in~\cref{eq:cabfit}, we get
\begin{equation}
   \left| \frac{\lambda_u^{*2}}{|\lambda_u|^2}\,
\left(\hi_{1u}^* \hi_{1d} \, \hi_{2u}\hi_{2s}^* \,\frac{\vw^4}{a^2\mB^4} \right)^2 \right| \, 
  \frac{f(a)\mB^2 }{m_W^2}
\,\approx\, \left(10^{-6}\right)^2 \;  \frac{f(a)\mB^2 }{m_W^2}
\,\lesssim\, 5\times 10^{-10}\,,
\end{equation}
which is well below the experimental constraints
of~\cref{eq:N2DeltaK}.
We can inspect the scenario in which
the mixing angle can be large, for instance $\theta_{45}\approx\pi/4$, with $|1-a|\lesssim 10^{-3}$.
In this case we would have
\begin{equation}
\begin{aligned}
 &M_{12,\text{NP}}^K \frac{\lambda_u^2}{|\lambda_u|^2}\approx 
\frac{1}{3}m_{K}f_{K}^{2} \,  \frac{G_F^2m_W^2}{4\pi^2} \frac{\lambda_u^2}{|\lambda_u|^2}
\\ & \qquad \qquad \times
\Bigg[
\left(\,\tilde{s}^*_{45}c_{45}\hi_{1d}^*\hi_{2s} \frac{\vw^2}{a\mB^2}  \right)^2 
\left[ 
F\left(\frac{M_{T^{\prime\prime}}^2}{m_W^2} \right) +F\left(\frac{M_{T^\prime}^2}{m_W^2}\right)
-2\, F\left(\frac{M_{T^{\prime\prime}}^2}{m_W^2},\frac{M_{T^\prime}^2}{m_W^2}  \right)
\right]
\\ & \qquad \qquad 
+2\big(c_{45}^3\tilde{s}_{45}^*\hi_{1u}  \, \hi_{2u}^*\hi_{1d}^{*2}\hi_{2s}^{2}  \big)\frac{\vw^6}{a^3\mB^6}
\left(F\left(\frac{M_{T^{\prime\prime}}^2}{m_W^2}\right)-F\left(\frac{M_{T^{\prime\prime}}^2}{m_W^2},\frac{M_{T^\prime}^2}{m_W^2}  \right)\right)
\\ &\qquad \qquad
+\big(c_{45}^2 \hi_{1u} \hi_{1d}^* \, \hi_{2u}^*\hi_{2s}  \big)^2\frac{\vw^8}{a^4\mB^8}
F\left(\frac{M_{T^{\prime\prime}}^2}{m_W^2}\right)
\Bigg]
\,,
\end{aligned}
\end{equation}
(the first two lines would disappear in the case of mass degeneracy). Taking into account the values in~\cref{eq:cabfit} we have
\begin{equation}\label{eq:KKs45}
\begin{aligned}
\left| \frac{\lambda_u^{*2}}{|\lambda_u|^2}\,
\left(\tilde{s}_{45}c_{45}\hi_{1d}\hi_{2s}^*  \right)^2 \right| \frac{\vw^4}{a^2\mB^4}   
\frac{\mB^2\, \tilde{f}(\tilde{a}) }{m_W^2}
\,&\approx\, \frac{\left(10^{-6}\right)^2}{4} \frac{\mB^4}{\vw^4} \; \frac{\mB^2\, \tilde{f}(\tilde{a}) }{m_W^2}
\,\lesssim\, 10^{-9}\,,
\\
\left|\frac{\lambda_u^{*2}}{|\lambda_u|^2}\big(c_{45}^2\hi_{1u}^* \hi_{1d} \, \hi_{2u}\hi_{2s}^*  \big)^2\right|\frac{\vw^8}{a^4\mB^8}
\frac{M_{T^{\prime\prime}}^2}{4m_W^2}
&\approx \frac{\left(10^{-6}\right)^2}{4} \frac{M_{T^{\prime\prime}}^2}{4m_W^2}
\,\lesssim\,  10^{-9}\,,
\end{aligned}
\end{equation}
below the experimental constraints of~\cref{eq:N2DeltaK}. The first inequality follows since, for $|1-a|\lesssim 10^{-3}$, the combination of Inami-Lim functions in the first line of~\cref{eq:KKs45} can be written as 
$\tilde{f}(\tilde{a})\mB^2/m_W^2\approx 0.3 (\tilde{a}-1)^2 \mB^2/m_W^2$, 
$\tilde{a}=\max(M_{T^{\prime\prime}}/M_{T^{\prime}},\, M_{T^{\prime}}/M_{T^{\prime\prime}})$, which
gives a value $<2\times 10^{-3}$ taking into account $|\tilde{a}-1|\approx |\hi_{1(2)u}|^2\vw^2/\mB^2$ and~\cref{eq:Msplit,eq:Zdecay-constr} and $\hi_{\alpha}\lesssim 1$.

\clearpage

\bibliographystyle{JHEPwithnote}
\bibliography{bibliography}

\providecommand{\noopsort}[1]{}\providecommand{\singleletter}[1]{#1}%

\providecommand{\href}[2]{#2}\begingroup\raggedright\begin{thebibliography}{100}

\bibitem{Gursey:1975ki}
F.~Gursey, P.~Ramond and P.~Sikivie, \emph{{A Universal Gauge Theory Model
  Based on $E_6$}},
  \href{https://doi.org/10.1016/0370-2693(76)90417-2}{\emph{Phys. Lett. B}
  {\bfseries 60} (1976) 177}.

\bibitem{Achiman:1978vg}
Y.~Achiman and B.~Stech, \emph{{Quark Lepton Symmetry and Mass Scales in an E6
  Unified Gauge Model}},
  \href{https://doi.org/10.1016/0370-2693(78)90584-1}{\emph{Phys. Lett. B}
  {\bfseries 77} (1978) 389}.

\bibitem{Berezhiani:1989bd}
Z.~G. Berezhiani and G.~R. Dvali, \emph{{Possible solution of the hierarchy
  problem in supersymmetrical grand unification theories}}, {\emph{Bull.
  Lebedev Phys. Inst.} {\bfseries 5} (1989) 55}.

\bibitem{Barbieri:1994kw}
R.~Barbieri, G.~R. Dvali, A.~Strumia, Z.~Berezhiani and L.~J. Hall,
  \emph{{Flavor in supersymmetric grand unification: A Democratic approach}},
  \href{https://doi.org/10.1016/0550-3213(94)90593-2}{\emph{Nucl. Phys. B}
  {\bfseries 432} (1994) 49}
  [\href{https://arxiv.org/abs/hep-ph/9405428}{{\ttfamily hep-ph/9405428}}].

\bibitem{Berezhiani:1995dt}
Z.~Berezhiani, \emph{{SUSY SU(6) GIFT for doublet-triplet splitting and fermion
  masses}}, \href{https://doi.org/10.1016/0370-2693(95)00705-P}{\emph{Phys.
  Lett. B} {\bfseries 355} (1995) 481}
  [\href{https://arxiv.org/abs/hep-ph/9503366}{{\ttfamily hep-ph/9503366}}].

\bibitem{Randall:1999ee}
L.~Randall and R.~Sundrum, \emph{{A Large mass hierarchy from a small extra
  dimension}}, \href{https://doi.org/10.1103/PhysRevLett.83.3370}{\emph{Phys.
  Rev. Lett.} {\bfseries 83} (1999) 3370}
  [\href{https://arxiv.org/abs/hep-ph/9905221}{{\ttfamily hep-ph/9905221}}].

\bibitem{Arkani-Hamed:2001nha}
N.~Arkani-Hamed, A.~G. Cohen and H.~Georgi, \emph{{Electroweak symmetry
  breaking from dimensional deconstruction}},
  \href{https://doi.org/10.1016/S0370-2693(01)00741-9}{\emph{Phys. Lett. B}
  {\bfseries 513} (2001) 232}
  [\href{https://arxiv.org/abs/hep-ph/0105239}{{\ttfamily hep-ph/0105239}}].

\bibitem{Arkani-Hamed:2002sdy}
N.~Arkani-Hamed, A.~G. Cohen, T.~Gregoire and J.~G. Wacker,
  \emph{{Phenomenology of electroweak symmetry breaking from theory space}},
  \href{https://doi.org/10.1088/1126-6708/2002/08/020}{\emph{JHEP} {\bfseries
  08} (2002) 020} [\href{https://arxiv.org/abs/hep-ph/0202089}{{\ttfamily
  hep-ph/0202089}}].

\bibitem{Perelstein:2003wd}
M.~Perelstein, M.~E. Peskin and A.~Pierce, \emph{{Top quarks and electroweak
  symmetry breaking in little Higgs models}},
  \href{https://doi.org/10.1103/PhysRevD.69.075002}{\emph{Phys. Rev. D}
  {\bfseries 69} (2004) 075002}
  [\href{https://arxiv.org/abs/hep-ph/0310039}{{\ttfamily hep-ph/0310039}}].

\bibitem{Han:2003wu}
T.~Han, H.~E. Logan, B.~McElrath and L.-T. Wang, \emph{{Phenomenology of the
  little Higgs model}},
  \href{https://doi.org/10.1103/PhysRevD.67.095004}{\emph{Phys. Rev. D}
  {\bfseries 67} (2003) 095004}
  [\href{https://arxiv.org/abs/hep-ph/0301040}{{\ttfamily hep-ph/0301040}}].

\bibitem{Fajfer:2013wca}
S.~Fajfer, A.~Greljo, J.~F. Kamenik and I.~Mustac, \emph{{Light Higgs and
  Vector-like Quarks without Prejudice}},
  \href{https://doi.org/10.1007/JHEP07(2013)155}{\emph{JHEP} {\bfseries 07}
  (2013) 155} [\href{https://arxiv.org/abs/1304.4219}{{\ttfamily 1304.4219}}].

\bibitem{Berezhiani:1983hm}
Z.~G. Berezhiani, \emph{{The Weak Mixing Angles in Gauge Models with Horizontal
  Symmetry: A New Approach to Quark and Lepton Masses}},
  \href{https://doi.org/10.1016/0370-2693(83)90737-2}{\emph{Phys. Lett. B}
  {\bfseries 129} (1983) 99}.

\bibitem{Dimopoulos:1983rz}
S.~Dimopoulos, \emph{{Natural Generation of Fermion Masses}},
  \href{https://doi.org/10.1016/0370-2693(83)90132-6}{\emph{Phys. Lett. B}
  {\bfseries 129} (1983) 417}.

\bibitem{Berezhiani:1985in}
Z.~G. Berezhiani, \emph{{Horizontal Symmetry and Quark-Lepton Mass Spectrum:
  The $SU(5) \otimes SU(3)_H$ Model}},
  \href{https://doi.org/10.1016/0370-2693(85)90164-9}{\emph{Phys. Lett. B}
  {\bfseries 150} (1985) 177}.

\bibitem{Berezhiani:1991ds}
Z.~G. Berezhiani and R.~Rattazzi, \emph{{Universal seesaw and radiative quark
  mass hierarchy}},
  \href{https://doi.org/10.1016/0370-2693(92)91851-Y}{\emph{Phys. Lett. B}
  {\bfseries 279} (1992) 124}.

\bibitem{Berezhiani:1992pj}
Z.~G. Berezhiani and R.~Rattazzi, \emph{{Inverse hierarchy approach to fermion
  masses}}, \href{https://doi.org/10.1016/0550-3213(93)90057-V}{\emph{Nucl.
  Phys. B} {\bfseries 407} (1993) 249}
  [\href{https://arxiv.org/abs/hep-ph/9212245}{{\ttfamily hep-ph/9212245}}].

\bibitem{Berezhiani:2000cg}
Z.~Berezhiani and A.~Rossi, \emph{{Predictive grand unified textures for quark
  and neutrino masses and mixings}},
  \href{https://doi.org/10.1016/S0550-3213(00)00653-2}{\emph{Nucl. Phys. B}
  {\bfseries 594} (2001) 113}
  [\href{https://arxiv.org/abs/hep-ph/0003084}{{\ttfamily hep-ph/0003084}}].

\bibitem{Berezhiani:1990be}
Z.~G. Berezhiani and J.~L. Chkareuli, \emph{{Low-energy horizontal symmetry of
  $SU(3)_H \times U(1)_H$ and B anti-B oscillation. (In Russian)}}, {\emph{Sov.
  J. Nucl. Phys.} {\bfseries 52} (1990) 383}.

\bibitem{Berezhiani:1996ii}
Z.~Berezhiani, \emph{{Unified picture of the particle and sparticle masses in
  SUSY GUT}}, \href{https://doi.org/10.1016/S0370-2693(97)01359-2}{\emph{Phys.
  Lett. B} {\bfseries 417} (1998) 287}
  [\href{https://arxiv.org/abs/hep-ph/9609342}{{\ttfamily hep-ph/9609342}}].

\bibitem{Anselm:1996jm}
A.~Anselm and Z.~Berezhiani, \emph{{Weak mixing angles as dynamical degrees of
  freedom}}, \href{https://doi.org/10.1016/S0550-3213(96)00597-4}{\emph{Nucl.
  Phys. B} {\bfseries 484} (1997) 97}
  [\href{https://arxiv.org/abs/hep-ph/9605400}{{\ttfamily hep-ph/9605400}}].

\bibitem{Berezhiani:2001mh}
Z.~Berezhiani and A.~Rossi, \emph{{Flavor structure, flavor symmetry and
  supersymmetry}},
  \href{https://doi.org/10.1016/S0920-5632(01)01527-4}{\emph{Nucl. Phys. B
  Proc. Suppl.} {\bfseries 101} (2001) 410}
  [\href{https://arxiv.org/abs/hep-ph/0107054}{{\ttfamily hep-ph/0107054}}].

\bibitem{Kim:1979if}
J.~E. Kim, \emph{{Weak Interaction Singlet and Strong CP Invariance}},
  \href{https://doi.org/10.1103/PhysRevLett.43.103}{\emph{Phys. Rev. Lett.}
  {\bfseries 43} (1979) 103}.

\bibitem{Berezhiani:1989fp}
Z.~G. Berezhiani and M.~Y. Khlopov, \emph{{Cosmology of Spontaneously Broken
  Gauge Family Symmetry}}, \href{https://doi.org/10.1007/BF01570798}{\emph{Z.
  Phys. C} {\bfseries 49} (1991) 73}.

\bibitem{Nelson:1983zb}
A.~E. Nelson, \emph{{Naturally Weak CP Violation}},
  \href{https://doi.org/10.1016/0370-2693(84)92025-2}{\emph{Phys. Lett. B}
  {\bfseries 136} (1984) 387}.

\bibitem{Barr:1984qx}
S.~M. Barr, \emph{{Solving the Strong CP Problem Without the Peccei-Quinn
  Symmetry}}, \href{https://doi.org/10.1103/PhysRevLett.53.329}{\emph{Phys.
  Rev. Lett.} {\bfseries 53} (1984) 329}.

\bibitem{Babu:1989rb}
K.~S. Babu and R.~N. Mohapatra, \emph{{A Solution to the Strong {CP} Problem
  Without an Axion}},
  \href{https://doi.org/10.1103/PhysRevD.41.1286}{\emph{Phys. Rev. D}
  {\bfseries 41} (1990) 1286}.

\bibitem{Berezhiani:1990vp}
Z.~G. Berezhiani, \emph{{On the possibility of a solution to the strong CP
  problem without axion in a $SU(3)_H$ family symmetry model}},
  \href{https://doi.org/10.1142/S0217732391002864}{\emph{Mod. Phys. Lett. A}
  {\bfseries 6} (1991) 2437}.

\bibitem{Berezhiani:1992pq}
Z.~G. Berezhiani, R.~N. Mohapatra and G.~Senjanovic, \emph{{Planck scale
  physics and solutions to the strong CP problem without axion}},
  \href{https://doi.org/10.1103/PhysRevD.47.5565}{\emph{Phys. Rev. D}
  {\bfseries 47} (1993) 5565}
  [\href{https://arxiv.org/abs/hep-ph/9212318}{{\ttfamily hep-ph/9212318}}].

\bibitem{Vecchi:2014hpa}
L.~Vecchi, \emph{{Spontaneous CP violation and the strong CP problem}},
  \href{https://doi.org/10.1007/JHEP04(2017)149}{\emph{JHEP} {\bfseries 04}
  (2017) 149} [\href{https://arxiv.org/abs/1412.3805}{{\ttfamily 1412.3805}}].

\bibitem{Kuchimanchi:2023imj}
R.~Kuchimanchi, \emph{{P and CP solution of the strong CP puzzle}},
  \href{https://doi.org/10.1103/PhysRevD.108.095023}{\emph{Phys. Rev. D}
  {\bfseries 108} (2023) 095023}
  [\href{https://arxiv.org/abs/2306.03039}{{\ttfamily 2306.03039}}].

\bibitem{Branco:1986my}
G.~C. Branco and L.~Lavoura, \emph{{On the Addition of Vector Like Quarks to
  the Standard Model}},
  \href{https://doi.org/10.1016/0550-3213(86)90060-X}{\emph{Nucl. Phys. B}
  {\bfseries 278} (1986) 738}.

\bibitem{delAguila:1989rq}
F.~del Aguila, L.~Ametller, G.~L. Kane and J.~Vidal, \emph{{Vector Like Fermion
  and Standard Higgs Production at Hadron Colliders}},
  \href{https://doi.org/10.1016/0550-3213(90)90655-W}{\emph{Nucl. Phys. B}
  {\bfseries 334} (1990) 1}.

\bibitem{Nir:1990yq}
Y.~Nir and D.~J. Silverman, \emph{{$Z$ Mediated Flavor Changing Neutral
  Currents and Their Implications for {CP} Asymmetries in $\B^0$ Decays}},
  \href{https://doi.org/10.1103/PhysRevD.42.1477}{\emph{Phys. Rev. D}
  {\bfseries 42} (1990) 1477}.

\bibitem{delAguila:2000rc}
F.~del Aguila, M.~Perez-Victoria and J.~Santiago, \emph{{Observable
  contributions of new exotic quarks to quark mixing}},
  \href{https://doi.org/10.1088/1126-6708/2000/09/011}{\emph{JHEP} {\bfseries
  09} (2000) 011} [\href{https://arxiv.org/abs/hep-ph/0007316}{{\ttfamily
  hep-ph/0007316}}].

\bibitem{Barenboim:2001fd}
G.~Barenboim, F.~J. Botella and O.~Vives, \emph{{Constraining Models with
  Vector-Like Fermions from FCNC in $K$ and $B$ Physics}},
  \href{https://doi.org/10.1016/S0550-3213(01)00390-X}{\emph{Nucl. Phys. B}
  {\bfseries 613} (2001) 285}
  [\href{https://arxiv.org/abs/hep-ph/0105306}{{\ttfamily hep-ph/0105306}}].

\bibitem{Cacciapaglia:2010vn}
G.~Cacciapaglia, A.~Deandrea, D.~Harada and Y.~Okada, \emph{{Bounds and Decays
  of New Heavy Vector-like Top Partners}},
  \href{https://doi.org/10.1007/JHEP11(2010)159}{\emph{JHEP} {\bfseries 11}
  (2010) 159} [\href{https://arxiv.org/abs/1007.2933}{{\ttfamily 1007.2933}}].

\bibitem{Botella:2012ju}
F.~J. Botella, G.~C. Branco and M.~Nebot, \emph{{The Hunt for New Physics in
  the Flavour Sector with up vector-like quarks}},
  \href{https://doi.org/10.1007/JHEP12(2012)040}{\emph{JHEP} {\bfseries 12}
  (2012) 040} [\href{https://arxiv.org/abs/1207.4440}{{\ttfamily 1207.4440}}].

\bibitem{Ishiwata:2015cga}
K.~Ishiwata, Z.~Ligeti and M.~B. Wise, \emph{{New Vector-Like Fermions and
  Flavor Physics}}, \href{https://doi.org/10.1007/JHEP10(2015)027}{\emph{JHEP}
  {\bfseries 10} (2015) 027}
  [\href{https://arxiv.org/abs/1506.03484}{{\ttfamily 1506.03484}}].

\bibitem{Wang:2016mjr}
W.~Wang, Z.-H. Xiong and X.-Y. Zhao, \emph{{General scan in flavor parameter
  space in models with vector quark doublets and an enhancement in the $B\to
  X_s\gamma$ process}},
  \href{https://doi.org/10.1088/1674-1137/40/9/093102}{\emph{Chin. Phys. C}
  {\bfseries 40} (2016) 093102}
  [\href{https://arxiv.org/abs/1603.05756}{{\ttfamily 1603.05756}}].

\bibitem{Biekotter:2016kgi}
A.~Biek\"otter, J.~L. Hewett, J.~S. Kim, M.~Kr\"amer, T.~G. Rizzo, K.~Rolbiecki
  et~al., \emph{{Complementarity of Resonant Scalar, Vector-Like Quark and
  Superpartner Searches in Elucidating New Phenomena}},
  \href{https://doi.org/10.1142/S0217751X17500324}{\emph{Int. J. Mod. Phys. A}
  {\bfseries 32} (2017) 1750032}
  [\href{https://arxiv.org/abs/1608.01312}{{\ttfamily 1608.01312}}].

\bibitem{Bobeth:2016llm}
C.~Bobeth, A.~J. Buras, A.~Celis and M.~Jung, \emph{{Patterns of Flavour
  Violation in Models with Vector-Like Quarks}},
  \href{https://doi.org/10.1007/JHEP04(2017)079}{\emph{JHEP} {\bfseries 04}
  (2017) 079} [\href{https://arxiv.org/abs/1609.04783}{{\ttfamily
  1609.04783}}].

\bibitem{Botella:2016ibj}
F.~J. Botella, G.~C. Branco, M.~Nebot, M.~N. Rebelo and J.~I. Silva-Marcos,
  \emph{{Vector-like Quarks at the Origin of Light Quark Masses and Mixing}},
  \href{https://doi.org/10.1140/epjc/s10052-017-4933-3}{\emph{Eur. Phys. J. C}
  {\bfseries 77} (2017) 408}
  [\href{https://arxiv.org/abs/1610.03018}{{\ttfamily 1610.03018}}].

\bibitem{Cacciapaglia:2018lld}
G.~Cacciapaglia, A.~Deandrea, N.~Gaur, D.~Harada, Y.~Okada and L.~Panizzi,
  \emph{{The LHC potential of Vector-like quark doublets}},
  \href{https://doi.org/10.1007/JHEP11(2018)055}{\emph{JHEP} {\bfseries 11}
  (2018) 055} [\href{https://arxiv.org/abs/1806.01024}{{\ttfamily
  1806.01024}}].

\bibitem{Belfatto:2019swo}
B.~Belfatto, R.~Beradze and Z.~Berezhiani, \emph{{The CKM unitarity problem: A
  trace of new physics at the TeV scale?}},
  \href{https://doi.org/10.1140/epjc/s10052-020-7691-6}{\emph{Eur. Phys. J. C}
  {\bfseries 80} (2020) 149}
  [\href{https://arxiv.org/abs/1906.02714}{{\ttfamily 1906.02714}}].

\bibitem{Belfatto:2021jhf}
B.~Belfatto and Z.~Berezhiani, \emph{{Are the CKM anomalies induced by
  vector-like quarks? Limits from flavor changing and Standard Model precision
  tests}}, \href{https://doi.org/10.1007/JHEP10(2021)079}{\emph{JHEP}
  {\bfseries 10} (2021) 079}
  [\href{https://arxiv.org/abs/2103.05549}{{\ttfamily 2103.05549}}].

\bibitem{Branco:2021vhs}
G.~C. Branco, J.~T. Penedo, P.~M.~F. Pereira, M.~N. Rebelo and J.~I.
  Silva-Marcos, \emph{{Addressing the CKM unitarity problem with a vector-like
  up quark}}, \href{https://doi.org/10.1007/JHEP07(2021)099}{\emph{JHEP}
  {\bfseries 07} (2021) 099}
  [\href{https://arxiv.org/abs/2103.13409}{{\ttfamily 2103.13409}}].

\bibitem{Balaji:2021lpr}
S.~Balaji, \emph{{Asymmetry in flavour changing electromagnetic transitions of
  vector-like quarks}},
  \href{https://doi.org/10.1007/JHEP05(2022)015}{\emph{JHEP} {\bfseries 05}
  (2022) 015} [\href{https://arxiv.org/abs/2110.05473}{{\ttfamily
  2110.05473}}].

\bibitem{Botella:2021uxz}
F.~J. Botella, G.~C. Branco, M.~N. Rebelo, J.~I. Silva-Marcos and J.~F. Bastos,
  \emph{{Decays of the heavy top and new insights on $\epsilon _K$ in a one-VLQ
  minimal solution to the CKM unitarity problem}},
  \href{https://doi.org/10.1140/epjc/s10052-022-10299-9}{\emph{Eur. Phys. J. C}
  {\bfseries 82} (2022) 360}
  [\href{https://arxiv.org/abs/2111.15401}{{\ttfamily 2111.15401}}], [Erratum:
  Eur.Phys.J.C 82, 423 (2022)].

\bibitem{Belfatto:2023tbv}
B.~Belfatto and S.~Trifinopoulos, \emph{{Cabibbo angle anomalies and oblique
  corrections: The remarkable role of the vectorlike quark doublet}},
  \href{https://doi.org/10.1103/PhysRevD.108.035022}{\emph{Phys. Rev. D}
  {\bfseries 108} (2023) 035022}
  [\href{https://arxiv.org/abs/2302.14097}{{\ttfamily 2302.14097}}].

\bibitem{Cepedello:2024qmq}
R.~Cepedello, F.~Esser, M.~Hirsch and V.~Sanz, \emph{{Faking ZZZ vertices at
  the LHC}}, \href{https://doi.org/10.1007/JHEP12(2024)098}{\emph{JHEP}
  {\bfseries 12} (2024) 098}
  [\href{https://arxiv.org/abs/2409.06776}{{\ttfamily 2409.06776}}].

\bibitem{Cheung:2020vqm}
K.~Cheung, W.-Y. Keung, C.-T. Lu and P.-Y. Tseng, \emph{{Vector-like Quark
  Interpretation for the CKM Unitarity Violation, Excess in Higgs Signal
  Strength, and Bottom Quark Forward-Backward Asymmetry}},
  \href{https://doi.org/10.1007/JHEP05(2020)117}{\emph{JHEP} {\bfseries 05}
  (2020) 117} [\href{https://arxiv.org/abs/2001.02853}{{\ttfamily
  2001.02853}}].

\bibitem{Endo:2020tkb}
M.~Endo and S.~Mishima, \emph{{Muon $g-2$ and CKM unitarity in extra lepton
  models}}, \href{https://doi.org/10.1007/JHEP08(2020)004}{\emph{JHEP}
  {\bfseries 08} (2020) 004}
  [\href{https://arxiv.org/abs/2005.03933}{{\ttfamily 2005.03933}}].

\bibitem{Crivellin:2020ebi}
A.~Crivellin, F.~Kirk, C.~A. Manzari and M.~Montull, \emph{{Global Electroweak
  Fit and Vector-Like Leptons in Light of the Cabibbo Angle Anomaly}},
  \href{https://doi.org/10.1007/JHEP12(2020)166}{\emph{JHEP} {\bfseries 12}
  (2020) 166} [\href{https://arxiv.org/abs/2008.01113}{{\ttfamily
  2008.01113}}].

\bibitem{Crivellin:2022rhw}
A.~Crivellin, M.~Kirk, T.~Kitahara and F.~Mescia, \emph{{Global fit of modified
  quark couplings to EW gauge bosons and vector-like quarks in light of the
  Cabibbo angle anomaly}},
  \href{https://doi.org/10.1007/JHEP03(2023)234}{\emph{JHEP} {\bfseries 03}
  (2023) 234} [\href{https://arxiv.org/abs/2212.06862}{{\ttfamily
  2212.06862}}].

\bibitem{Dcruz:2022rjg}
R.~Dcruz and K.~S. Babu, \emph{{Resolving W boson mass shift and CKM unitarity
  violation in left-right symmetric models with a universal seesaw mechanism}},
  \href{https://doi.org/10.1103/PhysRevD.108.095011}{\emph{Phys. Rev. D}
  {\bfseries 108} (2023) 095011}
  [\href{https://arxiv.org/abs/2212.09697}{{\ttfamily 2212.09697}}].

\bibitem{Alves:2023ufm}
J.~M. Alves, G.~C. Branco, A.~L. Cherchiglia, C.~C. Nishi, J.~T. Penedo,
  P.~M.~F. Pereira et~al., \emph{{Vector-like singlet quarks: A roadmap}},
  \href{https://doi.org/10.1016/j.physrep.2023.12.004}{\emph{Phys. Rept.}
  {\bfseries 1057} (2024) 1}
  [\href{https://arxiv.org/abs/2304.10561}{{\ttfamily 2304.10561}}].

\bibitem{delAguila:2000aa}
F.~del Aguila, M.~Perez-Victoria and J.~Santiago, \emph{{Effective description
  of quark mixing}},
  \href{https://doi.org/10.1016/S0370-2693(00)01071-6}{\emph{Phys. Lett. B}
  {\bfseries 492} (2000) 98}
  [\href{https://arxiv.org/abs/hep-ph/0007160}{{\ttfamily hep-ph/0007160}}].

\bibitem{Aguilar-Saavedra:2013qpa}
J.~A. Aguilar-Saavedra, R.~Benbrik, S.~Heinemeyer and M.~P\'erez-Victoria,
  \emph{{Handbook of vectorlike quarks: Mixing and single production}},
  \href{https://doi.org/10.1103/PhysRevD.88.094010}{\emph{Phys. Rev. D}
  {\bfseries 88} (2013) 094010}
  [\href{https://arxiv.org/abs/1306.0572}{{\ttfamily 1306.0572}}].

\bibitem{Cirigliano:2023nol}
V.~Cirigliano, W.~Dekens, J.~de~Vries, E.~Mereghetti and T.~Tong,
  \emph{{Anomalies in global SMEFT analyses. A case study of first-row CKM
  unitarity}}, \href{https://doi.org/10.1007/JHEP03(2024)033}{\emph{JHEP}
  {\bfseries 03} (2024) 033}
  [\href{https://arxiv.org/abs/2311.00021}{{\ttfamily 2311.00021}}].

\bibitem{Bernabeu:1986fc}
J.~Bernabeu, G.~C. Branco and M.~Gronau, \emph{{CP Restrictions on Quark Mass
  Matrices}}, \href{https://doi.org/10.1016/0370-2693(86)90659-3}{\emph{Phys.
  Lett. B} {\bfseries 169} (1986) 243}.

\bibitem{Gronau:1986xb}
M.~Gronau, A.~Kfir and R.~Loewy, \emph{{Basis Independent Tests of {CP}
  Violation in Fermion Mass Matrices}},
  \href{https://doi.org/10.1103/PhysRevLett.56.1538}{\emph{Phys. Rev. Lett.}
  {\bfseries 56} (1986) 1538}.

\bibitem{Olechowski:1989ny}
M.~Olechowski and S.~Pokorski, \emph{{Some Useful Invariants of Quark Mass
  Matrices}}, \href{https://doi.org/10.1016/0370-2693(89)90132-9}{\emph{Phys.
  Lett. B} {\bfseries 231} (1989) 165}.

\bibitem{delAguila:1996pa}
F.~del Aguila and J.~A. Aguilar-Saavedra, \emph{{Invariant formulation of CP
  violation for four quark families}},
  \href{https://doi.org/10.1016/0370-2693(96)00939-2}{\emph{Phys. Lett. B}
  {\bfseries 386} (1996) 241}
  [\href{https://arxiv.org/abs/hep-ph/9605418}{{\ttfamily hep-ph/9605418}}].

\bibitem{delAguila:1997vn}
F.~del Aguila, J.~A. Aguilar-Saavedra and G.~C. Branco, \emph{{CP violation
  from new quarks in the chiral limit}},
  \href{https://doi.org/10.1016/S0550-3213(97)00708-6}{\emph{Nucl. Phys. B}
  {\bfseries 510} (1998) 39}
  [\href{https://arxiv.org/abs/hep-ph/9703410}{{\ttfamily hep-ph/9703410}}].

\bibitem{Branco:2011aa}
G.~C. Branco and J.~I. Silva-Marcos, \emph{{Invariants, Alignment and the
  Pattern of Fermion Masses and Mixing}},
  \href{https://doi.org/10.1016/j.physletb.2012.07.064}{\emph{Phys. Lett. B}
  {\bfseries 715} (2012) 315}
  [\href{https://arxiv.org/abs/1112.1631}{{\ttfamily 1112.1631}}].

\bibitem{Albergaria:2022zaq}
F.~Albergaria, G.~C. Branco, J.~F. Bastos and J.~I. Silva-Marcos, \emph{{CP-odd
  and CP-even weak-basis invariants in the presence of vector-like quarks}},
  \href{https://doi.org/10.1088/1361-6471/acc349}{\emph{J. Phys. G} {\bfseries
  50} (2023) 055001} [\href{https://arxiv.org/abs/2210.14248}{{\ttfamily
  2210.14248}}].

\bibitem{Bento:2023owf}
M.~P. Bento, J.~P. Silva and A.~Trautner, \emph{{The basis invariant flavor
  puzzle}}, \href{https://doi.org/10.1007/JHEP01(2024)024}{\emph{JHEP}
  {\bfseries 01} (2024) 024}
  [\href{https://arxiv.org/abs/2308.00019}{{\ttfamily 2308.00019}}].

\bibitem{deLima:2024vrn}
E.~L.~F. de~Lima and C.~C. Nishi, \emph{{Flavor invariants for the SM with one
  singlet vector-like quark}},
  \href{https://doi.org/10.1007/JHEP11(2024)157}{\emph{JHEP} {\bfseries 11}
  (2024) 157} [\href{https://arxiv.org/abs/2408.10325}{{\ttfamily
  2408.10325}}].

\bibitem{Branco:1999fs}
G.~C. Branco, L.~Lavoura and J.~P. Silva, \emph{{CP Violation}}, vol.~103.
  Oxford University Press, 1999.

\bibitem{Chen:2015uza}
C.-Y. Chen, S.~Dawson and Y.~Zhang, \emph{{Higgs CP Violation from Vectorlike
  Quarks}}, \href{https://doi.org/10.1103/PhysRevD.92.075026}{\emph{Phys. Rev.
  D} {\bfseries 92} (2015) 075026}
  [\href{https://arxiv.org/abs/1507.07020}{{\ttfamily 1507.07020}}].

\bibitem{Bastos:2024afz}
J.~F. Bastos and J.~I. Silva-Marcos, \emph{{Reducing Complex Phases and other
  Subtleties of CP Violation}},
  \href{https://arxiv.org/abs/2407.07158}{{\ttfamily 2407.07158}}.

\bibitem{Silva-Marcos:2002upu}
J.~I. Silva-Marcos, \emph{{On the reduction of CP violation phases}},
  \href{https://arxiv.org/abs/hep-ph/0212089}{{\ttfamily hep-ph/0212089}}.

\bibitem{Jarlskog:1985ht}
C.~Jarlskog, \emph{{Commutator of the Quark Mass Matrices in the Standard
  Electroweak Model and a Measure of Maximal CP Nonconservation}},
  \href{https://doi.org/10.1103/PhysRevLett.55.1039}{\emph{Phys. Rev. Lett.}
  {\bfseries 55} (1985) 1039}.

\bibitem{Jarlskog:1985cw}
C.~Jarlskog, \emph{{A Basis Independent Formulation of the Connection Between
  Quark Mass Matrices, CP Violation and Experiment}},
  \href{https://doi.org/10.1007/BF01565198}{\emph{Z. Phys. C} {\bfseries 29}
  (1985) 491}.

\bibitem{ATLAS:2024zlo}
{\scshape ATLAS} collaboration, G.~Aad et~al., \emph{{Search for pair-produced
  vectorlike quarks coupling to light quarks in the lepton plus jets final
  state using 13~TeV pp collisions with the ATLAS detector}},
  \href{https://doi.org/10.1103/PhysRevD.110.052009}{\emph{Phys. Rev. D}
  {\bfseries 110} (2024) 052009}
  [\href{https://arxiv.org/abs/2405.19862}{{\ttfamily 2405.19862}}].

\bibitem{ATLAS:2022hnn}
{\scshape ATLAS} collaboration, G.~Aad et~al., \emph{{Search for
  pair-production of vector-like quarks in pp collision events at s=13 TeV with
  at least one leptonically decaying Z boson and a third-generation quark with
  the ATLAS detector}},
  \href{https://doi.org/10.1016/j.physletb.2023.138019}{\emph{Phys. Lett. B}
  {\bfseries 843} (2023) 138019}
  [\href{https://arxiv.org/abs/2210.15413}{{\ttfamily 2210.15413}}].

\bibitem{CMS:2022fck}
{\scshape CMS} collaboration, A.~Tumasyan et~al., \emph{{Search for pair
  production of vector-like quarks in leptonic final states in proton-proton
  collisions at $ \sqrt{s} $ = 13 TeV}},
  \href{https://doi.org/10.1007/JHEP07(2023)020}{\emph{JHEP} {\bfseries 07}
  (2023) 020} [\href{https://arxiv.org/abs/2209.07327}{{\ttfamily
  2209.07327}}].

\bibitem{CMS:2020ttz}
{\scshape CMS} collaboration, A.~M. Sirunyan et~al., \emph{{A search for
  bottom-type, vector-like quark pair production in a fully hadronic final
  state in proton-proton collisions at $\sqrt{s} =$ 13 TeV}},
  \href{https://doi.org/10.1103/PhysRevD.102.112004}{\emph{Phys. Rev. D}
  {\bfseries 102} (2020) 112004}
  [\href{https://arxiv.org/abs/2008.09835}{{\ttfamily 2008.09835}}].

\bibitem{Chau:1984fp}
L.-L. Chau and W.-Y. Keung, \emph{{Comments on the Parametrization of the
  Kobayashi-Maskawa Matrix}},
  \href{https://doi.org/10.1103/PhysRevLett.53.1802}{\emph{Phys. Rev. Lett.}
  {\bfseries 53} (1984) 1802}.

\bibitem{ParticleDataGroup:2022pth}
{\scshape Particle Data Group} collaboration, R.~L. Workman et~al.,
  \emph{{Review of Particle Physics}},
  \href{https://doi.org/10.1093/ptep/ptac097}{\emph{PTEP} {\bfseries 2022}
  (2022) 083C01}.

\bibitem{Botella:1985gb}
F.~J. Botella and L.-L. Chau, \emph{{Anticipating the Higher Generations of
  Quarks from Rephasing Invariance of the Mixing Matrix}},
  \href{https://doi.org/10.1016/0370-2693(86)91468-1}{\emph{Phys. Lett. B}
  {\bfseries 168} (1986) 97}.

\bibitem{Glashow:1970gm}
S.~L. Glashow, J.~Iliopoulos and L.~Maiani, \emph{{Weak Interactions with
  Lepton-Hadron Symmetry}},
  \href{https://doi.org/10.1103/PhysRevD.2.1285}{\emph{Phys. Rev. D} {\bfseries
  2} (1970) 1285}.

\bibitem{Glashow:1976nt}
S.~L. Glashow and S.~Weinberg, \emph{{Natural Conservation Laws for Neutral
  Currents}}, \href{https://doi.org/10.1103/PhysRevD.15.1958}{\emph{Phys. Rev.
  D} {\bfseries 15} (1977) 1958}.

\bibitem{Paschos:1976ay}
E.~A. Paschos, \emph{{Diagonal Neutral Currents}},
  \href{https://doi.org/10.1103/PhysRevD.15.1966}{\emph{Phys. Rev. D}
  {\bfseries 15} (1977) 1966}.

\bibitem{Seng:2018qru}
C.~Y. Seng, M.~Gorchtein and M.~J. Ramsey-Musolf, \emph{{Dispersive evaluation
  of the inner radiative correction in neutron and nuclear $\beta$ decay}},
  \href{https://doi.org/10.1103/PhysRevD.100.013001}{\emph{Phys. Rev. D}
  {\bfseries 100} (2019) 013001}
  [\href{https://arxiv.org/abs/1812.03352}{{\ttfamily 1812.03352}}].

\bibitem{Seng:2018yzq}
C.-Y. Seng, M.~Gorchtein, H.~H. Patel and M.~J. Ramsey-Musolf, \emph{{Reduced
  Hadronic Uncertainty in the Determination of $V_{ud}$}},
  \href{https://doi.org/10.1103/PhysRevLett.121.241804}{\emph{Phys. Rev. Lett.}
  {\bfseries 121} (2018) 241804}
  [\href{https://arxiv.org/abs/1807.10197}{{\ttfamily 1807.10197}}].

\bibitem{Hardy:2020qwl}
J.~C. Hardy and I.~S. Towner, \emph{{Superallowed $0^+ \to 0^+$ nuclear $\beta$
  decays: 2020 critical survey, with implications for V$_{ud}$ and CKM
  unitarity}}, \href{https://doi.org/10.1103/PhysRevC.102.045501}{\emph{Phys.
  Rev. C} {\bfseries 102} (2020) 045501}.

\bibitem{Gorchtein:2018fxl}
M.~Gorchtein, \emph{{\ensuremath{\gamma}W Box Inside Out: Nuclear
  Polarizabilities Distort the Beta Decay Spectrum}},
  \href{https://doi.org/10.1103/PhysRevLett.123.042503}{\emph{Phys. Rev. Lett.}
  {\bfseries 123} (2019) 042503}
  [\href{https://arxiv.org/abs/1812.04229}{{\ttfamily 1812.04229}}].

\bibitem{UCNt:2021pcg}
{\scshape UCN\ensuremath{\tau}} collaboration, F.~M. Gonzalez et~al.,
  \emph{{Improved neutron lifetime measurement with UCN$\tau$}},
  \href{https://doi.org/10.1103/PhysRevLett.127.162501}{\emph{Phys. Rev. Lett.}
  {\bfseries 127} (2021) 162501}
  [\href{https://arxiv.org/abs/2106.10375}{{\ttfamily 2106.10375}}].

\bibitem{Ezhov:2014tna}
V.~F. Ezhov et~al., \emph{{Measurement of the neutron lifetime with ultra-cold
  neutrons stored in a magneto-gravitational trap}},
  \href{https://doi.org/10.1134/S0021364018110024}{\emph{JETP Lett.} {\bfseries
  107} (2018) 671} [\href{https://arxiv.org/abs/1412.7434}{{\ttfamily
  1412.7434}}].

\bibitem{Pattie:2017vsj}
R.~W. Pattie, Jr. et~al., \emph{{Measurement of the neutron lifetime using a
  magneto-gravitational trap and in situ detection}},
  \href{https://doi.org/10.1126/science.aan8895}{\emph{Science} {\bfseries 360}
  (2018) 627} [\href{https://arxiv.org/abs/1707.01817}{{\ttfamily
  1707.01817}}].

\bibitem{Serebrov:2017bzo}
A.~P. Serebrov et~al., \emph{{Neutron lifetime measurements with a large
  gravitational trap for ultracold neutrons}},
  \href{https://doi.org/10.1103/PhysRevC.97.055503}{\emph{Phys. Rev. C}
  {\bfseries 97} (2018) 055503}
  [\href{https://arxiv.org/abs/1712.05663}{{\ttfamily 1712.05663}}].

\bibitem{Arzumanov:2015tea}
S.~Arzumanov, L.~Bondarenko, S.~Chernyavsky, P.~Geltenbort, V.~Morozov, V.~V.
  Nesvizhevsky et~al., \emph{{A measurement of the neutron lifetime using the
  method of storage of ultracold neutrons and detection of inelastically
  up-scattered neutrons}},
  \href{https://doi.org/10.1016/j.physletb.2015.04.021}{\emph{Phys. Lett. B}
  {\bfseries 745} (2015) 79}.

\bibitem{Steyerl:2012zz}
A.~Steyerl, J.~M. Pendlebury, C.~Kaufman, S.~S. Malik and A.~M. Desai,
  \emph{{Quasielastic scattering in the interaction of ultracold neutrons with
  a liquid wall and application in a reanalysis of the Mambo I neutron-lifetime
  experiment}}, \href{https://doi.org/10.1103/PhysRevC.85.065503}{\emph{Phys.
  Rev. C} {\bfseries 85} (2012) 065503}.

\bibitem{Pichlmaier:2010zz}
A.~Pichlmaier, V.~Varlamov, K.~Schreckenbach and P.~Geltenbort, \emph{{Neutron
  lifetime measurement with the UCN trap-in-trap MAMBO II}},
  \href{https://doi.org/10.1016/j.physletb.2010.08.032}{\emph{Phys. Lett. B}
  {\bfseries 693} (2010) 221}.

\bibitem{Serebrov:2004zf}
A.~Serebrov et~al., \emph{{Measurement of the neutron lifetime using a
  gravitational trap and a low-temperature Fomblin coating}},
  \href{https://doi.org/10.1016/j.physletb.2004.11.013}{\emph{Phys. Lett. B}
  {\bfseries 605} (2005) 72}
  [\href{https://arxiv.org/abs/nucl-ex/0408009}{{\ttfamily nucl-ex/0408009}}].

\bibitem{Mund:2012fq}
D.~Mund, B.~Maerkisch, M.~Deissenroth, J.~Krempel, M.~Schumann, H.~Abele
  et~al., \emph{{Determination of the Weak Axial Vector Coupling from a
  Measurement of the Beta-Asymmetry Parameter A in Neutron Beta Decay}},
  \href{https://doi.org/10.1103/PhysRevLett.110.172502}{\emph{Phys. Rev. Lett.}
  {\bfseries 110} (2013) 172502}
  [\href{https://arxiv.org/abs/1204.0013}{{\ttfamily 1204.0013}}].

\bibitem{UCNA:2017obv}
{\scshape UCNA} collaboration, M.~A.~P. Brown et~al., \emph{{New result for the
  neutron $\beta$-asymmetry parameter $A_0$ from UCNA}},
  \href{https://doi.org/10.1103/PhysRevC.97.035505}{\emph{Phys. Rev. C}
  {\bfseries 97} (2018) 035505}
  [\href{https://arxiv.org/abs/1712.00884}{{\ttfamily 1712.00884}}].

\bibitem{Markisch:2018ndu}
B.~M\"arkisch et~al., \emph{{Measurement of the Weak Axial-Vector Coupling
  Constant in the Decay of Free Neutrons Using a Pulsed Cold Neutron Beam}},
  \href{https://doi.org/10.1103/PhysRevLett.122.242501}{\emph{Phys. Rev. Lett.}
  {\bfseries 122} (2019) 242501}
  [\href{https://arxiv.org/abs/1812.04666}{{\ttfamily 1812.04666}}].

\bibitem{FlavourLatticeAveragingGroupFLAG:2021npn}
{\scshape Flavour Lattice Averaging Group (FLAG)} collaboration, Y.~Aoki
  et~al., \emph{{FLAG Review 2021}},
  \href{https://doi.org/10.1140/epjc/s10052-022-10536-1}{\emph{Eur. Phys. J. C}
  {\bfseries 82} (2022) 869}
  [\href{https://arxiv.org/abs/2111.09849}{{\ttfamily 2111.09849}}].

\bibitem{FlavourLatticeAveragingGroupFLAG:2024oxs}
{\scshape Flavour Lattice Averaging Group (FLAG)} collaboration, Y.~Aoki
  et~al., \emph{{FLAG Review 2024}},
  \href{https://arxiv.org/abs/2411.04268}{{\ttfamily 2411.04268}}.

\bibitem{Seng:2021nar}
C.-Y. Seng, D.~Galviz, W.~J. Marciano and U.-G. Mei\ss{}ner, \emph{{Update on
  $|V_{us}|$ and $|V_{us}/V_{ud}|$ from semileptonic kaon and pion decays}},
  \href{https://doi.org/10.1103/PhysRevD.105.013005}{\emph{Phys. Rev. D}
  {\bfseries 105} (2022) 013005}
  [\href{https://arxiv.org/abs/2107.14708}{{\ttfamily 2107.14708}}].

\bibitem{Seng:2022wcw}
C.-Y. Seng, D.~Galviz, M.~Gorchtein and U.-G. Mei\ss{}ner, \emph{{Complete
  theory of radiative corrections to K$_{\ell 3}$ decays and the V$_{us}$
  update}}, \href{https://doi.org/10.1007/JHEP07(2022)071}{\emph{JHEP}
  {\bfseries 07} (2022) 071}
  [\href{https://arxiv.org/abs/2203.05217}{{\ttfamily 2203.05217}}].

\bibitem{Moulson:2017ive}
M.~Moulson, \emph{{Experimental determination of $V_{us}$ from kaon decays}},
  \href{https://doi.org/10.22323/1.291.0033}{\emph{PoS} {\bfseries CKM2016}
  (2017) 033} [\href{https://arxiv.org/abs/1704.04104}{{\ttfamily
  1704.04104}}].

\bibitem{Marciano:2004uf}
W.~J. Marciano, \emph{{Precise determination of $|V_{us}|$ from lattice
  calculations of pseudoscalar decay constants}},
  \href{https://doi.org/10.1103/PhysRevLett.93.231803}{\emph{Phys. Rev. Lett.}
  {\bfseries 93} (2004) 231803}
  [\href{https://arxiv.org/abs/hep-ph/0402299}{{\ttfamily hep-ph/0402299}}].

\bibitem{Cirigliano:2011tm}
V.~Cirigliano and H.~Neufeld, \emph{{A note on isospin violation in $P_{\ell
  2(\gamma)}$ decays}},
  \href{https://doi.org/10.1016/j.physletb.2011.04.038}{\emph{Phys. Lett. B}
  {\bfseries 700} (2011) 7} [\href{https://arxiv.org/abs/1102.0563}{{\ttfamily
  1102.0563}}].

\bibitem{DiCarlo:2019thl}
M.~Di~Carlo, D.~Giusti, V.~Lubicz, G.~Martinelli, C.~T. Sachrajda,
  F.~Sanfilippo et~al., \emph{{Light-meson leptonic decay rates in lattice
  QCD+QED}}, \href{https://doi.org/10.1103/PhysRevD.100.034514}{\emph{Phys.
  Rev. D} {\bfseries 100} (2019) 034514}
  [\href{https://arxiv.org/abs/1904.08731}{{\ttfamily 1904.08731}}].

\bibitem{Boyle:2022lsi}
P.~Boyle et~al., \emph{{Isospin-breaking corrections to light-meson leptonic
  decays from lattice simulations at physical quark masses}},
  \href{https://doi.org/10.1007/JHEP02(2023)242}{\emph{JHEP} {\bfseries 02}
  (2023) 242} [\href{https://arxiv.org/abs/2211.12865}{{\ttfamily
  2211.12865}}].

\bibitem{Grossman:2019bzp}
Y.~Grossman, E.~Passemar and S.~Schacht, \emph{{On the Statistical Treatment of
  the Cabibbo Angle Anomaly}},
  \href{https://doi.org/10.1007/JHEP07(2020)068}{\emph{JHEP} {\bfseries 07}
  (2020) 068} [\href{https://arxiv.org/abs/1911.07821}{{\ttfamily
  1911.07821}}].

\bibitem{Coutinho:2019aiy}
A.~M. Coutinho, A.~Crivellin and C.~A. Manzari, \emph{{Global Fit to Modified
  Neutrino Couplings and the Cabibbo-Angle Anomaly}},
  \href{https://doi.org/10.1103/PhysRevLett.125.071802}{\emph{Phys. Rev. Lett.}
  {\bfseries 125} (2020) 071802}
  [\href{https://arxiv.org/abs/1912.08823}{{\ttfamily 1912.08823}}].

\bibitem{Crivellin:2020lzu}
A.~Crivellin and M.~Hoferichter, \emph{{\ensuremath{\beta} Decays as Sensitive
  Probes of Lepton Flavor Universality}},
  \href{https://doi.org/10.1103/PhysRevLett.125.111801}{\emph{Phys. Rev. Lett.}
  {\bfseries 125} (2020) 111801}
  [\href{https://arxiv.org/abs/2002.07184}{{\ttfamily 2002.07184}}].

\bibitem{Capdevila:2020rrl}
B.~Capdevila, A.~Crivellin, C.~A. Manzari and M.~Montull, \emph{{Explaining
  $b\to s\ell^+\ell^-$ and the Cabibbo angle anomaly with a vector triplet}},
  \href{https://doi.org/10.1103/PhysRevD.103.015032}{\emph{Phys. Rev. D}
  {\bfseries 103} (2021) 015032}
  [\href{https://arxiv.org/abs/2005.13542}{{\ttfamily 2005.13542}}].

\bibitem{Kirk:2020wdk}
M.~Kirk, \emph{{Cabibbo anomaly versus electroweak precision tests: An
  exploration of extensions of the Standard Model}},
  \href{https://doi.org/10.1103/PhysRevD.103.035004}{\emph{Phys. Rev. D}
  {\bfseries 103} (2021) 035004}
  [\href{https://arxiv.org/abs/2008.03261}{{\ttfamily 2008.03261}}].

\bibitem{Manzari:2020eum}
C.~A. Manzari, A.~M. Coutinho and A.~Crivellin, \emph{{Modified lepton
  couplings and the Cabibbo-angle anomaly}},
  \href{https://doi.org/10.22323/1.382.0242}{\emph{PoS} {\bfseries LHCP2020}
  (2021) 242} [\href{https://arxiv.org/abs/2009.03877}{{\ttfamily
  2009.03877}}].

\bibitem{Alok:2020jod}
A.~K. Alok, A.~Dighe, S.~Gangal and J.~Kumar, \emph{{The role of non-universal
  $Z$ couplings in explaining the $V_{us}$ anomaly}},
  \href{https://doi.org/10.1016/j.nuclphysb.2021.115538}{\emph{Nucl. Phys. B}
  {\bfseries 971} (2021) 115538}
  [\href{https://arxiv.org/abs/2010.12009}{{\ttfamily 2010.12009}}].

\bibitem{Crivellin:2020oup}
A.~Crivellin, C.~A. Manzari, M.~Alguero and J.~Matias, \emph{{Combined
  Explanation of the Z\textrightarrow{}bb\textasciimacron{} Forward-Backward
  Asymmetry, the Cabibbo Angle Anomaly, and
  \ensuremath{\tau}\textrightarrow{}\ensuremath{\mu}\ensuremath{\nu}\ensuremath{\nu}
  and b\textrightarrow{}s\ensuremath{\ell}+\ensuremath{\ell}- Data}},
  \href{https://doi.org/10.1103/PhysRevLett.127.011801}{\emph{Phys. Rev. Lett.}
  {\bfseries 127} (2021) 011801}
  [\href{https://arxiv.org/abs/2010.14504}{{\ttfamily 2010.14504}}].

\bibitem{Crivellin:2020klg}
A.~Crivellin, F.~Kirk, C.~A. Manzari and L.~Panizzi, \emph{{Searching for
  lepton flavor universality violation and collider signals from a singly
  charged scalar singlet}},
  \href{https://doi.org/10.1103/PhysRevD.103.073002}{\emph{Phys. Rev. D}
  {\bfseries 103} (2021) 073002}
  [\href{https://arxiv.org/abs/2012.09845}{{\ttfamily 2012.09845}}].

\bibitem{Crivellin:2021njn}
A.~Crivellin, M.~Hoferichter and C.~A. Manzari, \emph{{Fermi Constant from Muon
  Decay Versus Electroweak Fits and Cabibbo-Kobayashi-Maskawa Unitarity}},
  \href{https://doi.org/10.1103/PhysRevLett.127.071801}{\emph{Phys. Rev. Lett.}
  {\bfseries 127} (2021) 071801}
  [\href{https://arxiv.org/abs/2102.02825}{{\ttfamily 2102.02825}}].

\bibitem{Marzocca:2021azj}
D.~Marzocca and S.~Trifinopoulos, \emph{{Minimal Explanation of Flavor
  Anomalies: B-Meson Decays, Muon Magnetic Moment, and the Cabibbo Angle}},
  \href{https://doi.org/10.1103/PhysRevLett.127.061803}{\emph{Phys. Rev. Lett.}
  {\bfseries 127} (2021) 061803}
  [\href{https://arxiv.org/abs/2104.05730}{{\ttfamily 2104.05730}}].

\bibitem{Fischer:2021sqw}
O.~Fischer et~al., \emph{{Unveiling hidden physics at the LHC}},
  \href{https://doi.org/10.1140/epjc/s10052-022-10541-4}{\emph{Eur. Phys. J. C}
  {\bfseries 82} (2022) 665}
  [\href{https://arxiv.org/abs/2109.06065}{{\ttfamily 2109.06065}}].

\bibitem{Pich:2021yll}
A.~Pich and A.~Rodr\'\i{}guez-S\'anchez, \emph{{SU(3) analysis of four-quark
  operators: $K\to\pi\pi$ and vacuum matrix elements}},
  \href{https://doi.org/10.1007/JHEP06(2021)005}{\emph{JHEP} {\bfseries 06}
  (2021) 005} [\href{https://arxiv.org/abs/2102.09308}{{\ttfamily
  2102.09308}}].

\bibitem{Blum:2012uk}
T.~Blum et~al., \emph{{Lattice determination of the $K \to (\pi\pi)_{I=2}$
  Decay Amplitude $A_2$}},
  \href{https://doi.org/10.1103/PhysRevD.86.074513}{\emph{Phys. Rev. D}
  {\bfseries 86} (2012) 074513}
  [\href{https://arxiv.org/abs/1206.5142}{{\ttfamily 1206.5142}}].

\bibitem{RBC:2015gro}
{\scshape RBC, UKQCD} collaboration, Z.~Bai et~al., \emph{{Standard Model
  Prediction for Direct CP Violation in
  K\textrightarrow{}\ensuremath{\pi}\ensuremath{\pi} Decay}},
  \href{https://doi.org/10.1103/PhysRevLett.115.212001}{\emph{Phys. Rev. Lett.}
  {\bfseries 115} (2015) 212001}
  [\href{https://arxiv.org/abs/1505.07863}{{\ttfamily 1505.07863}}].

\bibitem{RBC:2020kdj}
{\scshape RBC, UKQCD} collaboration, R.~Abbott et~al., \emph{{Direct CP
  violation and the $\Delta I=1/2$ rule in $K\to\pi\pi$ decay from the standard
  model}}, \href{https://doi.org/10.1103/PhysRevD.102.054509}{\emph{Phys. Rev.
  D} {\bfseries 102} (2020) 054509}
  [\href{https://arxiv.org/abs/2004.09440}{{\ttfamily 2004.09440}}].

\bibitem{Bertolini:2013noa}
S.~Bertolini, A.~Maiezza and F.~Nesti,
  \emph{{K\textrightarrow{}\ensuremath{\pi}\ensuremath{\pi} hadronic matrix
  elements of left-right current-current operators}},
  \href{https://doi.org/10.1103/PhysRevD.88.034014}{\emph{Phys. Rev. D}
  {\bfseries 88} (2013) 034014}
  [\href{https://arxiv.org/abs/1305.5739}{{\ttfamily 1305.5739}}].

\bibitem{Cirigliano:2016yhc}
V.~Cirigliano, W.~Dekens, J.~de~Vries and E.~Mereghetti, \emph{{An $\epsilon'$
  improvement from right-handed currents}},
  \href{https://doi.org/10.1016/j.physletb.2017.01.037}{\emph{Phys. Lett. B}
  {\bfseries 767} (2017) 1} [\href{https://arxiv.org/abs/1612.03914}{{\ttfamily
  1612.03914}}].

\bibitem{Chen:2008kt}
P.~Chen, H.~Ke and X.~Ji, \emph{{Direct CP Violation in K-decay and Minimal
  Left-Right Symmetry Scale}},
  \href{https://doi.org/10.1016/j.physletb.2009.04.016}{\emph{Phys. Lett. B}
  {\bfseries 677} (2009) 157}
  [\href{https://arxiv.org/abs/0810.2576}{{\ttfamily 0810.2576}}].

\bibitem{ParticleDataGroup:2024cfk}
{\scshape Particle Data Group} collaboration, S.~Navas et~al., \emph{{Review of
  particle physics}},
  \href{https://doi.org/10.1103/PhysRevD.110.030001}{\emph{Phys. Rev. D}
  {\bfseries 110} (2024) 030001}.

\bibitem{Gisbert:2017vvj}
H.~Gisbert and A.~Pich, \emph{{Direct CP violation in $K^0\to\pi\pi$: Standard
  Model Status}}, \href{https://doi.org/10.1088/1361-6633/aac18e}{\emph{Rept.
  Prog. Phys.} {\bfseries 81} (2018) 076201}
  [\href{https://arxiv.org/abs/1712.06147}{{\ttfamily 1712.06147}}].

\bibitem{Cirigliano:2019ani}
V.~Cirigliano, H.~Gisbert, A.~Pich and A.~Rodr\'\i{}guez-S\'anchez,
  \emph{{Theoretical status of $\varepsilon'/\varepsilon$}},
  \href{https://doi.org/10.1088/1742-6596/1526/1/012011}{\emph{J. Phys. Conf.
  Ser.} {\bfseries 1526} (2020) 012011}
  [\href{https://arxiv.org/abs/1912.04736}{{\ttfamily 1912.04736}}].

\bibitem{Cirigliano:2019cpi}
V.~Cirigliano, H.~Gisbert, A.~Pich and A.~Rodr\'\i{}guez-S\'anchez,
  \emph{{Isospin-violating contributions to $\epsilon'/\epsilon$}},
  \href{https://doi.org/10.1007/JHEP02(2020)032}{\emph{JHEP} {\bfseries 02}
  (2020) 032} [\href{https://arxiv.org/abs/1911.01359}{{\ttfamily
  1911.01359}}].

\bibitem{Abel:2020pzs}
C.~Abel et~al., \emph{{Measurement of the Permanent Electric Dipole Moment of
  the Neutron}},
  \href{https://doi.org/10.1103/PhysRevLett.124.081803}{\emph{Phys. Rev. Lett.}
  {\bfseries 124} (2020) 081803}
  [\href{https://arxiv.org/abs/2001.11966}{{\ttfamily 2001.11966}}].

\bibitem{Sahoo:2016zvr}
B.~K. Sahoo, \emph{{Improved limits on the hadronic and semihadronic $CP$
  violating parameters and role of a dark force carrier in the electric dipole
  moment of $^{199}$Hg}},
  \href{https://doi.org/10.1103/PhysRevD.95.013002}{\emph{Phys. Rev. D}
  {\bfseries 95} (2017) 013002}
  [\href{https://arxiv.org/abs/1612.09371}{{\ttfamily 1612.09371}}].

\bibitem{Alioli:2017ces}
S.~Alioli, V.~Cirigliano, W.~Dekens, J.~de~Vries and E.~Mereghetti,
  \emph{{Right-handed charged currents in the era of the Large Hadron
  Collider}}, \href{https://doi.org/10.1007/JHEP05(2017)086}{\emph{JHEP}
  {\bfseries 05} (2017) 086}
  [\href{https://arxiv.org/abs/1703.04751}{{\ttfamily 1703.04751}}].

\bibitem{Cirigliano:2022qdm}
V.~Cirigliano, W.~Dekens, J.~de~Vries, E.~Mereghetti and T.~Tong,
  \emph{{Beta-decay implications for the W-boson mass anomaly}},
  \href{https://doi.org/10.1103/PhysRevD.106.075001}{\emph{Phys. Rev. D}
  {\bfseries 106} (2022) 075001}
  [\href{https://arxiv.org/abs/2204.08440}{{\ttfamily 2204.08440}}].

\bibitem{Inami:1980fz}
T.~Inami and C.~S. Lim, \emph{{Effects of Superheavy Quarks and Leptons in
  Low-Energy Weak Processes $K_L \to \mu \bar{\mu}$, $K^+ \to \pi^+ \nu
  \bar{\nu}$ and $K_0 \leftrightarrow \overline{K_0}$}},
  \href{https://doi.org/10.1143/PTP.65.297}{\emph{Prog. Theor. Phys.}
  {\bfseries 65} (1981) 297} [Erratum: Prog.Theor.Phys. 65, 1772 (1981)].

\bibitem{Buchalla:1995vs}
G.~Buchalla, A.~J. Buras and M.~E. Lautenbacher, \emph{{Weak decays beyond
  leading logarithms}},
  \href{https://doi.org/10.1103/RevModPhys.68.1125}{\emph{Rev. Mod. Phys.}
  {\bfseries 68} (1996) 1125}
  [\href{https://arxiv.org/abs/hep-ph/9512380}{{\ttfamily hep-ph/9512380}}].

\bibitem{Bai:2014cva}
Z.~Bai, N.~H. Christ, T.~Izubuchi, C.~T. Sachrajda, A.~Soni and J.~Yu,
  \emph{{$K_L-K_S$ Mass Difference from Lattice QCD}},
  \href{https://doi.org/10.1103/PhysRevLett.113.112003}{\emph{Phys. Rev. Lett.}
  {\bfseries 113} (2014) 112003}
  [\href{https://arxiv.org/abs/1406.0916}{{\ttfamily 1406.0916}}].

\bibitem{Bai:2018mdv}
Z.~Bai, N.~H. Christ and C.~T. Sachrajda, \emph{{The $K_L-K_S$ Mass
  Difference}}, \href{https://doi.org/10.1051/epjconf/201817513017}{\emph{EPJ
  Web Conf.} {\bfseries 175} (2018) 13017}.

\bibitem{Garron:2016mva}
{\scshape RBC/UKQCD} collaboration, N.~Garron, R.~J. Hudspith and A.~T. Lytle,
  \emph{{Neutral Kaon Mixing Beyond the Standard Model with $n_f=2+1$ Chiral
  Fermions Part 1: Bare Matrix Elements and Physical Results}},
  \href{https://doi.org/10.1007/JHEP11(2016)001}{\emph{JHEP} {\bfseries 11}
  (2016) 001} [\href{https://arxiv.org/abs/1609.03334}{{\ttfamily
  1609.03334}}].

\bibitem{Buras:2001ra}
A.~J. Buras, S.~Jager and J.~Urban, \emph{{Master formulae for $\Delta F=2$ NLO
  QCD factors in the standard model and beyond}},
  \href{https://doi.org/10.1016/S0550-3213(01)00207-3}{\emph{Nucl. Phys. B}
  {\bfseries 605} (2001) 600}
  [\href{https://arxiv.org/abs/hep-ph/0102316}{{\ttfamily hep-ph/0102316}}].

\bibitem{Bobeth:2017xry}
C.~Bobeth, A.~J. Buras, A.~Celis and M.~Jung, \emph{{Yukawa enhancement of
  $Z$-mediated new physics in $\Delta S = 2$ and $\Delta B = 2$ processes}},
  \href{https://doi.org/10.1007/JHEP07(2017)124}{\emph{JHEP} {\bfseries 07}
  (2017) 124} [\href{https://arxiv.org/abs/1703.04753}{{\ttfamily
  1703.04753}}].

\bibitem{Bona:2022zhn}
M.~Bona et~al., \emph{{Unitarity Triangle global fits beyond the Standard
  Model: UTfit 2021 NP update}},
  \href{https://doi.org/10.22323/1.398.0500}{\emph{PoS} {\bfseries EPS-HEP2021}
  (2022) 500}.

\bibitem{Littenberg:1989ix}
L.~S. Littenberg, \emph{{The CP Violating Decay $K_L^0 \to \pi^0 \nu
  \bar{\nu}$}}, \href{https://doi.org/10.1103/PhysRevD.39.3322}{\emph{Phys.
  Rev. D} {\bfseries 39} (1989) 3322}.

\bibitem{Kayser:1996sv}
B.~Kayser, \emph{{CP violation in the K and B systems}},  in \emph{{ICTP Summer
  School in High-energy Physics and Cosmology}}, pp.~432--478, 11, 1996,
  \href{https://arxiv.org/abs/hep-ph/9702264}{{\ttfamily hep-ph/9702264}}.

\bibitem{Grossman:1997sk}
Y.~Grossman and Y.~Nir, \emph{{$K_L \to \pi^0 \nu \nu$ beyond the standard
  model}}, \href{https://doi.org/10.1016/S0370-2693(97)00210-4}{\emph{Phys.
  Lett. B} {\bfseries 398} (1997) 163}
  [\href{https://arxiv.org/abs/hep-ph/9701313}{{\ttfamily hep-ph/9701313}}].

\bibitem{Buchalla:1998ux}
G.~Buchalla and G.~Isidori, \emph{{The CP conserving contribution to $K_L \to
  \pi^0 \nu \bar{\nu}$ in the standard model}},
  \href{https://doi.org/10.1016/S0370-2693(98)01088-0}{\emph{Phys. Lett. B}
  {\bfseries 440} (1998) 170}
  [\href{https://arxiv.org/abs/hep-ph/9806501}{{\ttfamily hep-ph/9806501}}].

\bibitem{Buras:2015qea}
A.~J. Buras, D.~Buttazzo, J.~Girrbach-Noe and R.~Knegjens, \emph{{$ {K}^{+}\to
  {\pi}^{+}\nu \overline{\nu} $ and $ {K}_L\to {\pi}^0\nu \overline{\nu} $ in
  the Standard Model: status and perspectives}},
  \href{https://doi.org/10.1007/JHEP11(2015)033}{\emph{JHEP} {\bfseries 11}
  (2015) 033} [\href{https://arxiv.org/abs/1503.02693}{{\ttfamily
  1503.02693}}].

\bibitem{Pich:2018ltt}
A.~Pich, \emph{{Effective Field Theory with Nambu-Goldstone Modes}},
  \href{https://doi.org/10.1093/oso/9780198855743.003.0003}{\emph{Les Houches
  Lect.Notes} {\bfseries 108} (2020) 137}
  [\href{https://arxiv.org/abs/1804.05664}{{\ttfamily 1804.05664}}].

\bibitem{Santamaria:1993ah}
A.~Santamaria, \emph{{Masses, mixings, Yukawa couplings and their symmetries}},
  \href{https://doi.org/10.1016/0370-2693(93)91110-9}{\emph{Phys. Lett. B}
  {\bfseries 305} (1993) 90}
  [\href{https://arxiv.org/abs/hep-ph/9302301}{{\ttfamily hep-ph/9302301}}].

\bibitem{Berger:2008zq}
J.~Berger and Y.~Grossman, \emph{{Parameter counting in models with global
  symmetries}},
  \href{https://doi.org/10.1016/j.physletb.2009.04.050}{\emph{Phys. Lett. B}
  {\bfseries 675} (2009) 365}
  [\href{https://arxiv.org/abs/0811.1019}{{\ttfamily 0811.1019}}].

\bibitem{deBlas:2017xtg}
J.~de~Blas, J.~C. Criado, M.~Perez-Victoria and J.~Santiago, \emph{{Effective
  description of general extensions of the Standard Model: the complete
  tree-level dictionary}},
  \href{https://doi.org/10.1007/JHEP03(2018)109}{\emph{JHEP} {\bfseries 03}
  (2018) 109} [\href{https://arxiv.org/abs/1711.10391}{{\ttfamily
  1711.10391}}].

\bibitem{Crivellin:2022fdf}
A.~Crivellin, M.~Kirk, T.~Kitahara and F.~Mescia, \emph{{Large $t\to cZ$ as a
  sign of vectorlike quarks in light of the $W$ mass}},
  \href{https://doi.org/10.1103/PhysRevD.106.L031704}{\emph{Phys. Rev. D}
  {\bfseries 106} (2022) L031704}
  [\href{https://arxiv.org/abs/2204.05962}{{\ttfamily 2204.05962}}].

\bibitem{Aebischer:2023nnv}
L.~Allwicher et~al., \emph{{Computing tools for effective field theories:
  SMEFT-Tools 2022 Workshop Report, 14\textendash{}16th September 2022,
  Z\"urich}}, \href{https://doi.org/10.1140/epjc/s10052-023-12323-y}{\emph{Eur.
  Phys. J. C} {\bfseries 84} (2024) 170}
  [\href{https://arxiv.org/abs/2307.08745}{{\ttfamily 2307.08745}}].

\bibitem{Carmona:2021xtq}
A.~Carmona, A.~Lazopoulos, P.~Olgoso and J.~Santiago, \emph{{Matchmakereft:
  automated tree-level and one-loop matching}},
  \href{https://doi.org/10.21468/SciPostPhys.12.6.198}{\emph{SciPost Phys.}
  {\bfseries 12} (2022) 198}
  [\href{https://arxiv.org/abs/2112.10787}{{\ttfamily 2112.10787}}].

\bibitem{Fuentes-Martin:2022jrf}
J.~Fuentes-Mart\'\i{}n, M.~K\"onig, J.~Pag\`es, A.~E. Thomsen and F.~Wilsch,
  \emph{{A proof of concept for matchete: an automated tool for matching
  effective theories}},
  \href{https://doi.org/10.1140/epjc/s10052-023-11726-1}{\emph{Eur. Phys. J. C}
  {\bfseries 83} (2023) 662}
  [\href{https://arxiv.org/abs/2212.04510}{{\ttfamily 2212.04510}}].

\bibitem{Jenkins:2017jig}
E.~E. Jenkins, A.~V. Manohar and P.~Stoffer, \emph{{Low-Energy Effective Field
  Theory below the Electroweak Scale: Operators and Matching}},
  \href{https://doi.org/10.1007/JHEP03(2018)016}{\emph{JHEP} {\bfseries 03}
  (2018) 016} [\href{https://arxiv.org/abs/1709.04486}{{\ttfamily
  1709.04486}}].

\end{thebibliography}\endgroup

\end{document}